# WAR-ALGORITHM ACCOUNTABILITY

Dustin A. Lewis, Gabriella Blum, and Naz K. Modirzadeh

Harvard Law School Program on International Law and Armed Conflict

Research Briefing + Appendices

August 2016

# WAR-ALGORITHM ACCOUNTABILITY

**DUSTIN A. LEWIS, GABRIELLA BLUM, AND NAZ K. MODIRZADEH**

HARVARD LAW SCHOOL PROGRAM ON
INTERNATIONAL LAW AND ARMED CONFLICT

*Research Briefing*

August 2016

# EXECUTIVE SUMMARY

Across many areas of modern life, "authority is increasingly expressed algorithmically."[1] War is no exception.

In this briefing report, we introduce a new concept—war algorithms—that elevates algorithmically-derived "choices" and "decisions" to a, and perhaps *the*, central concern regarding technical autonomy in war. We thereby aim to shed light on and recast the discussion regarding "autonomous weapon systems."

In introducing this concept, our foundational technological concern is the capability of a constructed system, without further human intervention, to help make and effectuate a "decision" or "choice" of a war algorithm. Distilled, the two core ingredients are an algorithm expressed in computer code and a suitably capable constructed system. Through that lens, we link international law and related accountability architectures to relevant technologies. We sketch a three-part (non-exhaustive) approach that highlights traditional and unconventional accountability avenues. By not limiting our inquiry only to weapon systems, we take an expansive view, showing how the broad concept of war algorithms might be susceptible to regulation—and how those algorithms might already fit within the existing regulatory system established by international law.

\* \* \*

Warring parties have long expressed authority and power through algorithms. For decades, algorithms have helped weapons systems—first at sea and later on land—to identify and intercept inbound missiles. Today, military systems are increasingly capable of navigating novel environments and surveilling faraway populations, as well as identifying targets, estimating harm, and launching direct attacks—all with fewer humans at the switch. Indeed, in recent years, commercial and military developments in algorithmically-derived autonomy have created diverse benefits for the armed forces in terms of "battlespace awareness," protection, "force application," and logistics. And those are by no means the exhaustive set of applications.

---

1. Frank Pasquale, The Black Box Society: The Secret Algorithms That Control Money and Society 8 (2015), *citing* Clay Shirky, *A Speculative Post on the Idea of Algorithmic Authority*, Clay Shirky (November 15, 2009, 4:06 PM), http://www.shirky.com/weblog/2009/11/a-speculative-post-on-the-idea-of-algorithmic-authority (referencing Shirky's definition of "algorithmic authority" as "the decision to regard as authoritative an unmanaged process of extracting value from diverse, untrustworthy sources, without any human standing beside the result saying 'Trust this because you trust me.'"). All further citations for sources underlying this Executive Summary are available in the full-text version of the briefing report.

Much of the underlying technology—often developed initially in commercial or academic contexts—is susceptible to both military and non-military use. Most of it is thus characterized as "dual-use," a shorthand for being capable of serving a wide array of functions. Costs of the technology are dropping, often precipitously. And, once the technology exists, the assumption is usually that it can be utilized by a broad range of actors.

Driven in no small part by commercial interests, developers are advancing relevant technologies and technical architectures at a rapid pace. The potential for those advancements to cross a moral Rubicon is being raised more frequently in international forums and among technical communities, as well as in the popular press.

Some of the most relevant advancements involve constructed systems through which huge amounts of data are quickly gathered and ensuing algorithmically-derived "choices" are effectuated. "Self-driving" or "autonomous" cars are one example. Ford, for instance, mounts four laser-based sensors on the roof of its self-driving research car, and collectively those sensors "can capture 2.5 million 3-D points per second within a 200-foot range." Legal, ethical, political, and social commentators are casting attention on—and vetting proposed standards and frameworks to govern—the life-and-death "choices" made by autonomous cars.

Among the other relevant advancements is the potential for learning algorithms and architectures to achieve more and more human-level performance in previously-intractable artificial-intelligence (AI) domains. For instance, a computer program recently achieved a feat previously thought to be at least a decade away: defeating a human professional player in a full-sized game of Go. In March 2016, in a five-game match, AlphaGo—a computer program using an AI technique known as "deep learning," which "allows computers to extract patterns from masses of data with little human hand-holding"—won four games against Go expert Lee Sedol. Google, Amazon, and Baidu use the same AI technique or similar ones for such tasks as facial recognition and serving advertisements on websites. Following AlphaGo's series of wins, computer programs have now outperformed humans at chess, backgammon, "Jeopardy!", and Go.

Yet even among leading scientists, uncertainty prevails as to the technological limits. That uncertainty repels a consensus on the current capabilities, to say nothing of predictions of what might be likely developments in the near- and long-term (with those horizons defined variously).

The stakes are particularly high in the context of political violence that reaches the level of "armed conflict." That is because international law admits of far more lawful death, destruction, and disruption in war than in peace. Even for responsible parties who are committed to the rule of law, the legal regime contemplates the deployment of lethal and destructive technologies on a wide scale. The use of advanced technologies—to say nothing of the failures, malfunctioning, hacking, or spoofing of those technologies—might therefore entail far more significant consequences in relation to war than to peace. We focus here largely on international law because it is the only normative regime that purports—in key respects but with important caveats—to be both universal and uniform. In this way, international law is different from the myriad domestic legal systems, administrative rules, or industry codes that govern the development and use of technology in all other spheres.

Of course, the development and use of advanced technologies in relation to war have long generated ethical, political, and legal debates. There is nothing new about the general desire and the need to discern whether the use of an emerging technological capability would comport with or violate the law. Today, however, emergent technologies sharpen— and, to a certain extent, recast—that enduring endeavor. A key reason is that those technologies are seen as presenting an inflection point at which human judgment might be "replaced" by algorithmically-derived "choices." To unpack and understand the implications of that framing requires, among other things, technical comprehension, ethical awareness, and legal knowledge. Understandably if unfortunately, competence across those diverse domains has so far proven difficult to achieve for the vast majority of states, practitioners, and commentators.

Largely, the discourse to date has revolved around a concept that so far lacks a definitional consensus: "autonomous weapon systems" (AWS). Current conceptions of AWS range enormously. On one end of the spectrum, an AWS is an automated component of an existing weapon. On the other, it is a platform that is itself capable of sensing, learning, and launching resulting attacks. Irrespective of how it is defined in a particular instance, the AWS framing narrows the discourse to weapons, excluding the myriad other functions, however benevolent, that the underlying technologies might be capable of.

What autonomous weapons mean for legal responsibility and for broader accountability has generated one of the most heated recent debates about the law of war. A constellation of factors has shaped the discussion.

Perceptions of evolving security threats, geopolitical strategy, and accompanying developments in military doctrine have led governments to prioritize the use of unmanned and increasingly autonomous systems (with "autonomous" defined variously) in order to gain and maintain a qualitative edge. By 2013, leadership in the U.S. Navy and Department of Defense (DoD) had identified autonomy in unmanned systems as a "high priority." In March 2016, the Ministries of Foreign Affairs and Defense of the Netherlands affirmed their belief that "if the Dutch armed forces are to remain technologically advanced, autonomous weapons will have a role to play, now and in the future." A growing number of states hold similar views.

At the same time, human-rights advocates and certain technology experts have catalyzed initiatives to promote a ban on "fully autonomous weapons" (which those advocates and experts also call "killer robots"). The primary concerns are couched in terms of delegating decisions about lethal force away from humans—thereby "dehumanizing" war—and, in the process, of making wars easier to prosecute. Following the release in 2012 of a report by Human Rights Watch and the International Human Rights Clinic at Harvard Law School, the Campaign to Stop Killer Robots was launched in April 2013 with an explicit goal of fostering a "pre-emptive ban on fully autonomous weapons." The rationale is that such weapons will, pursuant to this view, never be capable of comporting with international humanitarian law (IHL) and are therefore *per se* illegal. In July 2015, thousands of prominent AI and robotics experts, as well as other scientists, endorsed an "Open Letter" on autonomous weapons, arguing that "[t]he key question for humanity today is whether to start a global AI arms race or to prevent it from starting." Those endorsing the letter "believe that AI has great potential to benefit humanity in many ways, and that the goal of the field should be to do so." But, they cautioned, "[s]tarting a military AI arms race is a bad idea, and should be prevented by a ban on offensive autonomous weapons beyond meaningful human control."

Meanwhile, a range of commentators has argued in favor of regulating autonomous weapon systems, primarily through existing international law rules and provisions. In general, these voices focus on grounding the discourse in terms of the capability of existing legal norms—especially those laid down in IHL—to regulate the design, development, and use, or to prohibit the use, of emergent technologies. In doing so, these commentators often emphasize that states have already developed a relatively thick set of international law rules that guide decisions about life and death in war. Even if there is no specific treaty addressing a particular weapon, they argue, IHL regulates the use of all weapons through general rules and principles governing the conduct of hostilities that apply irrespective of the weapon used. A number of these voices also aver that—for political, military, commercial, or other reasons—states are unlikely to agree on a preemptive ban on fully autonomous weapons, and therefore a better use of resources would be to focus on regulating the technologies and monitoring their use. In addition, these commentators often emphasize the modularity of the technology and raise concerns about foreclosing possible beneficial applications in the service of an (in their eyes, highly unlikely) prohibition on fully autonomous weapons.

Over all, the lack of consensus on the root classification of AWS and on the scope of the resulting discussion make it difficult to generalize. But the main contours of the ensuing "debate" often cast a purportedly unitary "ban" side versus a purportedly unitary "regulate" side. As with many shorthand accounts, this formulation is overly simplistic. An assortment of thoughtful contributors does not fit neatly into either general category. And, when scrutinized, those wholesale categories—of "ban" vs. "regulate"— disclose fundamental flaws, not least because of the lack of agreement on what, exactly, is meant to be prohibited or regulated. Be that as it may, a large portion of the resulting discourse has been captured in these "ban"- vs.-"regulate" terms.

Underpinning much of this debate are arguments about decision-making in war, and who is better situated to make life-and-death decisions—humans or machines. There is also a disagreement over the benefits and costs of distancing human combatants from the battlefield and whether the possible life-saving benefits of AWS are offset by the fact that war also becomes, in certain respects, easier to conduct. There are also different understandings of and predictions about what machines are and will be capable of doing.

With the rise of expert and popular interest in AWS, states have been paying more public attention to the issue of regulating autonomy in war. But the primary venue at which they are doing so functionally limits the discussion to weapons. Since 2014, informal expert meetings on "lethal autonomous weapons systems" have been convened on an annual basis at the United Nations Office in Geneva. These meetings take place within the structure of the 1980 Convention on Prohibitions or Restrictions on the Use of Certain Conventional Weapons which may be deemed to be Excessively Injurious or to have Indiscriminate Effects (CCW). That treaty is set up as a framework convention: through it, states may adopt additional instruments that pertain to the core concerns of the baseline agreement (five such protocols have been adopted). Alongside the CCW, other arms-control treaties address specific types of weapons, including chemical weapons, biological weapons, anti-personnel landmines, cluster munitions, and others. The CCW is the only existing regime, however, that is ongoing and open-ended and is capable of being used as a framework to address additional types of weapons.

The original motivation to convene states as part of the CCW was to propel a protocol banning fully autonomous weapons. The most recent meeting (which was convened in April 2016) recommended that the Fifth Review Conference of states parties to the CCW (which is scheduled to take place in December 2016) "may decide to establish an open-ended Group of Governmental Experts (GGE)" on AWS. In the past, the establishment of a GGE has led to the adoption of a new CCW protocol (one banning permanently-blinding lasers). Whether states parties establish a GGE on AWS—and, if so, what its mandate will be—are open questions. In any event, at the most recent meetings, about two-dozen states endorsed the notion—the contours of which remain undefined so far—of "meaningful human control" over autonomous weapon systems.

Zooming out, we see that a pair of interlocking factors has obscured and hindered analysis of whether the relevant technologies can and should be regulated.

One factor is the sheer technical complexity at issue. Lack of knowledge of technical intricacies has hindered efforts by non-experts to grasp how the core technologies may either fit within or frustrate existing legal frameworks.

This is not a challenge particular to AWS, of course. The majority of IHL professionals are not experts in the inner workings of the numerous technologies related to armed conflict. Most IHL lawyers could not detail the technical specifications, for instance, of various armaments, combat vehicles, or intelligence, surveillance, and reconnaissance (ISR) systems. But in general that lack of technical knowledge would not necessarily impede at least a provisional analysis of the lawfulness of the use of such a system. That is because an initial IHL analysis is often an exercise in identifying the relevant rule and beginning to apply it in relation to the applicable context. Yet the widely diverse conceptions of AWS and the varied technologies accompanying those conceptions pose an as-yet-unresolved set of classification challenges. Without a threshold classification, a general legal analysis cannot proceed.

The other, related factor is that states—as well as lawyers, technologists, and other commentators—disagree in key respects on what should be addressed. The headings so far include "lethal autonomous robots," "lethal autonomous weapons systems," "autonomous weapons systems" more broadly, and "intelligent partnerships" more broadly still. And the possible standards mentioned include "meaningful human control" (including in the "wider loop" of targeting operations), "meaningful state control," and "appropriate levels of human judgment." More basically, there is no consensus on whether to include only weapons or, additionally, systems capable of involvement in other armed conflict-related functions, such as transporting and guarding detainees, providing medical care, and facilitating humanitarian assistance.

Against this backdrop, the AWS framing has largely precluded meaningful analysis of whether it (whatever "it" entails) can be regulated, let alone whether and how it should be regulated. In this briefing report, we recast the discussion by introducing the concept of "war algorithms." We define "war algorithm" as any algorithm that is expressed in computer code, that is effectuated through a constructed system, and that is capable of operating in relation to armed conflict. Those algorithms seem to be a—and perhaps *the*—key ingredient of what most people and states discuss when they address AWS. We expand the purview beyond weapons alone (important as those are) because the technological capabilities are rarely, if ever, limited to use only as weapons and because other war functions involving algorithmically-derived autonomy should be considered for regulation as well. Moreover, given the modular nature of much of the technology, a focus on weapons alone might thwart attempts at regulation.

Algorithms are a conceptual and technical building block of many systems. Those systems include self-learning architectures that today present some of the sharpest questions about "replacing" human judgment with algorithmically-derived "choices." Moreover, algorithms form a foundation of most of the systems and platforms—and even the "systems of systems"— often discussed in relation to AWS. Absent an unforeseen development, algorithms are likely to remain a pillar of the technical architectures.

The constructed systems through which these algorithms are effectuated differ enormously. So do the nature, forms, and tiers of human control and governance over them. Existing constructed systems include, among many others, stationary turrets, missile systems, and manned or unmanned aerial, terrestrial, or marine vehicles.

All of the underlying algorithms are developed by programmers and are expressed in computer code. But some of these algorithms—especially those capable of "self-learning" and whose "choices" might be difficult for humans to anticipate or unpack—seem to challenge fundamental and interrelated concepts that underpin international law pertaining to armed conflict and related accountability frameworks. Those concepts include attribution, control, foreseeability, and reconstructability.

At their core, the design, development, and use of war algorithms raise profound questions. Most fundamentally, those inquiries concern who, or what, should decide—and what it means to decide—matters of life and death in relation to war. But war algorithms also bring to the fore an array of more quotidian, though also important, questions about the benefits and costs of human judgment and "replacing" it with algorithmically-derived systems, including in such areas as logistics.

We ground our analysis by focusing on war-algorithm accountability. In short, we are primarily interested in the "duty to account … for the exercise of power" over—in other words, holding someone or some entity answerable for—the design, development, or use (or a combination thereof) of a war algorithm. That power may be exercised by a diverse assortment of actors. Some are obvious, especially states and their armed forces. But myriad other individuals and entities may exercise power over war algorithms, too. Consider the broad classes of "developers" and "operators," both within and outside of government, of such algorithms and their related systems. Also think of lawyers, industry bodies, political authorities, members of organized armed groups—and many, many others. Focusing on war algorithms encompasses them all.

We draw on the extensive—and rapidly growing—amount of scholarship and other analytical analyses that have addressed related topics. To help illuminate the discussion, we outline what technologies and weapon systems already exist, what fields of international law might be relevant, and what regulatory avenues might be available. As noted above, because international law is the touchstone normative framework for accountability in relation to war, we focus on public international law sources and methodologies. But as we show, other norms and forms of governance might also merit attention.

Accountability is a broad term of art. We adapt—from the work of an International Law Association Committee in a different context (the accountability of international organizations)—a three-part accountability approach. Our framework outlines three axes on which to focus initially on war algorithms.

The first axis is state responsibility. It concerns state responsibility arising out of acts or omissions involving a war algorithm where those acts or omissions constitute a breach of a rule of international law. State responsibility entails discerning the content of the rule, identifying a breach of the rule, assigning attribution for that breach to a state, determining available excuses (if any), and imposing measures of remedy.

The second axis is a form of individual responsibility under international law. In particular, it concerns individual responsibility under international law for international crimes—such as war crimes—involving war algorithms. This form of individual responsibility entails establishing the commission of a crime under the relevant jurisdiction, assessing the existence of a justification or excuse (if any), and, upon conviction, imposing a sentence.

The third and final axis is scrutiny governance. Embracing a wider notion of accountability, it concerns the extent to which a person or entity is and should be subject to, or should exercise, forms of internal or external scrutiny, monitoring, or regulation (or a combination thereof) concerning the design, development, or use of a war algorithm. Scrutiny governance does not hinge on—but might implicate—potential and subsequent liability or responsibility (or both). Forms of scrutiny governance include independent monitoring, norm (such as legal) development, adopting non-binding resolutions and codes of conduct, normative design of technical architectures, and community self-regulation.

Following an introduction that highlights the stakes, we proceed with a section outlining pertinent considerations regarding algorithms and constructed systems. We highlight recent advancements in artificial intelligence related to learning algorithms and architectures. We also examine state approaches to technical autonomy in war, focusing on five such approaches—those of Switzerland, the Netherlands, France, the United States, and the United Kingdom. Finally, to ground the often-theoretical debate pertaining to autonomous weapon systems, we describe existing weapon systems that have been characterized by various commentators as AWS.

The next section outlines the main fields of international law that war algorithms might implicate. There is no single branch of international law dedicated solely to war algorithms. So we canvass how those algorithms might fit within or otherwise implicate various fields of international law. We ground the discussion by outlining the main ingredients of state responsibility. To help illustrate states' positions concerning AWS, we examine whether an emerging norm of customary international law specific to AWS may be discerned. We find that one cannot (at least not yet). So we next highlight how the design, development, or use (or a combination thereof) of a war algorithm might implicate more general principles and rules found in various fields of international law. Those fields include the *jus ad bellum*, IHL, international human rights law, international criminal law (ICL), and space law. Because states and commentators have largely focused on AWS to date, much of our discussion here relates to the AWS framing.

The subsequent section elaborates a (non-exhaustive) war-algorithm accountability approach. That approach focuses on state responsibility for an internationally wrongful act, on individual responsibility under international law for international crimes, and on wider forms of scrutiny, monitoring, and regulation. We highlight existing accountability actors and architectures under international law that might regulate war algorithms. These include war reparations as well as international and domestic tribunals. We then turn to less conventional accountability avenues, such as those rooted in normative design of technical architectures (including maximizing the auditability of algorithms) and community self-regulation.

In the conclusion, we return to the deficiencies of current discussions of AWS and emphasize the importance of addressing the wide and serious concerns raised by AWS with technical proficiency, legal expertise, and non-ideological commitment to a genuine and inclusive inquiry. On the horizon, we see that two contradictory trends may be combining into a new global climate that is at once enterprising and anxious. Militaries see myriad technological triumphs that will transform warfighting. Yet the possibility of "replacing" human judgment with algorithmically-derived "decisions"—especially in war—threatens what many consider to define us as humans.

To date, the lack of demonstrated technical knowledge by many states and commentators, the unwillingness of states to share closely-held national-security technologies, and an absence of a definitional consensus on what is meant by autonomous weapon systems have impeded regulatory efforts on AWS. Moreover, uncertainty about which actors would benefit most from advances in AWS and for how long such benefits would yield a meaningful qualitative edge over others seems likely to continue to inhibit efforts at negotiating binding international rules on the development and deployment of AWS. In this sense, efforts at reaching a dedicated international regime to address AWS may follow the same frustrations as analogous efforts to address cyber warfare. True, unlike with the early days of cyber warfare, there has been greater state engagement on regulation of AWS. In particular, the concept of "meaningful human control" over AWS has already been endorsed by over two-dozen states. But much remains up in the air as states decide whether to establish a Group of Governmental Experts on AWS at the upcoming Fifth Review Conference of the CCW.

The current crux, as we see it, is whether advances in technology—especially those capable of "self-learning" and of operating in relation to war and whose "choices" may be difficult for humans to anticipate or unpack or whose "decisions" are seen as "replacing" human judgment—are susceptible to regulation and, if so, whether and how they should be regulated. One way to think about the core concern which vaults over at least some of the impediments to the discussion on AWS is the new concept we raise: war algorithms. War algorithms include not only those algorithms capable of being used in weapons but also in any other function related to war.

More war algorithms are on the horizon. Two months ago, the Defense Science Board, which is connected with the U.S. Department of Defense, identified five "stretch problems"—that is, goals that are "hard-but-not-too-hard" and that have a purpose of accelerating the process of bringing a new algorithmically-derived capability into widespread application:

- Generating "future loop options" (that is, "using interpretation of massive data including social media and rapidly generated strategic options");

- Enabling autonomous swarms (that is, "deny[ing] the enemy's ability to disrupt through quantity by launching overwhelming numbers of low-cost assets that cooperate to defeat the threat");

- Intrusion detection on the Internet of Things (that is, "defeat[ing] adversary intrusions in the vast network of commercial sensors and devices by autonomously discovering subtle indicators of compromise hidden within a flood of ordinary traffic");

- Building autonomous cyber-resilient military vehicle systems (that is, "trust[ing] that … platforms are resilient to cyber-attack through autonomous system integrity validation and recovery"); and

- Planning autonomous air operations (that is, "operat[ing] inside adversary timelines by continuously planning and replanning tactical operations using autonomous ISR analysis, interpretation, option generation, and resource allocation").

What this trajectory toward greater algorithmic autonomy in war—at least among more technologically-sophisticated armed forces and even some non-state armed groups—means for accountability purposes seems likely to stay a contested issue for the foreseeable future.

In the meantime, it remains to be authoritatively determined whether war algorithms will be capable of making the evaluative decisions and value judgments that are incorporated into IHL. It is currently not clear, for instance, whether war algorithms will be capable of formulating and implementing the following IHL-based evaluative decisions and value judgments:

- The presumption of civilian status in case of "doubt";

- The assessment of "excessiveness" of expected incidental harm in relation to anticipated military advantage;

- The betrayal of "confidence" in IHL in relation to the prohibition of perfidy; and

- The prohibition of destruction of civilian property except where "imperatively" demanded by the necessities of war.

*   *   *

Two factors may suggest that, at least for now, the most immediate ways to regulate war algorithms specifically and to pursue accountability over them might be to follow not only traditional paths but also less conventional ones. As illustrated above, the latter might include relatively formal avenues—such as states making, applying, and enforcing war-algorithm rules of conduct within and beyond their territories—or less formal avenues—such as coding law into technical architectures and community self-regulation. First, even where the formal law may seem sufficient, concerns about practical enforcement abound. Second, the proliferation of increasingly advanced technical systems based on self-learning and distributed control raises the question of whether the model of individual responsibility found in ICL might pose conceptual challenges to regulating AWS and war algorithms.

In short, individual responsibility for international crimes under international law remains one of the vital accountability avenues in existence today, as do measures of remedy for state responsibility. Yet in practice responsibility along either avenue is unfortunately relatively rare. And thus neither path, on its own or in combination, seems to be sufficient to effectively address the myriad regulatory concerns pertaining to war algorithms—at least not until we better understand what is at issue. These concerns might lead those seeking to strengthen accountability of war algorithms to pursue not only traditional, formal avenues but also less formal, softer mechanisms.

In that connection, it seems likely that attempts to change governments' approaches to technical autonomy in war through social pressure (at least for those governments that might be responsive to that pressure) will continue to be a vital avenue along which to pursue accountability. But here, too, there are concerns. Numerous initiatives already exist. Some of them are very well informed; others less so. Many of them are motivated by ideological, commercial, or other interests that—depending on one's

viewpoint—might strengthen or thwart accountability efforts. And given the paucity of formal regulatory regimes, some of these initiatives may end up having considerable impact, despite their shortcomings.

Stepping back, we see that technologies of war, as with technologies in so many areas, produce an uneasy blend of promise and threat. With respect to war algorithms, understanding these conflicting pulls requires attention to a century-and-a-half-long history during which war came to be one of the most highly regulated areas of international law. But it also requires technical know-how. Thus those seeking accountability for war algorithms would do well not to forget the essentially political work of IHL's designers—nor to obscure the fact that today's technology is, at its core, designed, developed, and deployed by humans. Ultimately, war-algorithm accountability seems unrealizable without sufficient competence in technical architectures and in legal frameworks, coupled with ethical, political, and economic awareness.

Finally, we also include a Bibliography and Appendices. The Bibliography contains over 400 analytical sources, in various languages, pertaining to technical autonomy in war. The Appendices contain detailed charts listing and categorizing states' statements at the 2015 and 2016 Informal Meetings of Experts on Lethal Autonomous Weapons Systems convened within the framework of the CCW.

# CREDITS

## About PILAC

The Program on International Law and Armed Conflict (PILAC) is a platform at Harvard Law School that provides a space for research on critical challenges facing the various fields of public international law related to armed conflict, including the *jus ad bellum*, the *jus in bello* (international humanitarian law/the law of armed conflict), international human rights law, international criminal law, and the law of state responsibility. Its mode is critical, independent, and rigorous. PILAC's methodology fuses traditional public international law research with targeted analysis of changing security environments. The Program does not engage in advocacy. While its contributors may express a range of views on contentious legal and policy debates, PILAC does not take institutional positions on these matters.

## About the Authors

Dustin A. Lewis is a Senior Researcher at the Harvard Law School Program on International Law and Armed Conflict (PILAC). Gabriella Blum, the Faculty Director of PILAC, is the Rita E. Hauser Professor of Human Rights and Humanitarian Law at Harvard Law School. And Naz K. Modirzadeh, the Director of PILAC, is a Professor of Practice at Harvard Law School.

## Acknowledgments and Disclaimers

The authors extend their thanks to: Adam Broza, Jessica Burniske, Molly Doggett, Joshua Kestin, and Katie King for research assistance; Jessica Burniske and Katie King for editorial assistance; Adam Broza, Jiawei He, Katie King, Francesco Romani, Svitlana Starosvit, and Anton Vallélian for translation assistance; Jiawei He and the Chinese Initiative on International Law for logistical and translation support; Jennifer Allison, PILAC Liaison to the Harvard Law School Library (HLSL), and the staff of the HLSL for research support; participants at events featuring early PILAC research at Fudan University, Shanghai (May 2016), at China University of Political Science and Law, Beijing (May 2016), and at the Berkman Klein Center for Internet and Society of Harvard University, Cambridge (July 2016) for critical feedback and comments; and Peter Krafft, Claudia Pérez D'Arpino, and Merel A. C. Ekelhof for technical assistance and critical feedback.

Adam Broza and Molly Doggett produced the Bibliography. Jessica Burniske and Joshua Kestin drafted the sub-section of Section 2 on examples of purported autonomous weapon systems. Katie King and Joshua Kestin produced Appendices I and II and provided extensive research assistance concerning the sub-section in section 3 regarding customary international law and autonomous weapon systems.

This Briefing Report has been produced, in part, with financial assistance from the Pierre and Pamela Omidyar Fund, which is an advised fund of the Silicon Valley Community Foundation (SVCF). PILAC also receives generous support from the Swiss Federal Department of Foreign Affairs (FDFA). The views expressed in this Briefing Report should not be taken, in any way, to reflect the official opinion of the SVCF or of the Swiss FDFA. PILAC is grateful for the support the SVCF and Swiss FDFA provide for independent research and analysis. The research undertaken by the authors of this Briefing Report was completely independent;

the views and opinions reflected in this Briefing Report are those solely of the authors; and the authors alone are responsible for any errors in this Briefing Report.

## License

Creative Commons Attribution-NonCommercial-ShareAlike 4.0 International license (CC BY-NC-SA 4.0).

## Web

This Briefing Report is available free of charge at http://pilac.law.harvard.edu.

# CONTENTS

# 1

# INTRODUCTION

Across many areas of modern life, "authority is increasingly expressed algorithmically."[1] War is no exception.

Complex algorithms help determine a person's creditworthiness.[2] They suggest what movies to watch. They detect healthcare fraud. And they are used to trade stocks at speeds far faster than humans are capable of. (Sometimes, algorithms contribute to market crashes[3] or form a basis for anti-trust prosecutions.[4])

Warring parties express authority and power through algorithms, too. For decades, algorithms have helped weapons systems—first at sea and later on land—to identify and intercept inbound missiles.[5] Today, military systems are increasingly capable of navigating novel environments and surveilling faraway populations, as well as identifying targets, estimating harm, and launching direct attacks—all with fewer humans at the switch.[6] Indeed, in recent years, commercial and military developments in algorithmically-derived autonomy[7] have created diverse benefits for the armed forces in terms of "battlespace awareness,"[8] protection,[9] "force application,"[10] and logistics.[11] And those are by no means the exhaustive set of applications. Meanwhile, other algorithmically-derived war functions may not be far off— and, indeed, might already exist. Consider the provision of medical care to the wounded and sick *hors de combat* (such as certain combatants rendered incapable of fighting and who are therefore "outside of the battle"[12]) or the capture, transfer, and detention of enemy fighters.

---

1. Frank Pasquale, The Black Box Society: The Secret Algorithms That Control Money and Society 8 (2015), *citing* Clay Shirky, *A Speculative Post on the Idea of Algorithmic Authority*, Clay Shirky (November 15, 2009, 4:06 PM), http://www.shirky.com/weblog/2009/11/a-speculative-post-on-the-idea-of-algorithmic-authority (referencing Shirky's definition of "algorithmic authority" as "the decision to regard as authoritative an unmanaged process of extracting value from diverse, untrustworthy sources, without any human standing beside the result saying 'Trust this because you trust me.'").

2. On the examples in this paragraph, *see generally* Pasquale, *supra* note 1.

3. *See generally* U.S. Commodity Futures Trading Commission & U.S. Securities & Exchange Commission, Findings Regarding the Market Events of May 6, 2010: Report of the Staffs of the CFTF and SEC to the Joint Advisory Committee on Emerging Regulatory Issues (2010), https://www.sec.gov/news/studies/2010/marketevents-report.pdf.

4. *See, e.g.,* Jill Prulick, *When Bots Collude*, New Yorker, April 25, 2015, http://www.newyorker.com/business/currency/when-bots-collude.

5. The use of artificial intelligence and other forms of algorithmic systems in relation to war is far from new. For examples from nearly three decades ago, *see* Defense Applications of Artificial Intelligence (Stephen J. Andriole & Gerald W. Hopple eds., 1988).

---

6.   *See generally*, *e.g.*, Paul J. Springer, Military Robots and Drones: A Reference Handbook (2013); *see also infra* Section 2: Examples of Purported Autonomous Weapon Systems.

7.   In a recent report, the Defense Science Board uses a definition of autonomy that implies the use of one or more algorithms: "To be autonomous, a system must have the capability to independently compose and select among different courses of action to accomplish goals based on its knowledge and understanding of the world, itself, and the situation." Defense Science Board, Summer Study on Autonomy 4 (June 2016) (noting that "[d]efinitions for intelligent system, autonomy, automation, robots, and agents can be found in L.G. Shattuck, *Transitioning to Autonomy: A human systems integration perspective*, p. 5. Presentation at *Transitioning to Autonomy: Changes in the role of humans in air transportation* [March 11, 2015]. Available at http://human-factors.arc.nasa.gov/workshop/autonomy/download/presentations/ Shaddock%20.pdf."). *Id.* at n.1.

8.   E.g., autonomous agents to improve cyber-attack indicators and warnings; onboard autonomy for sensing; and time-critical intelligence from seized media. *See* Defense Science Board, *supra* note 7, at 46–53.

9.   E.g., dynamic spectrum management for protection missions; unmanned underwater vehicles (UUVs) to autonomously conduct sea-mine countermeasures missions; and automated cyber-response. *See* Defense Science Board, *supra* note 7, at 53–60.

10.   E.g., cascaded UUVs for offensive maritime mining, and organic tactical unmanned aircraft to support ground forces. *See* Defense Science Board, *supra* note 7, at 60–68. The term "force application" is defined in the report as "the ability to integrate the use of maneuver and engagement in all environments to create the effects necessary to achieve mission objectives." *Id.* at 60.

11.   E.g., predictive logistics and adaptive planning, and adaptive logistics for rapid deployment. *See* Defense Science Board, *supra* note 7, at 69–75.

12.   Under international humanitarian law (IHL), a person is *hors de combat* if (i) she is in the power of an adverse party, (ii) she clearly expresses an intention to surrender, or (iii) she has been rendered unconscious or is otherwise incapable of defending herself, provided that in any of these cases she abstains from any hostile act and does not attempt to escape; shipwrecked persons cannot be excluded from the construct of *hors de combat*. This formulation is derived from the Protocol Additional to the Geneva Conventions of 12 August 1949, and Relating to the Protection of Victims of International Armed Conflicts art. 41(2), June 8, 1977, 1125 U.N.T.S. 3 [hereinafter AP I]; *see also*, *e.g.*, Yoram Dinstein, Non-International Armed Conflicts in International Law 164 (2014).

Much of the underlying technology—often developed initially in commercial or academic contexts—is susceptible to both military and non-military use. Most of it is thus characterized as "dual-use," a shorthand for being capable of serving a wide array of functions. Costs of the technology are dropping, often precipitously. And, once the technology exists, the assumption is usually that it can be utilized by a broad range of actors.

Driven in no small part by commercial interests, developers are advancing relevant technologies and technical architectures at a rapid pace. The potential for those advancements—often in consumer-facing computer science and robotics fields—to be used to cross a moral Rubicon if unscrupulously adapted for belligerent purposes is being raised more frequently in international forums and among technical communities, as well as in the popular press.

Some of the most relevant advancements involve constructed systems through which huge amounts of data are quickly gathered and ensuing algorithmically-derived "choices" are effectuated. "Self-driving" or "autonomous" cars are one example. Ford, for instance, mounts four laser-based sensors on the roof of its self-driving research car, and collectively those sensors "can capture 2.5 million 3-D points per second within a 200-foot range."[13] Legal, ethical, political, and social commentators are casting attention on—and vetting proposed standards and frameworks to govern—the life-and-death "choices" made by autonomous cars.

Among the other relevant advancements is the potential for learning algorithms and architectures to achieve more and more human-level performance in previously-intractable artificial-intelligence (AI) domains. For instance, a computer program recently achieved a feat previously thought to be at least a decade away: defeating a human professional player in a full-sized game of Go.[14] In March 2016, in a five-game match, AlphaGo—a computer program using an AI technique known as "deep learning," which "allows computers to extract patterns from masses of data with little human

---

13.  Ucilia Wang, *Driverless Cars Are Data Guzzlers*, Wall Street Journal, March 23, 2014, http://www.wsj.com/articles/SB10001424052702304815004579417441475998338.

14.  David Silver et al., *Mastering the Game of Go with Deep Neural Networks and Tree Search*, 529 Nature 484, 488 (2016). Go is a board game pitting two players in a contest to surround more territory than each other's opponent; it is played on a grid of black lines, with game pieces played on the lines' intersections. A full-sized board is 19 by 19. Part of the reason Go presents such a difficult computational challenge is because its search space is so large. "After the first two moves of a Chess game," for instance, "there are 400 possible next moves. In Go, there are close to 130,000." Danielle Muoio, *Why Go is So Much Harder for AI to Beat Than Chess*, Tech Insider, March 10, 2016, http://www.techinsider.io/why-google-ai-game-go-is-harder-than-chess-2016-3.

hand-holding"—won four games against Go expert Lee Sedol.[15] Google, Amazon, and Baidu use the same AI technique or similar ones for such tasks as facial recognition and serving advertisements on websites. Following AlphaGo's series of wins, computer programs have now outperformed humans at chess, backgammon, "Jeopardy!", and Go.[16]

Yet even among leading scientists, uncertainty prevails as to the technological limits. That uncertainty repels a consensus on the current capabilities, to say nothing of predictions of what might be likely developments in the near- and long-term (with those horizons defined variously).

The stakes are particularly high in the context of political violence that reaches the level of "armed conflict." That is because international law admits of far more lawful death, destruction, and disruption in war than in peace.[17] Even for responsible parties who are committed to the rule of law, the legal regime contemplates the deployment of lethal and destructive technologies on a wide scale. The use of advanced technologies—to say nothing of the failures, malfunctioning, hacking, or spoofing of those technologies— might therefore entail far more significant consequences in relation to war than to peace.[18] We focus here largely on international law because it is the only normative regime that purports—in key respects but with important caveats—to be both universal and uniform. In this way, international law is different from the myriad domestic legal systems, administrative rules, or industry codes that govern the development and use of technology in all other spheres.

Of course, the development and use of advanced technologies in relation to war have long generated ethical, political, and legal debates. There is nothing new about the general desire and the need to discern whether the use of an emerging technological capability would comport with or violate the law. Today, however, emergent technologies sharpen—and, to a certain extent,

---

15.  *A Game-Changing Result*, The Economist, March 19, 2016, http://www.economist.com/news/science-and-technology/21694883-alphagos-masters-taught-it-game-electrifying-match-shows-what.

16.  *Id.*

17.  In this report, while recognizing certain distinctions and overlaps between them, we use the terms "war" and "armed conflict" interchangeably to denote an armed conflict (whether of an international or a non-international character) as defined in international law and a state of war in the legal sense. *See, e.g.*, Jann Kleffner, *Scope of Application of International Humanitarian Law*, *in* The Handbook of International Humanitarian Law (Dieter Fleck ed., 3rd ed. 2013).

18.  *See, e.g.*, Marten Zwanenburg et al., *Humans, Agents and International Humanitarian Law: Dilemmas in Target Discrimination*, BNAIC 408 (2005) (examining the destruction of a commercial airliner by the USS Vincennes to illustrate legal and ethical dilemmas involving the use of autonomous agents).

recast—that enduring endeavor. A key reason is that those technologies are seen as presenting an inflection point at which human judgment might be "replaced" by algorithmically-derived "choices." To unpack and understand the implications of that framing requires, among other things, technical comprehension, ethical awareness, and legal knowledge. Understandably if unfortunately, competence across those diverse domains has so far proven difficult to achieve for the vast majority of states, practitioners, and commentators.

Largely, the discourse to date has revolved around a concept that so far lacks a definitional consensus: "autonomous weapon systems" (AWS).[19] Current conceptions of AWS range enormously. On one end of the spectrum, an AWS is an automated component of an existing weapon. On the other, it is a platform that is itself capable of sensing, learning, and launching resulting attacks. Irrespective of how it is defined in a particular instance, the AWS framing narrows the discourse to *weapons*, excluding the myriad other functions, however benevolent, that the underlying technologies might be capable of.

What autonomous weapons mean for legal responsibility and for broader accountability has generated one of the most heated recent debates about the law of war. A constellation of factors has shaped the discussion.

Perceptions of evolving security threats, geopolitical strategy, and accompanying developments in military doctrine have led governments to prioritize the use of unmanned and increasingly autonomous systems (with "autonomous" defined variously) in order to gain and maintain a qualitative edge. The systems are said to present manifold military advantages—in short, a "seductive combination of distance, accuracy, and lethality."[20] By 2013,

---

19. Among states and commentators, there is no agreement on whether to refer to "autonomous weapons," "autonomous weapon systems," or "autonomous weapons systems," among many other formulations. Throughout this report, where referring to the views of a particular state(s) or commentator(s), we adopt that entity's or person's framing. Otherwise, for ease of reference, we adopt the "autonomous weapon system(s)" framing.

20. Rebecca Crootof, *War Torts: Accountability for Autonomous Weapons Systems*, 164 U. Penn. L. Rev. (forthcoming June 2016), http://ssrn.com/abstract=2657680 [hereinafter Crootof, *War Torts*]. In June 2016, the Defense Science Board highlighted six categories of how autonomy can benefit (Department of Defense) DoD missions:

- Required decision speed: more autonomy is valuable when decisions must be made quickly (*e.g.*, cyber operations and missile defense);
- Heterogeneity and volume of data: more autonomy is valuable with high volume data and variety of data types (*e.g.*, imagery; intelligence data analysis; intelligence, surveillance, reconnaissance (ISR) data integration);
- Quality of data links: more autonomy is valuable when communication is intermittent (*e.g.*, times of contested communications, unmanned undersea operations);
- Complexity of action: more autonomy is valuable when activity is multimodal (*e.g.*, an

leadership in the U.S. Navy and Department of Defense (DoD) had identified autonomy in unmanned systems as a "high priority."[21] A few months ago, the Ministries of Foreign Affairs and Defense of the Netherlands affirmed their belief that "if the Dutch armed forces are to remain technologically advanced, autonomous weapons will have a role to play, now and in the future."[22] A growing number of states hold similar views.

At the same time, human-rights advocates and certain technology experts have catalyzed initiatives to promote a ban on "fully autonomous weapons" (which those advocates and experts also call "killer robots"). The primary concerns are couched in terms of delegating decisions about lethal force away from humans—thereby "dehumanizing" war—and, in the process, of making wars easier to prosecute.[23] Following the release in 2012 of a report by Human Rights Watch and the International Human Rights Clinic at Harvard Law School,[24] the Campaign to Stop Killer Robots was launched in April 2013 with an explicit goal of fostering a "pre-emptive ban on fully autonomous weapons."[25] The rationale is that such weapons will, pursuant to this view, never be capable of comporting with international humanitarian law (IHL) and are therefore *per se* illegal. In July 2015, thousands of prominent AI and robotics experts, as well as other scientists, endorsed an "Open Letter" on autonomous weapons, arguing that "[t]he key question for humanity today is whether to start a global AI arms race or to prevent it from starting."[26] Those

---

air operations center, multi-mission operations);

- Danger of mission: more autonomy can reduce the number of warfighters in harm's way (*e.g.*, in contested operations; chemical, biological, radiological, or nuclear attack cleanup); and

- Persistence and endurance: more autonomy can increase mission duration (*e.g.*, enabling unmanned vehicles, persistent surveillance).

*See* Defense Science Board, *supra* note 7, at 45 (June 2016).

21. U.S. Dep't of Defense, Unmanned Systems Integrated Roadmap: FY2013–2038, at 67 (2013), http://www.defense.gov/Portals/1/Documents/pubs/DOD-USRM-2013.pdf.

22. Gov't (Neth.), Government Response to AIV/CAVV Advisory Report no. 97, Autonomous Weapon Systems: The Need for Meaningful Human Control (2016), http://aiv-advice.nl/8gr#government-responses [hereinafter Dutch Government, Response to AIV/CAVV Report]. At the same time, however, the Dutch government "reject[ed] outright the possibility of developing and deploying fully autonomous weapons." *Id.*

23. *See, e.g.*, Mary Ellen O'Connell, *Banning Autonomous Killing*, *in* The American Way of Bombing: Changing Ethical and Legal Norms, from Flying Fortresses to Drones (Matthew Evangelista & Henry Shue eds., 1st ed. 2014).

24. Human Rights Watch and the Harvard Law School International Human Rights Clinic, Losing Humanity: The Case against Killer Robots (2012), https://www.hrw.org/report/2012/11/19/losing-humanity-case-against-killer-robots.

25. *See, e.g.*, *Act*, Campaign to Stop Killer Robots, https://www.stopkillerrobots.org/act (last visited Aug. 23, 2016).

26. *Autonomous Weapons: An Open Letter from AI & Robotics Researchers*, Future of Life

endorsing the letter "believe that AI has great potential to benefit humanity in many ways, and that the goal of the field should be to do so." But, they cautioned, "[s]tarting a military AI arms race is a bad idea, and should be prevented by a ban on offensive autonomous weapons beyond meaningful human control."[27]

Meanwhile, a range of commentators has argued in favor of regulating AWS, primarily through existing international law rules and provisions. In general, these voices focus on grounding the discourse in terms of the capability of existing legal norms—especially those laid down in IHL—to regulate the design, development, and use, or to prohibit the use, of emergent technologies. In doing so, these commentators often emphasize that states have already developed a relatively thick set of international law rules that guide decisions about life and death in war. Even if there is no specific treaty addressing a particular weapon, they argue, IHL regulates the use of all weapons through general rules and principles governing the conduct of hostilities that apply irrespective of the weapon used. A number of these voices also aver that—for political, military, commercial, or other reasons—states are unlikely to agree on a preemptive ban on fully autonomous weapons, and therefore a better use of resources would be to focus on regulating the technologies and monitoring their use. In addition, these commentators often emphasize the modularity of the technology and raise concerns about foreclosing possible beneficial applications in the service of an (in their eyes, highly unlikely) prohibition on fully autonomous weapons.

Over all, the lack of consensus on the root classification of AWS and on the scope of the resulting discussion make it difficult to generalize. But the main contours of the ensuing "debate" often cast a purportedly unitary "ban" side versus a purportedly unitary "regulate" side. As with many shorthand accounts, this formulation is overly simplistic. An assortment of thoughtful contributors does not fit neatly into either general category. And, when scrutinized, those wholesale categories—of "ban" vs. "regulate"—disclose fundamental flaws, not least because of the lack of agreement on what, exactly, is meant to be prohibited or regulated. Be that as it may, a large portion of the resulting discourse has been captured in these "ban"-vs.-"regulate" terms.

Underpinning much of this debate are arguments about decision-making in war, and who is better situated to make life-and-death decisions—humans or machines. There is also a disagreement over the benefits and costs of distancing human combatants from the battlefield and whether the possible life-saving benefits of AWS are offset by the fact that war also becomes, in

---

Institute (July 28, 2015), http://futureoflife.org/open-letter-autonomous-weapons.
27.   *Id.*

certain respects, easier to conduct. There are also different understandings of and predictions about what machines are and will be capable of doing.

With the rise of expert and popular interest in AWS, states have been paying more public attention to the issue of regulating autonomy in war. But the primary venue at which they are doing so functionally limits the discussion to weapons.[28] Since 2014, informal expert meetings on "lethal autonomous weapons systems" have been convened on an annual basis at the United Nations Office in Geneva. These meetings take place within the structure of the 1980 Convention on Prohibitions or Restrictions on the Use of Certain Conventional Weapons which may be deemed to be Excessively Injurious or to have Indiscriminate Effects (CCW). That treaty is set up as a framework convention: through it, states may adopt additional instruments that pertain to the core concerns of the baseline agreement (five such protocols have been adopted). Alongside the CCW, other arms-control treaties address specific types of weapons, including chemical weapons, biological weapons, anti-personnel landmines, cluster munitions, and others. The CCW is the only existing regime, however, that is ongoing and open-ended and is capable of being used as a framework to address additional types of weapons.

The original motivation to convene states as part of the CCW was to propel a protocol banning fully autonomous weapons. The most recent meeting (which was convened in April 2016) recommended that the Fifth Review Conference of states parties to the CCW (which is scheduled to take place in December 2016) "may decide to establish an open-ended Group of Governmental Experts (GGE)" on AWS. In the past, the establishment of a GGE has led to the adoption of a new CCW protocol (one banning permanently-blinding lasers). Whether states parties establish a GGE on AWS—and, if so, what its mandate will be—are open questions. In any event, at the most recent meetings, about two-dozen states endorsed the notion—the contours of which remain undefined so far—of "meaningful human control" over autonomous weapon systems.[29]

Zooming out, we see that a pair of interlocking factors has obscured and hindered analysis of whether the relevant technologies can and should be regulated.

One factor is the sheer technical complexity at issue. Lack of knowledge of technical intricacies has hindered efforts by non-experts to grasp how the core technologies may either fit within or frustrate existing legal frameworks.

---

28.   AWS have also been raised at the U.N. Human Rights Council, though without the thematic focus given to them in the context of the Convention on Certain Conventional Weapons (CCW). *See, e.g.*, Christof Heyns (Special Rapporteur on Extrajudicial, Summary or Arbitrary Executions), *Rep. to Human Rights Council*, ¶¶ 142–45, UN Doc. A/HRC/26/36 (Apr. 1, 2014).
29.   *See infra* Section 3: International Law pertaining to Armed Conflict — Customary International Law concerning AWS.

This is not a challenge particular to AWS, of course. The majority of IHL professionals are not experts in the inner workings of the numerous technologies related to armed conflict. Most IHL lawyers could not detail the technical specifications, for instance, of various armaments, combat vehicles, or intelligence, surveillance, and reconnaissance (ISR) systems. But in general that lack of technical knowledge would not necessarily impede at least a provisional analysis of the lawfulness of the use of such a system. That is because an initial IHL analysis is often an exercise in identifying the relevant rule and beginning to apply it in relation to the applicable context. Yet the widely diverse conceptions of AWS and the varied technologies accompanying those conceptions pose an as-yet-unresolved set of classification challenges. And without a threshold classification, a general legal analysis cannot proceed.

The other, related factor is that states—as well as lawyers, technologists, and other commentators—disagree in key respects on what should be addressed. The headings so far include "lethal autonomous robots," "lethal autonomous weapons systems," "autonomous weapons systems" more broadly, and "intelligent partnerships" more broadly still. And the possible standards mentioned include "meaningful human control" (including in the "wider loop" of targeting operations), "meaningful state control," and "appropriate levels of human judgment."[30] More basically, there is no consensus on whether to include only weapons or, additionally, systems capable of involvement in *other* armed conflict-related functions, such as transporting and guarding detainees, providing medical care, and facilitating humanitarian assistance.

Against this backdrop, the AWS framing has largely precluded meaningful analysis of whether it (whatever "it" entails) *can* be regulated, let alone whether and how it *should* be regulated.[31] In this briefing report, we recast the discussion by introducing the concept of "war algorithms."[32] We define

---

30. *See infra* Appendices I and II.

31. On various formal and informal models of regulating new technologies, *see generally* Benjamin Wittes & Gabriella Blum, The Future of Violence: Robots and Germs, Hackers and Drones—Confronting A New Age of Threat (2015); with respect to autonomous military robots, *see* Gary E. Marchant et al., *International Governance of Autonomous Military Robots*, 12 Colum. Sci. & Tech. L. Rev. 272 (2011).

32. Our concept of "war algorithms" should be distinguished from the "WAR algorithm" concept that has been developed in relation to evaluating environmental impacts. *See* Environmental Protection Agency, *Waste Reduction Algorithm: Chemical Process Simulation for Waste Reduction*, https://www.epa.gov/chemical-research/waste-reduction-algorithm-chemical-process-simulation-waste-reduction (last visited Aug. 27, 2016) (explaining that "[t]raditionally chemical process designs, focus on minimizing cost, while the environmental impact of a process is often overlooked. This may in many instances lead to the production of large quantities of waste materials. It is possible to reduce the generation of these wastes and their environmental impact by modifying the design of the process. The WAste Reduction

"war algorithm" as any algorithm[33] that is expressed in computer code, that is effectuated through a constructed system, and that is capable of operating in relation to armed conflict. Those algorithms seem to be a—and perhaps *the*—key ingredient of what most people and states discuss when they address AWS. We expand the purview beyond weapons alone (important as those are) because the technological capabilities are rarely, if ever, limited to use only as weapons and because other war functions involving algorithmically-derived autonomy should be considered for regulation as well. Moreover, given the modular nature of much of the technology, a focus on weapons alone might thwart attempts at regulation.

Algorithms are a conceptual and technical building block of many systems. Those systems include self-learning architectures that today present some of the sharpest questions about "replacing" human judgment with algorithmically-derived "choices." Moreover, algorithms form a foundation of most of the systems and platforms—and even the "systems of systems"—often discussed in relation to AWS. Absent an unforeseen development, algorithms are likely to remain a pillar of the technical architectures.

The constructed systems through which these algorithms are effectuated differ enormously. So do the nature, forms, and tiers of human control and governance over them. Existing constructed systems include, among many others, stationary turrets, missile systems, and manned or unmanned aerial, terrestrial, or marine vehicles.[34]

All of the underlying algorithms are developed by programmers and are expressed in computer code. But some of these algorithms—especially those capable of "self-learning" and whose "choices" might be difficult for humans to anticipate or unpack—seem to challenge fundamental and interrelated concepts that underpin international law pertaining to armed conflict and related accountability frameworks. Those concepts include attribution, control, foreseeability, and reconstructability.

At their core, the design, development, and use of war algorithms raise profound questions. Most fundamentally, those inquiries concern who, or what, should decide—and what it means to decide—matters of life and death in relation to war. But war algorithms also bring to the fore an array of more quotidian, though also important, questions about the benefits and

---

(WAR) algorithm was developed so that environmental impacts of designs could easily be evaluated. The goal of WAR is to reduce environmental and related human health impacts at the design stage.")

33.  *See infra* Section 2: Technology Concepts and Developments (on general definitions of "algorithm").

34.  *See infra* Section 2: Examples of Purported Autonomous Weapon Systems.

costs of human judgment and "replacing" it with algorithmically-derived systems, including in such areas as logistics.

We ground our analysis by focusing on war-algorithm accountability. In doing so, we sketch a three-axis accountability approach for those algorithms: state responsibility for a breach of a rule of international law, individual responsibility under international law for international crimes, and a broad notion of scrutiny governance. This is not an exhaustive list of possible types of accountability. But the axes we outline offer a flavor of how accountability, in general, could be conceptualized in the context of war algorithms.

In short, we are primarily interested in the "duty to account … for the exercise of power"[35] over—in other words, holding someone or some entity answerable for—the design, development, or use (or a combination thereof) of a war algorithm.[36] That power may be exercised by a diverse assortment of actors. Some are obvious, especially states and their armed forces. But myriad other individuals and entities may exercise power over war algorithms, too. Consider the broad classes of "developers" and "operators," both within and outside of government, of such algorithms and their related systems. Also think of lawyers, industry bodies, political authorities, members of organized armed groups—and many, many others. Focusing on war algorithms encompasses them all.

# OBJECTIVE, APPROACH, AND METHODOLOGY

In this briefing report, our objective is not to argue whether international law, as it currently exists, sufficiently addresses the plethora of issues raised by autonomous weapon systems. Rather, we aim to shed light on and recast the discussion in terms of a new concept: war algorithms. Through that lens, we link international law and related accountability architectures to relevant technologies. We sketch a three-part (non-exhaustive) approach that highlights traditional and unconventional accountability avenues. By not limiting our inquiry only to weapon systems, we take an expansive view, showing how the broad category of war algorithms might be susceptible to regulation (and how those algorithms might already fit within the existing regulatory system established by international law).

---

35.  Drawn from the discussion of INTERNATIONAL LAW ASSOCIATION, COMMITTEE ON ACCOUNTABILITY OF INTERNATIONAL ORGANIZATIONS, BERLIN CONFERENCE: FINAL REPORT 5 (2004), http://www.ila-hq.org/en/committees/index.cfm/cid/9 *in* JAMES CRAWFORD, STATE RESPONSIBILITY: THE GENERAL PART 85 (2013).

36.  In principle, the *threat* of use of a war algorithm may (also) give rise to legal implications; however, we focus on the design, development, and use of those algorithms.

We draw on the extensive—and rapidly growing—amount of scholarship and other analytical analyses that have addressed related topics.[37] To help illuminate the discussion, we outline what technologies and weapon systems already exist, what fields of international law might be relevant, and what regulatory avenues might be available. As noted above, because international law is the touchstone normative framework for accountability in relation to war, we focus on public international law sources and methodologies. But as we show, other norms and forms of governance might also merit attention.

Accountability is a broad term of art. We adapt—from the work of an International Law Association Committee in a different context (the accountability of international organizations)—a three-part accountability approach.[38] Our framework outlines three axes on which to focus initially on war algorithms.

The first axis is *state responsibility*. It concerns state responsibility arising out of acts or omissions involving a war algorithm where those acts or omissions constitute a breach of a rule of international law. State responsibility entails discerning the content of the rule, identifying a breach of the rule, assigning attribution for that breach to a state, determining available excuses (if any), and imposing measures of remedy.

The second axis is a form of *individual responsibility* under international law. In particular, it concerns individual responsibility under international law for international crimes—such as war crimes—involving war algorithms. This form of individual responsibility entails establishing the commission of a crime under the relevant jurisdiction, assessing the existence of a justification or excuse (if any), and, upon conviction, imposing a sentence.

The third and final axis is *scrutiny governance*. Embracing a wider notion of accountability, it concerns the extent to which a person or entity is and should be subject to, or should exercise, forms of internal or external scrutiny, monitoring, or regulation (or a combination thereof) concerning the design, development, or use of a war algorithm. Scrutiny governance does not hinge on—but might implicate—potential and subsequent liability or responsibility (or both). Forms of scrutiny governance include independent monitoring, norm (such as legal) development, adopting non-binding resolutions and codes of conduct, normative design of technical architectures, and community self-regulation.

---

37.  *See infra* Bibliography.

38.  Our approach is derived in part from International Law Association, *supra* note 35, at 5.

# OUTLINE

In **Section 2,** we outline pertinent considerations regarding algorithms and constructed systems. We then highlight recent advancements in artificial intelligence related to learning algorithms and architectures. We next examine state approaches to technical autonomy in war, focusing on five such approaches. Finally, to ground the often-theoretical debate pertaining to autonomous weapon systems, we describe existing weapon systems that have been characterized by various commentators as AWS.

In **Section 3,** we outline the main fields of international law that war algorithms might implicate. There is no single branch of international law dedicated solely to war algorithms. So we canvass how those algorithms might fit within or otherwise implicate various fields of international law. We ground the discussion by outlining the main ingredients of state responsibility: attribution, breach, excuses, and consequences. Then, to help illustrate states' positions concerning AWS, we examine whether an emerging norm of customary international law specific to AWS may be discerned. We find that one cannot (at least not yet). So we next highlight how the design, development, or use (or a combination thereof) of a war algorithm might implicate more general principles and rules found in various fields of international law. Those fields include the *jus ad bellum*, IHL, international human rights law, international criminal law, and space law. Because states and commentators have largely focused on AWS to date, much of our discussion here relates to the AWS framing.

In **Section 4,** we elaborate a (non-exhaustive) war-algorithm accountability approach. That approach focuses on state responsibility for an internationally wrongful act, on individual responsibility under international law for international crimes, and on wider forms of scrutiny, monitoring, and regulation. We highlight existing accountability actors and architectures under international law that might regulate war algorithms. These include war reparations as well as international and domestic tribunals. We then turn to less conventional accountability avenues, such as those rooted in normative design of technical architectures (including maximizing the auditability of algorithms) and community self-regulation.

In the **Conclusion,** we return to the deficiencies of current discussions of AWS and emphasize the importance of addressing the wide and serious concerns raised by AWS with technical proficiency, legal expertise, and non-ideological commitment to a genuine and inclusive inquiry.

We also attach a **Bibliography** and **Appendices**. The Bibliography contains over 400 analytical sources, in various languages, pertaining to technical autonomy in war. The Appendices contain detailed charts listing and categorizing states' statements at the 2015 and 2016 Informal Meetings of Experts on Lethal Autonomous Weapons Systems convened within the framework of the CCW.

# CAVEATS

The bulk of the secondary-source research was conducted in English. Moreover, none of us is an expert in computer science or robotics. We consulted specialists in these fields, but we alone are responsible for any remaining errors. In any event, given the rapid pace of development, the technologies discussed in this briefing report may soon be eclipsed—if they have not been already.

# 2

# TECHNOLOGY CONCEPTS AND DEVELOPMENTS

This section sketches key technology concepts and developments, as well as certain states' understandings of autonomy in relation to war. We set the stage by discussing algorithms and constructed systems. We then outline recent advancements in the AI field of deep learning. Next, we highlight five states' approaches to technical autonomy in war. In doing so, we also note accompanying standards that states and commentators are actively vetting, such as "meaningful human control" over AWS. Finally, we describe some of the main technologies that various commentators have addressed in relation to autonomous weapon systems.

## TWO KEY INGREDIENTS

In this briefing report, our foundational technological concern is the capability of a constructed system, without further human intervention, to help make and effectuate a "decision" or "choice" of a war algorithm. Distilled, the two core ingredients are an algorithm expressed in computer code and a suitably capable constructed system.

### ALGORITHM

An algorithm has been defined informally as "any well-defined computational procedure that takes some value, or set of values, as ***input*** and produces some value, or set of values, as *output*."[39] Accordingly, an algorithm is "a sequence of computational steps that transform the input into the output."[40] Yet "[w]e can also view an algorithm as a tool for solving a well-specified *computational problem*."[41] In this second approach, "[t]he statement of the problem specifies in general terms the desired input/output relationship. The algorithm describes a specific computational procedure for achieving that input/output relationship."[42] Here, we are most concerned with algorithms that are expressed in computer code and that can be conceptualized as making "decisions" or "choices" along the computational pathway undertaken in light of the input and in accordance with programmed parameters.

The relevant algorithms may vary enormously in terms of their sophistication and complexity. But, at base, they all are conceived and coded initially by humans to take some input and produce some output or to describe a specific computational procedure for achieving a defined desirable input/ output relationship.

By limiting our inquiry to war algorithms, we narrow the types of algorithms at issue to those that fulfill three conditions: algorithms (1) that are expressed in computer code; (2) that are effectuated through a constructed system; and (3) that are capable of operating in relation to armed conflict. Not all weapons or systems that have been characterized as "AWS" meet these criteria. But most do. And, more to the point, we see these algorithms as a key ingredient in what most commentators and states mean when they address notions of autonomy.

We predicate our definition on the algorithm being *capable* of operating in relation to armed conflict, even if it is not initially designed for such use. We thus do not limit our classification to algorithms that are in fact *used* in armed conflict (though the broader category of capability would subsume those that are actually used). A critique of this approach might be that it is over-inclusive because it does not distinguish between algorithms and the relevant constructed systems that are intended for use in relation to war from the vast array of other such algorithms and systems that might be adapted for such use. Yet one reason to focus on capability—instead of intent—is that much of the underlying technology is modular and can therefore be adapted for use in relation to war even if it was not initially designed and developed to do so. Moreover, with respect to accountability, focusing on capability sweeps in not only those who are in a position to choose to deploy or to operate war

---

algorithms but also those involved in the design and development of those algorithms. The emphasis on capability thereby helps account for the diverse assortment of actors—whether in government, commercial, academic, or other contexts—who might exercise power over, and thus who might be held answerable for, the design, development, or use of war algorithms.

## CONSTRUCTED SYSTEM

"Robot" is not a legal term of art under international law. One oft-cited, decades-old definition comes from the Robot Institute of America, a trade association of robot manufacturers and users: "a reprogrammable, multifunctional manipulator designed to move material, parts, tools, or specialized devices through various programmed motions for the performance of a variety of tasks."[43] Others draw different definitional boundaries. Alan Winfield, for instance, defines a robot as "an artificial device that can sense its environment and purposefully act on or in that environment."[44] Neil Richards and William Smart argue that a robot is "a constructed system that displays both physical and mental agency but is not alive in the biological sense."[45] And the Oxford English Dictionary Online defines a robot in the modern sense[46] as "[a]n intelligent artificial being typically made of metal and resembling in some way a human or other animal."[47]

We sidestep some of the definitional quandaries attending "robot" by focusing instead on *constructed systems*. For our purposes, a constructed system is a manufactured machine, apparatus, plant, or platform that is capable both of being used to gather information and of effectuating a "choice" or "decision" which is, in whole or in part, derived through an algorithm expressed in computer code but that is not alive in the biological sense. By limiting our inquiry to systems that are not alive in the biological sense, we also circumvent the subject of biologically engineered agents.

43. *Robotics Today*, RIA News, Spring 1980, at 7, *cited in* Robotics and the Economy: A Staff Study, Prepared for the Use of the Subcommittee on Monetary and Fiscal Policy of the Joint Economic Committee, Congress of the United States 4 n.3 (1982).

44. Robohub Editors, *Robohub Roundtable: Why Is It So Difficult to Define Robot?*, Robohub, April 29, 2016, http://robohub.org/robohub-roundtable-why-is-it-so-difficult-to-define-robot.

45. Neil Richards & William Smart, *How Should the Law Think About Robots?*, *in* Robot Law 3, 6 (Ryan Calo, Michael Froomkin & Ian Kerr eds., 2016).

46. The now-historical sense of the term "robot" denotes "[a] central European system of serfdom, by which a tenant's rent was paid in forced labour or service." *See Robot* n.1, Oxford English Dictionary (online ed.) (2016).

47. *Robot* n.2, Oxford English Dictionary (online ed.) (2016) (noting that, originally, this sense of the term was used "with reference to the mass-produced workers in Karel Čapek's play *R.U.R.: Rossum's Universal Robots* (1920) which are assembled from artificially synthesized organic material.").

Among the most common sensors used to gather information in "constructed systems" include methods to detect how far away objects are by transmitting certain waves and monitoring their reflections, such as radar (radio waves), sonar (sound waves), and lidar (light waves), as well as cameras. The system may be tele-operated (also known as remotely operated)—or not. It may have a manipulator (used loosely here to denote a component providing the capability to interact in the built environment)—or not. However, if it does not have a manipulator, the system needs, to meet our definition, another avenue to effectuate the algorithmically-derived "choice" or "decision."

The constructed systems may come in a diverse array of forms,[48] such as marine, terrestrial, aerial, or space vehicles; missile systems; or biped or quadruped robots.[49] They may operate collaboratively—including as so-called "swarms"[50]—or individually. They may use a range of power sources, such as batteries or internal combustion engines to generate electricity or to power hydraulic or pneumatic actuators. And their costs may run the gamut from the budget of a tinkerer to industrial or governmental-scale programs.

# A.I. ADVANCEMENTS

Recently published advancements in AI—especially machine learning and a class of techniques called deep learning—underscore the rapid pace of technical development.[51] Those advancements reach into many areas of modern digital life, underlying "web searches to content filtering on social networks to recommendations on e-commerce websites."[52]

---

48.  *See infra* Section 2: Examples of Purported Autonomous Weapon Systems.

49.  *See, e.g.*, Boston Dynamics, *Introducing SpotMini*, YOUTUBE (June 23, 2016), https://www.youtube.com/watch?v=tf7IEVTDjng [https://perma.cc/LNV5-3SCH] (video of Boston Dynamic's SpotMini robot, which purports to "perform[] some tasks autonomously, but often uses a human for high-level guidance.").

50.  *See, e.g.*, Michael Rubenstein, Alejandro Cornejo & Radhika Nagpal, *Programmable Self-Assembly in a Thousand-Robot Swarm*, 345 SCIENCE 795, 796 (2014) ("We demonstrate a thousand-robot swarm capable of large-scale, flexible self-assembly of two-dimensional shapes entirely through programmable local interactions and local sensing, achieving highly complex collective behavior. The approach involves the design of a collective algorithm that relies on the composition of basic collective behaviors and cooperative monitoring for errors to achieve versatile and robust group behavior, combined with an unconventional physical robot design that enabled the creation of more than 1000 autonomous robots."). In respect of this large-scale robotic swarm, the extent to which the robots "can be fully autonomous" is measured in terms of being "capable of computation, locomotion, sensing, and communication." *Id.* at 796.

51.  For an excellent analysis of some of the key technologies in relation to AWS, see Peter Margulies, *Making Autonomous Weapons Accountable: Command Responsibility for Computer-Guided Lethal Force in Armed Conflicts*, *in* RESEARCH HANDBOOK ON REMOTE WARFARE (Jens David Ohlin ed., forthcoming 2016).

52.  Yann LeCun, Yoshua Bengio & Geoffrey Hinton, *Deep Learning*, 521 NATURE 436, 436 (2015).

For many years, "[c]onventional machine-learning techniques were limited in their ability to process natural data in their raw form."[53] For decades, for instance, "constructing a pattern-recognition or machine-learning system required careful engineering and considerable domain expertise to design a feature extractor that transformed the raw data … into a suitable internal representation or feature vector from which the learning subsystem, often a classifier, could detect or classify patterns in the input."[54] An advance came with *representational learning*, which "is a set of methods that allows a machine to be fed with raw data and to automatically discover the representations needed for detection or classification."[55]

*Deep learning*—including deep neural networks—marked another advance. (A deep neural network can be thought of as "a network of hardware and software that mimics the web of neurons in the human brain."[56]) Deep-learning methods have been explained as "representation-learning methods with multiple levels of representation, obtained by composing simple but non-linear modules that each transform the representation at one level (starting with the raw input) into a representation at a higher, slightly more abstract level."[57] As experts have explained, "[w]ith the composition of enough such transformations, very complex functions can be learned."[58] The gist is that, "[f]or classification tasks, higher layers of representation amplify aspects of the input that are important for discrimination and suppress irrelevant variations."[59]

Consider the example of a digital image. It

> comes in the form of an array of pixel values, and the learned features in the first layer of representation typically represent the presence or absence of edges at particular orientations and locations in the image. The second layer typically detects motifs by spotting particular arrangements of edges, regardless of small variations in the edge positions. The third layer may assemble motifs into larger combinations that correspond to parts of familiar objects, and subsequent layers would detect objects as combinations of these parts.[60]

Through deep-learning techniques, "these layers of features are not designed

---

by human engineers: they are learned from data using a general-purpose learning procedure."[61]

Already, "[d]eep learning is making major advances in solving problems that have resisted the best attempts of the artificial intelligence community for many years."[62] Those include beating records in image recognition and speech recognition, as well as beating other machine-learning techniques at, for example, predicting the activity of drug molecules.[63] Writing in 2015, some experts "think that deep learning will have many more successes in the near future because it requires very little engineering by hand, so it can easily take advantage of increases in the amount of available computation and data."[64] In line with this view, "[n]ew learning algorithms and architectures that are currently being developed for deep neural networks will only accelerate this progress."[65]

One mark of that progress came late last year when a computer program, AlphaGo, achieved a feat previously thought to be at least a decade away: defeating a human professional player in a full-sized game of Go.[66] (A few months later, AlphaGo won four of five matches against Lee Sedol, who, as one of the top players in the world, had achieved the highest rank of nine dan.[67]) The system designers introduced a new approach based on deep convolutional neural networks that used "value networks" to evaluate board decisions and "policy networks" to select moves. (Convolutional neural networks—the typical architecture of which is structured as a series of stages—"are designed to process data that come in the form of multiple arrays."[68] In other words, these networks "use many layers of neurons, each arranged in overlapping tiles, to construct increasingly abstract, localized representations of an image."[69]) For AlphaGo, those deep neural networks were "trained by a novel combination of supervised learning from human expert games, and reinforcement learning from games of self-play."[70] AlphaGo developers also introduced a new search algorithm—which was designed in part to encourage exploration on its own—that combines a sophisticated

simulation technique (called Monte Carlo tree search) with the value and policy networks.[71]

By grounding our discussion in algorithms expressed in computer code and effectuated through constructed systems, we sidestep some of the doctrinal debates on what constitutes "artificial intelligence" and "artificial general intelligence"—and on whether the latter may be realistically achievable or is more the stuff of science fiction. These questions are outside of the scope of this briefing report, but they are nonetheless vitally important. In any event, it merits emphasis that existing learning algorithms and architectures already have remarkable capabilities that, at least, seem to approach aspects of human "decision-making."

For their part, creators of AlphaGo have characterized Go as "exemplary in many ways of the difficulties faced by artificial intelligence: a challenging decision-making task, an intractable search space, and an optimal solution so complex it appears infeasible to directly approximate using a policy or value function."[72] In the eyes of its designers, AlphaGo provides "hope that human-level performance can now be achieved in other seemingly intractable artificial intelligence domains."[73]

# APPROACHES TO TECHNICAL AUTONOMY IN WAR

As noted above, there is no agreement on what "autonomy" means in the context of the discussion to date on autonomous weapon systems.

Commentators' views on what constitutes "autonomy" in this context range enormously. Some, for instance, focus on whether the system navigates with a human on board ("manned") or without one ("unmanned"). Others emphasize geography, such as whether the weapon is operated by a human remotely or proximately. Some hold that the "autonomy" in AWS should be reserved only for "critical functions" in the conduct-of-hostilities targeting cycle. Still others argue that it is the capability of a system, once launched, to sense, think, learn, and act all without further human intervention. A number of definitions combine various components of these notions. But depending on the definition and classification, it is beyond doubt that some existing military systems contain at least a degree of autonomy. (In the last sub-section of this section, we profile examples of weapons, weapon systems, and weapon platforms that some commentators have characterized as AWS.)

---

71.  Silver et al., *supra* note 14, at 486.

72.  *Id*. at 489 (citations omitted).

73.  *Id*.

In this sub-section, we focus on the positions of states, because discerning states' positions and practices is one of the key steps in illuminating the scope of international law as it currently stands (*lex lata*) and distinguishing that from nascent norms and from the law as it should be (*lex ferenda*). A handful of states have considered or formally adopted definitions relevant to AWS, whether while focusing on weapon systems or unmanned aerial systems. Below, we summarize five of the most elaborate sets of these considerations and definitions—those by Switzerland, France, the Netherlands, the United States, and the United Kingdom.

## SWITZERLAND

In the lead-up to the 2016 Informal Meeting of Experts on Lethal Autonomous Weapons Systems, Switzerland published an "Informal Working Paper" titled "Towards a 'compliance-based' approach to LAWS." The paper proposes "to initially describe autonomous weapons systems (AWS) simply as" follows:

> [W]eapons systems that are capable of carrying out tasks governed by IHL in partial or full replacement of a human in the use of force, notably in the targeting cycle.[74]

According to the paper, "[s]uch a working definition is inclusive, accounts for a wide array of system configurations, and allows for a debate that is differentiated, compliance-based, and without prejudice to the question of appropriate regulatory response."[75] In the view of Switzerland, "the working definition proposed is not conceived in any way to single out only those systems which could be seen as legally objectionable."[76] The authors note that "[a]t one end of the spectrum of systems falling within that working definition, States may find some subcategories to be entirely unproblematic, while at the other end of the spectrum, States may find other subcategories unacceptable."[77] Finally, the paper notes, "[a]s discussions advance, this working definition could and probably should evolve to become more specific and purposeful."[78]

---

74. Gov't of Switz., Towards a "Compliance-Based" Approach to LAWS [Lethal Autonomous Weapons Systems] 1 (March 30, 2016) (informal working paper), http://www.unog.ch/80256EDD006B8954/(httpAssets)/D2D66A9C427958D6C1257F8700415473/$file/2016_LAWS+MX_CountryPaper+Switzerland.pdf [hereinafter Swiss, "Compliance-Based" Approach].
75. *Id.*
76. *Id.* at 1–2.
77. *Id.* at 2.
78. *Id.*

# THE NETHERLANDS

On April 7, 2015, the Netherlands Ministries of Foreign Affairs and of Defense requested a report from the Advisory Council on International Affairs (AIV) and the Advisory Committee on Issues of Public International Law (CAVV) addressing five sets of questions concerning autonomous weapon systems:

> 1. What role can autonomous weapons systems (and autonomous functions within weapons systems) fulfil in the context of military action now and in the future?

> 2. What changes might occur in the accountability mechanism for the use of fully or semi-autonomous weapons systems in the light of associated ethical issues? What role could the concept of 'meaningful human control' play in this regard, and what other concepts, if any, might be helpful here?

> 3. In its previous advisory report, the CAVV states that the deployment of any weapons system, whether or not it is wholly or partly autonomous, remains subject to the same legal framework. As far as the CAVV is concerned, there is no reason to assume that the existing international legal framework is inadequate to regulate the deployment of armed drones. Does the debate on fully or semi-autonomous weapons systems give cause to augment or amend this position?

> 4. How do the AIV and the CAVV view the UN Special Rapporteur's call for a moratorium on the development of fully autonomous weapons systems?

> 5. How can the Netherlands best contribute to the international debate on this issue?

A joint committee of the AIV and the CAVV prepared a report, which the AIV adopted on October 2, 2015 and the CAVV adopted on October 12, 2015.[79] On March 2, 2016, the government responded to the report. (We use the term "government" in this context interchangeably with reference to the Ministries of Foreign Affairs and of Defense of the Netherlands.) The main conclusion of the report, in the words of the government's response, "is that meaningful human control is required in the deployment of autonomous weapon systems"—a view with which the government concurs.[80]

The government—while noting "[t]here is as yet no internationally agreed definition of an autonomous weapon system"—supports the working

---

79.  Advisory Council on International Affairs, Autonomous Weapon Systems: The Need for Meaningful Human Control 7 (Advisory Report No. 97, 2015), http://aiv-advice. nl/8gr [hereinafter AIV].

80.  Dutch Government, Response to AIV/CAVV Report, *supra* note 22.

definition of AWS which the advisory committee adopted:[81]

> A weapon that, without human intervention, selects and engages targets matching certain predetermined criteria, following a human decision to deploy the weapon on the understanding that an attack, once launched, cannot be stopped by human intervention.[82]

Underlying this definition is the notion of the "wider loop" of the decision-making process, which plays a prominent role in the Dutch government's understanding of accountability concerning AWS. In the view of the Dutch government, with respect to AWS humans are involved in that "wider loop" because humans "play a prominent role in programming the characteristics of the targets that are to be engaged and in the decision to deploy the weapon."[83] That means, in short, "that humans continue to play a crucial role in the wider targeting process. An autonomous weapon as defined above is therefore only deployed after human consideration of aspects such as target selection, weapon selection and implementation planning, including an assessment of potential collateral damage."[84] In addition, the government notes, "the autonomous weapon is programmed to perform specific functions within pre-programmed conditions and parameters. Its deployment is followed by a human assessment of the effects. Assessments of potential collateral damage (proportionality) and accountability under international humanitarian law are of key importance in this respect."[85]

As summarized by the Dutch government, "[t]he advisory committee states that if the deployment of an autonomous weapon system takes place in accordance with the process described above, there is meaningful human control. In such cases, humans make informed, conscious choices regarding the use of weapons, based on adequate information about the target, the weapon in question and the context in which it is to be deployed."[86] For its part, "[t]he advisory committee sees no immediate reason to draft new or additional legislation for the concept of meaningful human control."[87] Instead, "[t]he concept should be regarded as a standard deriving from existing legislation and practices (such as the targeting process)."[88] Over all, the government expressly affirms that it "supports the definition given above of an autonomous weapon system, including the concept of meaningful human control, and agrees that no new legislation is required."[89]

---

81. *Id.*

82. *Id.*

83. Dutch Government, Response to AIV/CAVV Report, *supra* note 22.

84. *Id.*

85. *Id.*

86. *Id.*

87. *Id.*

88. *Id.*

# FRANCE

In a "non-paper" circulated in the context of the 2016 Informal Meeting of Experts on Lethal Autonomous Weapons Systems, France articulated the following considerations with respect to such systems:

> France considers that LAWS [Lethal Autonomous Weapons Systems] share the following characteristics:
>
> - Lethal autonomous weapons systems are **fully autonomous systems.** LAWS are future systems: they do not currently exist.
>
> - **Remotely operated weapons systems and supervised weapons systems should not be regarded as LAWS** since a human operator remains involved, in particular during the targeting and firing phases. **Existing automatic systems are not LAWS either**[.]
>
> - **LAWS should be understood as implying a total absence of human supervision,** meaning there is absolutely no link (communication or control) with the military chain of command.
>
> - **The delivery platform of a LAWS would be capable of moving, adapting to its land, marine or aerial environments and targeting and firing a lethal effector (bullet, missile, bomb, etc.) without any kind of human intervention or validation."**[90]

Compared to most other states that have put forward working definitions, France articulates a relatively narrow definition of what constitutes a lethal autonomous weapons system in the context of the CCW. Most striking, perhaps, is the condition that there be "a total absence of human supervision, meaning there is absolutely no link (communication or control) with the military chain of command." Moreover, France clarifies that, in its view, the definition of a "lethal autonomous weapons system" includes only a delivery "platform" that "would be capable of moving, adapting to its land, marine or aerial environments and targeting and firing a lethal effector … without any kind of human intervention or validation." This formulation combines autonomy in navigation and maneuver with autonomy in certain key elements of the targeting cycle.

---

89.  Though the government agrees with the advisory committee "that definitions should be agreed on (in accordance with recommendation no. 4)." DUTCH GOVERNMENT, RESPONSE TO AIV/CAVV REPORT, *supra* note 22. As noted above, the Dutch government "reject[ed] outright the possibility of developing and deploying fully autonomous weapons." *Id.*

90.  GOV'T OF FR., Characterization of a LAWS (April 11–15, 2016) (non-paper), http://www.unog.ch/80256EDD006B8954/(httpAssets)/5FD844883B46FEACC1257F8F00401FF6/$file/2016_LAWSMX_CountryPaper_France+CharacterizationofaLAWS.pdf (bold in the original).

## *UNITED STATES*

In a series of directives and other documents, the U.S. Department of Defense (DoD) has elaborated one of the most technically specific state approaches to autonomy in relation to weapon systems.

A central document is DoD Directive 3000.09 (2012). It "[e]stablishes DoD policy and assigns responsibilities for the development and use of autonomous and semi-autonomous functions in weapon systems, including manned and unmanned platforms."[91] The directive is applicable to certain DoD actors and related organizational entities.[92] It concerns "[t]he design, development, acquisition, testing, fielding, and employment of autonomous and semi-autonomous weapon systems, including guided munitions that can independently select and discriminate targets," as well as "[t]he application of lethal or non-lethal, kinetic or non-kinetic, force by autonomous or semi-autonomous weapon systems."[93] However, the directive expressly "does not apply to autonomous or semi-autonomous cyberspace systems for cyberspace operations; unarmed, unmanned platforms; unguided munitions; munitions manually guided by the operator (e.g., laser- or wire-guided munitions); mines; or unexploded explosive ordnance."[94] Among the relevant terms defined in the glossary of Directive 3000.09 are the following:

> *Autonomous weapon system*: "A weapon system that, once activated, can select and engage targets without further intervention by a human operator. This includes human-supervised autonomous weapon systems that are designed to allow human operators to override operation of the weapon system, but can select and engage targets without further human input after activation."[95]

> *Human-supervised autonomous weapon system*: "An autonomous weapon system that is designed to provide human operators with the ability to intervene and terminate engagements, including in the event of a weapon system failure, before unacceptable levels of damage occur."[96]

---

91.  U.S. Dep't of Def., Dir. 3000.09, Autonomy in Weapon Systems ¶ 1 (Nov. 21, 2012) [hereinafter DOD AWS Dir.].

92.  *Id*. at ¶ 2.

93.  *Id*.

94.  *Id*.

95.  *Id*. at 13–14.

96.  *Id*.at 14.

> *Semi-autonomous weapon system*: "A weapon system that, once activated, is intended to only engage individual targets or specific target groups that have been selected by a human operator. This includes: [s]emi-autonomous weapon systems that employ autonomy for engagement-related functions including, but not limited to, acquiring, tracking, and identifying potential targets; cueing potential targets to human operators; prioritizing selected targets; timing of when to fire; or providing terminal guidance to home in on selected targets, provided that human control is retained over the decision to select individual targets and specific target groups for engagement."[97]

Directive 3000.09 establishes that, as a matter of policy, "[a]utonomous and semi-autonomous weapon systems shall be designed to allow commanders and operators to exercise appropriate levels of human judgment over the use of force."[98] More specifically, "[s]ystems will go through rigorous hardware and software verification and validation … and realistic system developmental and operational test and evaluation … in accordance with" certain guidelines.[99] In addition, "[t]raining, doctrine, and tactics, techniques, and procedures … will be established."[100] In particular, those measures will ensure that autonomous and semi-autonomous weapon systems will, first, "[f]unction as anticipated in realistic operational environments against adaptive adversaries." Second, they will ensure that those systems will "[c]omplete engagements in a timeframe consistent with commander and operator intentions and, if unable to do so, terminate engagements or seek additional human operator input before continuing the engagement." And third, they will ensure that those systems "[a]re sufficiently robust to minimize failures that could lead to unintended engagements or to loss of control of the system to unauthorized parties."[101]

The directive also establishes that "[c]onsistent with the potential consequences of an unintended engagement or loss of control of the system to unauthorized parties, physical hardware and software will be designed with appropriate: … Safeties, anti-tamper mechanisms, and information assurance in accordance with [another relevant DoD directive]. … Human-machine interfaces and controls."[102] Furthermore, "[i]n order for operators to make informed and appropriate decisions in engaging targets," the directive establishes that "the interface between people and machines for autonomous and semi-autonomous weapon systems shall" have three characteristics. First, they shall "[b]e readily understandable to trained operators." Second, they shall "[p]rovide traceable feedback on system status." And third, they shall "[p]rovide clear procedures for trained operators to activate and deactivate system functions."[103]

---

97.  *Id.*
98.  *Id.* at ¶ 4.
99.  *Id.*
100.  *Id.*
101.  *Id.*
102.  *Id.*

Directive 3000.09 further lays down, also as a matter of policy, that "[p]ersons who authorize the use of, direct the use of, or operate autonomous and semi-autonomous weapon systems must do so with appropriate care and in accordance with the law of war, applicable treaties, weapon system safety rules, and applicable rules of engagement (ROE)."[104] The directive establishes that autonomous and semi-autonomous weapon systems intended to be used in a manner that falls within three certain sets of policies will be considered for approval in accordance with enumerated approval procedures and other applicable policies and issuances.[105] The first such policy set establishes that "[s]emi-autonomous weapon systems (including manned or unmanned platforms, munitions, or sub-munitions that function as semi-autonomous weapon systems or as subcomponents of semi-autonomous weapon systems) may be used to apply lethal or non-lethal, kinetic or non-kinetic force." Further pursuant to that policy set, "[s]emi-autonomous weapon systems that are onboard or integrated with unmanned platforms must be designed such that, in the event of degraded or lost communications, the system does not autonomously select and engage individual targets or specific target groups that have not been previously selected by an authorized human operator." The second policy set lays down that "[h]uman-supervised autonomous weapon systems may be used to select and engage targets, with the exception of selecting humans as targets, for local defense to intercept attempted time-critical or saturation attacks" for static defense of manned installations and for onboard defense of manned platforms. Finally in this connection, the third policy set establishes that autonomous weapon systems "may be used to apply non-lethal, non-kinetic force, such as some forms of electronic attack, against materiel targets in accordance with" a separate DoD directive.[106]

Directive 3000.09 further provides that "[a]utonomous or semi-autonomous weapon systems intended to be used in a manner that falls outside" those three sets of policies must be approved by the Under Secretary of Defense for Policy, the Under Secretary of Defense for Acquisition, Technology, and Logistics, and the Chairman of the Joint Chiefs of Staff "before formal development and again before fielding in accordance with" enclosed guidelines and other applicable policies and issuances.[107] In addition, Directive 3000.09 lays down, also as a matter of policy, that "[i]nternational sales or transfers of autonomous and semi-autonomous weapon systems will be approved in accordance with existing technology security and foreign disclosure requirements and processes, in accordance with" an enumerated memorandum.[108] Enclosures to the directive further explain certain references; further elaborate verification and validation as well as testing and evaluation of autonomous and semi-autonomous weapon systems; set down guidelines for review of certain such systems; elaborate responsibilities; and provide definitions in a glossary.[109]

---

103.  *Id.*
104.  *Id.*
105.  *Id.*
106.  *Id.*

For its part, the U.S. DoD *Law of War Manual* gives examples of two ways that some weapons may have autonomous functions. First, "mines may be regarded as rudimentary autonomous weapons because they are designed to explode by the presence, proximity, or contact of a person or vehicle, rather than by the decision of the operator."[110] And second, "[o]ther weapons may have more sophisticated autonomous functions and may be designed such that the weapon is able to select targets or to engage targets automatically after being activated by the user."[111] The *Manual* authors give the example that "the United States has used weapon systems for local defense with autonomous capabilities designed to counter time-critical or saturation attacks. These weapon systems have included the Aegis ship defense system and the Counter-Rocket, Artillery, and Mortar (C-RAM) system."[112]

## UNITED KINGDOM

The United Kingdom Ministry of Defence (MoD) has addressed autonomy primarily in relation to unmanned aircraft systems. The MoD promulgated the key document—*Joint Doctrine Note 2/11: The UK Approach to Unmanned Aircraft Systems* (Joint Doctrine Note)—on March 30, 2011.[113] That document's "purpose is to identify and discuss policy, conceptual, doctrinal and technology issues that will need to be addressed if such systems are to be

---

107.  *Id.*

108.  *Id.*

109.  *Id.* at 5–15.

110.  U.S. Dep't of Def., Law of War Manual § 6.5.9.1 (2016) (internal reference omitted) [hereinafter Law of War Manual].

111.  *Id.*

112.  *Id.*

113.  U.K. Ministry of Def., Joint Doctrine Note 2/11: The UK Approach to Unmanned Aircraft Systems, (2011), https://www.gov.uk/government/uploads/system/uploads/attachment_data/file/33711/20110505JDN_211_UAS_v2U.pdf.

successfully developed and integrated into future operations."[114]

In the section on definitions, the authors discuss "automation" and "autonomy," emphasizing that, confusingly, the two "terms are often used interchangeably even when referring to the same platform; consequently, companies may describe their systems to be autonomous even though they would not be considered as such under the military definition."[115] Noting that "[i]t would be impossible to produce definitions that every community would agree to," the Joint Doctrine Note authors chose the following definitions in order to be "as simple as possible, while making clear the essential differences in meaning between them":[116]

> *Automated system*: "In the unmanned aircraft context, an automated or automatic system is one that, in response to inputs from one or more sensors, is programmed to logically follow a pre-defined set of rules in order to provide an outcome. Knowing the set of rules under which it is operating means that its output is predictable."

> *Autonomous system*: "An autonomous system is capable of understanding higher level intent and direction. From this understanding and its perception of its environment, such a system is able to take appropriate action to bring about a desired state. It is capable of deciding a course of action, from a number of alternatives, without depending on human oversight and control, although these may still be present. Although the overall activity of an autonomous unmanned aircraft will be predictable, individual actions may not be."[117]

Based on those definitions, the Joint Doctrine Note authors deduce four sets of points. The basic notion of the first set is that "[a]ny or none of the functions involved in the operation of an unmanned aircraft may be automated."[118] In a related footnote, it is stated that "[f]or major functions such as target detection, only some of the sub-functions may be automated, requiring human input to deliver the overall function."[119]

The main idea guiding the second set of points is that "[a]utonomous systems will, in effect, be self-aware and their response to inputs indistinguishable from, or even superior to, that of a manned aircraft."[120] As such, according to the authors, those autonomous systems "must be capable of achieving the same level of situational understanding as a human."[121] At the time of publication (2011), the authors stated, "[t]his level of technology is not yet achievable and so, by the definition of autonomy in this JDN, none of the currently fielded or in-development unmanned aircraft platforms can be correctly described as autonomous."[122]

---

114.  *Id.* at iii.

115.  *Id.* at 2-2.

116.  *Id.* at 2-2–2-3.

117.  *Id.* at 2-3.

118.  *Id.*

119.  *Id.* at 2-3 n.5 (giving examples of "take-off and landing; navigation/route following; pre-programmed response to events such as loss of a command and communication link; and automated target detection and recognition").

120.  *Id.* at 2-3.

The third set of points concerns the importance of "[t]he distinction between autonomous and automated … as there are moral, ethical and legal implications regarding the use of autonomous unmanned aircraft."[123] Those issues are discussed in another part of the Joint Doctrine Note.[124] The fourth and final set of points deduced by the authors concerns "an over-arching principle that, whatever the degree of automation, an unmanned aircraft should provide at least the same, or better, safety standard as a manned platform carrying out the same task."[125]

In addressing accountability, the Joint Doctrine Note states that "[l]egal responsibility for any military activity remains with the last person to issue the command authorising a specific activity."[126] The Joint Doctrine Note authors recognize, however, that "[t]his assumes that a system's basic principles of operation have, as part of its release to service, already been shown to be lawful, but that the individual giving orders for use will ensure its continued lawful employment throughout any task."[127] An assumption underlying this process is "that a system will continue to behave in a predictable manner after commands are issued," yet, the authors note, "clearly this becomes problematical as systems become more complex and operate for extended periods."[128] Indeed, according to the authors, "[i]n reality, predictability is likely to be inversely proportional to mission and environmental complexity. For long-endurance missions engaged in complex scenarios, the authorised entity that holds legal responsibility will be required to exercise some level of supervision throughout."[129] If that is the case, in the view of the authors, "this implies that any fielded system employing weapons will have to maintain a 2-way data link between the aircraft and its controlling authority."[130]

---

121.  *Id.*

122.  *Id.* at 2-3–2-4 (further stating in this connection that "[a]s computing and sensor capability increases, it is likely that many systems, using very complex sets of control rules, will appear and be described as autonomous systems, but as long as it can be shown that the system logically follows a set of rules or instructions and is not capable of human levels of situational understanding, then they should only be considered to be automated").

123.  *Id.* at 2-4.

124.  *Id.* at 5-1–5-12. *See also infra* Section 3.

125.  *Id.* at 2-4 (citation omitted).

126.  *Id.* at 5-5.

127.  *Id.*

128.  *Id.*

129.  *Id.*

# EXAMPLES OF PURPORTED AUTONOMOUS WEAPON SYSTEMS

This section profiles weapons, weapon systems, and weapon platforms that have been couched, by various commentators, as autonomous weapon systems—such as by exhibiting or reflecting varying levels, forms, or notions of autonomy or automation, in relation to navigation or maneuvering or the targeting cycle. The inclusion of a weapon here is not meant to indicate our evaluation that the weapon, system, or platform has or does not have autonomous capabilities or that it fits within a legally relevant definition of autonomy. Most, but not all, of the weapons, systems, and platforms described here operate based, at least in part, on a war algorithm.

## *MINES*

### Anti-Personnel Mines

Anti-personnel mines are designed to "reroute or push back foot soldiers from a given geographic area," and can kill or injure foot soldiers[131] (in contrast to, for example, naval mines, which are designed to destroy ships).[132] They are typically activated "by direct pressure from above, by pressure put on a wire or filament attached to a pull switch, by a radio signal or other remote firing method, or even simply by the proximity of a person within a predetermined distance."[133] For these reasons, anti-personnel mines do not discriminate among potential targets, as they are not capable of independently tracking different targets and choosing among them.

### Underwater Mines

Naval Mines — General

Naval mines are capable of being detonated by either seismic sensors that sense vibrations in the water as a ship approaches[134] or acoustic sensors

---

that detect sounds generated by passing ships.[135] Some modern mines use a combination of seismic, acoustic, electric, and magnetic sensors to detect nearby ships.[136] Naval mines explode when triggered, without a proximate human directing them to detonate. Naval mines do not discriminate among potential targets; if something triggers its detonation, a naval mine explodes without any independent decision-making process in which it might "choose" whether to detonate.

## MK-60 CAPTOR (United States)

The MK-60 EnCAPsulated TORpedo (CAPTOR), manufactured by Alliant Techsystems, is a sophisticated anti-submarine weapon. It is a deep-water mine that, when triggered, launches a torpedo at hostile targets. It is anchored to the ocean floor and uses a surveillance system known as Reliable Acoustic Path (RAP) sound propagation to track vessels above it.[137] Vessels traveling on or very close to the surface are labeled as ships and are not attacked. Vessels traveling far enough below the surface are labeled as submarines. When it senses a submarine that does not have a "friendly" acoustic signature, the MK-60 launches a torpedo at the target.[138] It therefore has autonomy in its functions in terms of not requiring human authorization to unleash a specific attack. Yet the MK-60 is not capable of "choosing" whether to attack an enemy submarine; if it detects an enemy submarine, it launches the torpedo with no (further) "decision-making" process involved.

# *UNMANNED VEHICLES AND SYSTEMS*

## Unmanned Vehicles — General

### Unmanned Aerial Vehicles

Unmanned Aerial Vehicles (UAVs), also called drones, comprise a broad category and refer to any aircraft without a human pilot onboard. Their functions can span from surveillance and reconnaissance to military attacks. Unmanned Combat Aerial Vehicles (UCAVs) are a subset of UAVs. Different models operate with varying degrees of autonomy across different functions. Traditionally, pilots have operated drones remotely, but drones

---

19, 2014), http://www.popsci.com/blog-network/shipshape/terrible-thing-waits-under-ocean.

135.   Guillermo C. Gaunaurd, *Acoustic Mine*, ACCESS SCIENCE (2014), http://www. accessscience.com/content/006000.

136.   LaGrone, *supra* note 134.

137.   *MK 60 Encapsulated Torpedo (CAPTOR)*, FAS MILITARY ANALYSIS NETWORK (Dec. 13, 1998), http://fas.org/man/dod-101/sys/dumb/mk60.htm.

138.   *Id.*

are becoming increasingly capable of certain autonomous functions. Models such as the nEUROn (which has been referred to as a UCAV; see below) can in key respects fly autonomously,[139] compensating for unexpected events like changing weather patterns, and the X-47B (see below) can even refuel itself in mid-air at its carrier.[140] The technological capability of certain UAVs, once launched, to select and attack targets, without further human intervention, seems to exist, but most drones require human authorization or guidance before deploying lethal force. The Harpy (see below)—a "fire and forget, fully autonomous" so-called "loitering munition"—is one notable exception.[141]

## Unmanned Surface Vehicle

Unmanned Surface Vehicles (USVs) broadly refer to any watercraft that operates on the surface of the water without an onboard crew. They have a wide range of commercial and military functions. The U.S. Navy often uses them for minesweeping, for surveillance and reconnaissance, and to detect submarines.[142] Like UAVs, USVs might operate with various degrees of autonomy across different functions, spanning a range from remote-controlled operation to autonomy in navigation and maneuver.[143]

## Unmanned Maritime Vehicles

Unmanned Maritime Vehicles include both USVs and Autonomous Underwater Vehicles (AUVs). Both USVs and AUVs generally perform similar functions like surveillance and minesweeping.[144] Different models operate with various degrees of autonomy across different functions.[145]

# Unmanned Vehicles and Systems — Specific

## Dominator (United States)

Currently under development by Boeing, the Dominator aims to incorporate a "long-endurance, autonomous UAV for intelligence, surveillance, and reconnaissance missions and potentially for strike capability."[146] According to Boeing, the Dominator will employ "autonomous flight using small-diameter bomb avionics," and can be deployed from a variety of artillery and vehicles, including unmanned aircraft.[147] Boeing will also examine the potential to incorporate "Textron Defense System's Common Smart Submunition (CSS)" to differentiate and deploy against both fixed and moving targets.[148]

---

139.  *See, e.g.,* Ryan Gallagher, *Military Moves Closer to Truly Autonomous Drones,* Slate (Jan. 16, 2013), http://www.slate.com/blogs/future_tense/2013/01/16/taranis_neuron_militaries_moving_closer_to_truly_autonmoous_drones.html.

140.  *X-47B UCAS Makes Aviation History…Again!,* Northrop Grumman, http://www.northropgrumman.com/Capabilities/x47bucas/Pages/default.aspx (last visited Aug. 24, 2016).

141.  *Loitering with Intent,* Jane's Int'l Def. Rev. (Nov. 27, 2015).

142.  *See generally* U.S. Dep't of the Navy, The Navy Unmanned Surface Vehicle Master Plan (2007), http://www.navy.mil/navydata/technology/usvmppr.pdf.

143.  *See, e.g.,* Autonomous Surface Vehicles Ltd., *Unmanned Marine Systems,* Unmanned Systems Technology, http://www.unmannedsystemstechnology.com/company/autonomous-surface-vehicles-ltd (last visited Aug. 24, 2016).

144.  Denise Crimmins & Justin Manley, *What Are AUVs and Why Do We Use Them?,* National Oceanic and Atmospheric Administration (2008), http://oceanexplorer.noaa.gov/explorations/08auvfest/background/auvs/auvs.html.

145.  *Autonomous Underwater Vehicles,* Woods Hole Oceanographic Institution, http://www.whoi.edu/main/auvs (last visited Aug. 24, 2016).

## Guardium (Israel)

The Guardium system, developed by G-NIUS, Israel Aerospace Industries, and Elbit Systems, includes both manned and unmanned ground vehicles (UGVs) and is used by the Israel Defense Forces.[149] According to the chief executive officer for G-NIUS, the latest design of Guardium displayed at a weapons exhibition in 2015 has the capability of serving a variety of purposes, including carrying missiles, loitering munitions, or UAV for reconnaissance missions.[150] The Guardium vehicles have "varying degrees" of autonomy: for instance, the vehicles are capable of responding to various obstacles, "automatically deploy[ing] subsystems," and patrolling Israel's border with Gaza,[151] yet human operators may override or intervene to control the vehicle's functions.[152]

## K-MAX Helicopter (United States)

Lockheed Martin designed the K-MAX helicopter, which is capable of deploying in a variety of environments, including cargo delivery in combat, firefighting, and humanitarian aid.[153] While the K-MAX helicopter has the capability to seat a pilot onboard, it is capable of being operated remotely to allow the system to function in a variety of high-risk environments.[154]

146.  Bill Carey, *Boeing Phantom Develops 'Dominator' UAV*, AIN ONLINE (Nov. 2, 2012), http://www.ainonline.com/aviation-news/defense/2012-11-02/boeing-phantom-works-develops-dominator-uav.

147.  *Id.*

148.  London Huw Williams, *Boeing to Evaluate CSS for Dominator*, JANE'S INT'L DEF. REV., (Oct. 31, 2012).

149.  London Huw Williams, *IAI to Offer Broad UGV Portfolio*, JANE'S INT'L DEF. REV. (July 8, 2016).

150.  Damian Kemp, *AUSA 2015: G-NIUS Displays Loitering Munition-Equipped Guardium Concept*, JANE'S INT'L DEF. REV. (Oct. 13, 2015).

151.  London Huw Williams, *G-NIUS Reveals Its Plans for Guardium Development*, JANE'S INT'L DEF. REV. (June 25, 2008).

152.  *Id.*

153.  *K-MAX*, LOCKHEED MARTIN, http://www.lockheedmartin.com/us/products/kmax.html (last visited Aug. 24, 2016).

154.  *K-MAX Unmanned Aircraft System*, LOCKHEED MARTIN, http://www.lockheedmartin.com/content/dam/lockheed/data/ms2/documents/K-MAX-brochure.pdf (last visited Aug. 24,

## Knifefish (United States)

The Knifefish, designed as an unmanned underwater vehicle (UUV), is used to locate mines,[155] including those buried in so-called "high clutter environments."[156] General Dynamics Mission Systems and Bluefin Robotics have been developing various models to be used by the U.S. Navy, possibly beginning in 2018 or 2019.[157] The Knifefish operates with autonomy in its function to sweep for mines in various underwater environments.[158]

## Lijian (China)

China launched a prototype of Lijian, meaning "sharp sword," on November 20, 2013.[159] Shenyang Aircraft Company and the Hongdu Aircraft Industries Corporation reportedly designed and manufactured the unmanned combat aerial vehicle (UCAV).[160] Other than its similar configuration to the X-47B, little is known about the UCAV or its capabilities.[161] Notably, it did not appear at Airshow China in 2014; however, the China Aerospace Science and Technology Corporation has "insinuated" that the Lijian program is "alive and well."[162] Because little, if any, information about the Lijian's capabilities is publicly known, it remains unclear whether the Lijian employs autonomy in its system. More generally, the release of information about China's air forces indicates that China aims to develop an air force "capable of conducting both offensive and defensive operations," to include "the enhancement of reconnaissance and strategic projection capabilities."[163]

## nEUROn (France, Greece, Italy, Spain, Sweden, Switzerland)

The nEUROn is an unmanned combat air vehicle (UCAV) being developed by

Dassault Aviation and several European nations.[164] The nEUROn is designed to perform reconnaissance and combat missions. The various countries involved in the nEUROn program have been testing its capabilities, assessing, among other things, the "detection, localization, and reconnaissance of ground targets in autonomous modes."[165] Testing of the nEUROn, which is designed as a demonstrator of current technologies, will also evaluate its capability to "drop…Precision Guided Munitions through the internal weapon bay."[166]

## Platform M (Russia)

According to Russian media, Platform M is a "remote-controlled robotic unit" developed by the Progress Scientific Research Technological Institute of Izhevsk.[167] Reportedly, Platform M has the capability to "destroy targets in automatic or semiautomatic control systems."[168] Its "targeting mechanism works automatically without human assistance," according to news reports.[169]

## Pluto Plus (Italy)

The Pluto and Pluto Plus remotely operated vehicles (ROVs), also referred to as unmanned underwater vehicles (UUVs),[170] operate underwater to identify mines using features such as "sonar sensors for navigation, search, obstacle avoidance and identification," as well as the capability to relay information, including video imagery, to the operator.[171] The Italian company Gaymarine developed the Pluto and Pluto Plus models, which are used in conjunction with other mine-countermeasure vehicles (MCMVs) by various navies throughout the world, including Italy, Nigeria, Norway, South Korea, Spain, and Thailand.[172] A pilot operates the Pluto Plus above the water, using a

---

164.   Nicholas de Larrinaga, *France Begins Naval Testing of Neuron UCAV*, Jane's Defence Wkly. (May 19, 2016).

165.   Berenice Baker, *Taranis vs. nEUROn – Europe's Combat Drone Revolution*, Airforce-Technology.com (May 6, 2014), http://www.airforce-technology.com/features/featuretaranis-neuron-europe-combat-drone-revolution-4220502.

166.   David Cenciotti, *First European Experimental Stealth Combat Drone Rolled Out: The Neuron UCAV Almost Ready for Flight*, The Aviationist (Jan. 20, 2012), https://theaviationist.com/2012/01/20/neuron-roll-out.

167.   *Russia's Platform-M Combat Robot on Display in Sevastopol*, RT News (July 22, 2015, 8:20 AM), https://www.rt.com/news/310291-russia-military-robot-sevastopol.

168.   *Id.*

169.   Franz-Stefan Gady, *Meet Russia's New Killer Robot*, The Diplomat (July 21, 2015), http://thediplomat.com/2015/07/meet-russias-new-killer-robot.

170.   Gary Martinic, *Unmanned Maritime Surveillance and Weapons Systems*, Australian Naval Institute (July 8, 2014), http://navalinstitute.com.au/unmanned-maritime-surveillance-and-weapons-systems.

171.   Casandra Newell, *Egypt Orders Pluto Plus ROVs*, Jane's Navy Int'l (June 19, 2009).

172.   *Briefing: Rolling in the Deep*, Jane's Def. Wkly. (March 6, 2011).

"remote control console" to maneuver the vehicle.[173]

## Protector USV (Israel)

Developed and manufactured by Rafael Advanced Defense Systems, the 11m version of the Protector USV contains an "enhanced remotely controlled water can[n]on system for non-lethal and firefighting capabilities."[174] It includes an unmanned boat, a tactical control system, and mission modules.[175] The 11m model includes features that will reportedly enable the USV to engage in "surveillance, reconnaissance, mine warfare, and anti-submarine warfare."[176] The 11m model, as with earlier models of the Protector, employs two operators that work remotely from a dual-console station, controlling both the boat and the payload.[177]

## Sea Hunter (United States)

In 2016, the Defense Advanced Research Projects Agency (DARPA), a U.S. government agency, designed a prototype of an autonomous surface vessel named Sea Hunter, which was manufactured by Leidos.[178] According to DARPA, the vessel can "robustly track quiet diesel electric submarines,"[179] with the ability to travel up to several months and for considerable distances; developers anticipate that it has the capability to perform other functions as well.[180] Sea Hunter is capable of autonomy in certain functions in two ways. First, it is capable of navigating and maneuvering independently without colliding with other ships.[181] Second, it is capable of locating and tracking diesel electric submarines, which can be extremely quiet and difficult to detect, within a range of two miles.[182] A human can take control of the vessel

---

173. *Columbia Group to Supply Pluto Plus UUVs to Egyptian Navy*, Def. Industry Daily (June 21, 2009), http://www.defenseindustrydaily.com/Columbia-Group-to-Supply-Pluto-Plus-UUVs-to-Egyptian-Navy-05530/.

174. *Protector Unmanned Surface Vehicle (USV), Israel*, Naval-Technology.com, http://www.naval-technology.com/projects/protector-unmanned-surface-vehicle/ (last visited Aug. 24, 2016).

175. London Huw Williams, *Rafael Looks to Extend Protector USV Control Range*, Jane's Int'l Def. Rev. (Aug. 8, 2013).

176. *Id.*

177. Richard Scott, *New Protector USV Variant Detailed*, Jane's Int'l Def. Rev., Nov. 12, 2012.

178. Rachel Courtland, *DARPA's Self-Driving Submarine Hunter Steers Like a Human*, IEEE Spectrum (Apr. 7, 2016), http://spectrum.ieee.org/automaton/robotics/military-robots/darpa-actuv-self-driving-submarine-hunter-steers-like-a-human.

179. Scott Littlefield, *Anti-Submarine Warfare (ASW) Continuous Trail Unmanned Vessel (ACTUV)*, Defense Advanced Research Projects Agency, http://www.darpa.mil/program/anti-submarine-warfare-continuous-trail-unmanned-vessel (last visited Aug. 24, 2016).

180. Courtland, *supra* note 178.

181. Littlefield, *supra* note 179.

182. Rick Stella, *Ghost Ship: Stepping aboard Sea Hunter, the Navy's Unmanned Drone Ship*,

if necessary, but it is designed to perform its functions without any proximate human direction.[183]

## Skat (Russia)

In 2013, the developer MiG reportedly signed an agreement to develop an unmanned combat air vehicle (UCAV) called Skat.[184] According to a Russian news agency, Skat would "carry out strike missions on stationary targets, especially air defense systems in high-threat areas, as well as mobile land and sea targets."[185] Also according to a Russian news agency, Skat would "navigate in autonomous modes."[186] More recent reports, however, note it is "unclear" whether Russia has continued to develop this kind of technology, stating that Russia cancelled plans to develop Skat.[187]

## Taranis (United Kingdom)

Taranis is an unmanned aerial combat stealth drone being developed by the British company BAE Systems to demonstrate current technologies.[188] It is capable of performing surveillance and reconnaissance, and also serving in combat missions. According to BAE Systems, the company is attempting to determine whether the Taranis can "strike targets 'with real precision at long range, even in another continent.'"[189] Taranis is theoretically capable of flying autonomously (although during test flights, it has always been controlled remotely by a human operator).[190] A remote human operator must give authorization before Taranis is capable of attacking any target, although the drone identifies potential targets and, once an attack has been authorized, it aims at those targets.[191]

## X-47B (United States)

The X-47B is an unmanned aerial combat stealth drone that was developed by the United States, built by Northrop Grumman, and designed as a "test

DIGITAL TRENDS (Apr. 11, 2016), http://www.digitaltrends.com/cool-tech/darpa-officially-christens-the-actuv-in-portland.

183.  *Id.*

184.  John Reed, *Meet Skat, Russia's Stealthy Drone*, FOREIGN POLICY, June 3, 2013, http://foreignpolicy.com/2013/06/03/meet-skat-russias-stealthy-drone.

185.  *Id.*

186.  *Id.*

187.  Andrew White, *Unmanned Ambitions: European UAV Developments*, JANE'S DEF. WKLY., (Oct. 27, 2015).

188.  Guia Marie Del Prado, *This Drone Is One of the Most Secretive Weapons in the World*, TECH INSIDER (Sep. 29, 2015), http://www.techinsider.io/british-taranis-drone-first-autonomous-weapon-2015-9.

189.  Gallagher, *supra* note 139.

190.  *Id.*

191.  *Id.*

and development vehicle for advancing control technologies and systems necessary for operating [UAVs] in and around aircraft carriers."[192] According to the U.S. Navy, it developed the X-47B as a "demonstrator" to showcase current capabilities; although the X-47B has not been armed, it is capable of carrying two 2,000-pound bombs.[193] While the X-47B reportedly has autonomy in certain functions,[194] an operator can take control of the X-47B via a Control Display Unit.[195] The X-47B pioneered several autonomous flight maneuvers, including the "first autonomous landing on an aircraft carrier and the first mid-air refueling by a [UAV]."[196] In principle, human authorization is required before the X-47B could be used to intentionally deploy deadly force, but the precise way in which the human operator fits into this equation is not publicly reported.[197]

## MISSILE SYSTEMS

### Missile Systems — General

#### "Fire and Forget" Missile Systems

"Fire and forget" missiles are capable, once launched, of reaching their target with no further human assistance. With older missile systems, the operator who fired the missile had to help guide the missile towards its target by, for example, continuing to track the target and transmitting "corrective commands" to the missile.[198] Newer "fire and forget" missiles, such as the FMG-148 Javelin (discussed below), are capable, once fired, of independently tracking their targets without outside guidance or control.[199] They are also capable of navigating certain difficult terrain on their own, and some, like the Brimstone and Brimstone 2 (discussed below), are capable of locating their target even when it was not initially in the line of sight of the launch location.

---

## Missile Systems — Specific

### Brimstone and Brimstone 2 (United Kingdom)

Brimstone is an anti-armor, "fire and forget" missile first used in 2005, and developed initially by GEC-Marconi Radar and Defense Systems (later MBDA UK).[200] The Royal Air Force (RAF) began using the Brimstone in Iraq and Afghanistan during 2008 and 2009.[201] Brimstone 2, which entered service in 2016,[202] incorporates a number of improvements from the initial Brimstone model.[203] Brimstone included "embedded algorithms" and could strike both land and naval targets.[204] Brimstone 2 introduced "an improved set of targeting algorithms," as well as "autopilot and seeker enhancements."[205] It is a "fire and forget" missile that is capable of autonomy in navigating terrain as it travels toward its target and in certain respects of independently locating a particular target by discriminating among potential candidates.[206] Once launched, Brimstone is capable of "sweeping" a large target area, searching for a specific type of target, the details of which can be pre-programmed into each individual missile prior to launch. For example, a Brimstone missile is capable of being programmed to target only an armored vehicle, ignoring other objects.[207]

### FMG-148 Javelin (United States)

The Javelin is a "fire and forget" anti-tank missile developed by the United States with a range of 2,500 meters.[208] Multiple countries have purchased the Javelin, including Australia, Bahrain, the Czech Republic, France, Ireland, Jordan, Lithuania, New Zealand, Norway, Oman, the United Arab Emirates, the United Kingdom, and the United States.[209] The United States has also recently approved sales of the missile to other countries, including Qatar.[210] Both Raytheon and Lockheed Martin manufacture the Javelin.[211] Two human

---

200.  London Hughes, *Reign of Fire: UK RAF Readies for Brimstone 2*, Jane's Int'l Def. Rev. (Sept. 4, 2014).

201.  *Id.*

202.  Nicholas de Larrinaga, *Farnborough 2016: Brimstone 2 Enters Service, Begins Apache Trials*, Jane's Def. Wkly (July 14, 2016).

203.  Hughes, *supra* note 200.

204.  *Id.*

205.  *Id.*

206.  *Brimstone Advanced Anti-Armour Missile*, Army Technology.com, http://www.army-technology.com/projects/brimstone (last visited Aug. 24, 2016).

207.  *Id.*

208.  *Raytheon*, *supra* note 199.

209.  *Id.*

210.  Jeremy Binnie, *U.S. Clears More Javelins for Qatar*, Jane's Def. Wkly (May 27, 2016).

211.  *Raytheon*, *supra* note 199.

operators carry and launch the Javelin.[212] A human operator must select the Javelin's target; however, the missile guides itself to the target, allowing the human operators to leave the launch site before the missile strikes. Operators are capable of identifying targets "either directly [in] line-of-sight or with help from the missile's guidance capability."[213]

## Harpy (Israel)

Developed by Israel Aerospace Industries and used principally by China, India, South Korea, Turkey, and Israel, the Harpy is a "transportable, canister-launched, fire-and-forget, fully autonomous" system,[214] which is also called a "loitering munition."[215] Harop, a variant of the Harpy developed in 2009, has the capability to "engage time-critical, high-value, relocatable targets," and is also capable of being launched from both land and naval-based canisters.[216]

## Joint Strike Missile (Norway)

The recently-developed Joint Strike Missile builds on the technology of the Naval Strike Missile.[217] Norway has funded the development of the missile, which is manufactured by Kongsberg.[218] It is designed to be integrated into the F-35 Joint Strike Fighters and to attack both naval and land targets.[219] In 2015, the Joint Strike Missile was deployed successfully in a test run, and further testing and developments are scheduled through 2017.[220] The Joint Strike Missile is not capable of choosing an initial target. It is also incapable of locating a hidden target; however, it does include a Global Positioning System/Inertial Navigation System to help it autonomously navigate close to terrain towards a preselected target. It is also programmed to automatically fly in unpredictable patterns to make it harder to intercept.[221]

---

212.  *Id.*

213.  *Id.*

214.  *Loitering with Intent*, *supra* note 141.

215.  *Id.*

216.  *Id.*

217.  Richard Scott, *Joint Strike Missile Starts Flight Test Programme*, Jane's Missiles & Rockets (Nov. 16, 2015), http://www.janes.com/article/55989/joint-strike-missile-starts-flight-test-programme.

218.  *Id.*

219.  *Kongsberg's NSM/JSM Anti-Ship & Strike Missile Attempts to Fit in Small F-35 Stealth Bay*, Defense Industry Daily (Nov. 12, 2015), http://www.defenseindustrydaily.com/norwegian-contract-launches-nsm-missile-03417 [hereinafter *Kongsberg*].

220.  Franz-Stefan Gady, *F-35's Joint Strike Missile Successfully Completes Flight Test in US*, The Diplomat (Nov. 13, 2015), http://thediplomat.com/2015/11/f-35s-joint-strike-missile-successfully-completes-flight-test-in-us.

221.  *Kongsberg*, *supra* note 219.

# STATIONARY SYSTEMS, INCLUDING CLOSE-IN WEAPON SYSTEMS

## Aegis Combat System (United States)

The Aegis Combat System, manufactured by Lockheed Martin,[222] is a weapons control system capable of identifying, tracking, and attacking hostile targets.[223] Several countries use the system, including Australia, Japan, Norway, South Korea, Spain, and the United States.[224] Aegis has many more capabilities than a standalone Phalanx CIWS (see below). Like the Phalanx, Aegis relies on radar to identify possibly hostile targets.[225] Unlike the Phalanx, Aegis is capable of engaging over 100 targets simultaneously.[226] The Aegis Combat System is capable of being operated autonomously[227] in terms of the computer interface tracking various targets, determining their threat levels, and, in certain respects, independently determining whether to attack them.

## AK-630 CIWS (Russia)

The AK-630 Close-In Weapons System (CIWS) gun turret is "designed to engage manned and unmanned aerial targets, small-size surface targets, soft-skinned coastal targets, and floating mines."[228] Multiple countries have used the AK-630, including Bulgaria, Croatia, Greece, Lithuania, Poland, Romania, and Ukraine.[229]

## Centurion (United States)

The Centurion Weapons System, manufactured by Raytheon, uses a "radar-guided gun" against "incoming rocket and mortar fire."[230] The Centurion has been described as a "land-based version" of the Phalanx CIWS (see below).[231] In addition to the United States, the United Kingdom also uses the Centurion.

---

222. *Aegis Combat System*, Lockheed Martin, http://www.lockheedmartin.com/us/products/aegis.html (last visited Aug. 24, 2016).

223. *Aegis Weapon System*, America's Navy: United States Navy Fact File (Jan. 5, 2016), http://www.navy.mil/navydata/fact_display.asp?cid=2100&tid=200&ct=2.

224. Paul Scharre & Michael C. Horowitz, An Introduction to Autonomy in Weapons Systems 21 (Feb. 2015) (working paper), http://www.cnas.org/sites/default/files/publications-pdf/Ethical%20Autonomy%20Working%20Paper_021015_v02.pdf.

225. *Id.*

226. *Id.*

227. Scharre & Horowitz, *supra* note 224, at 21.

228. *Thales Targets AK-630 Users for Fire Control*, Jane's Navy Int'l (Apr. 13, 2005).

229. *Id.*

230. Nathan Hodge, *Raytheon Ramps Up Centurion Production*, Jane's Def. Wkly. (March 20, 2008).

231. *Id.*

The Centurion uses the same capabilities as the Phalanx CIWS, including automatically tracking and destroying incoming fire.[232]

## Counter Rocket, Artillery, and Mortar (C-RAM) (United States)

C-RAM, manufactured by Northrop Grumman and Raytheon, is a missile defense system designed to intercept hostile projectiles before they reach their intended targets. Its central component is a revised version of the U.S. Navy's Phalanx CIWS (see below), as well as existing radar systems, adapted for on-land use.[233] Australia and the United Kingdom have purchased the system from the United States.[234] C-RAM reportedly has autonomy in its operations in terms of "intercept[ing] incoming munitions at speeds too quick for a human to react."[235]

## GDF (Switzerland)

The Oerlikon GDF is an anti-aircraft cannon initially developed in the late 1950s and currently used by over 30 countries.[236] Once activated, the GDF-005 model is capable, without further human intervention, of operating using radar to identify targets, attacking them, and reloading.[237]

## Goalkeeper CIWS (The Netherlands)

The Goalkeeper CIWS, manufactured by the Thales Group, includes a gun with "missile-piercing ammunition" that enables the system to "destroy missile warheads."[238] The navies of Belgium, Chile, the Netherlands, Portugal, Qatar, South Korea, the United Arab Emirates, and the United Kingdom use the system.[239] According to information provided by Thales, the Goalkeeper system "automatically performs the entire process from surveillance and detection to destruction, including selection of the next priority target."[240]

---

232. *Centurion C-RAM Counter-Rocket, Artillery, and Mortar Weapon System*, Army Recognition, http://www.armyrecognition.com/united_states_us_army_artillery_vehicles_system_uk/centurion_c-ram_land-based_weapon_system_phalanx_technical_data_sheet_specifications_pictures_video.html (last visited Aug. 24, 2016).

233. *Counter Rocket, Artillery and Mortar (C-RAM)*, GlobalSecurity.org (July 7, 2011), http://www.globalsecurity.org/military/systems/ground/cram.htm.

234. Kristin Horitski, *Counter-Rocket, Artillery, Mortar (C-RAM)*, Missile Defense Advocacy Alliance (March 2016), http://missiledefenseadvocacy.org/missile-defense-systems-2/missile-defense-systems/u-s-deployed-intercept-systems/counter-rocket-artillery-mortar-c-ram.

235. Heather M. Roff, *Killer Robots on the Battlefield*, Slate (April 7, 2016), http://www.slate.com/articles/technology/future_tense/2016/04/the_danger_of_using_an_attrition_strategy_with_autonomous_weapons.html.

236. *GDF*, Weapons Systems.net, http://weaponsystems.net/weaponsystem/EE02%20-%20GDF.html (last visited Aug. 24, 2016).

237. Noah Shachtman, *Robot Cannon Kills 9, Wounds 14*, Wired (Oct. 18, 2007), https://www.wired.com/2007/10/robot-cannon-ki.

238. *Goalkeeper – Close-In Weapon System*, Thales, https://www.thalesgroup.com/en/goalkeeper-close-weapon-system# (last visited Aug. 24, 2016).

239. *Id.*

## Iron Dome (Israel)

The Iron Dome is manufactured by Raytheon and seeks to "detect, assess, and intercept incoming rockets, artillery, and mortars."[241] The Iron Dome has autonomy in some of its functions. It locates potential targets using radar and calculates their expected trajectory. If a rocket would hit a populated area, the Iron Dome is capable of launching a Tamir interceptor missile at the rocket. A human operator must authorize the launch, and she must often make the decision very quickly, sometimes in a matter of minutes.[242] Once a launch is authorized, the computer system will independently aim the Tamir and determine when to launch it. Once close enough to the hostile rocket, the Tamir explodes, destroying both projectiles. The computer algorithm, not the human operator, determines when to detonate the Tamir.

## Kashtan CIWS (Russia)

Manufactured by KBP Instrument Design Bureau and used by China, India, and Russia,[243] the Kashtan Close-In Weapon System (CIWS) "can engage up to six targets simultaneously," and includes gun and missile armaments.[244] The Kashtan system has been described as a human-supervised system with certain autonomous functions.[245]

## MANTIS (Germany)

The Modular, Automatic, and Network-Capable Targeting and Interception System, or MANTIS, manufactured by Rheinmetall and used by German forces, is capable of quickly acquiring a target and firing 1,000 rounds a minute.[246] An operator must first activate the MANTIS, but, once activated, "the system is fully automated, although a man in the loop allows for engagement to be

---

240.  *Id.*

241.  *Iron Dome Weapon System*, Raytheon, http://www.raytheon.com/capabilities/products/irondome (last visited Aug. 24, 2016).

242.  Raoul Heinrichs, *How Israel's Iron Dome Anti-Missile System Works*, Business Insider, July 30, 2014, http://www.businessinsider.com/how-israels-iron-dome-anti-missile-system-works-2014-7.

243.  Scharre & Horowitz, *supra* note 224, at 21.

244.  *India – Kashtan Self-Defence System for Retrofit*, Jane's Int'l Def. Rev. (May 1, 2001).

245.  Scharre & Horowitz, *supra* note 224, at 21.

246.  Nicholas Fiorenza, *Luftwaffe Receives MANTIS C-RAM System*, Jane's Def. Wkly. (Nov. 28, 2012).

overruled if needed."[247]

## MK 15 Phalanx CIWS (United States)

Manufactured by Raytheon[248] and used by at least 25 countries,[249] MK 15 Phalanx Close-In Weapons System (CIWS) is a "fast-reaction, detect-through-engage, radar guided, 20-millimeter gun weapon system" used to explode anti-ship missiles (ASMs) and other approaching threats, such as aircraft and unmanned aerial systems (UASs).[250] The Phalanx CIWS can be operated manually or in an autonomous mode.[251] The Phalanx CIWS uses radar to track nearby projectiles, and it is capable of independently determining whether they pose a threat based on their speed and direction.[252] When it is programmed to operate autonomously, the Phalanx CIWS automatically fires at incoming missiles without further human direction.[253]

## MK-60 Griffin Missile System (United States)

Used by the U.S. Navy and manufactured by Raytheon, the MK-60 Griffin Missile System enables ships to defend themselves against "small boat threats" by employing a "surface-to-surface missile system."[254] The MK-60 Griffin Missile System includes at least two variants: Griffin A, an unmanned aircraft system (UAS), and Griffin B, an unmanned aerial vehicle (UAV).[255] The Griffin B model uses GPS guidance to help identify a target, while the human operator is capable of controlling the type of detonation, as well as of changing the target location after the missile has been launched.[256]

## Patriot Missile (United States)

The Patriot System, manufactured by Raytheon, is a surface-to-air missile defense system that uses radar to detect and identify hostile incoming missiles and fires missiles to intercept them.[257] Multiple countries use the Patriot system, including Egypt, Germany, Greece, Israel, Japan, Kuwait, the Netherlands, Saudi Arabia, South Korea, Spain, the United Arab Emirates, and the United States.[258] The Patriot's radar system is responsible for automatically detecting and tracing incoming projectiles. When operating semi-autonomously, the Patriot computer system requires a human operator to authorize a launch.[259] When operating in a mode of heightened autonomy, the Patriot computer itself chooses whether or not to launch, based upon the speed and direction of the approaching projectile.[260]

## SeaRAM (United States)

The SeaRAM anti-ship missile defense system, used by the U.S. Navy, combines features of the Phalanx and rolling airframe missile (RAM) guided weapons systems.[261] According to the manufacturer Raytheon, the SeaRAM can "identify and destroy approaching supersonic and subsonic threats, such as cruise missiles, drones, small boats, and helicopters."[262] The RAM "fire and forget" missile contains some autonomy in its features, including a "dual-mode passive radio frequency system."[263]

## Sentry Robot (Russia)

In 2014, the Russian Strategic Missile Forces announced that they were planning to release armed sentry robots that could exhibit autonomy in identifying and attacking targets.[264] Little else is publicly known about the specific features of these machines because the prototypes have not yet been released. Uralvagonzavod, a Russian defense firm, anticipates that it will be able to demonstrate prototypes by 2017.[265] In December 2015, U.S. Defense

---

257.  Andreas Parsch, *Raytheon MIM-104 Patriot*, Directory of Military US Rockets and Missiles (Dec. 3, 2002), http://www.designation-systems.net/dusrm/m-104.html.

258.  Scharre & Horowitz, *supra* note 224, at 21–22.

259.  *Patriot Missiles (PAC-1, PAC-2, PAC-3)*, Missile Threat (Dec. 22, 2013), http://missilethreat.com/defense-systems/patriot-pac-1-pac-2-pac-3.

260.  Marshall Brain, *How Patriot Missiles Work*, How Stuff Works (March 28, 2003), http://science.howstuffworks.com/patriot-missile.htm.

261.  *SeaRAM Anti-Ship Missile Defense System*, Raytheon, http://www.raytheon.com/capabilities/products/searam (last visited Aug. 24, 2016).

262.  *Id.*

263.  *SeaRAM Anti-Ship Missile Defence System, United States of America*, Naval-Technology. com, http://www.naval-technology.com/projects/searam-anti-ship-missile-defence-system (last visited Aug. 24, 2016).

264.  Patrick Tucker, *The Pentagon Is Nervous about Russian and Chinese Killer Robots*, Defense One (Dec. 14, 2015), http://www.defenseone.com/threats/2015/12/pentagon-nervous-about-russian-and-chinese-killer-robots/124465.

265.  *Producer of Russia's Armata T-14 Plans to Create Army of AI Robots*, RT International

Department officials expressed alarm at the development of the "highly capable autonomous combat robots" that would be "capable of independently carrying out military operations."[266]

## Sentry Tech (Israel)

Manufactured by Rafael Advanced Defense Systems, the Sentry Tech system "consists of a lineup of remote-controlled weapon stations integrated with security and intelligence sensors…providing an infiltration alert via ground and airborne sensors" to provide operators with information on whether to fire weapons.[267] The system is mainly used by Israel along the Gaza border.[268] Sentry Tech does not operate with autonomy in its features; rather, it is a remote-controlled weapon station. Once a potential target has been identified, an operator remotely controls the Sentry Tech to track the target and is capable of choosing to attack the target with the Sentry's machine gun turret.[269]

## SGR A1 Sentry Gun (South Korea)

The SGR A1 is a stationary robot that operates a machine-gun turret, originally designed by the Korea University and the Samsung Techwin Company. The robot guards the Demilitarized Zone (DMZ) between North and South Korea. It uses an infrared camera surveillance system to identify potential intruders. When an individual comes within ten meters of the robot, the SGR A1 demands the necessary access code and uses voice recognition to determine whether the intruder has provided the correct code. If the intruder fails to do so, the SGR A1 has three options: ring an alarm bell, fire rubber bullets, or fire its turreted machine gun.[270] The SGR A1 normally operates with remote human authorization required to enable the SGR A1 to fire.[271] Central to this decision is whether the target has appeared to "surrender." The

---

robot is programmed to recognize that a human with its arms held high in the air is attempting to surrender.[272]

## Super aEgis II (South Korea)

The Super aEgis II is a robot sentry with certain automated features manufactured by DoDAAM. It incorporates a machine gun turret, which is used primarily by South Korea in the Demilitarized Zone (DMZ).[273] It uses a combination of digital cameras and thermal imaging to identify potential targets, allowing it to operate in the dark.[274] The Super aEgis II requires a human to authorize any use of lethal force. Before firing, it automatically emits a warning, advising potential targets to "turn back or we will shoot" (in Korean).[275] If the target continues to advance, a remote human operator enters a password to enable the aEgis to shoot the target.[276]

# CYBER CAPABILITIES

## Stuxnet (United States and Israel)

Reportedly, Stuxnet is a cyberweapon that was used to attack Iran's nuclear-enrichment operations in 2009 and 2010. The specifics of the malware are uncertain, but it was reportedly developed by the United States and Israel in a mission codenamed "Olympic Games."[277] Allegedly, Stuxnet caused computers in Natanz (Iran's nuclear enrichment facility) to malfunction, reprogramming the centrifuges to spin too fast and damaging delicate pieces of the machinery.[278] It is believed to have damaged 1,000 of Iran's 6,000 centrifuges in 2010.[279] Since it was intended to operate in the Iranian nuclear enrichment facility, a computer system that is "air-gapped" (disconnected from the internet and other computer networks), Stuxnet was designed to operate, once launched, without (further) external human direction or input.[280] Stuxnet's code was written to ensure that once connected to the nuclear facility's computer network, it would begin sabotaging the centrifuge software immediately and to continue doing so without further outside guidance.

---

272.   Sentry Guard, *supra* note 270.

273.   Simon Parkin, *Killer Robots: The Soldiers That Never Sleep*, BBC Future (July 16, 2015), http://www.bbc.com/future/story/20150715-killer-robots-the-soldiers-that-never-sleep. Other countries, however, such as the United Arab Emirates and Qatar, also use the system.

274.   *Id.*

275.   *Id.*

276.   *Id.*

277.   Ellen Nakashima & Joby Warrick, *Stuxnet Was Work of U.S. and Israeli Experts, Officials Say*, Wash. Post (June 2, 2012), https://www.washingtonpost.com/world/national-security/stuxnet-was-work-of-us-and-israeli-experts-officials-say/2012/06/01/gJQAlnEy6U_story.html.

278.   David E. Sanger, *Obama Order Sped Up Wave of Cyberattacks Against Iran*, N.Y. Times (May 31, 2012), http://www.nytimes.com/2012/06/01/world/middleeast/obama-ordered-wave-of-cyberattacks-against-iran.html.

279.   Kim Zetter, *An Unprecedented Look at Stuxnet, the World's First Digital Weapon*, Wired (Nov. 3, 2013), https://www.wired.com/2014/11/countdown-to-zero-day-stuxnet.

---

280.   *See* Dorothy E. Denning, *Stuxnet: What Has Changed?*, 4 Future Internet 672, 674 (2012).

# 3

# INTERNATIONAL LAW PERTAINING TO ARMED CONFLICT

In this section, we outline key fields, concepts, and rules relating to international law pertaining to armed conflict. We do so to identify some of the fundamental substantive norms that may be relevant to war algorithms in general and to our three-part accountability approach in particular.[281] *State responsibility* entails, among other things, identifying the content of the underlying obligation. *Individual responsibility* entails, among other things, identifying the elements of the crime and the mode of responsibility under international law. Finally, *scrutiny governance* entails detecting— and potentially surpassing—a baseline of relevant normative regimes, and international law may provide a foundational normative framework concerning regulation of war algorithms.

This section is divided into two parts. We first set the stage with an introduction of state responsibility. Then, in the bulk of the section, we highlight relevant considerations in the substantive law of obligations. Part of the focus is on AWS, since that has been the main framing states have addressed to date. We examine whether a customary international law norm pertaining to AWS in particular has crystallized. We find that one has not, at least not yet. So we then outline some of the main international law rules of a more general nature. We focus here primarily on rules that may relate to AWS, but we also note a number of rules that may (otherwise or also) implicate war algorithms.

---

281.  *See infra* Section 4.

With respect to AWS, most commentators and states focus primarily on international humanitarian law and international criminal law. In this section, we raise concerns not only in those fields but also in some of the other regimes of international law that might apply with respect to war algorithms. The section, however, is not meant to be exhaustive.[282] We note that some states—including Switzerland, the United States, and the United Kingdom—have articulated much more detailed analyses of how AWS might relate to a particular rule or field of international law; in light of our interest in discerning state practice, we focus, in part, on those states' positions and practices.

# STATE RESPONSIBILITY

State responsibility underpins international law. To grasp the broader accountability architecture governing the design, development, or use (or a combination thereof) of war algorithms, therefore, it is necessary to have at least a basic understanding of the conceptual framework of state responsibility.

## *UNDERLYING CONCEPTS*

The underlying concepts of state responsibility, which are general in character, are attribution, breach, excuses, and consequences.[283] Attribution concerns the circumstances under which an act may be attributed to a state.[284] Breach concerns the conditions under which an act (or omission) may qualify as an internationally wrongful act.[285] Excuses concern the general defenses that may be available to a state in relation to an internationally wrongful act.[286] And consequences concern the forms of liability that may arise in relation to an internationally wrongful act. As James Crawford explains, "[i]ndividual treaties or rules may vary these underlying concepts in some respect; otherwise they are assumed and apply unless excluded."[287]

---

282. One field of international law that we do not address but that might merit attention is international trade law, perhaps especially to the extent that it is used as a framework for developing technology-related standards and procedures at the national and international levels.

283. *See* James R. Crawford, *State Responsibility*, *in* Max Planck Encyclopedia of Public International Law ¶ 3 (2006).

284. *See* Draft Articles on Responsibility of States for Internationally Wrongful Acts with Commentaries arts. 4–11, Report of the International Law Commission, 53d Sess., Apr. 23-June 1, July 2-Aug. 10, 2001, U.N. Doc. A/56/10, U.N. GAOR 56th Sess., Supp. No. 10 (2001), http://legal.un.org/ilc/texts/instruments/english/commentaries/9_6_2001.pdf [hereinafter Draft Articles].

285. *Id*. at arts. 12–15.

Conduct may be attributed to a state under a variety of circumstances. These circumstances include the conduct of any state organ, such as the armed forces.[288] They also include the conduct of a person or entity empowered by the law of the state to exercise elements of governmental authority (so long as the person or entity is acting in that capacity in a particular instance),[289] and the conduct of an organ placed at the disposal of a state by another state so long as that "organ is acting in the exercise of elements of the governmental authority of the State at whose disposal it is placed."[290] The conduct of these organs, persons, and entities where acting in those capacities shall be considered an act of the state under international law even if that conduct exceeds its authority or contravenes instructions.[291] Furthermore, "[t]he conduct of a person or group of persons shall be considered an act of a State under international law if the person or group of persons is in fact acting on the instructions of, or under the direction or control of, that State in carrying out the conduct."[292] And "[t]he conduct of a person or group of persons shall be considered an act of a State under international law if the person or group of persons is in fact exercising elements of the governmental authority in the absence or default of the official authorities and in circumstances such as to call for the exercise of those elements of authority."[293] Also, "[t]he conduct of an insurrectional movement which becomes the new Government of a State shall be considered an act of that State under international law."[294] And, finally, "[c]onduct which is not attributable to a State under the preceding [circumstances] shall nevertheless be considered an act of that State under international law if and to the extent that the State acknowledges and adopts the conduct in question as its own."[295]

---

286. *Id*. at arts. 20–25.

287. Crawford, *supra* note 283, at ¶ 3.

288. Draft Articles, *supra* note 284, at art. 4(1).

289. *Id*. at art. 5.

290. *Id*. at art. 6.

291. *Id*. at art. 7.

292. *Id*. at art. 8.

293. *Id*. at art. 9.

294. *Id*. at art. 10(1); *see also id*. at art. 10(2)–(3).

295. *Id*. at art. 11.

In general, a consequence of state responsibility is the liability to make reparation.[296] As noted by Pietro Sullo and Julian Wyatt, "[t]he principle that States have to provide reparations to other States to redress wrongful acts they have committed is undisputed under international law and is confirmed by other instruments of international law."[297] Those authors explain that "[t]he primary function of reparations in international law is the re-establishment of the situation that would have existed if an internationally wrongful act had not been committed and the forms that such reparation may take are various."[298]

# SUBSTANTIVE LAW OF OBLIGATIONS

While state responsibility provides the basic framework, the substantive law of obligations fleshes out the relevant rules and procedures. The substantive law of obligations may be found in a relevant branch or branches of public international law. The operation of a specific branch may have implications for particular forms of attribution, breach, excuses, and consequences. IHL, for instance, contains specific provisions on what may constitute a "serious violation" and what consequences may arise with respect to certain rule breaches.

The two sources of the substantive law of obligations most relevant to war algorithms are treaties and customary international law. Treaties are often defined as international agreements between two or more states.[299] And customary international law is often defined as being made up of the "rules of international law that derive from and reflect a general practice accepted as law."[300] Below, we first explore whether there is a specific customary rule

296.  *See* Rosalyn Higgins, Problems and Process: International Law and How We Use It 162 (1995).

297.  Pietro Sullo & Julian Wyatt, *War Reparations, in* Max Planck Encyclopedia of Public International Law ¶ 5 (2015) (citing to the 2001 International Law Commission Draft Articles on Responsibility of States for Internationally Wrongful Acts (art. 31 and arts. 34–37)).

298.  Sullo & Wyatt, *supra* note 297, at ¶ 5.

299.   *See, e.g.*, Vienna Convention on the Law of Treaties art. 2(1)(a), May 23, 1969, 1155 U.N.T.S. 133; Restatement (Third) of the Foreign Relations Law of the United States § 301(1) (1987).

300.   Michael Wood (Special Rapporteur), Int'l Law Comm'n, Second Report on Identification of Customary International Law, at 20, U.N. Doc. A/CN.4/672 (2014), http://daccess-ods.un.org/access.nsf/Get?Open&DS=A/CN.4/672&Lang=E [hereinafter Wood, Second Report]. Though the International Law Commission (ILC) Drafting Committee ultimately did not include this definition in its subsequent report, this exclusion was related to concerns about redundancy, not objections to its content. *See* Gilberto Saboia (Chairman of the Drafting Committee), Int'l pertaining to AWS in particular. (We focus on AWS here and not on war algorithms more broadly because, to date, the bulk of the state practice pertains to AWS.) Answering in the negative, we then highlight treaty provisions (and corresponding customary rules) of a more general character that may relate to AWS and war algorithms. These provisions stretch across an array of fields of international law—not only IHL and international criminal law, but also space law, telecommunications law, and others.

## CUSTOMARY INTERNATIONAL LAW CONCERNING AWS[301]

Customary international law has two constituent elements: state practice and *opinio juris sive necessitates* (shorthand: *opinio juris*).[302] State practice has recently been formulated as the "conduct of the State, whether in the exercise of executive, legislative, judicial or any other functions of the State."[303] And *opinio juris* has recently been formulated as "the belief that [a practice] is obligatory under a rule of law."[304] In other words, a state following a particular practice merely as a matter of policy or out of habit, not out of a sense of legal obligation, does not qualify as *opinio juris*.[305]

It seems fair to say that statements made by official state representatives at the 2015 and 2016 Convention on Certain Conventional Weapons (CCW) Informal Meetings of Experts on Lethal Autonomous Weapons Systems could qualify as state practice or *opinio juris*. (Though those statements probably should not be counted as both.) Such gatherings are "informal implementation mechanism[s],"[306] not formal gatherings of state parties. But these meetings

---

Law Comm'n, Identification of Customary International Law, at 4 (2014), http://legal.un.org/ilc/sessions/66/pdfs/english/dc_chairman_statement_identification_of_custom.pdf.

301.   Katie King and Joshua Kestin provided extensive research assistance for this section.

302.   *See, e.g.*, Int'l Law Comm'n, Identification of Customary International Law: Text of the Draft Conclusions Provisionally Adopted by the Drafting Committee, draft conclusion 2, U.N. Doc. A/CN.4/L.869 (2015), https://documents-dds-ny.un.org/doc/UNDOC/LTD/G15/156/93/PDF/G1515693.pdf?OpenElement; Wood, Second Report, *supra* note 300, at 9, 21–27.

303.   Int'l Law Comm'n, *supra* note 302, at draft conclusion 5.

304.   Wood, Second Report, *supra* note 302, at 24 (quoting the explanation of various states). *See also* Michael Wood (Special Rapporteur), Int'l Law Comm'n, Third Report on Identification of Customary International Law, at 13, U.N. Doc. A/CN.4/682 (2015), https://documents-dds-ny.un.org/doc/UNDOC/GEN/N15/088/91/PDF/N1508891.pdf?OpenElement      [hereinafter Wood, Third Report]; Int'l Law Comm'n, *supra* note 302, at draft conclusion 9 ("The requirement, as a constituent element of customary international law, that the general practice be accepted as law (*opinio juris*) means that the practice in question must be undertaken with a sense of legal right or obligation").

305.   *See, e.g.*, *id.*

306.   U.N. Office at Geneva, *2010 Meeting of Experts*, Disarmament, http://www.unog.ch/80256EE600585943/(httpPages)/701141247B6C85E7C12576F200587847?OpenDocument

nevertheless involved the sort of public pronouncements that, when conducted by state agents, are capable of comprising evidence of the elements of customary international law. In at least some cases, states' presentations at meetings of experts have been considered as state practice for the purposes of assessing customary international law.[307] Whether a particular statement is evidence depends in part on its content. For example, a state merely implying or expressing a *desire* that something become illegal would not be evidence of state practice.[308]

So far, it appears that there is not enough consensus among these statements for any clear customary international law to have emerged due to state practice or *opinio juris*. Be that as it may, the 2016 meeting revealed

---

(last visited March 12, 2016).

307. *See*, *e.g.*, Customary International Humanitarian Law 1338, 3164 (Jean-Marie Henckaerts and Louise Doswald-Beck eds., 2005), https://www.icrc.org/eng/assets/files/other/customary-international-humanitarian-law-ii-icrc-eng.pdf (citing remarks at a meeting of experts as evidence related to state practice on deception and a Colombian Ministry of Foreign Affairs working paper presented at a meeting of experts as evidence of state practice). The same International Committee of the Red Cross (ICRC) study also took statements at CCW conferences as evidence of state practice, both when at official States Parties conferences, *see*, *e.g.*, *id.* at 1965 (citing China's remarks about blinding lasers; however, since these remarks were made a year after China adopted the protocol banning blinding lasers and are generally an endorsement of that protocol, it is not clear what added value they have), and in preparatory or implementation gatherings, *see*, *e.g.*, *id.* at 1966 (noting India's statement at the Third Preparatory Committee for the Second Review Conference of States Parties to the CCW that it "fully supported the idea of expanding the scope of the CCW to cover armed internal conflicts"). Even if one is not willing to accept the ICRC's assessment of what qualifies as state practice, *see*, *e.g.*, John Bellinger & William Haynes, *A U.S. Government Response to the International Committee of the Red Cross Study* Customary International Humanitarian Law, 89 Int'l. Rev. Red Cross 443, 444–46 (2007), https://www.icrc.org/eng/assets/files/other/irrc_866_bellinger.pdf, international tribunals like the International Tribunal for the Former Yugoslavia have accepted states' remarks before the United Nations General Assembly as state practice, *see* Prosecutor v. Tadic, Case No. IT-94-1-I, Decision on Defence Motion for Interlocutory Appeal on Jurisdiction, para. 120 (Int'l Cri. Trib. For the Former Yugoslavia Oct. 2, 1995), as well as statements before national legislatures, *see id.* at para. 100. Statements at a meetings of experts are similarly public, recorded, and made by state representatives in an official capacity. Further, at least one International Court of Justice judge has also declared that "the positions taken up by the delegates of States in international organizations and conferences…naturally form part of State practice." Barcelona Traction, Light and Power Company Limited (Belgium v. Spain), Judgment, 3 I.C.J. Rep 286, para. 302 (Feb. 5, 1970) (Ammoun, J., separate opinion), http://www.icj-cij.org/docket/index.php?p1=3&p2=3&case=50&p3=4. Statements at the Meeting of the Experts would fulfill that description.

308. *See* Henrik Meijers, *On International Customary Law in the Netherlands*, *in* On the Foundations and Sources of International Law 77, 85 (Ige F. Dekker & Harry H.G. Post eds., 2003) (A "declaration by a state which implies no more than that it is in favor of a proposed rule becoming law, does not contribute to the formation of…custom" because "[i]f one declares to be in favour of something happening in [the] future, that 'something' has not yet taken place in the present, and no present practice (relating to that something) can have been formed yet").

relatively wide agreement on some important points. First, nearly all states that explicitly addressed the issue concurred that "fully" autonomous weapon systems do not yet exist (although some maintained that such systems will never exist, whereas others seemed to assume that they inevitably will). Second, there was wide agreement on the need for further discussion or monitoring (or both). Nearly every state mentioned the importance of continuing the dialogue. Third, most states indicated their belief that the current definitions of "autonomous weapon systems" are inadequate, impeding the progress that international society can make in assessing legal concerns.

In terms of taking a concrete position concerning the legality of "lethal autonomous weapons systems," at the 2016 Meeting the greatest agreement was on the importance or relevance of the review process under Article 36 of the first Additional Protocol to the Geneva Conventions (described in more detail below) and on the need for "meaningful human control" over AWS. In statements at the 2016 Meeting, thirteen states referenced the importance or relevance of Article 36—more than twice as many as at the 2015 Meeting. Also at the 2016 Meeting, thirteen states expressly referenced the need for "meaningful human control." However, as in 2015, this agreement was undercut by the lack of clarity as to what "meaningful human control" means. (Some states seemed to think that something akin to a human override capability would be sufficient, while others disagreed.[309]) Given the disparities in how different states interpret the concept, some states expressed skepticism about the usefulness of the notion of "meaningful human control."[310]

When comparing the 2015 and 2016 CCW Informal Meetings of Experts, it is important to bear in mind that the participating states are not identical. The differences between the meetings may simply reflect the altered composition of participating states, not necessarily a coherent shift in position among the same group of states. Nonetheless, the growing number of states that referenced Article 36 reviews might reflect a growing recognition that the category "autonomous weapon systems" involves a broad spectrum of weapons and may require review on a case-by-case basis.

---

309.  *See, e.g.*, Statement of Israel, Characteristics of LAWS (Part II), http://www.unog. ch/80256EDD006B8954/(httpAssets)/AB30BF0E02AA39EAC1257E29004769F3/$file/2015_ LAWS_MX_Israel_characteristics.pdf ("During the discussions, delegations have made use of various phrases referring to the appropriate degree of human involvement in respect to LAWS. Several States mentioned the phrase 'meaningful human control'. Several other States did not express support for this phrase. Some of them thought that it was too vague, and the alternative phrasing 'appropriate levels of human judgment' was suggested. We have also noted, that even those who did choose to use the phrase 'meaningful human control', had different understandings of its meaning. Some of its proponents had in mind human control or oversight of each targeting action in real-time, while others thought that, at least from a perspective of ensuring compliance with IHL, the preset by a human of certain limitations on the way a lethal autonomous system would operate, may also amount to meaningful human control. In our view, it is safe to assume that human judgment will be an integral part of any process to introduce LAWS, and will be applied throughout the various phases of research, development, programming, testing, review, approval, and decision to employ them. LAWS will not actually be making decisions or exercising judgment by themselves, but will operate as designed and programmed by humans").

310.  *See* Appendices I and II.

Another consideration in the evaluation of customary international law that may be relevant to AWS concerns "specially affected" states. The basic idea is that the practice of "specially affected" states[311]—that is, states that are "affected or interested to a higher degree than other states with regard to the rule in question"—"should weigh heavily (to the extent that, in appropriate circumstances, it may prevent a rule from emerging)."[312] For example, with respect to the rights associated with a state's territorial sea, the practices of states with a coastline have been considered as more significant than those of landlocked states.[313] There is some dispute over the determination and role of "specially affected" states in customary international humanitarian law.[314] Yet the position of the majority of commentators seems to be that "[i]f an emerging rule in respect to the use of sophisticated weaponry is considered then the practice of only a few states technically capable of production may suffice."[315]

---

311.   North Sea Continental Shelf Cases (Germany v. Denmark; Germany v. Netherlands), Judgment, 1969 I.C.J. Rep. 3, para. 73 (Feb. 1969) ("State practice, including that of States whose interests are specially affected, should have been both extensive and virtually uniform in the sense of the provision invoked;—and should moreover have occurred in such a way as to show a general recognition that a rule of law or legal obligation is involved").

312.   Wood, Second Report, *supra* note 300, at 38–39 (internal citations omitted).

313.   *See*, *e.g.*, Yoram Dinstein, The Interaction between Customary International Law and Treaties 288–89 (2007).

314.   *See*, *e.g.*, Ward Ferdinandusse, Book Review, 53 Netherlands Int'l L. Rev. 502, 504 (2006) ("it may be asked whether there are specially affected states in IHL at all. It is easy to see how the concept of specially affected states is useful when discussing delimitation of the continental shelf: some states have a continental shelf to delimit while other states do not and, one may assume, never will. There is an aspect of permanency there which is lacking in IHL. Belligerent states, one may hope, are the peace makers of tomorrow. Occupied states may be the occupiers of tomorrow. Customary rules develop slowly and should be stable enough to withstand such changing of positions. Moreover, one would think that it is irreconcilable with the very character of IHL to grant specially affected status to the manufacturers of certain dubious weapons, just as it would have been problematic at least to grant South-Africa specially affected status with regard to the question of apartheid"). *See also* Richard Price, *Emerging Customary Norms and Anti-Personnel Landmines*, *in* The Politics of International Law 106, 120–21 (Christian Reus-Smit ed., 2004); Jean-Marie Henckaerts, *Customary International Humanitarian Law: Taking Stock of the ICRC Study*, 78 Nordic J. Int'l L. 435, 446 (2010).

315.   Harry H.G. Post, *The Role of State Practice in the Formation of Customary International Humanitarian Law*, *in* On the Foundations and Sources of International Law 129,

If this view is accurate, then the practice of states that are more technologically advanced in the weapons arena—such as the United States, Israel, and South Korea, which are reportedly some of the states furthest along in the development of relevant technologies[316]—would be particularly important for any customary rules about AWS. So far, these and other similar states have largely favored continuing to monitor or discuss the development of such weapons. Indeed, these states mostly refrain from deciding on their *per se* legality while offering hints that they have apprehensions about bans that they view as potentially premature or restricting civilian technological development.[317]

Yet another line of reasoning suggests that states in whose territory *where* autonomous weapons might be deployed (regardless of whether the territorial state grants consent) may also be considered "specially affected." Along these lines, Pakistan's statements about the illegality of lethal autonomous weapons systems would also receive a privileged status.[318] This claim might have some value as *lex ferenda* (the law as it should be). But, as mentioned above, existing scholarly commentary tends to focus on the *weapons-possessors*, not on the *places where the weapons may be used*, as the "specially affected" states.

142 (Ige F. Dekker & Harry H.G. Post eds., 2003). *See also* Dinstein, *supra* note 313, at 293; Customary International Humanitarian Law, *supra* note 307, at xliv–xlv ("Concerning the question of the legality of the use of blinding laser weapons, for example, 'specially affected States' include those identified as having been in the process of developing such weapons"). *Cf.* H.W.A. Thirlway, International Customary Law and Codification: An Examination of the Continuing Role of Custom in the Present Period of Codification of International Law 71–72 (stating that, in relation to laws for outer space, specially affected states would be those "actually or potentially in control of the economic and scientific assets necessary for the exploration of space," and that it might even be unnecessary to look beyond those states to determine the relevant state practice).

316.   *See Kenneth Anderson & Matthew Waxman, Law and Ethics for Autonomous Weapon Systems: Why a Ban Won't Work and How the Laws of War Can*, American University Washington College of Law Research Paper No. 2013-11, at 1 (2013), http://ssrn.com/abstract=2250126.

317.   At least in 2015, Germany did somewhat differentiate itself, drawing a "red line" about the need for meaningful human control and calling for states to "take care to closely monitor the development and introduction of any new weapon system to guarantee that there will be no transgression."

318.   This sort of argument would not be too far removed from some states' claims before the International Court of Justice (ICJ) that the potentially world-affecting damage nuclear weapons could create should mean that all states qualify as specially affected, *see* Hugh Thirlway, *The Sources of International Law*, *in* International Law 91, 99 (Malcolm D. Evans ed., 2014). The ICJ did not weigh in on the validity of this claim. Still, if anything, the sort of argument outlined above would be less extreme than the nuclear-weapons claim, since, it seems, AWS might be capable of being more geographically limited than nuclear weapons. That argument would nevertheless rely on states believing that they could accurately predict where AWS would be used, if the customary law was to precede their development.

## Summary of States' Positions as Reflected by Their Statements at the 2015 and 2016 CCW Meetings of Experts

*Charts containing the relevant quotations, caveats, and explanations are in Appendices I and II.*

**Position:**[319] *Currently unacceptable, unallowable, or unlawful*
**States reflecting this position:** Austria,[320] Chile,[321] Costa Rica, Ecuador, Germany,[322] Mexico, Pakistan, Poland,[323] and Zambia

**Position:** *Need to monitor or continue to discuss*
**States reflecting this position:** Algeria, Austria, Australia, Canada, Chile, Colombia, Croatia, Costa Rica, Czech Republic, Ecuador, Finland, France, Germany, India, Ireland, Israel, Italy, Japan, Korea, Mexico, Morocco, Netherlands, New Zealand, Pakistan, Poland, Sierra Leone, South Africa, Spain, Sri Lanka, Sweden, Switzerland, Turkey, United Kingdom, United States of America, and Zambia

319.   When states advocate the need to regulate AWS, the need for meaningful human control, or the need for an Article 36 review, they are not necessarily suggesting that any of these steps, on their own, would adequately address the issues presented by autonomous weapons. Rather, states often presented these actions as necessary but not sufficient steps to effectively dealing with AWS. Additionally, this table is not intended to and does not necessarily represent a comprehensive, accurate list of all states' current positions on AWS. One reason for this fact is that it represents states' positions as assessed through both the 2015 and 2016 meetings; a state's position could have changed between 2015 and 2016, but both the 2015 and 2016 positions would be listed here. Also, the table generally excludes states' remarks outside of the written statements they offered at these two meetings. There are several exceptions, which are noted through footnotes.

320.   In this context, Austria concludes only that the technology as it currently stands is unlawful; though concerned about future versions also being unlawful, Austria does not categorically state that lawfulness would be impossible.

321.   Chile's position on this issue is slightly ambiguous. Some of its statements clearly indicate that it believes that fully autonomous weapons are unlawful, but some of its other statements seem to suggest that those weapons should simply be regulated. (This raises the question whether Chile believes that AWS would become lawful if we simply regulated their use.)

322.   In this context, Germany never explicitly uses the word "unlawful." Nevertheless, Germany has given strong indications that it considers the use of lethal force by fully autonomous weapon systems to be illegitimate. Not only does Germany explicitly state that it is "not acceptable" for a weapon system to have control over life and death, but Germany portrays its current stance as a repetition of the stance that it took in last year's meeting. (In last year's meeting, Germany stated that it considered AWS to be unlawful.)

323.   In this context, Poland indicated only that a *fully* autonomous weapon system would not be allowed, but it was very careful to indicate that it believes that such weapon systems do not yet exist. Therefore, Poland does not believe that any autonomous weapon systems, as they currently exist, are unlawful. But its Human Rights and Ethical Issues Statement does suggest that if a fully autonomous weapon system were to be developed in the future, it would "not be allowed." (As with Germany, however, Poland does not explicitly use the word "unlawful," though Poland's statement that fully autonomous weapon systems would "not be allowed" seems to suggest that such systems would indeed be illegal.)

**Position:** *Need to regulate*[324]

**States reflecting this position:** Austria, Chile, Colombia, Czech Republic, Netherlands, Poland, Sri Lanka, Sweden, and Zambia

**Position:** *Need to ban (or favorably disposed towards the idea)*[325]

**States reflecting this position:** Algeria, Bolivia,[326] Chile, Costa Rica, Croatia,[327] Cuba, Ecuador, Egypt,[328] Ghana, Mexico,[329] Nicaragua,[330]

---

324.   Scholarly debates about AWS are often framed as a choice between regulation and a ban. However, when states at the 2015 and 2016 CCW Informal Meeting of Experts have discussed regulation, it is not clear that they were implying regulation was to be preferred over a ban; often, those endorsing regulation seemed to be conceiving of the act as distinguished from doing nothing, not in contrast to a ban.

325.   The Holy See has also spoken in favor of a ban (for example, in a written statement for the 2015 CCW Meeting of Experts). However, as it is not a state, *see* Gerd Westdickenberg, *Holy See*, *in* Max Planck Encyclopedia of Public International Law (James R. Crawford, ed., 2006) ("The Holy See is neither a State nor only an abstract entity like an international organization….The international personality the Holy See enjoys as a unique entity and the sovereignty it exercises are different from those of other subjects of international law, be it States, international organizations like the International Committee of the Red Cross (ICRC), or [other] subject[s] of international law…[Its] international legal personality can best be defined as being '*sui generis*'"), the Holy See has not been included in this table or any of the ones that follow in Appendices I and II.

326.   Bolivia did not express its desire for a ban via a written statement at the 2015 or 2016 CCW Meeting of Experts, but it did reportedly offer an oral statement favoring a ban at the 2015 CCW Meeting of Experts. *See* Campaign to Stop Killer Robots, Report on Activities: Convention on Conventional Weapons Second Informal Meeting of Experts on Lethal Autonomous Weapons Systems 25 (2015), http://www.stopkillerrobots.org/wp-content/uploads/2013/03/KRC_CCWx2015_Report_4June2015_uploaded.pdf ("Bolivia made a late statement—its first on the matter—that called for a ban on fully autonomous weapons systems, citing concerns that the right to life should not be delegated and doubts that international humanitarian and human rights law is sufficient to deal with the challenges posed"). Bolivia's position has been included here to more fully represent states' attitudes on an important issue.

327.   In 2015, Croatia did not necessarily endorse a ban on all AWS but seemed to at least indicate it would be favorably inclined toward efforts to ban any AWS that did not involve "meaningful human control;" Croatia also repeatedly indicated that the option of a ban or moratorium should still be on the table. *See* Appendices I and II for more.

328.   At the 2015 or 2016 CCW Meeting of Experts, Egypt did not express its desire for a ban via a written statement. It has, however, orally indicated a preference for a moratorium on the development of AWS until more debate has occurred. *See* Appendices I and II for more. Egypt's position has been included here to more fully represent states' attitudes on an important issue.

329.   At the 2015 or 2016 CCW Meeting of Experts, Mexico did not express its desire for a ban via a written statement. It did, however, orally indicate a preference for a ban during the 2016 meeting. *See* Appendices I and II for more. Mexico's position has been included here to more fully represent states' attitudes on an important issue.

330.   At the 2015 or 2016 CCW Meeting of Experts, Nicaragua did not express its desire for a ban via a written statement. It did, however, orally indicate a preference for a ban during the 2016 meeting. *See* Appendices I and II for more. Nicaragua's position has been included here to more fully represent states' attitudes on an important issue.

Pakistan, Sierra Leone,[331] Palestine,[332] Zambia,[333] and Zimbabwe[334]

**Position:** *Need for meaningful human control*
**States reflecting this position:** Argentina, Austria, Australia, Canada, Chile, Colombia, Croatia, Czech Republic, Denmark, Ecuador, Germany, Greece, Ireland, Korea, Morocco, Netherlands, Pakistan, Poland, South Africa, Sweden, Switzerland, Turkey, United Kingdom, Zambia, and Zimbabwe

**Position:** *AP I Article 36 weapons review (defined below) necessary*[335]
**States reflecting this position:** Australia, Austria, Canada, Cuba, Finland, France, Germany, Netherlands, Sierra Leone, South Africa,[336] Sri Lanka, Sweden, Switzerland, United Kingdom, and Zambia

**Position:** *Refers to legal principles while remaining undecided on* per se *legality of AWS*
**States reflecting this position:** Algeria, Argentina, Australia, Austria, Canada, Chile, Czech Republic, Denmark, Ecuador, Finland, France, Germany, Greece, India, Ireland, Israel, Italy, Japan, New Zealand, Poland, Sierra Leone, South Africa, Spain, Sri Lanka, Sweden, Switzerland, Turkey, United Kingdom, United States of America, and Zambia

---

331.   Sierra Leone did not explicitly call for a ban but is seemingly against any AWS not under human control. *See* Appendices I and II for more.

332.   At the 2015 or 2016 CCW Meeting of Experts, Palestine did not express its desire for a ban via a written statement (it did offer a written statement for the 2015 meeting, but it is not available online, and no press reports cite that 2015 statement as announcing Palestine favored a ban). Palestine did, however, orally indicate a preference for a ban during the 2015 CCW meeting (*not* the Meeting of Experts). *See* Appendices I and II for more. Palestine's position has been included here to more fully represent states' attitudes on an important issue.

333.   Zambia believes a prohibition on the use of AWS should be "on the CCW agenda." *See* Appendices I and II for more.

334.   At the 2015 or 2016 CCW Meeting of Experts, Zimbabwe did not express its desire for a ban via a written statement. It did, however, orally indicate a preference for a ban during the 2016 CCW meeting (*not* the Meeting of Experts). *See* Appendices I and II for more. Zimbabwe's position has been included here to more fully represent states' attitudes on an important issue.

335.   Other states spoke about the importance of proper national review but did not necessarily frame it in terms of an international legal obligation or, more specifically, an obligation derived from Article 36 of AP I.

336.   South Africa's position on Article 36 is somewhat ambiguous. South Africa does not explicitly state that an Article 36 review is necessary, nor does South Africa discuss how it would plan to implement it. But South Africa's General Statement directly quotes the language of Article 36 when discussing compliance with international law, strongly implying that an Article 36 review is important or relevant to assessing the legality of AWS.

# TREATY PROVISIONS AND CUSTOMARY RULES NOT SPECIFIC TO AWS

Having established that a rule of customary international law specific to AWS has not crystallized (at least not yet),[337] we turn to treaty provisions and customary rules that might nonetheless govern the design, development, or use (or a combination thereof) of an AWS or, more generally, a war algorithm. The following section is not meant to be exhaustive but rather to highlight some of the main rules that might be implicated by AWS or war algorithms.

## Jus ad Bellum

The *jus ad bellum* (also known as the *jus contra bellum*) is the field of public international law governing the threat of force or the use of force by a state in its international relations. Current international law establishes a general prohibition on such threats of force and such uses of force unless undertaken pursuant to a lawful exception to that prohibition. Recognized exceptions include an enforcement action pursuant to a mandate of the U.N. Security Council, an exercise of lawful self-defense conforming to the principles of necessity and proportionality, and lawful consent.[338]

At least two concerns arise with respect to war algorithms as a matter of the *jus ad bellum*. The first is whether the determination of a breach of a rule of the *jus ad bellum* is independent of the type of weapon used.[339] For instance, some commentators have debated the use of so-called "predecessors of AWS," such as UAVs, in the context of obviating threats of terrorism as a matter of the *jus ad bellum*.[340] Others find those contributions "misguided,"[341] arguing instead that "[t]he use of AWS does not render an operation illegal under rules of *ius ad bellum*."[342]

---

337. This conclusion aligns with the statement in the U.S. DoD *Law of War Manual* that "[t]he law of war does not prohibit the use of autonomy in weapon systems." Law of War Manual, *supra* note 110, at § 6.5.9; *see also id.* at § 6.9.5.2 ("The law of war does not specifically prohibit or restrict the use of autonomy to aid in the operation of weapons").

338. On Security Council authorizations and self-defense, *see, e.g.*, Oliver Dörr, *Use of Force, Prohibition of, in* Max Planck Encyclopedia of Public International Law ¶¶ 38, 40–42 (2015).

339. *See* Markus Wagner, *Autonomous Weapon Systems, in* Max Planck Encyclopedia of Public International Law ¶ 11 (2016) (arguing that "[w]hether a breach of a rule of *ius ad bellum* has occurred is a determination that is independent from the type of weapon that has been used….").

340. *Id.*

341. *Id.*

342. *Id.*

The second concern is whether a particular use of a war algorithm in relation to the use of force in international relations falls under the category of prohibited "force." The most pertinent analogue might be a computer network attack. Oliver Dörr notes that, so far, such attacks against the information systems of another state have not been treated in practice under the principle of the non-use of force.[343] However, Dörr argues, "current and future State practice may, in this respect, lead to a different interpretation, given the weapon-like destructive potential which some attacks by means of information technology may develop: computer network attacks intended to directly cause physical damage to property or injury to human beings in another State may reasonably be considered armed force."[344]

## International Humanitarian Law

IHL is the primary field of international law governing armed conflict. It applies only in relation to armed conflict. Under international law, armed conflicts may be either international or non-international in character. IHL binds all of the parties to the armed conflict (whether states or non-state organized armed groups), as well as individuals.[345] And, where applicable, the law of neutrality also binds neutral states or other states not party to the armed conflict.[346]

The discussion on AWS and war algorithms enters into a number of preexisting debates in IHL. Those concern such issues as the contours of civilian "direct participation in hostilities,"[347] the geographic and temporal scope of armed conflict, and the relationship of IHL to international human rights law. The AWS discourse to date has largely revolved around IHL provisions concerning the conduct of hostilities, given the focus on autonomous *weapon* systems. Here we highlight the major considerations concerning AWS as *weapons*, though we note some other areas of IHL that might be relevant for war algorithms more broadly.

### Suppression of Acts Contrary to the Geneva Conventions

As a framework matter, states parties to the Geneva Conventions of 1949 have a general obligation to "undertake to respect and to ensure respect for the … Convention[s] in all circumstances."[348] More broadly, each state party

---

343.  *See, e.g.*, Dörr, *supra* note 338, at ¶ 12.

344.  *Id.* (citations omitted).

345.  *See, e.g.,* Jann Kleffner, *supra* note 17.

346.  *See, e.g.*, Michael Bothe, *Law of Neutrality*, *in* The Handbook of International Humanitarian Law (Dieter Fleck ed., 3rd ed. 2013).

347.  *See generally* the Forum *in* 42 N.Y.U. J. Int'l Law & Pol. 3, 637 *et seq.* (2010).

348.  *See* Geneva Convention for the Amelioration of the Condition of the Wounded and Sick in Armed Forces in the Field art. 1, Aug. 12, 1949, T.I.A.S. 3362 [hereinafter GC I]; Geneva Convention for the Amelioration of the Condition of Wounded, Sick and Shipwrecked

"shall take measures necessary for the suppression of all acts contrary to the provisions of the" Geneva Conventions of 1949 other than grave breaches.[349] (States are required to take certain other, more exacting measures with respect to grave breaches, as noted below.)

## Classification: Weapons (or Weapon Systems) or Combatants?

An initial issue is whether under IHL the relevant AWS (however defined) is considered a weapon (or a weapon system) or should be classified as something else, such as a combatant. The bulk of states and commentators focus on AWS in the sense of *weapons*.[350] But others, such as Hin-Yan Liu, raise the prospect that an AWS may be considered a combatant where, for instance, the focus is on the system's decision-making capability. Liu adopts the U.S. DoD Law of War Working Group's approach to differentiating between the terms "weapon" and "weapon systems."[351] The former refers to "all arms, munitions, materiel, instruments, mechanisms, or devices that have an intended effect of injuring, damaging, destroying or disabling personnel or property," while the latter is more broadly conceived to include "the weapon itself and those components required for its operation, including new, advanced or emerging technologies."[352]

For Liu, "the *capacity for autonomous decision-making* pushes these technologically advanced systems to the boundary of the notion of 'combatant.'"[353] As an indicator of the "potential for the confusion between means and methods of warfare and combatants," Liu points to the German military manual, which provides that "combatants are persons who may take a direct part in hostilities, i.e., participate in the use of a weapon or a weapon-system in an indispensable function."[354] Liu notes that "this characterization was used in the context of differentiating categories of non-combatants who are members of the armed forces," yet his broader point is that "the circularity of this definition illustrates precisely the

---

Members of Armed Forces at Sea art. 1, Aug. 12, 1949, T.I.A.S. 3363 [hereinafter GC II]; Geneva Convention Relative to the Treatment of Prisoners of War art. 1, Aug. 12, 1949, T.I.A.S. 3364 [hereinafter GC III]; Geneva Convention Relative to the Protection of Civilian Persons in Time of War art. 1, Aug. 12, 1949, T.I.A.S. 3365 [hereinafter GC IV].

349.   GC I, *supra* note 348, at art. 59; GC II, *supra* note 348, at art. 50; GC III, *supra* note 348, at art. 129; GC IV, *supra* note 348, at art. 146.

350.   On the conflation between weapons and "means and methods of warfare," at least in the context of Article 36 AP I weapons reviews, *see generally* Hin-Yan Liu, *Categorization and Legality of Autonomous and Remote Weapons Systems*, 94 Int'l Rev. Red Cross 627, 636 (2012).

351.   *Id*. at 635.

352.   *Id*. (citations omitted).

353.   *Id*. at 636 (italics added).

354.   *Id*. at 637.

difficulties associated with defining 'weapon' and 'weapons system.'"[355]

## Weapons: Reviews

As noted relatively frequently at the 2016 CCW Informal Expert Meeting on Lethal Autonomous Weapons Systems, Article 36 of Additional Protocol I imposes an obligation on states parties concerning "the study, development, acquisition or adoption of a new weapon, means or method of warfare." In particular, states parties are obliged to determine "whether [the] employment [of a new weapon, means or method of warfare] would, in some or all circumstances, be prohibited by" AP I or by any other rule of international law applicable to the state party.

With respect to AWS, Christopher Ford argues that "[t]he complexity of the weapons review will be a function of the sophistication of the technology, the geographic and temporal scope of use, and the nature of the environment in which the system is expected to be used."[356] He puts forward four "best practices" to consider in all such reviews. First, "[t]he weapons review should either be a multi-disciplinary process or include attorneys who have the technical expertise to understand the nature and results of the testing process." Second, "[r]eviews should delineate the planned and normal circumstances of use for which the weapon was reviewed." Third, "[t]he review should provide a clear delineation between expected human and system roles." And fourth, "optimally, the review should occur at three points in time." Those points are: "when the proposal is made to transition a weapon from research to development"; before the weapon is fielded; and, after fielding, "based upon feedback on how the weapon is functioning." The latter "would necessitate the establishment of a clear feedback loop which provides information from the developer to the reviewer to the user, and back again."

## Weapons: Grounds for Unlawfulness

Under IHL, a weapon or its use may be considered unlawful under two sets of circumstances.[357] First, the weapon may be considered unlawful *per se* (in and of itself), either because the weapon has been expressly prohibited in applicable international law or because the weapon is not capable of being used in a manner that comports with IHL. Second, the weapon may be considered unlawful based on a particular use. In relation to this factor, only that unlawful use of the weapon, not the weapon itself, would be illegal.

---

355. *Id.*

356. Lt. Col. Christopher M. Ford, Stockton Center for the Study of International Law, Remarks at the 2016 Informal Meeting of Experts, at 4, UN Office in Geneva (April 2016), http://www.unog.ch/80256EDD006B8954/(httpAssets)/D4FCD1D20DB21431C1257F9B0050B318/$file/2016_LAWS+MX_presentations_challengestoIHL_fordnotes.pdf; *see also* U.K. Ministry of Def., *supra* note 113, at 5-3 (discussing factors concerning legal review and situation awareness of manned vs. unmanned aircraft systems).

357. This sub-section on weapons and IHL draws extensively on William H. Boothby, *Prohibited Weapons*, *in* Max Planck Encyclopedia of Public International Law (2015).

## Weapons: Unlawful *Per Se* Due to Applicable Prohibition

A number of IHL treaties prohibit or restrict the use of certain weapons. The prohibitions in IHL treaties concerning specific weapons that might be relevant to war algorithms or AWS (or both) include:

- Pursuant to the Hague Convention on the Laying of Automatic Submarine Contact Mines (1907 Hague Convention VIII),[358] it is prohibited to lay unanchored automatic contact mines, except when they are so constructed as to become harmless one hour at most after the person who laid them ceases to control them;[359] it is also prohibited to lay anchored automatic contact mines that do not become harmless as soon as they have broken loose from their moorings and to use torpedoes that do not become harmless when they have missed their mark;[360] finally, it is also forbidden to lay automatic contact mines off the coast and ports of the enemy with the sole object of intercepting commercial shipping.[361]

- The Convention on the Prohibition of Military or any other Hostile Use of Environmental Modification Techniques (1977)[362] prohibits, among other things, military or other hostile use of environmental modification techniques if these would have widespread, long-lasting, or severe effects as the means of destruction, damage, or injury to another state party.[363]

- The Convention on Prohibitions or Restrictions on the Use of Certain Conventional Weapons which may be Deemed to be Excessively Injurious or to have Indiscriminate Effects (1980)[364] "facilitates the negotiation of protocols which can address particular weapons or types of weapon technology."[365] Under the aegis of the CCW, the following weapons prohibitions, among others, have been adopted:

o   Pursuant to the Protocol on Non-Detectable Fragments (Protocol I, 1980),[366] it is prohibited to use any weapon "the primary effect of which is to injure by fragments which in the human body escape detection by x-rays";[367]

o   Pursuant to the Protocol on Prohibitions or Restrictions on the Use of Mines, Booby-Traps and other Devices (Protocol II, as amended, 1996),[368] it is prohibited to use booby-traps in the form of apparently harmless portable objects specifically designed and constructed to contain explosive material and to detonate when they are disturbed or approached[369] (note that the U.S. DoD *Law of War Manual* states that "to the extent a weapon system with autonomous functions falls within the definition of a 'mine' in the CCW Amended Mines Protocol, it would be regulated as such."[370]);

o   Pursuant to the Protocol on Prohibitions or Restrictions on the Use of Incendiary Weapons (Protocol III, 1980),[371] it is prohibited to make any military objective located within a concentration of civilians the object of attack by air-delivered incendiary weapons;[372]

o   Pursuant to the Protocol on Blinding Laser Weapons (Protocol IV, 1995),[373] it is prohibited to employ laser-weapons

---

358.   Convention No. VIII Relative to the Laying of Automatic Submarine Contact Mines, Oct. 18, 1907, 36 Stat. 2332.

359.   *Id.* at art. 1.

360.   *Id.*

361.   *Id.* at art. 2.

362.   Convention on the Prohibition of Military or Any Other Hostile Use of Environmental Modification Techniques, May 18, 1977, 31 U.S.T. 333, 1108 U.N.T.S. 15.

363.   *See id.* at art. 1. *See also* AP I, *supra* note 12, at arts. 35(3) and 55.

364.   Convention on Prohibitions or Restrictions on the Use of Certain Conventional Weapons Which May Be Deemed to Be Excessively Injurious or to Have Indiscriminate Effects, Oct. 10, 1980, 1342 U.N.T.S. 137, 19 I.L.M. 1523 [hereinafter CCW].

---

365.   Boothby, *supra* note 357, at ¶ 16.

366.   Protocol [I to the Convention on Prohibitions on the Use of Certain Conventional Weapons Which May Be Deemed to Be Excessively Injurious or to Have Indiscriminate Effects] on Non-Detectable Fragments, Oct. 10, 1980, 1342 U.N.T.S. 168.

367.   *Id.*

368.   Protocol [II to the Convention on Prohibitions or Restrictions on the Use of Certain Conventional Weapons Which May Be Deemed to Be Excessively Injurious or to Have Indiscriminate Effects] on Prohibitions or Restrictions on the Use of Mines, Booby-Traps and Other Devices, Oct. 10, 1980, 1342 U.N.T.S. 168.

369.   *Id.* at art. 2-3.

370.   Law of War Manual, *supra* note 110, at § 6.5.9.2 (internal reference omitted).

371.   Protocol [III to the Convention on Prohibitions or Restrictions on the Use of Certain Conventional Weapons Which May Be Deemed to Be Excessively Injurious or to Have Indiscriminate Effects] on Prohibitions or Restrictions on the Use of Incendiary Weapons art. 2(2), Oct. 10, 1980, 1342 U.N.T.S. 171.

372.   *Id.* at art. 2.

373.   Protocol [IV to the Convention on Prohibitions or Restrictions on the Use of Certain Conventional Weapons Which May Be Deemed to Be Excessively Injurious or to Have specifically designed, as their sole combat function or as one of their combat functions, to cause permanent blindness to unenhanced vision, that is, to the naked eye or to the eye with corrective eyesight devices.[374]

- The Convention on the Prohibition of the Use, Stockpiling, Production and Transfer of Anti-Personnel Mines and on Their Destruction (1997)[375] prohibits the use, development, production, acquisition, stockpiling, retention, or transfer of anti-personnel landmines and provides for their destruction.[376]

- The Biological Weapons Convention (1972)[377] prohibits the development, production, stockpiling, acquisition, or retention of microbial or other biological agents or toxins where the types or quantities are such that there is no justification for prophylactic, protective, or other peaceful purposes.

- The Chemical Weapons Convention (1993)[378] prohibits the development, production, acquisition, stockpiling, retention, direct or indirect transfer, or use of chemical weapons, preparing for their use or assisting, encouraging, or inducing any person to do any of these things.

- The Convention on Cluster Munitions (2008)[379] prohibits the use, development, production, acquisition, stockpiling, retention, and direct or indirect transfer of cluster munitions and forbids assistance, encouragement, or inducement of any of these activities.[380]

As noted above, whether AWS (however defined) should be the subject of a preemptive prohibition remains an area of discussion and debate. As of August 2016, 16 states have stated that there is a need for a ban on fully autonomous

---

Indiscriminate Effects] on Blinding Laser Weapons art. 1, Oct. 13, 1995, 1380 U.N.T.S. 370.

374.　*Id.* at art. 1.

375.　Convention on the Prohibition of the Use, Stockpiling, Production and Transfer of Anti-Personnel Mines and on Their Destruction, Sept. 18, 1997, 2056 U.N.T.S. 211, 242.

376.　*Id.* at art. 1.

377.　Convention on the Prohibition of the Development, Production, and Stockpiling of Bacteriological (Biological) and Toxin Weapons and on Their Destruction, art. 1, Apr. 10, 1972, 1015 U.N.T.S. 163.

378.　Convention on the Prohibition of the Development, Production, Stockpiling and Use of Chemical Weapons and on their Destruction art. 1, Jan. 13, 1993, 1974 U.N.T.S. 317.

379.　Convention on Cluster Munitions art. 1, May 30, 2008, 48 I.L.M. 357.

380.　*Id.* at art. 1.

weapons or have made statements indicating that they are favorably disposed toward the idea.[381]

Some advocates of a preemptive ban have pointed to the development of the Protocol on Blinding Lasers (CCW Protocol IV) as a relevant precedent. However, commentators have noted a number of distinguishing factors between permanently-blinding lasers and AWS. The combined analyses of two scholars suggest that, in general, a weapons ban is more likely to be successful where:

- The weapon is ineffective;

- Other means exist for accomplishing a similar military objective;

- The weapon is not novel: it is easily analogized to other weapons, and its usages and effects are well understood;

- The weapon or similar weapons have been previously regulated;

- The weapon is unlikely to cause social or military disruption;

- The weapon has not already been integrated into a state's armed forces;

- The weapon causes superfluous injury or suffering in relation to prevailing standards of medical care;

- The weapon is inherently indiscriminate;

- The weapon is or is perceived to be sufficiently notorious to galvanize public concern and spur civil society activism;

- There is sufficient state commitment in enacting regulations;

- The scope of the ban is clear and narrowly tailored; or

- Violations can be identified.[382]

According to one of those scholars, "[o]f these, only a single factor – civil society engagement – supports the likelihood of a successful ban on autonomous weapon systems; the others are irrelevant, inconclusive, or imply that autonomous weapon systems will resist regulation."[383] The extent

---

to which states agree or disagree with these arguments seems likely to shape whether states will take more concrete steps towards a preemptive ban concerning AWS.

## Weapons: Unlawful *Per Se* — Of a Nature to Cause Superfluous Injury or Unnecessary Suffering

Pursuant to Article 35(2) of AP I, "[i]t is prohibited to employ weapons, projectiles and material and methods of warfare *of a nature to cause superfluous injury or unnecessary suffering*."[384] According to Bill Boothby, "[t]his is now a customary rule of law that binds all States in all types of armed conflict."[385] Accordingly, to not be unlawful, a war algorithm must not be of a nature to cause superfluous injury or unnecessary suffering.

## Weapons: Unlawful *Per Se* — Indiscriminate by Nature

In addition to the customary superfluous-injury principle, "[t]he second, equally important, customary weapons law principle holds that weapons that are *indiscriminate by nature* are prohibited."[386] The principle is derived in part from Article 51(4) of AP I. That provision prohibits indiscriminate attacks that are defined as including attacks "which employ a method or means of combat which *cannot be directed at a specific military objective*; or … which *employ a method or means of combat the effects of which cannot be limited*" as required by AP I and which consequently are of a nature to strike military objectives and civilians or civilian objects without distinction.[387] Thus, according to Switzerland, "in order for an AWS to be lawful under this rule [prohibiting indiscriminate-by-nature weapons], it must be possible to ensure that its operation will not result in unlawful outcomes with respect to the principle of distinction."[388]

---

com/why-prohibition-permanently-blinding-lasers-poor-precedent-ban-autonomous-weapon-systems.

384.   AP I, *supra* note 12, at art. 35(2) (emphasis added). *See also* Regulations Concerning the Laws and Customs of War on Land art. 23(e), annexed to Hague Convention (IV) Respecting the Laws and Customs of War on Land, Oct. 18, 1907, T.S. 539 [hereinafter 1907 Hague Regulations].

385.   Boothby, *supra* note 357, at ¶ 10; *see also, e.g.,* LAW OF WAR MANUAL, *supra* note 110, at § 6.5.9.2 (stating that "[i]n addition, the general rules applicable to all weapons would apply to weapons with autonomous functions. For example, autonomous weapon systems must not be calculated to cause superfluous injury ….") (internal reference omitted).

386.   Boothby, *supra* note 357, at ¶ 11; *see also, e.g.,* LAW OF WAR MANUAL, *supra* note 110, at § 6.5.9.2 (stating that "[i]n addition, the general rules applicable to all weapons would apply to weapons with autonomous functions. For example, autonomous weapon systems must not … be inherently indiscriminate.") (internal reference omitted).

387.   AP I, *supra* note 12, at art. 51 (emphasis added).

388.   Swiss, "Compliance-Based" Approach, *supra* note 74, at 3.

## Weapons: Unlawful by Use — Failure to Conform to Principles Governing Conduct of Hostilities

As noted above, where a weapon is not unlawful *per se* it may nonetheless be considered unlawful based on a particular use. In relation to this factor, only that unlawful use of the weapon, not the weapon itself, would be illegal. To avoid contravening IHL, in an armed conflict a direct attack using a weapon that is not unlawful *per se* must comport with IHL principles governing the conduct of hostilities.

The three such principles most frequently cited in discussions of AWS are distinction, proportionality, and precautionary measures. Each of these principles has IHL treaty roots and customary cognates. According to Switzerland, the basic guidelines in relation to AWS are as follows:

> Most notably, in order to lawfully use an AWS for the purpose of attack, belligerents must: (1 - Distinction) distinguish between military objectives and civilians or civilian objects and, in case of doubt, presume civilian status; (2 - Proportionality) evaluate whether the incidental harm likely to be inflicted on the civilian population or civilian objects would be excessive in relation to the concrete and direct military advantage anticipated from that particular attack; (3 - Precaution) take all feasible precautions to avoid, and in any event minimize, incidental harm to civilians and damage to civilian objects; and cancel or suspend the attack if it becomes apparent that the target is not a military objective, or that the attack may be expected to result in excessive incidental harm.[389]

With respect to the principle of proportionality and AWS, the U.S. DoD *Law of War Manual* states that "in the situation in which a person is using a weapon that selects and engages targets autonomously, that person must refrain from using that weapon where it is expected to result in incidental harm that is excessive in relation to the concrete and direct military advantage expected to be gained."[390]

Regarding precautions in attack, the wording of Article 57(2) of AP I raises the question of whether some of the precautionary-measures obligations laid down therein may be carried out, as a matter of treaty law, *only by humans* (compared with other obligations therein, which are reposed *in the party* to the armed conflict). Consider how Article 57(2)(a) of AP I lays down obligations of "*those* who plan or decide upon an attack."[391] But Article 57(2)(b)–(c) of AP I frames the obligations, respectively, as "an attack shall be

---

389. *Id.*

390. Law of War Manual, *supra* note 110, at § 6.5.9.3 (internal reference omitted).

391. AP I, *supra* note 12, art. 57(2)(a).

cancelled or suspended"[392] and "effective advance warning shall be given."[393]

For their part, the authors of the U.S. DoD *Law of War Manual* emphasize their view that "[t]he law of war rules on conducting attacks (such as the rules relating to discrimination and proportionality) impose obligations *on persons*. These rules do not impose obligations on the weapons themselves; of course, an inanimate object could not assume an 'obligation' in any event."[394] According to this view, "the obligation on the person using the weapon to take feasible precautions in order to reduce the risk of civilian casualties may be more significant when the person uses weapon systems with more sophisticated autonomous functions."[395] As an example, the *Manual* authors state that "such feasible precautions a person is obligated to take may include monitoring the operation of the weapon system or programming or building mechanisms for the weapon to deactivate automatically after a certain period of time."[396]

The UK MoD Joint Doctrine Note on unmanned aircraft systems discusses the obligations laid down in Additional Protocol I on the constant care that must be "taken in the conduct of military operations to spare civilians and civilian objects. This means that any system, before an attack is made, must verify that targets are military entities, take all feasible precautions to minimise civilian losses and ensure that attacks do not cause disproportionate incidental losses."[397] The Joint Doctrine Note authors state that "[f]or automated systems, operating in anything other than the simplest of scenarios, this process will provide a severe technological challenge for some years to come."[398]

While not focusing on AWS in particular, the UK MoD Joint Doctrine Note also addresses a situation where "a mission may require an unmanned aircraft to carry out surveillance or monitoring of a given area, looking for a particular target type, before reporting contacts to a supervisor when found."[399] According to the Joint Doctrine Note authors, "[a] human-authorised subsequent attack would be no different to that by a manned aircraft and would be fully compliant with the LOAC [law of armed conflict], provided the human believed that, based on the information available, the

---

attack met LOAC requirements and extant ROE [rules of engagement]."[400] The Joint Doctrine Note authors elaborate this line of reasoning, noting that, "[f]rom this position, it would be only a small technical step to enable an unmanned aircraft to fire a weapon based solely on its own sensors, or shared information, and without recourse to higher, human authority."[401] This would be entirely legal, the Joint Doctrine Note concludes, "[p]rovided it could be shown that the controlling system appropriately assessed the LOAC principles (military necessity; humanity; distinction and proportionality) and that ROE were satisfied…."[402] Yet the authors highlight a number of additional factors to consider:

> In practice, such operations would present a considerable technological challenge and the software testing and certification for such a system would be extremely expensive as well as time consuming. Meeting the requirement for proportionality and distinction would be particularly problematic, as both of these areas are likely to contain elements of ambiguity requiring sophisticated judgement. Such problems are particularly difficult for a machine to solve and would likely require some form of artificial intelligence to be successful.[403]

Finally in this connection, the Joint Doctrine Note notes that "the MOD currently has no intention to develop systems that operate without human intervention in the weapon command and control chain, but it is looking to increase levels of automation where this will make systems more effective."[404]

According to the U.S. DoD *Law of War Manual*, "in many cases, the use of autonomy could *enhance* the way law of war principles are implemented in military operations. For example, some munitions have homing functions that enable the user to strike military objectives with greater discrimination and less risk of incidental harm."[405] The *Manual* authors also note that "some munitions have mechanisms to self-deactivate or to self-destruct, which helps reduce the risk they may pose generally to the civilian population or after the munitions have served their military purpose."[406]

In a similar connection, the UK MoD Joint Doctrine Note on unmanned aircraft systems states that "[s]ome fully automated weapon systems have already entered service, following legal review, and contributing factors – such as required timeliness of response – can make compliance with LOAC easier to demonstrate."[407] The authors give an example of "the Phalanx and Counter-Rocket, Artillery and Mortar (C-RAM) systems that are already employed in Afghanistan," arguing that "it can be clearly shown that there is insufficient time for a human initiated response to counter incoming fire."[408] According to this view, "[t]he potential damage caused by not using C-RAM in its automatic mode justifies the level of any anticipated collateral damage."[409]

---

400.  *Id.*

401.  *Id.*

402.  *Id.*

403.  *Id.*

404.  *Id.*

405.  Law of War Manual, *supra* note 110, at § 6.5.9.2.

406.  *Id.* (internal reference omitted).

Other potentially relevant conduct-of-hostilities considerations raised in relation to AWS include principles concerning prohibitions on the denial of quarter and on the protection of persons *hors de combat* (such as the wounded and sick *hors de combat*). For instance, in relation to denial of quarter, in the view of Switzerland, "[a]ny reliance on AWS would need to preserve a reasonable possibility for adversaries to surrender. A general denial of this possibility would violate the prohibition of ordering that there shall be no survivors or of conducting hostilities on this basis (denial of quarter)."[410]

Stepping back, we see that, where a war algorithm is capable of being used in relation to the conduct of hostilities in connection with an armed conflict, that possible use is already regulated by a number of IHL rules and principles. Few states, however, have offered detailed views on what implications may arise for such uses of war algorithms.

## Other Functions in relation to Armed Conflict

IHL governs far more than just weapons and the conduct of hostilities. As the primary normative framework regulating armed conflict, IHL also lays down rules concerning such activities as capture, detention, and transfer of enemies; medical care to the wounded and sick *hors de combat*; and humanitarian access and assistance to civilian populations in need. Switzerland has noted, for instance, that it is conceivable that AWS "could be used to perform other tasks governed by IHL, such as the guarding and transport of persons deprived of their liberty or tasks related to crowd control and public security in occupied territories."[411]

---

407. U.K. Ministry of Def., *supra* note 113, at 5-2.

408. *Id.*

409. *Id.*

410. Swiss, "Compliance-Based" Approach, *supra* note 74, at 3 (citation omitted).

411. *Id.* (citation omitted) (noting that "[a]dditional specific rules need to be taken into consideration if AWS were to be relied for such activities").

## Martens Clause

With respect to AWS, the IHL "Martens clause" would, according to Switzerland, afford "an important fallback protection in as much as the 'laws of humanity and the requirements of the public conscience' need to be referred to if IHL is not sufficiently precise or rigorous."[412] Pursuant to this line of reasoning, "not everything that is not explicitly prohibited can be said to be legal if it would run counter [to] the principles put forward in the Martens clause. Indeed, the Martens clause may be said to imply positive obligations where contemplated military action would result in untenable humanitarian consequences."[413]

## Seizure of Private Property Susceptible of Direct Military Use

In a situation of belligerent occupation (a type of international armed conflict), the Occupying Power may seize, among other things, "all kinds of munitions of war … even if they belong to private persons."[414] Items so seized "must be restored and compensation fixed when peace is made."[415] With respect to AWS, this provision may implicate, for example, the private property—including the software and hardware components involved in developing AWS—of individuals or commercial entities subject to a belligerent occupation.[416]

# International Criminal Law

International criminal law (ICL) is a framework through which individual responsibility arises for international crimes. Under certain circumstances, the design, development, or use (or a combination thereof) of a war algorithm may form part of the conduct underlying an international crime. Recognized categories of international crimes include war crimes, genocide, and crimes against humanity. Each international crime is made up of a prohibited act or acts (the *actus reus* or *actus rei*) and the prohibited mental state (the *mens rea*). War crimes may arise only in relation to armed conflict. Genocide and crimes against humanity may arise outside of situations of armed conflict (though they often do in fact arise in relation to armed conflict). Here, we focus on the Statute of the International Criminal Court (ICC),[417] though we note that other ICL rules—those derived from applicable treaties or customary international law—also may be relevant.

---

412.   *See, e.g., id.* at 4 (citing to CCW, *supra* note 364, at preamble and AP I, *supra* note 12, at art. 1(2), and noting that "[i]n its 1996 Advisory Opinion on the legality of the threat or use of nuclear weapons, the International Court of Justice held that the clause 'proved to be an effective means of addressing the rapid evolution of military technology' (§78)").

413.   *Id.* at 3 (citing respectively, to AP I, *supra* note 12, at art. 57(2)(a) and to GCs I–IV, *supra* note 348, at arts. 49, 50, 129, 146 (respectively); AP I, *supra* note 12, at Section III.

414.   1907 Hague Regulations, *supra* note 384, at art. 53(2).

415.   *Id.*

416.   According to the U.S. DoD *Law of War Manual*, "[p]rivate property susceptible of direct military use includes cables, telephone and telegraph facilities, radio, television, telecommunications and computer networks and equipment, motor vehicles, railways, railway plants, port facilities, ships in port, barges and other watercraft, airfields, aircraft, depots of arms (whether military or sporting), documents connected with the conflict, all varieties of military equipment (including that in the hands of manufacturers), component parts of, or material suitable only for use in, the foregoing, and, in general, all kinds of war material." Law of War Manual, *supra* note 110, at § 11.18.6.2, *citing to* U.S. Dep't of the Army, The Law of Land Warfare, 1956 FM 27-10 ¶410a (Change No. 1 1976).

Various states and commentators disagree on whether ICL, especially in relation to war crimes, sufficiently addresses the design, development, and use of AWS. The discussion is hampered by lack of agreement on the definition of AWS, on the technological capabilities of AWS, and on the nature of the relationship between the various actors involved in the development and operation of AWS. These disagreements implicate underlying legal concepts of attribution, control, foreseeability, and reconstructability.

Much of the debate on AWS in relation to ICL revolves around modes of responsibility for international crimes and the mental element of international crimes.[418] Those arguing that ICL is sufficient to address AWS concerns typically emphasize that, ultimately, a single person—often, the commander or superior—may and should be held responsible where, in connection with an armed conflict, the design, development, or use of an AWS gives rise to an international crime.[419] Those arguing that ICL may not be sufficient typically emphasize that the ICL modes of command and superior responsibility are predicated on relationships between humans, not on relationships between humans and machines or constructed systems. (The ICC Statute establishes jurisdiction for individual responsibility only over natural persons, thereby excluding legal entities such as corporations.) They also note that it might not be possible, due to a lack of a temporal nexus to an armed conflict, to prosecute a developer who, before the war began, coded an AWS to function in a way that later gives rise to a war crime.[420] Critics also argue that due to the distributed nature of technical and physical control over an operation involving an AWS, it may not be possible to establish the relevant intent and knowledge of a particular perpetrator. Or, they assert, even if it is possible to establish the mental element, a perpetrator may argue to exclude criminal responsibility due to a mistake of fact, given how complex the operation of an AWS may be.

---

417.   Rome Statute of the International Criminal Court, July 17, 1998, 2187 U.N.T.S. 90 [hereinafter ICC Statute].

418.   *Id.* at arts. 25(3) and 28.

419.   *See, e.g.*, Dutch Government, Response to AIV/CAVV Report, *supra* note 22.

420.   *See* Tim McFarland & Tim McCormack, *Mind the Gap: Can Developers of Autonomous Weapons Systems Be Liable for War Crimes?*, 90 Int'l L. Stud. 361 (2014).

## Arms-Transfer Law

The Arms Trade Treaty of 2013 (ATT)[421] may implicate war algorithms that form part of the conventional arms and certain other items covered by that instrument. It may do so not only with respect to exporting and importing states parties but also in connection with trans-shipment states parties.

The ATT regulates certain activities of the international trade in arms— in particular, "export, import, transit, trans-shipment and brokering," all of which fall under the umbrella term of "transfer."[422] Many of the arms and related items covered by the treaty already use war algorithms. In relation to states parties, the treaty applies in respect of all conventional arms within eight categories: battle tanks, armored combat vehicles, large-caliber artillery systems, combat aircraft, attack helicopters, warships, missiles and missile launchers, and small arms and light weapons.[423] The ATT also regulates the export of "ammunition/munitions fired, launched or delivered by"[424] such conventional weapons, as well as of "parts and components where the export is in a form that provides the capability to assemble the [relevant] conventional arms."[425] (The ATT expressly does "not apply to the international movement of conventional arms by, or on behalf of, a State Party for its use provided that the conventional arms remain under that State Party's ownership."[426])

As part of the regulatory system established by the ATT, a state party is prohibited from authorizing any transfer of conventional arms or other covered items in three situations. First, the state party may not authorize such a transfer if it "would violate its obligations under measures adopted by the United Nations Security Council acting under Chapter VII of the Charter of the United Nations, in particular arms embargoes."[427] Second, an authorization is prohibited if the transfer "would violate its relevant international obligations under international agreements to which it is a Party, in particular those relating to the transfer of, or illicit trafficking in, conventional arms."[428] And third, an authorization is prohibited if the state party "has knowledge at the time of authorization that the arms or items would be used in the commission of … grave breaches of the Geneva Conventions of 1949, attacks directed against civilian objects or civilians protected as such, or other war crimes as defined by international agreements to which it is a Party."[429]

---

421. Arms Trade Treaty, Apr. 2, 2013, U.N. Doc. A/RES/67/234B [hereinafter ATT].

422. *Id.*, at art. 2(2).

423. *Id.* at art. 1.

424. *Id.* at art. 3.

425. *Id.* at art. 4.

426. *Id* at art. 2(3).

427. *Id.* at art. 6(a).

428. *Id.* at art. 6(b).

Even if the export is not prohibited under one of those stipulations, the ATT imposes an obligation not to authorize the export where the state party determines "that there is an overriding risk of any of the negative consequences" identified in a provision of the treaty.[430] Those consequences include the potential that the conventional arms or other covered items:

(a) would contribute to or undermine peace and security;

(b) could be used to:

(i) commit or facilitate a serious violation of international humanitarian law;

(ii) commit or facilitate a serious violation of international human rights law;

(iii) commit or facilitate an act constituting an offence under international conventions or protocols relating to terrorism to which the exporting State is a Party; or

(iv) commit or facilitate an act constituting an offence under international conventions or protocols relating to transnational organized crime to which the exporting State is a Party.[431]

Also, pursuant to the ATT, each export state party "shall make available appropriate information about the authorization in question, upon request, to the importing State Party and to the transit or trans-shipment States Parties, subject to its national laws, practices or policies."[432] Finally, each state party "involved in the transfer of conventional arms covered under Article 2 (1) [of the ATT] shall take measures to prevent their diversion."[433]

The upshot is that, under the ATT, a detailed and somewhat expansive regime exists to regulate the transfer of war algorithms where those algorithms form part of certain conventional weapons and related items.

## International Human Rights Law

While IHL traces its roots to the regulation of interstate wars, international human rights law (IHRL) arose out of an attempt to regulate, as a matter of international law and policy, the relationship between the state—through its governmental authority—and its population. Unlike the relatively narrow war-related field of IHL, IHRL spans a seemingly ever-growing range of dealings an individual, community, or nation may have with the state.

---

429.  *Id*. at art. 6(c).

430.  *Id*. at art. 7(3).

431.  *Id*. at art. 7(1).

432.  *Id*. at art. 7(6).

433.  *Id*. at art. 11(1).

In recent decades, the connection between IHL and IHRL has been the subject of increased jurisprudential treatment and interpretation by states. The precise links between the two branches of public international law have also merited extensive academic commentary. The debate on this relationship is largely over three issues. First, whether IHRL applies extraterritorially such that states bring all, some, or none of their obligations with them when they fight wars under IHL outside of their territories. Second, whether organized armed groups have IHRL obligations (or, at least, responsibilities). And third, what is the apposite interpretive procedure or principle to use when discerning the content of a particular right under the relevant framework(s).

With these considerations in mind, IHRL may impose substantive obligations on a state party to an armed conflict concerning the design, development, or use of a war algorithm. These obligations may range, for instance, from violations of the right to privacy to the right not to be arbitrarily deprived of life. That is, of course, not an exhaustive list, but it demonstrates the wide array of rights under IHRL that a war algorithm might implicate. IHRL might also implicate state obligations in relation to the design, development, and use of war algorithms during times of peace.

## Law of the Sea

As illustrated in section 2, many of the existing weapon systems with autonomous functions operate in the sea.[434] A number of provisions of the 1982 U.N. Convention on the Law of the Sea (UNCLOS),[435] "many of which are recognised as stating customary international law, ... apply to ships with mounted autonomous weapon systems and possibly to independent seafaring autonomous weapon systems."[436] Among these are the UNCLOS articles outlining "state obligations to protect and preserve both the marine environment generally and specific areas, such as the seabed and ocean floor,"[437] as well as the general prohibition on the threat of force or the use of force.[438] Furthermore, "[i]n addition to providing that the high seas 'shall be reserved for peaceful purposes', UNCLOS sets forth a number of prohibitions applicable to ships equipped with autonomous weapon systems that wish to exercise rights to innocent and transit passage."[439] Finally, "[w]hile automated and autonomous weapon systems have long been used on warships, future autonomous weapon systems may themselves be warships." Accordingly, "[s]hould they be granted warship status, such systems would gain certain rights and associated obligations."[440]

---

434.   The vast majority of scholars and states addressing AWS in relation to international law focus only on IHL and ICL; Rebecca Crootof has provided one of the most expansive analyses of various fields of public international law that might be implicated by AWS. Rebecca Crootof, *The Varied Law of Autonomous Weapon Systems, in* NATO ALLIED COMMAND TRANSFORMATION, AUTONOMOUS SYSTEMS: ISSUES FOR DEFENCE POLICY MAKERS 98, 109 (Andrew P. Williams & Paul D. Scharre eds., 2015) [hereinafter Crootof, *Varied*]. With respect to the law of the sea, space law, and international telecommunications law, we draw in part on her analysis.

435.   United Nations Convention on the Law of the Sea, Dec. 10, 1982, 1833 U.N.T.S. 397 [hereinafter UNCLOS].

436.   Crootof, *Varied*, *supra* note 434, at 109 (citation omitted).

## Space Law

Guidance concerning the design, use, and liability of war algorithms in outer space in relation to armed conflict may be found in the 1967 Outer Space Treaty,[441] other space-law treaties, and various U.N. General Assembly declarations.[442] Yet "aside from a few plain prohibitions," "the 'ceiling' of space law regulation is sky high … it allows for a wide range of potential extraterrestrial autonomous weapon systems"[443] and of war algorithms more broadly.

One such prohibition—laid down in the Outer Space Treaty, which may be binding as a codification of international law[444]—is on the use of space for destructive purposes. In particular, states parties to the Outer Space Treaty "undertake not to place in orbit around the Earth any objects carrying nuclear weapons or any other kinds of weapons of mass destruction, install such weapons on celestial bodies, or station such weapons in outer space in any other manner."[445] Among the other issues raised in this context include jurisdiction, control over objects launched into space, international responsibility for activities in space, and international liability for damage

---

437.   *Id.* (citing to UNCLOS, *supra* note 435, at art. 192–196).

438.   *Id.* (citing to UNCLOS, *supra* note 435, at art. 301)

439.   *Id.* at 110 (citing to UNCLOS, *supra* note 435, at art. 88).

440.   *Id.* (referring to the definition of "warship" in UNCLOS, *supra* note 435, at art. 29). *Id.* at 110 n.41.

441.   Treaty on Principles Governing the Activities of States in the Exploration and Use of Outer Space, Including the Moon and Other Celestial Bodies, Jan. 27, 1967, 18 U.S.T. 2410, 610 U.N.T.S. 205 [hereinafter OST].

442.   Crootof, *Varied*, *supra* note 434, at 111.

443.   *Id.*

444.   *Id.* (citation omitted).

445.   OST, *supra* note 441, at art. IV.

caused by space-based objects.[446]

## International Telecommunications Law

Constructed systems that use the electromagnetic spectrum or international telecommunications networks in effectuating war algorithms may be governed in part by telecommunications law. That law is regulated primarily by the International Telecommunications Union (ITU).[447] Scholars have already raised AWS in relation to telecommunications law,[448] including with respect to obligations to legislate against certain "harmful interference," preserving the secrecy of international correspondence and military radio installations, as well as exceptions concerning certain uses of military installations.[449]

---

446.   Crootof, *Varied*, *supra* note 434, at 112 (citations omitted).

447.   *See* Dietrich Westphal, *International Telecommunication Union*, *in* Max Planck Encyclopedia of Public International Law (2014).

448.   Crootof, *Varied*, *supra* note 434, at 113–114.

449.   *Id*. at 114 (citation omitted).

# 4

# ACCOUNTABILITY AXES

In this section, we outline three accountability axes that might be relevant to regulating war algorithms. We do not claim to be exhaustive but rather aim to provide examples of key accountability avenues. We adapt an accountability approach focusing on the regulation of war algorithms along three axes: state responsibility for internationally wrongful acts, individual responsibility under international law for international crimes, and a wider notion of scrutiny governance.[450]

Below, for each axis, we highlight existing and possible accountability actors, forums, and mechanisms. Some of these axes utilize existing formal legal regimes; others depend more on "soft law" or less formal codes, standards, guidelines, and the like. Regulation may arise, for instance, through direct or intermediary modes, as well as by setting rules to allocate risk and by defining rules of private interaction.[451] As noted above, we focus on international law in part because it is the only normative framework that purports, in key respects but with important caveats, to be universal and uniform.

## STATE RESPONSIBILITY

Along this axis, accountability is a matter of state responsibility arising out of acts or omissions involving a war algorithm where those acts or omissions constitute a breach of a rule of international law. State responsibility entails discerning the content of the rule, assigning attribution to a state, determining available excuses (if any), and imposing measures of remedy.

---

450.   Derived in part from INTERNATIONAL LAW ASSOCIATION, *supra* note 35, at 5.
451.   *See* WITTES & BLUM, *supra* note 31, at 203–206.

# MEASURES OF REMEDY

A range of consequences may arise where a war algorithm involved in an internationally wrongful act, not otherwise excused, is attributable to a state. In this sub-section, we highlight a main form of liability: war reparations. But we also note some of the other existing mechanisms and avenues through which state responsibility may be pursued, such as diplomatic channels, arbitration, judicial proceedings, weapons-control regimes, and an IHL fact-finding body.

## War Reparations to a State

As noted above, in general a consequence of state responsibility is the liability to make reparation. War reparations constitute one such form of liability. They "involve the transfer of legal rights, goods, property and, typically, money from one State to another in response to the injury caused by the use of armed force."[452] Historical practice favors, "[i]n the specific case of war reparations, … the use of restitution, monetary compensation, territorial guarantees, guarantees of non-repetition, and symbolic reparations."[453]

The Hague Convention concerning the Laws and Customs of War on Land (1907) and Additional Protocol I "establish an inter-State duty to pay compensation when a belligerent party violates the provisions of the Convention and ... Protocol I."[454] Thus, with respect to who can claim reparations, "a State's duty to provide inter-State reparations after the commission of an internationally wrongful act is certain."[455]

As a practical matter, war reparations are still the exception rather than the norm. When they do occur, the most common form of reparations, according to an assessment of practice up to 1995, was a lump sum at the end of the war.[456] Nonetheless, pursuant to Security Council resolutions the United Nations Compensation Commission (UNCC) was established to address damages incurred in the course of the Iraq-Kuwait War (1990–91).[457]

---

452.   Sullo & Wyatt, *supra* note 297, at ¶ 1.

453.   *Id*. at ¶ 4.

454.   *Id*. at ¶ 5 (referring to art. 3 Hague Peace Conferences [1899 and 1907]) and art. 91 AP I).

455.   *Id*. at ¶ 4.

456.   *Id*. ("Based on the analysis of practice until 1995, Lillich, Weston, and Bederman concluded that the settlement of international claims by lump sum agreements was by far the prevailing practice and the creation of arbitral tribunals such as the Iran-United States Claims Tribunal the exception.").

457.   *Id*. at ¶ 5.

And the governments of Eritrea and Ethiopia established a commission to deal with reparations claims concerning an armed conflict between those two states.[458]

Where a state party does not fulfill the obligation concerning suppression of acts contrary to the Geneva Conventions, another state party may also, for instance, pursue diplomatic channels to encourage the non-complying state to fulfill the obligation. That other state party may, where available, also pursue arbitration (if the transgressing state agrees) or institute judicial proceedings (if a relevant tribunal can assert its jurisdiction over the transgressing state).

## Weapons Monitoring, Inspection, and Verification Regimes

Weapons regimes may establish consequences for certain violations. Arms-control instruments range, in general, "from mere reporting duties and routine inspections (monitoring) to more invasive ad hoc inspections, sometimes so-called 'challenge inspections' at the request of a Member State (verification), up to compulsive methods in case of a determined breach (enforcement)."[459] Two of the main challenges of effective arms-control law are weak verification and limited enforcement mechanisms.

As noted above, the Arms Trade Treaty—which might cover various war algorithms—lays down a regulatory framework concerning the transfer of certain conventional weapons and related items. Through activities such as reporting and inspections, the Organization for the Prohibition of Chemical Weapons (OPCW) supervises the Chemical Weapons Convention. That treaty also provides for a challenge inspection procedure, "which is considered one of the most extensive verification procedures in the law of arms control, but has never been used, mainly due to political constraints."[460] In comparison, the supervisory mechanism of the Biological Weapons Convention is weaker, consisting mainly of review conferences every five years.

## International Fact-Finding Commission

Where certain rules of IHL are breached, the International Fact-Finding Commission (IFFC) established in Additional Protocol I may help provide measures of remedy. With respect to states parties to that treaty, the IFFC is competent, first, to enquire into any facts alleged to be a grave breach in or other serious violation of the Geneva Conventions of 1949 and Additional Protocol I.[461] Second, the IFFC is competent to "facilitate, through its good offices, the restoration of an attitude of respect for" the Geneva Conventions of 1949 and Additional Protocol I.[462] Where relevant, the design, development, or use of a war algorithm might implicate either or both of these competences. However, as practical matter, it bears emphasis that the IFFC has never been utilized for either competence.

---

458.   *Id.* (citing to Agreement between the Government of the State of Eritrea and the Government of the Federal Democratic Republic of Ethiopia, U.N. Doc. A/55/686-S/2000/1183 Annex).

459.   Adrian Loets, *Arms Control*, *in* Max Planck Encyclopedia of Public International Law ¶ 21 (2013).

460.   *Id.* at ¶ 23 (citation omitted).

## OTHER AVENUES

Certain other state accountability avenues may arise even where the design, development, or use of a war algorithm attributable to a state does not constitute an internationally wrongful act. Two such measures to consider are reparations to an individual pursuant to international human rights law, and a highly contentious form of domestic tort liability.

### Reparations to an Individual

As noted above, it is clear that a state may be provided reparations after the commission of an internationally wrongful act, including an applicable violation of IHL. Yet it is far less clear whether an individual right to reparation for victims of gross human rights violations has crystallized.[463] The U.N. General Assembly has adopted a resolution on the matter.[464] But that resolution has been characterized as falling into a category often referred to as "soft law": while "[t]hese documents do not have the formal status of legally binding instruments such as treaties, … they nonetheless reflect principles of justice and serve as tools for victim-oriented policies and practices at national and international levels."[465]

Nonetheless, to the extent it is applicable in relation to the design, development, or use of a war algorithm, IHRL may provide grounds for an individual to seek redress and reparation. The relevant violation would not be an internationally wrongful act vis-à-vis another state (or states) but rather a violation of an applicable provision of IHRL vis-à-vis an individual. For instance, "[t]he case-law developed in the jurisprudence of the [European Court of Human Rights] and the Inter-American Court of Human Rights … demonstrates an increasing readiness of these international (regional) adjudicative bodies to afford substantial reparative justice to victims, in particular in cases of gross violations of human rights."[466]

---

461.   AP I, *supra* note 12, at art. 90(2)(c)(i).

462.   *Id.* at art. 90(2)(c)(ii).

463.   *See* Sullo & Wyatt, *supra* note 297, at ¶ 4; *see generally* Christian Tomuschat, *State Responsibility and the Individual Right to Compensation Before National Courts*, *in* The Oxford Handbook of International Law in Armed Conflict (Andrew Clapham, Paola Gaeta & Tom Haeck eds., 2014).

464.   Basic Principles and Guidelines on the Right to a Remedy and Reparation for Victims of Gross Violations of International Human Rights Law and Serious Violations of International Humanitarian Law, adopted by UNGA Resolution 60/147, Dec. 16, 2005.

465.   Theo van Boven, *Victims' Rights*, *in* Max Planck Encyclopedia of Public International Law ¶ 19 (2007).

## Tortious Liability

Another state accountability avenue might arise in relation to a highly disputed form of tortious liability:[467] pecuniary compensation under domestic tort law for death or injury to the person, or damage to or loss of tangible property, caused by an act or omission which is alleged to be attributable under domestic law to a state other than the forum state and which involved a war algorithm. That compensation may be available only so long as the act or omission occurred in whole or in part in the territory of the forum state and so long as the author of the act or omission was present in the forum-state territory at the time of the act or omission.[468]

This notion of tortious liability requires discerning the content of applicable domestic law (including the relevant standard of care), attributing responsibility for the resulting harm to a state other than the forum state, confirming the presence of the author of the act in the forum state, determining the availability of immunity claims (if any), and imposing pecuniary compensation. This contested form of liability is derived from a purported "territorial tort" restriction to the applicability of state immunity found in the 2004 United Nations Convention on Jurisdictional Immunities

---

466.  *Id.* at ¶¶ 10–13.

467.  *Compare, e.g.,* Joanne Foakes & Roger O'Keefe, *Article 12, in* The United Nations Convention on Jurisdictional Immunities of States and Their Property: A Commentary 209, 209–224 (Roger O'Keefe, Christian J. Tams & Antonios Tzanakopoulos eds., 2013) *with* Tomuschat, *supra* note 463. As noted above, another form of pecuniary compensation—though one not framed in terms of tortious liability—may arise under IHRL.

468.  Another form of tortious liability—one that, in principle, establishes jurisdiction for serious violations of IHL to national courts in accordance with the principle of universal jurisdiction—may be relevant, though perhaps more in theory than in practice, at least under current interpretations. *See, e.g.,* Tomuschat, *supra* note 463. Under the Alien Tort Claims Act (ATCA), federal judges "shall have original jurisdiction of any civil action by an alien for a tort only, committed in violation of the law of nations or a treaty of the United States." 28 U.S.C. § 1350 (2012). Actions have been filed under the ATCA against foreign governments and foreign corporations, as well as against the U.S. government. Yet recent judicial interpretations have narrowed the statute's scope of application. *See, e.g.,* Ingrid Wuerth, *Kiobel v. Royal Dutch Petroleum Co.: The Supreme Court and the Alien Tort Statute,* 107 Am. J. Int'l L. 601 (2013).

of States and Their Property (UNCSI), which is not yet in force, and its customary analogue (if any).[469]

# INDIVIDUAL RESPONSIBILITY UNDER INTERNATIONAL LAW

As noted in section 3, a natural person may be held responsible under international law for committing an international crime connected with a war algorithm, including certain war crimes and crimes against humanity. To impose that liability, the judicial body would need to be able to understand the underlying war algorithm so as to adjudicate the legal parameters applicable in relation to it. Also as noted above, commentators have raised a number of concerns as to whether international law concerning individual responsibility for international crimes is suitable to address AWS, especially in relation to certain modes of responsibility, such as command and superior responsibility, and to mental elements (especially the requisite knowledge and intent).

This axis describes international and domestic avenues through which an individual may be held responsible for committing an international crime. We also briefly highlight another avenue—extraterritorial jurisdiction not in respect of internationally defined crimes—through which an individual may be held responsible in relation to the design, development, or use of a war algorithm.

## *INTERNATIONAL CRIMES*

### International Criminal Tribunals

As noted in section 3, where it has jurisdiction, an international criminal court or tribunal may impose individual responsibility for the commission of international crimes. The ICC—which operates pursuant to the principle of complementarity to national jurisdictions—is the first such court established on a permanent basis. Numerous war crimes under the ICC's jurisdiction may in principle be committed through the design, development, or use of war algorithms.

---

469.   *See generally* Foakes & O'Keefe, *supra* note 467. The form of pecuniary compensation here, which is based on a municipal tort law of the forum state, is distinguishable from the innovative "war tort" idea articulated by Rebecca Crootof, which is based on serious violations of IHL; however, the two might interface where a municipal tort is linked to a serious violation of IHL. *See* Crootof, *War Torts*, *supra* note 20, at 2. Crootof argues that "just as the Industrial Revolution fostered the development of modern tort law, autonomous weapon systems highlight the need for 'war torts': serious violations of international humanitarian law that give rise to state responsibility." *Id.* She believes that a "successful ban on autonomous weapon systems is unlikely (and possibly even detrimental)." *Id.* Instead, in her view, "what is needed is a complementary legal regime that holds states accountable for the injurious wrongs that are the side effects of employing these uniquely effective but inherently unpredictable and dangerous weapons." *Id.*

## Suppression of Grave Breaches

Under the Geneva Conventions of 1949, states parties are obliged "to enact any legislation necessary to provide effective penal sanctions for persons committing, or ordering to be committed, any of the grave breaches of the" relevant instrument.[470] In principle, a war algorithm may be involved in the commission of such a breach. Each state party is obliged "to search for persons alleged to have committed, or to have ordered to be committed, such grave breaches, and shall bring such persons, regardless of their nationality, before its own courts."[471] And each state party "may also, if it prefers, and in accordance with the provisions of its own legislation, hand such persons over for trial to another" state party, so long as that party has "made out a prima facie case."[472]

## Universal Jurisdiction

While "[s]tates generally do not have jurisdiction to define and punish crimes committed abroad by and against foreign nationals," pursuant to universal jurisdiction "any State has the right to try a person with regard to certain internationally defined crimes."[473] Originally, this "jurisdiction was recognized only with respect to piracy on the high seas."[474] But "[a]s the human rights content of international law expanded, universal adjudicative jurisdiction also expanded to embrace universally condemned crimes and may now apply to slavery, genocide, torture, and war crimes."[475] Such "[u]niversal jurisdiction to try these offences is not limited to situations in which they are committed on the high seas or in other areas outside the territory of any State, but generally confers no enforcement power to enter foreign territory or board a foreign ship without consent."[476] Nonetheless, "[a]lthough the laws of each State define the offences over which its courts may exercise universal jurisdiction, the scope of legislative jurisdiction is limited by the fact that the offences subject to universal jurisdiction are determined by treaty and international law."[477] As a practical matter, to date the exercise of domestic universal jurisdiction has arguably been the strongest form (even if not very strong over all) of enforcement of accountability for war crimes.

---

470.   GC I, *supra* note 348, at art. 49; GC II, *supra* note 348, at art. 50; GC III, *supra* note 348, at art. 129; GC IV, *supra* note 348, at art. 146. *See also* AP I, *supra* note 12, at art. 85.

471.   GC I, *supra* note 348, at art. 49; GC II, *supra* note 348, at art. 50; GC III, *supra* note 348, at art. 129; GC IV, *supra* note 348, at art. 146. *See also* AP I, *supra* note 12, at art. 85.

472.   GC I, *supra* note 348, at art. 49; GC II, *supra* note 348, at art. 50; GC III, *supra* note 348, at art. 129; GC IV, *supra* note 348, at art. 146. *See also* AP I, *supra* note 12, at art. 85.

473.   Bernard H. Oxman, *Jurisdiction of States*, *in* Max Planck Encyclopedia of Public International Law ¶ 37 (2007).

474.   *Id*. at ¶ 38.

475.   *Id*. at ¶ 39.

## OTHER AVENUES

Certain other individual accountability avenues might arise even where the design, development, or use of a war algorithm attributable to a natural person does not give rise to individual responsibility under international law for an international crime. One such avenue to consider is extraterritorial jurisdiction, which more and more states are turning to in order to protect their perceived interests.

### Extraterritorial Jurisdiction

Extraterritorial jurisdiction refers "to the competence of a State to make, apply and enforce rules of conduct in respect of persons, property or events beyond its territory."[478] Traditionally, the exercise of extraterritorial jurisdiction was viewed as available only in exceptional circumstances.[479] But today, more and more states are creating such regimes.

The background idea is that, with respect to conduct occurring beyond a state's territory, the state perceives the need to protect not only its own interests but also the interests of international society.[480] States have perceived those interests in such areas as anti-trust and competition law, anti-terrorism law, and anti-bribery law.

Certain characteristics of war algorithms—including that some of the underlying technologies are developed by transnational corporations and the modularity of the technology—might lead states to perceive strong interests in making, applying, and enforcing war-algorithm rules of conduct beyond their territories. Where states do so, it may be important to be attentive to the distinctions between the different ways that states may exercise extraterritorial jurisdiction. That is because some of those methods "are more likely to conflict with the competence of other States and therefore more likely to raise questions as to their compatibility with international law."[481]

---

476. *Id.*

477. *Id.*

478. Menno T. Kamminga, *Extraterritoriality*, *in* Max Planck Encyclopedia of Public International Law ¶ 1 (2012).

479. *See id.* at ¶ 3.

480. *See id.* at ¶ 4.

# SCRUTINY GOVERNANCE

Along this axis, accountability is framed in terms of the extent to which a person or entity is and should be subject to, or should exercise, forms of internal or external scrutiny, monitoring, or regulation concerning a war algorithm.[482] Notably, scrutiny governance does not hinge on—but might implicate—potential and subsequent liability or responsibility.[483] The basic notion is that there are a number of avenues—other than or alongside of legal responsibility—to hold oneself or others answerable for the exercise of war-algorithm power and authority. We highlight only a few of the various possible approaches: independent monitoring, norm (including legal) development, non-binding resolutions and codes of conduct, normative design of technical architectures, and community self-regulation.

## *INDEPENDENT MONITORING*

A vast array of institutions independently monitor compliance with law and regulations that may be relevant to war algorithms. Those institutions include bodies within international organizations, treaty-based weapons-control regimes, and non-governmental organizations. Note, however, that the existence of all of these institutions does not absolve any state from its independent duty to ensure its own compliance with international law in general and with IHL in particular. While the competence of these institutions is not explicitly stated in war-algorithm terms, their general purviews would encompass monitoring of at least certain elements of the development and operation of those algorithms. Included among those institutions are:

- The U.N. Security Council;[484]

- The U.N. General Assembly;[485]

---

481.  *Id.* at ¶ 1.

482.  Derived in part from International Law Association, *supra* note 35, at 5.

483.  The obligation to review weapons, means, and methods of warfare laid down in Article 36 of AP I and the customary law cognate (if any), discussed above, constitutes a form of required scrutiny that directly implicates legal responsibility.

484.  *See* U.N. Charter art. 25, 39–42.

485.  Under the U.N. Charter, "[t]he General Assembly may discuss any questions or any matters within the scope of the … Charter or relating to the powers and functions of any organs provided for in the … Charter, and, except as provided in Article 12, may make recommendations to the Members of the United Nations or to the Security Council or to both on any such questions or matters." U.N. Charter art. 10. Among its explicit competences laid

- The U.N. Secretariat, including the Secretary-General,[486] the Office of the United Nations High Commissioner for Human Rights (OHCHR), and the U.N. Office for Outer Space Affairs (UNOOSA);

- The Human Rights Council, including Special Procedures (Special Rapporteurs);[487]

- Treaty-based human-rights and weapons-monitoring bodies and mechanisms;[488] and

- Non-governmental organizations.[489]

# NORM DEVELOPMENT (INCLUDING OF INTERNATIONAL LAW)

Norms may be developed through formal or informal mechanisms.

With respect to international law, for instance, the U.N. "General Assembly shall initiate studies and make recommendations for the down in the U.N. Charter, "[t]he General Assembly may consider the general principles of co-operation in the maintenance of international peace and security, including the principles governing disarmament and *the regulation of armaments*, and may make recommendations with regard to such principles to the Members or to the Security Council or to both." U.N. Charter art. 11 (emphasis added). And "[t]he General Assembly may call the attention of the Security Council to situations which are likely to endanger international peace and security." *Id.*

486.   Pursuant to the U.N. Charter, "[t]he Secretary-General may bring to the attention of the Security Council any matter which in his opinion may threaten the maintenance of international peace and security." U.N. Charter art. 99. An inherent right to investigate in connection with this power has been invoked by several Secretaries-General. Katja Göcke & Hubertus von Mohr, *United Nations, Secretary-General*, *in* Max Planck Encyclopedia of Public International Law ¶ 18 (2013). The rationale is that "[s]ince it is necessary for the Secretary-General to have comprehensive knowledge of the situation in the conflict area before taking action, his authority [to bring any relevant matter to the attention of the Security Council] must encompass the right to conduct investigations and to implement preparatory fact-finding missions." *Id.* at ¶ 20. According to Katja Göcke and Hubertus von Mohr, this power has proven its value especially "since States may for various reasons be reluctant to bring certain matters before the Security Council…." *Id.* at ¶ 19.

487.   *See, e.g.*, Christof Heyns (Special Rapporteur on Extrajudicial, Summary or Arbitrary Executions), *Rep. to Human Rights Council*, UN Doc. A/HRC/23/47 (Apr. 9, 2013).

488.   *See, e.g.*, Human Rights Committee; Committee on Economic, Social and Cultural Rights (CESCR); Committee against Torture (CAT); Committee on the Rights of the Child (CRC); and Organisation for the Prohibition of Chemical Weapons (OPCW).

489.   *See, e.g.*, the Steering Committee of the Campaign to Stop Killer Robots (Human Rights Watch, Article 36, Association for Aid and Relief Japan, International Committee for Robot Arms Control, Mines Action Canada, Nobel Women's Initiative, PAX, Pugwash Conferences on Science & World Affairs, Seguridad Humana en América Latina y el Caribe, and Women's International League for Peace and Freedom). *About Us*, Campaign to Stop Killer Robots, https://www.stopkillerrobots.org/about-us/ (last visited Aug. 25, 2016).

purpose of … encouraging the progressive development of international law and its codification."[490] The U.N. General Assembly established its Legal Committee (Sixth Committee), which "is responsible for the UN General Assembly's role in encouraging the codification and progressive development of international law."[491] The workings of the Sixth Committee led to the establishment of the International Law Commission (ILC).[492] According to its Statute, the ILC is expected to bring onto its agenda only topics that are "necessary and desirable"[493]—or, "[i]n other words, only topics 'ripe' for codification and progressive development of international law are to be the subject of its work."[494] This criterion leaves some room for the ILC to consider various topics as possible candidates for its work. Broadly speaking, "a topic may be considered ripe if the subject-matter regulates the essential necessities of States or *the wider needs and/ or contemporary realities of the international community* or is one held central to the authority of international law, *notwithstanding any existing disagreements among States on the topic*."[495] In principle, war algorithms could arguably fit that definition.

Norms and accompanying standards relevant to war algorithms may also be developed at levels other than international law. Pursuant to their legislative jurisdiction, states may promulgate municipal laws.[496] Moreover, whether pursuant to domestic law or regulations or to less formal bases, agencies, regulatory bodies, and other standards-setting entities—governmental or non-governmental—may articulate guidelines, standards, and the like.[497]

---

490. U.N. Charter art. 13.

491. Huw Llewellyn, *United Nations, Sixth Committee, in* Max Planck Encyclopedia of Public International Law ¶ 1 (2012).

492. Pemmaraju Sreenivasa Rao, *International Law Commission (ILC), in* Max Planck Encyclopedia of Public International Law ¶ 3 (2013) (citing to G.A. Res. 174 (II) (November 1947)).

493. Statute of the International Law Commission, art. 18(2), GA Res. 174(II), UN Doc. A/519 (1947).

494. Rao, *supra* note 492, at ¶ 6.

495. *Id.* (emphasis added).

496. *See, e.g.,* Public Law 100-180, § 224 ("No agency of the Federal Government may plan for, fund, or otherwise support the development of command and control systems for strategic defense in the boost or post-boost phase against ballistic missile threats that would permit such strategic defenses to initiate the directing of damaging or lethal fire except by affirmative human decision at an appropriate level of authority."). *But see* Law of War Manual, *supra* note 110, at § 6.9.5.4 n.111 ("This statute may, however, be an unconstitutional intrusion on the President's authority, as Commander in Chief, to determine how weapons are to be used in military operations.").

497. *See, e.g.,* DOD AWS Dir., *supra* note 91; Hui-Min Huang et al., *Autonomy Levels for*

# NON-BINDING RESOLUTIONS AND DECLARATIONS, AND INTERPRETATIVE GUIDES

While not laying down legal obligations, non-binding resolutions and declarations, as well as codes of conduct or informal manuals, may also contribute to the development of the normative framework concerning war algorithms. This has already occurred in relation to AWS: a 2014 resolution of the European Parliament "[c]alls on the High Representative for Foreign Affairs and Security Policy, the Member States and the Council to ... ban the development, production and use of fully autonomous weapons which enable strikes to be carried out without human intervention."[498]

Moreover, at the 2016 CCW Informal Meeting of Experts, the Netherlands called "for the formulation of an interpretative guide that clarifies the current legal landscape with regard to the deployment of autonomous weapons."[499] In recent years, a number of "Manuals"[500] as well as an "Interpretive Guide"[501] on international law pertaining to armed conflict in relation to certain thematic areas have been drafted. It is unclear whether the initiative called for by the Netherlands will align with these approaches or might take another form. But based on the initial

---

*Unmanned Systems (ALFUS) Framework*, Volume II: Framework Models, NIST Special Publication 1011-II-1.0, Version 1.0 (2007), http://www.nist.gov/el/isd/ks/upload/ALFUS-BG.pdf; Jessie Y.C. Chen; Ellen C. Haas, Krishna Pillalamarri & Catherine N. Jacobson, "Human-Robot Interface: Issues in Operator Performance, Interface Design, and Technologies," U.S. Army Research Laboratory, ARL-TR-3834 (July 2006).

498. European Parliament Resolution on the Use of Armed Drones ¶ H.2(d) (2014/2567(RSP)) (Feb. 25, 2014),

499. Henk Cor van der Kwast, Perm. Rep. of Neth. to the Conference on Disarmament, Opening Statement at the 2016 Informal Meeting of Experts, at 4, UN Office in Geneva (April 11, 2016), http://www.unog.ch/80256EDD006B8954/(httpAssets)/FC2E59B32F14D791C1257F920057CAE6/$file/2016_LAWS+MX_GeneralExchange_Statements_Netherlands.pdf. *See also* Steven Groves, *A Manual Adapting the Law of Armed Conflict to Lethal Autonomous Weapons Systems* (Heritage Foundation, Special Report No. 183, 2016), http://www.heritage.org/research/reports/2016/04/a-manual-adapting-the-law-of-armed-conflict-to-lethal-autonomous-weapons-systems.

500. *E.g.*, Tallinn Manual on the International Law Applicable to Cyber Warfare (Michael Schmitt ed., 2013); Program on Humanitarian Policy and Conflict Research, Manual on International Law Applicable to Air and Missile Warfare (2009); International Institute of Humanitarian Law, San Remo Manual on International Law Applicable to Armed Conflicts at Sea (1995). *See also Project on a Manual on International Law Applicable to Military Uses of Outer Space* (MILAMOS), https://www.mcgill.ca/milamos/home (last visited Aug. 27, 2016).

501. Nils Melzer (ICRC), Interpretive Guidance on the Notion of Direct Participation in Hostilities Under International Humanitarian Law (2009).

articulation, it appears that the focus of the called-for "interpretative guide" will be on clarifying currently applicable law concerning the deployment of autonomous weapons.

# NORMATIVE DESIGN OF TECHNICAL ARCHITECTURES

Programmers, engineers, and others involved in the design, development, and use of war algorithms might take diverse measures to embed normative principles into those systems. The background idea is that code and technical architectures can function like a kind of law. Maximizing the auditability of that code—especially in light of legally-relevant concepts such as attribution and reconstructability—might help strengthen external and internal scrutiny mechanisms.

To increase the likelihood of being adopted, such normative-design approaches would likely need to be devised in a manner that takes due consideration of the tension between, on one side, external transparency, and, on the other, a state's interest in protecting classified technologies as well as the intellectual-property interests associated with those technologies. In addition, those thinking through ways to pursue war-algorithm accountability along this avenue should critically assess the experience of attempting to regulate cyber operations and cyber "warfare." So far, those areas have eluded a universal normative regime. Like war algorithms, cyber operations and cyber "warfare" raise concerns regarding intellectual-property interests, the modularity and dual-use nature of the technologies, transparency with external actors due to classification regimes, and maintaining a qualitative edge.

## Designing "Morally Responsible Engineering" and a "Partnership Architecture"

Some governments have recognized the importance of incorporating moral and ethical considerations into the engineering of systems that might be relevant to war algorithms.

In an October 2015 report on AWS, a Dutch "advisory committee advocates taking the interaction between humans and machines into account sufficiently in the design phase of autonomous weapon systems."[502] Furthermore, "[i]n light of the importance of attributing responsibility and accountability, the [advisory committee] believes that, when procuring autonomous weapons, the government should ensure that the concept of morally responsible engineering is applied during the design stage." For their part, the Ministries of Foreign Affairs and Defense consider that "recommendation to be an affirmation of existing policy,"[503] and emphasize that "the government and several of its knowledge partners are studying this theme."[504]

---

502.   DUTCH GOVERNMENT, RESPONSE TO AIV/CAVV REPORT, *supra* note 22.

Among the research programs funded by the Dutch government was a project entitled "Military Human Enhancement: Design for Responsibility and Combat Systems," which was carried out by Delft University of Technology. One of the articles published as part of that project put forward the idea of a "partnership architecture."[505] Two components undergird this idea. First, a mechanism is put forward through which both parties—the human and the machine—"do their job *concurrently*. In this way, each actor arrives at an own interpretation of the world thereby constructing a human representation of the world and a machine representation of the world at the same time."[506] Second, work agreements—"explicit contracts between the human and the machine about the division of work"—are used to "minimize[] the automation-human coordination asymmetry because working agreements define an a priori explicit contract [regarding] what [to] and what not to delegate[] to the automation."[507]

The main idea is that the resulting "partnership architecture can protect a commitment to responsibility within the armed forces."[508] On one hand, "operators will be responsible for the terms of their working agreements with their machine."[509] And on the other, working agreements may help "ensure that operators receive the morally relevant facts needed to make decisions that comply with IHL, as well as key moral principles."[510]

---

503. This approach aligns in certain respects with the focus on systems engineering discussed in the UK MoD Joint Doctrine Note on unmanned aircraft systems. The authors of that document state that "[i]n order to ensure that new unmanned aircraft systems adhere to present and future legal requirements, it is likely that a systems engineering approach will be the best model for developing the requirement and specification." U.K. MINISTRY OF DEF., *supra* note 113, at 5-2. Using such an approach, according to the Joint Doctrine Note authors, "the legal framework for operating the platform would simply form a list of capability requirements that would sit alongside the usual technical and operational requirements." *Id.* In turn, "[t]his would then inform the specification and design of various sub-systems, as well as informing the concept of employment." *Id.*

504. DUTCH GOVERNMENT, RESPONSE TO AIV/CAVV REPORT, *supra* note 22.

505. *See* Tjerk de Greef & Alex Leveringhaus, Design for Responsibility: Safeguarding Moral Perception via a Partnership Architecture, 17 COGNITION, TECHNOLOGY & WORK 319 (2015).

506. *Id*. at 326 (emphasis original).

507. *Id*. (citations omitted).

508. *Id*. at 327.

509. *Id*. The authors note that "[t]his raises issues about foresight, negligence and so on that we cannot tackle here." Rather, "[f]or now, it suffices to note that the operator remains firmly control of his machine—even if there is a physical distance between them or that the machines operates at increased levels of automation." *Id.*

## Coding Law

Software and hardware engineers, roboticists, and others involved in the development of war algorithms may consider taking a page from the internet playbook. The internet protocol suite (also known as TCP/IP) is a core set of protocols that define the way in which the internet functions. A fundamental choice at the heart of the internet's architecture concerned defining the flow of information by allowing ordinary computers connected to the internet to not only receive but also to send information. This was neither a necessary nor inevitable feature of the internet. (And whether one sees it today as a feature or a bug depends on one's vantage point.) The suite of protocols could have been designed in other ways—for instance, the system could have distributed packets from a centralized hub, precluding individual computers to communicate directly with each other.

Lawrence Lessig argues that, through that structuring, TCP/IP embeds some regulatory—perhaps normative—principles in the design of the system.[511] Put another way, in defining the way in which computers could share data and communicate with one another, TCP/IP also forecloses alternative methods of communication, thereby imposing, if implicitly, regulations on the way in which the internet functions. In this way, *code is a kind of law* because it *enables* computers to do certain things (such as exchange packets of information) but, in doing so, also *indirectly defines and narrows* the specific way in which that exchange is accomplished. (It merits mention that code functions as a type of law in this conception irrespective of whether that was the intention of the system's designers.)

At the 2016 CCW Informal Meeting of Experts on Lethal Autonomous Weapons Systems, Danièle Bourcier imported Lessig's general idea into the specific discussion on AWS where she raised the notion of designing "humanitarian law" into the relevant technical system.[512] What this might mean in practice is unclear. But in principle it might concern the design of the underlying algorithms as well as the constructed systems through which those algorithms are effectuated.

---

510. *Id.*

511. *See generally* LAWRENCE LESSIG, CODE AND OTHER LAWS OF CYBERSPACE (1999).

512. Danièle Bourcier, Centre national de la recherche scientifique, Artificial Intelligence & Autonomous Decisions: From Judgelike Robot to Soldier Robot, Address at the 2016 Informal Meeting of Experts, UN Office in Geneva (April 2016), available at http://www.unog.ch/80256EDD006B8954/(httpAssets)/338ABCC8C57BB09CC1257F9A0045197A/$file/2016_LAWS+MX+Presentations_HRandEthicalIssues_Daniele+Vourcier.pdf.

## Auditable Algorithms

Making war algorithms more auditable may help foster accountability over them. "Audit logs," for instance, record activity that takes place in an information architecture. In the U.S., national-security fusion centers "are supposed to employ audit logs that record the activity taking place in the information-sharing network, including 'queries made by users, the information accessed, information flows between systems, and date- and time-markers for those activities.'"[513] (A fusion center is designed to promote information-sharing and to streamline intelligence-gathering, not only at the federal level between various agencies but also among the U.S. military and state- and local-level government.) In addition to the national-security realm, audit logs or similar mechanisms are mandated with respect to certain credit-rating agencies, financial transactions, and healthcare software. To be effective, audit logs need to be immutable.[514] While not specifically addressing AWS, the UK MoD Joint Doctrine Note on unmanned aircraft systems states that "[a] complex weapon system is also likely to require an authorisation and decisions log, to provide an audit trail for any subsequent legal enquiry."[515]

## *COMMUNITY SELF-REGULATION*

A recent call for self-imposed regulation by a group of expert scientists in the domain of genetic engineering may provide a regulatory model for those involved in the development of war algorithms. The basic idea is that, even where there is no or little formal regulation, a community can choose, on its own initiative, to delineate what is and is not acceptable and to self-police the resulting boundaries.

The plea by some leading scientists partly concerned a relatively easy-to-use gene-editing technique called CRISPR/Cas9. (Gene-editing techniques, in short, "use enzymes called nucleases to snip DNA at specific points and then delete or rewrite the genetic information at those locations."[516]) CRISPR/Cas9 had "suddenly made it possible to cross [a]

---

513.   Pasquale, *supra* note 1, at 157 (citing to Markle Task Force on National Security in the Information Age, Implementing a Trusted Information Sharing Environment: Using Immutable Audit Logs to Increase Security, Trust, and Accountability, at I (2006), http://research.policyarchive.org/15551.pdf).

514.   *See* Pasquale, *supra* note 1, at 157; *see also id.* at 159 (stating that "[i]f immutable audit logs of fusion centers are regularly reviewed, misconduct might be discovered, wrongdoers might be held responsible, and similar misuses might be deterred") (citation omitted).

515.   U.K. Ministry of Def., *supra* note 113, at 5-6. *See also* DOD AWS Dir., *supra* note 91 (establishing audit-like requirements in DoD policy).

516.   David Cyranoski, *Ethics of Embryo Editing Divides Scientists*, 519 Nature 272, 272 (2015).

Rubicon": "[f]or decades, the ability to make changes that could be inherited in the human genome has been viewed as a fateful decision — but one that could be postponed because there was no safe and efficient way to edit the genome."[517] With CRISPR/Cas9, it has been said, "the long theoretical issue now requires practical decisions."[518]

In December 2015, the Organizing Committee for the International Summit on Human Gene Editing came to an agreement on "a recommendation not to stop human-gene-editing research outright, but to refrain from research and applications that use modified human embryos to establish a pregnancy."[519] More specifically, intensive basic and preclinical research should proceed, the Committee said, but that research should be "subject to appropriate legal and ethical rules and oversight, on (i) technologies for editing genetic sequences in human cells, (ii) the potential benefits and risks of proposed clinical uses, and (iii) understanding the biology of human embryos and germline cells."[520] And "[i]f, in the process of research, early human embryos or germline cells undergo gene editing," the Committee entreated, "the modified cells should not be used to establish a pregnancy."[521]

The Committee also called for an ongoing forum to address these issues. The push should be for "[t]he international community … [to] strive to establish norms concerning acceptable uses of human germline editing and to harmonize regulations, in order to discourage unacceptable activities while advancing human health and welfare."[522] Against this backdrop, the Committee called upon the national academies that co-hosted the summit "to take the lead in creating an ongoing international forum to discuss potential clinical uses of gene editing; help inform decisions by national policymakers and others; formulate recommendations and guidelines; and promote coordination among nations."[523] This forum, the Committee stated, "should be inclusive among nations and engage a wide range of perspectives and expertise," such as "biomedical scientists, social scientists, ethicists, health care providers, patients and their families,

---

people with disabilities, policymakers, regulators, research funders, faith leaders, public interest advocates, industry representatives, and members of the general public."[524]

Zooming out, the call for various forms of self-regulation by these scientists might be relevant for those involved in the design and development of war algorithms—another area where some are concerned about crossing a moral Rubicon. In addition to the broader point (that, alongside forms of legal responsibility, a community can raise the normative bar for itself), specific possible regulatory avenues emerge: setting boundaries on possible research and imposing moratoriums (where deemed necessary); defining legal and ethical rules and oversight mechanisms; committing to review existing regulations on an ongoing basis; and establishing forums to address enduring and emergent concerns.

---

524.  *Id.*

# 5

# CONCLUSION

Two contradictory trends may be combining into a new global climate that is at once enterprising and anxious. Militaries see myriad technological triumphs that will transform warfighting. Yet the possibility of "replacing" human judgment with algorithmically-derived "decisions"—especially in war—threatens what many consider to define us as humans.

To date, the lack of demonstrated technical knowledge by many states and commentators, the unwillingness of states to share closely-held national-security technologies, and an absence of a definitional consensus on what is meant by autonomous weapon systems have impeded regulatory efforts on AWS. Moreover, uncertainty about which actors would benefit most from advances in AWS and for how long such benefits would yield a meaningful qualitative edge over others seems likely to continue to inhibit efforts at negotiating binding international rules on the development and deployment of AWS. In this sense, efforts at reaching a dedicated international regime to address AWS may follow the same frustrations as analogous efforts to address cyber warfare. True, unlike with the early days of cyber warfare, there has been greater state engagement on regulation of AWS. In particular, the concept of "meaningful human control" over AWS has already been endorsed by over two-dozen states. But much remains up in the air as states decide whether to establish a Group of Governmental Experts on AWS at the upcoming Fifth Review Conference of the CCW.

We have shown that, with respect to armed conflict, the primary formal regulatory avenues under international law are state responsibility for internationally wrongful acts and individual criminal responsibility for international crimes. These fields are well established and offer many more avenues than are often considered in the relatively narrow AWS discourse to date. In sum, ICL and, especially, IHL already address many of the concerns raised in relation to AWS—but ICL and IHL may not be sufficient to address all of those concerns.

The current crux, as we see it, is whether advances in technology—especially those capable of "self-learning" and of operating in relation to war and whose "choices" may be difficult for humans to anticipate or unpack or whose "decisions" are seen as "replacing" human judgment—are susceptible to regulation and, if so, whether and how they should be regulated. One way to think about the core concern which vaults over at least some of the impediments to the discussion on AWS is the new concept we raise: war algorithms. War algorithms include not only those algorithms capable of being used in weapons but also in any other function related to war.

More war algorithms are on the horizon. Two months ago, the Defense Science Board, which is connected with the U.S. Department of Defense, identified five "stretch problems"—that is, goals that are "hard-but-not-too-hard" and that have a purpose of accelerating the process of bringing a new algorithmically-derived capability into widespread application:

- Generating "future loop options" (that is, "using interpretation of massive data including social media and rapidly generated strategic options");

- Enabling autonomous swarms (that is, "deny[ing] the enemy's ability to disrupt through quantity by launching overwhelming numbers of low‑cost assets that cooperate to defeat the threat");

- Intrusion detection on the Internet of Things (that is, "defeat[ing] adversary intrusions in the vast network of commercial sensors and devices by autonomously discovering subtle indicators of compromise hidden within a flood of ordinary traffic");

- Building autonomous cyber-resilient military vehicle systems (that is, "trust[ing] that … platforms are resilient to cyber‑attack through autonomous system integrity validation and recovery"); and

- Planning autonomous air operations (that is, "operat[ing] inside adversary timelines by continuously planning and replanning tactical operations using autonomous ISR analysis, interpretation, option generation, and resource allocation").[525]

---

525.  Defense Science Board, *supra* note 7, at 76–97.

What this trajectory toward greater algorithmic autonomy in war—at least among more technologically-sophisticated armed forces and even some non-state armed groups—means for accountability purposes seems likely to remain a contested issue for the foreseeable future.

In the meantime, it remains to be authoritatively determined whether war algorithms will be capable of making the evaluative decisions and value judgments that are incorporated into IHL. It is currently not clear, for instance, whether war algorithms will be capable of formulating and implementing the following IHL-based evaluative decisions and value judgments:[526]

- The presumption of civilian status in case of "doubt";[527]

- The assessment of "excessiveness" of expected incidental harm in relation to anticipated military advantage;

- The betrayal of "confidence" in IHL in relation to the prohibition of perfidy; and

- The prohibition of destruction of civilian property except where "imperatively" demanded by the necessities of war.[528]

*    *    *

Two factors may suggest that, at least for now, the most immediate ways to regulate war algorithms more broadly and to pursue accountability over them might be to follow not only traditional paths but also less conventional ones. As illustrated above, the latter might include relatively formal avenues—such as states making, applying, and enforcing war-algorithm rules of conduct within and beyond their territories—or less formal avenues—such as coding law into technical architectures and community self-regulation.

First, even where the formal law may seem sufficient, concerns about practical enforcement abound. Recently, for instance, states parties to the Geneva Conventions failed to muster the political support to establish a new IHL compliance forum.[529] There are a number of ways to interpret

526.   These concerns were raised in relation to autonomous weapon systems, but they are also implicated by war algorithms.

527.   Swiss, "Compliance-Based" Approach, *supra* note 74, *citing* art. 50(3) and art. 52(3) of Additional Protocol I to the Geneva Conventions. *See* AP I, *supra* note 12, at art. 50(3), 52(3).

528.   *Id.*, *citing* art. 23(g) of Hague Regulation IV, *see* Hague Convention (IV) Respecting the Laws and Customs of War on Land art. 23(g), Oct. 18, 1907, T.S. 539, and art. 53 of the Fourth Geneva Convention, *see* GC IV, *supra* note 349, at art. 53.

529.   *Compare* 32nd International Conference of the Red Cross and Red Crescent, *Draft "0"*

this refusal. But, at a minimum, it seems to point to a lack of political will among states to cast more light on IHL compliance. This suggests that even where existing IHL seems adequate as a regulatory regime for some aspects of the design, development, and use of AWS or war algorithms, it still lacks dependable enforcement as far as state conduct is concerned.

Second, the proliferation of increasingly advanced technical systems based on self-learning and distributed control raises the question of whether the model of individual responsibility found in ICL might pose conceptual challenges to regulating AWS and war algorithms. At a general level, this is not a wholly new concern, as distributed systems have been used in relation to war for a long time. But the design, development, and operation of those systems might be increasingly difficult to square with the foundational tenet of ICL—that "[c]rimes against international law are committed by men, not by abstract entities"[530]—as learning algorithms and architectures advance.[531]

In short, individual responsibility for international crimes under international law remains one of the vital accountability avenues in existence today, as do measures of remedy for state responsibility. Yet in practice responsibility along either avenue is unfortunately relatively

---

*Resolution on "Strengthening compliance with international humanitarian law"* (undated), https://www.icrc.org/en/download/file/13244/32ic-draft-0-resolution-on-ihl-compliance-20150915-en.pdf *with* 32nd International Conference of the Red Cross and Red Crescent, *Resolution 2* (Dec. 10, 2015), http://rcrcconference.org/wp-content/uploads/sites/3/2015/04/32IC-AR-Compliance_EN.pdf.

530.  1 Trial of the Major War Criminals Before the International Military Tribunal 223 (1947).

531.  In a related context, M.C. Elish has noted a dilemma in which "control has become distributed across multiple actors (human and nonhuman)," and yet "our social and legal conceptions of responsibility have remained generally about an individual." She thus "developed the term *moral crumple zone* to describe the result of this ambiguity within systems of distributed control, particularly automated and autonomous systems." The basic idea is that "[j]ust as the crumple zone in a car is designed to absorb the force of impact in a crash, the human in a highly complex and automated system may become simply a component—accidentally or intentionally—that bears the brunt of the moral and legal responsibilities when the overall system malfunctions." M.C. Elish, Moral Crumple Zones: Cautionary Tales in Human-Robot Interaction 3–4 (We Robot 2016 Working Paper) (March 20, 2016), http://dx.doi.org/10.2139/ssrn.2757236 (using "the terms autonomous, automation, machine and robot as related technologies on a spectrum of computational technologies that perform tasks previously done by humans" and discussing a framework for categorizing types of automation proposed by Parasuraman, Sheridan and Wickens, who "define automation specifically in the context of human-machine comparison and as 'a device or system that accomplishes (partially or fully) a function that was previously, or conceivably could be, carried out (partially or fully) by a human operator.'"). *Id.* at n.5 (citing to Parasuraman et al., "A Model for Types and Levels of Human Interaction with Automation," 30 IEEE Transactions on Systems, Man and Cybernetics 3 (2000). Elish notes that the term arose in her work with Tim Hwang. *Id.* at 3.

rare. And thus neither path, on its own or in combination, seems to be sufficient to effectively address the myriad regulatory concerns pertaining to war algorithms—at least not until we better understand what is at issue. These concerns might lead those seeking to strengthen accountability of war algorithms to pursue not only traditional, formal avenues but also less formal, softer mechanisms.

In that connection, it seems likely that attempts to change governments' approaches to technical autonomy in war through social pressure (at least for those governments that might be responsive to that pressure) will continue to be a vital avenue along which to pursue accountability. But here, too, there are concerns. Numerous initiatives already exist. Some of them are very well informed; others less so. Many of them are motivated by ideological, commercial, or other interests that—depending on one's viewpoint—might strengthen or thwart accountability efforts. And given the paucity of formal regulatory regimes, some of these initiatives may end up having considerable impact, despite their shortcomings.

Stepping back, we see that technologies of war, as with technologies in so many areas, produce an uneasy blend of promise and threat.[532] With respect to war algorithms, understanding these conflicting pulls requires attention to a century-and-a-half-long history during which war came to be one of the most highly regulated areas of international law. But it also requires technical know-how. Thus those seeking accountability for war algorithms would do well not to forget the essentially political work of IHL's designers—nor to obscure the fact that today's technology is, at its core, designed, developed, and deployed by humans. Ultimately, war-algorithm accountability seems unrealizable without competence in technical architectures and in legal frameworks, coupled with ethical, political, and economic awareness.

---

532.   On broader historical, social, and political forces that shape notions and experiences of technology, at least in the American context, *see, e.g.*, John M. Staudenmaier, *Technology*, *in* A Companion to American Thought 667–669 (Richard Wrightman Fox & James T. Kloppenberg eds., 1995).

# BIBLIOGRAPHY

## ONLINE RESOURCES

The Ethical Autonomy Project of the Center for a New American Security (CNAS) maintains a Bibliography at http://www.cnas.org/research/defense-strategies-and-assessments/20YY-Warfare-Initiative/Ethical-Autonomy/bibliography.

The United Nations Office in Geneva maintains web pages that host papers produced by and statements given by states—as well as materials produced by non-state commentators—in relation to "Lethal Autonomous Weapon Systems" in the context of the CCW at http://www.unog.ch/80256EE600585943/(httpPages)/8FA3C2562A60FF81C1257CE600393DF6?OpenDocument.

## SOURCES

# APPENDIX I

## STATES' MOST COMMON INTERNATIONAL-LAW REFERENCE POINTS AT THE 2015 AND 2016 CCW MEETINGS OF EXPERTS ON LAWS

| | AWS clearly violates | Concerns in relation to AWS | A consideration in relation to AWS |
|---|---|---|---|
| Reference to conduct of hostilities | Cuba: "No se podría emplear estas armas con plenas garantías de cumplimiento y observancia de las normas y principios del Derecho Internacional Humanitario (DIH). No podría garantizarse la distinción entre civiles y combatientes, ni la evaluación de proporcionalidad, entre otros principios básicos del DIH." (General statement; National Document/Position Paper also uses latter sentence) (2015)[*]<br><br>Ecuador: "Nos preocupan aspectos fundamentales que merecen ser analizados y discutidos en profundidad como… incumplimiento del Derecho Internacional Humanitario en cuanto a…las normas de la distinción, la proporcionalidad y las precauciones en los ataques…. La Constitución del Ecuador…prohíbe y condena el desarrollo y uso de armas de destrucción masiva y de armas de efectos indiscriminados violatorias del Derecho Internacional Humanitario como es el caso de los Drones armados y sería el caso de las Armas Letales Autónomas." (General statement) (2015)[†]<br><br>Pakistan: "LAWS would not distinguish between combatants and non-combatants…. Whilst automated weapons and automatic weapons have to some degree a 'human in the loop', autonomous implies no scope for such 'interference' by | Argentina: "Resulta obvio que los principios humanitarios del DIH sobre proporcionalidad y discriminación aplicado a los SALA se encuentran visiblemente comprometidos y con numerosas alternativas de incumplimiento según se comporten un número de variables que intervienen en su uso. Es conveniente que cualquier desarrollo de los SALA, este sujeto a que se demuestre en forma indubitable que las mismas poseen la capacidad de discriminar y de diferenciar la proporcionalidad conforme las instrumentos legales existentes." (General statement) (2015)[‡]<br><br>Austria: "To take just the example of the IHL principle of distinction: today, clearly only humans are capable to distinguish reliably between civilians and combatants in a real combat situation, thereby ensuring observance of the principle. Whether technology will be able to create at some future point machines with an equivalent capability seems to be a matter of speculation at this stage. In any case, the blurring of the fundamental distinction between the military and civilian spheres, between front and rear, as an ever more prominent feature of modern warfare, does not make this an easy task." (General statement) (2015) | Denmark: "All use of force - including the use of autonomous weapon systems - must be in compliance with international humanitarian law, i.e. the fundamental rules of distinction, proportionality and precautions in attack." (General statement) (2015)<br><br>France: "D'un point de vue juridique, je crois que de nombreuses délégations ont souligné l'importance du respect du DIH dans les phases de développement et d'emploi des SALA. La France estime que les principes du DIH s'appliquent pour encadrer le développement et l'emploi des SALA." (General statement) (2015)[§§]<br><br>France: "Il est aujourd'hui trop tôt pour savoir si l'on pourra un jour développer des SALA conformes dans leur emploi aux principes de discrimination et de proportionnalité du DIH, mais nous ne pouvons pas prévoir les progrès techniques à venir. Par ailleurs, comme cela a été rappelé plusieurs fois dans cette enceinte, tout dépend du milieu dans lequel ces systèmes seront déployés : l'incapacité présumée de ces systèmes à distinguer un civil d'un combattant ne pose problème que dans un environnement où la machine aura à faire cette distinction entre civils et combattants, ce qui n'est pas toujours le cas. Tous les champs de bataille ne |

---

[*] "It would not be possible to employ these weapons with full guarantees of the fulfillment and observance of the norms and principles of IInternational Humanitarian Law (IHL). It would not be possible to guarantee distinction between civilians and combatants or the evaluation of proportionality, among other basic principles of IHL."

[†] "We worry about fundamental aspects that deserve to be analyzed and discussed in depth like…the breach of IHL in regard to…the norms of distinction, proportionality, and precaution in attacks….The Constitution of Ecuador…prohibits and condemns the development and use of weapons of mass destruction and of arms which are in effect indisrriminatory in breach of International Humanitarian Law as true of armed drones and would be true of Lethal Autonomous Weapons."

[‡] "It is obvious that the humanitarian principles of IHL about proportionality and discrimination applied to LAWS would find themselves visibly compromised and with numerous other options in violation depending on the involvement of a number of variables that could be part of their use. It is advisable that whatever the development of LAWS, it is shown in indubitable form that LAWS have the capacity to discriminate and distinguish proportionality in conformance with existing legal instruments."  "From a legal point of view, I believe that numerous delegations have underlined the important of respect for IHL in the phases of development and employment of LAWS. France believes that IHL principles apply to frame [or regulate] the development and employment of LAWS."

[§§] "From a legal point of view, I believe that numerous delegations have underlined the important of respect for IHL in the phases of development and employment of LAWS. France believes that IHL principles apply to frame [or regulate] the development and employment of LAWS."

any human, calling into question the principles of IHL: distinction, proportionality, precaution, humanity and military necessity…. Besides depriving the combatants of the targeted state, the protection offered to them by the international law of armed conflict, LAWS would also risk the lives of civilians and non-combatants on both sides. It remains unclear as to how "combatants" will be defined in case of LAWS. Will targets be chosen based on an algorithm that recognizes certain physical characteristics, for example, "beards and turbans"? Also, there are questions of the protection of those who are not; or no longer, taking part in fighting: "hors de combat". How will LAWS distinguish between noncombatants from combatants or hors de combat? Can a machine be trusted to have the same or better discerning abilities as a human? These questions remain unanswered…. Like any other complex machine, LAWS can never be fully predictable or reliable. They could fail for a wide variety of reasons including human error, malfunctions, degraded communications, software failures, cyber attacks, jamming and spoofing, etc. There will always be a level of uncertainty about the way an autonomous weapon system will interact with the external environment." (General statement) (2015)

Pakistan: "LAWS cannot be programmed to comply with International Humanitarian Law (IHL), in particular its cardinal rules of distinction, proportionality, and precaution. These rules can be complex and entail subjective decision making requiring human judgment. The introduction of fully autonomous weapons in the battlefield would be a major leap backward on account of their profound implications on norms and behaviour that the world has painstakingly arrived at after centuries of warfare. We firmly believe that developments in future military technologies should follow

Austria: "This obligation covers obligations under treaties and customary international law, inter alia the prohibition of indiscriminate attacks as well as the prohibition to cause avoidable injury and unnecessary suffering. New weapons need to comply inter alia with the following three fundamental IHL principles, namely the principle of proportionality, distinction and precaution. Under the proportionality principle, the evaluation of military advantage has to be assessed against the potential incidental loss of civilian life, injury to civilians or damage to civilian objects that may be expected from an attack and that must not be excessive in relation to the concrete and direct military advantage. It is necessary to examine whether a reasonably wellinformed person in the circumstances of the actual perpetrator, making reasonable use of the information available to him or her could have expected excessive casualties to result from the attack. Proportionality thus requires a distinctively human judgement ("common sense", "reasonable military commander standard"). The assessment must be based on information reasonably available not only at the time of the planning of the attack, but need to remain valid throughout the weapon's use. The principle of proportionality requires therefore an immediate temporal link between the assessment and the factual deployment of the weapon. LAWS usually are programmed well before the weapon actually attacks. Such a correct evaluation under the proportionality principle can be a particularly challenging or impossible task in populated areas where the situation changes rapidly. Under these circumstances it would be impossible to weigh anticipated military advantage against the expected collateral harm. Whether an attack complies with the rule needs to be assessed on a case-by-case basis, depending on the specific context and considering the totality of circumstances and should be done in a temporal comprennent pas de civils. Les SALA sont donc soumis à une forte logique de milieu et leur déploiement dans les milieux spatiaux et sous-marins, par exemple, semble a priori poser moins de problems." (Ethics/Overarching issues statement) (2015)[*]

Greece: "For the sake of argument, let us suppose that in the future autonomous weapon systems are developed which can fully comply with IHL and its cardinal principles, such as distinction, proportionality and precautions in attack." (Challenges to IHL statement) (2015)

Israel: "Humans who intend to develop and employ a lethal autonomous weapon system, are responsible to do so in a way that ensures the system's operation in accordance with the rules of IHL. In this regard, the context – referring to the specific system and the specific scenario of use – is of utmost importance. The characteristics and capabilities of each system must be adapted to the complexity of its intended environment of use. Where deemed necessary, the system's operation would be limited by, for example, restricting the system's operation to a specific perimeter, during a limited timeframe, against specific types of targets, to conduct specific kinds of tasks, or other such limitations which are all set by a human. Likewise, for example, if necessary, a system could be programmed to refrain from action when facing complexities it cannot resolve." (Characteristics of LAWS statement) (2015)

Poland: "The main principles of IHL which are of interest to us would be: humanity, military necessity, discrimination and proportionality…. Looking at the present level of technological advancement, however, there are reasons for concern that the existing systems will not be able to meet

the established law and not vice versa." (General Statement) (2016)

Pakistan: "Besides depriving the combatants of the targeted state the protection offered to them by the international law of armed conflict, LAWS would also risk the lives of civilians and non-combatants. The unavailability of a legitimate human target of the LAWS user State on the ground could lead to reprisals on its civilians including through terrorist acts." (General Statement) (2016) proximity to the attack. IHL further prohibits attacks on persons hors de combat under the principle of distinction. Although the ability of LAWS to comply with this rule will depend on its recognition technology and the environment in which it is used, it seems problematic to leave the assessment of whether an individual is hors de combat to a robotic weapon. It does not seem realistic that LAWS could distinguish whether someone is wounded or whether a soldier is in the process of surrendering or interpret human behavior as would be necessary. Furthermore, the principle of precaution, requiring that an attack must be cancelled or suspended if it becomes apparent that the objective is not a military one or is subject to special protection or that it would violate the rule of proportionality, is also challenged by LAWS. Even if humans take feasible planning precautions, their plans will need to remain relevant when the system makes the decision to launch an attack. This seems unlikely to be realistic in dynamic environments and in the absence of human override.... The assessment of compliance with the existing standards and rules under IHL has to be taken in a contextual manner in the light of concrete circumstances. Circumstances in the battlefield are shifting and human control of a weapon is a necessary prerequisite. IHL does require that combatants can make an objective assessment of the facts when applying force and targeting an objective. This assessment must be made on a case-by-case basis, in view of the concrete circumstances. In this context, the concurrence and inter-action of the three principles of proportionality, distinction and precaution can be seen as the basis for what can be considered under IHL as a requirement to consider until when human control needs to be maintained. Such a concept implies that States have to use particular restrain before deciding about the development and the deployment of new weapons, even if the evaluation of each of these principles on their own may not necessarily lead to a negative compliance assessment." (Working paper) (2015) those principles. Hence the importance of developing further the MHC [meaningful human control] concept and its institutional extension - the idea of MSC [meaningful state control]. The presence of human control in the form of institutional framework guarantees itself a reference to certain standards - legal and related international customs. Human or institutional oversight upholds accountability, the rule of law and supports procedures through which our decisions may be verified." (Characteristics of LAWS/Meaningful human control statement) (2015)

South Africa: "The use of such a weapon systems would need to comply with the fundamental rule of International Humanitarian Law, including those of distinction, proportionality and military necessity." (General statement) (2015)

Switzerland: "Lors d'engagement de SALA dans des conflits armés, les règles du droit international humanitaire, y compris celles relatives à la conduite des hostilités, doivent être pleinement observées." (General statement) (2015)*

United Kingdom: "From our perspective, to discuss LAWS is to discuss means and methods of warfare. As such, international humanitarian law provides the appropriate paradigm for discussion." (General statement) (2015)

Australia: "We have observed and considered the various ways of framing the question [including] a legal approach, which asks how IHL applies to weaponisation of increasingly autonomous systems, whether lethal autonomous weapons systems would function in conformity with IHL rules, whether clarification or interpretation of existing law is required, or whether new rules need to be developed." (General Statement) (2016)

---

* "At the time of LAWS's use in an armed conflict, the rules of international humanitarian law, including those related to the conduct of hostilities, are clearly to be observed."

Chile: "Creemos que el punto actual en que se encuentra la evolución del Derecho Internacional Humanitario aún no da respuestas solidas a los desafíos que plantea un sistema autónomo que llegara a tomar la decisión de quitar la vida, con independencia completa de la orden de un humano. Este es un desafío legal que contiene vacíos necesarios de llenar como...la merma en la dignidad humana." (General statement)* (2015)

Chile: "[I]t is reasonable to think that the proportionality principle may be placed in jeopardy with the use of lethal force by autonomous machines, inasmuch as the prevailing legal interpretations of the said principle are explicitly grounded on concepts such as "common sense", "good faith" and the "rule of the reasonable military commander."" (Paper) (2015)

Spain: "Nuestro principal punto de partida en este empeño debe fundamentarse, como debe hacerlo además en relación con cualquier otro tipo de armas, en la necesidad del respeto más escrupuloso del Derecho Internacional Humanitario y del Derecho Internacional de los Derechos Humanos, cuya primacía entendemos irrenunciable, en particular en relación con los principios de necesidad, proporcionalidad, distinción y precaución. Para lograr este objetivo, es necesaria la capacidad de control y supervisión humana en la fase de selección del blanco militar, incluida la capacidad de abortar el proceso de lanzamiento del arma de que se trate." (General statement) (2015)†

Finland: "Finland highlights the importance of adhering to the rule of international humanitarian law in all situations. In our opinion IHL is fully applicable also in a situation where LAWS would be used as a means of warfare on the battleground. We further underline, that each and every state has the ultimate responsibility in every situation where norms of international humanitarian or human rights law are breached." (General Statement) (2016)

India: "In our view, a discussion on LAWS should include questions on their compatibility with international law including international humanitarian law." (General Statement) (2016)

New Zealand: "We also look forward to an informed debate on the challenges posed by LAWS for compliance with the norms and dictates of international humanitarian law. For New Zealand, the absolutely essential requirement is that the development and subsequent usage of any weapon system – including LAWS – must take place only in accordance with IHL. Compliance with IHL, and, as applicable, other aspects of international law, remains of the highest priority for New Zealand and will continue to be the determining factor in our approach to these issues." (General Statement) (2016)

United Kingdom: "The UK's clear position is that IHL is the applicable legal framework for the assessment and use of all weapons systems in armed conflict. Distinction, proportionality, military necessity and humanity are fundamental to compliance with IHL. Any weapon system, no matter what its specific technical characteristics or which or how many of its critical functions are autonomous, would

---

* "We believe that the actual point at which one finds the evolution of International Humanitarian Law still does not give solid answers to the challenges laid out by an autonomous system that will be become able to take the decision to take a life, with complete independence from a human order. This is a legal challenge that contains lacunae that must be filled like...the reduction in human dignity."

† "Our principle point of departure in this effort must be based, as must be done in relation to whatever type of weapons, on the need for the most scrupulous respect for International Humanitarian Law and International Human Rights Law, whose primacy we understand to be irrenoueable, in particular in relation to the principles of necessity, proportionality, distinction and precaution. To succeed in this goal, it is necessary to have human control and supervision in the phase of selection of the military target, including the ability to abort the launch process of the weapon in question."

Sri Lanka: "We agree that the use of LAWS could open up new challenges on compliance with IHL principles such as distinction, proportionality, precaution and military necessity." (General statement) (2015)

Switzerland: "IHL imposes manifold obligations which would have to be respected when using LAWS, in particular the principles governing the conduct of hostilities. For example, in order for LAWS to be lawfully employed in an armed conflict, challenging assessments are required to distinguish between civilian and military objectives or in evaluating whether the causation of unavoidable incidental harm to the civilian population can be justified in view of the concrete and direct military advantage anticipated from that particular attack. These fundamental principles must not be circumvented by the use of LAWS. These and other legal requirements are derived directly from longstanding principles of IHL and allow for no compromise. It is therefore clear that existing IHL sets the bar very high in terms of technological prerequisites for the lawful use of LAWS in armed conflict.... Any legal review process concerned with such systems would have to assess not only their international lawfulness under the rules of classic weapons law (such as the prohibition of indiscriminate weapons), but also their capability to reliably implement the targeting principles of distinction, precaution and proportionality without human intervention. This is not the case with conventional weapons systems, where the actual targeting is always conducted by a human operator." (Challenges to IHL statement) (2015)

United Kingdom: "The UK's clear position is that IHL is the applicable legal framework for the assessment and use of all weapons systems in armed conflict. Distinction,

have to comply with those principles to be capable of being used lawfully." (Challenges to IHL Paper) (2016)

Spain: "Debemos partir para ello del máximo respeto a la legalidad internacional, fundamentada en el Derecho Internacional Humanitario y el Derecho Internacional de los Derechos Humanos, contando con los principios de necesidad, distinción, proporcionalidad y precaución." (General Statement) (2016)[*]

Mexico: "Los desarrollos tecnológicos bélicos, incluidos los Sistemas de Armas Letales Autónomas (SALAS) deben cumplir con las normas del Derecho Internacional Humanitario (DIH), normas convencionales y consuetudinarias; en particular las normas de distinción, proporcionalidad y precauciones en el ataque." (General Statement) (2016)[†]

Mexcio: "Mi país considera que para cumplir con los requerimientos del DIH, los SALAS deben tener además la capacidad de distinguir entre combatientes activos y personal de las fuerzas armadas fuera de combate, civiles que participan directamente en las hostilidades, fuerzas de seguridad públicas, personal sanitario, entre otros." (General Statement) (2016)[‡]

---

[*] "We should begin with maximum respect for international law based on IHL and IHRL, the principles of necessity, distinction, proportionality and precaution."

[†] "The development of military technology, including LAWS, must comply with the norms of IHL, conventional and customary law; in particular, the rules of distinction, proportionality and precautions in attack."

[‡] "My country believes that in order to meet the requirements of IHL, LAWS must also have the capacity to distinguish between active combatants and [hors de combat], civilians who directly participate in hostilities, public security forces, health [/medical] personnel, among others."

proportionality, military necessity and humanity are fundamental to compliance with IHL. Any LAWS, no matter what its specific technical characteristics, would have to comply with those principles to be capable of being used lawfully. However, the UK position is that those principles, and the requirement for precautions in attack, are best assessed and applied by a human. Within that process, a human may of course be supported by a system that has the appropriate level of automation to assist the human to make informed decisions." (Challenges to IHL statement) (2015)

Australia: "We have observed and considered the various ways of framing the question [including] an ethical approach, which raises the fundamental question whether the principles of humanity and dictates of public conscience can ever allow machines to select, attack and kill human beings, entirely outside of human control."* (General Statement) (2016)

India: "Our aim should be to strengthen the CCW in terms of its objectives and purposes through increased systemic controls on international armed conflict in a manner that does not widen the technology gap amongst states or encourage the increased resort to military force in the expectation of lesser casualties or that use of lethal force can be shielded from the dictates of public conscience." (General Statement) (2016)

Sri Lanka: "Over and above these technological issues, there are underlying fundamental moral questions. Even if any of the IHL principles are found to be inapplicable, the test of public conscience and laws of humanity as referred to in the Martens Clause provide compelling reasons for basic guiding principles on the legality of the use of LAWS." (General Statement) (2016)

Ecuador: "Cláusula de Martens: Los civiles y combatientes quedan amparados bajo la protección de los principios

---

* Note that Australia's reference of "public conscience" here does not include mention of the Martens Clause.

| | | derivados de la costumbre, de los principios de humanidad y de los dictados de la conciencia pública. Esta disposición es relevante para la revisión de las armas emergentes. El Derecho, incluido el Derecho Internacional Humanitario, es escrito por y para los seres humanos que deben aplicarlo no sólo con la razón sino con todos los atributos humanos como la compasión, la piedad, el sentido de moralidad. Dejar la decisión de la vida o muerte a una máquina no es moral y contraviene la conciencia pública, deshumanizaría la guerra. Los SALA amenazan con violar varios derechos humanos, incluido el derecho a la vida." (General Statement) (2016)[*] | |
| **Reference to Martens Clause/public conscience[§§]** | Ecuador: "[E]stas nuevas tecnologías…pueden estar reñidas con el Derecho Internacional Humanitario, la ética, los principios de humanidad y los dictados de la conciencia pública …Nos preocupan aspectos fundamentales que merecen ser analizados y discutidos en profundidad como… la inobservancia de la ética y de los derechos humanos fundamentales, en particular de la cláusula de Martens." (General statement) (2015)[***] | Austria: "The underlying unity of international humanitarian law is grounded on the basic values of humanity shared by all civilizations. The idea of humanity plays a crucial role and is reflected in the Martens clause, which is a binding rule under IHL and demands the application of "the principle of humanity" in armed conflict. In the context of LAWS, an interesting parallel is sometimes drawn to landmines, which were banned because of the delegation of the decision to initiate lethal force from humans." (Working paper) (2015)<br><br>Sri Lanka: "As the Convention stipulates 'the civilian population and the combatants shall at all times remain under the protection and authority of the principles of international law derived from established custom, from the | Chile: "it is clear that, in our ongoing debate, the Martens Clause - based upon "the usages established between civilized nations, the laws of humanity and the requirements of the public conscience", is an analytical and legal resource applicable to LAWS, along with all subsequent legal and political developments of the International Humanitarian and Human Rights Law. This holds true at least at these initial stages of the analysis and diplomatic reflections we are undertaking. Already back in 1996, the International Court of Justice, in its advisory opinion on the legality of the threat or use of nuclear weapons, referred to the Martens Clause stating that "it has proved to be an effective means of addressing the rapid evolution of military technology". This is a valid criterium that should necessarily be applied to an |

---

[*] "The Martens Clause: Civilians and combatants are protected [/covered by] under the protection of principles derived from custom, the principles of humanity and the dictates of public conscience. This provision [/principle] is relevant to the review of emerging weapons. The Law [sic], including IHL, is written by and for human beings who apply it not only with reason but with all human attributes such as compassion [/mercy?], piety [?], [and] notions of morality. Leaving the decision over life or death to a machine is immoral and contravenes the public conscience, [and would] dehumanize war. LAWS threaten to violate several human rights, including the right to life."

[§§] The Martens Clause, taken from the preamble to the 1899 Hague Convention on the laws of war on land, states, "Until a more complete code of the laws of war is issued, the High Contracting Parties think it right to declare that in cases not included in the Regulations adopted by them, populations and belligerents remain under the protection and empire of the principles of international law, as they result from the usages established between civilized nations, from the laws of humanity and the requirements of the public conscience." See Rupert Ticehurst, The Martens Clause and the Laws of Armed Conflict, 317 INT'L REV. RED CROSS (1997), https://www.icrc.org/eng/resources/documents/misc/57jnhy.htm. Thus, it makes reference to humanity, a concept given a separate category in this table. Since many states reference "humanity" without referring to the Martens Clause, "humanity" was given its own category. "Public conscience," on the other hand, tended to be referenced in relation to the Martens Clause. Thus, references to the Martens Clause and public conscience were combined.

[***] "These new technologies…can be at odds with International Humanitarian Law, ethics, the principles of humanity and the dictates of public conscience…We are worried about fundamental aspects that deserve to be analyzed and discussed in depth like…the violation of ethics and fundamental human rights, in particular the Martens Clause."

principles of humanity and from the dictates of public conscience,' and we therefore need to be wary of allowing any level of autonomy in the use of weapons systems. The implications of LAWS becoming the moral-discerner in its own right, without human control, are far reaching to contend with." (General statement) (2015) emerging technology, whose consequences are hard to predict, although it would need to be consider in varying degrees with regard to nuclear weapons and to this new type of weapons." (Paper) (2015)

Greece: "For the sake of argument, let us suppose that in the future autonomous weapon systems are developed which can fully comply with IHL and its cardinal principles, such as distinction, proportionality and precautions in attack; a weapon operating with better precision than being under human control. We are not there yet; indeed we are far from that juncture, however, for the sake of our debate, let us envisage such a hypothetical scenario. In such a case, one may ask oneself what would the legal basis be to justify their prohibition...[T]o argue that LAWs comply or do not comply with IHL at this stage would amount to an oracle of Delphi. What is left then is basically an ethical question, not a legal one. It boils down to the fundamental question of whether humans should delegate life and death decisions to machines and definitely Greece, like others, does not feel comfortable with such a prospect. Or as Germany stated on Monday, full autonomy is a line that should not be crossed, the line being when there is no longer any human oversight, as the delegate from the United Kingdom remarked earlier. The question which then arises is how does one operationalize this ethical concern into a legal provision. The only legal principle which comes to mind is the Martens Clause, given its dependence on the dictates of public conscience. Does though such a general principle suffice to lead to the codification in the future of a new set of legally binding rules? We have our doubts. Indeed, should we isolate this issue to its legal parameters, then- in our view- there is no other logical conclusion than the one made by Dr. Boothby earlier, that is, that a thorough and systematic weapons review is the only practical solution, at least at the present stage, to address the issue of LAWS from a legal angle. The discussion, however, takes a very different dimension when it is addressed ethically or politically, bringing to the fore the question of 'meaningful human control', but this is not a legal

| | | | norm. Hence, we should in our view be clear about what it is we are discussing and avoid a conflation which makes things even more complicated." (Challenges to IHL statement) (2015)<br><br>India: "A discussion on LAWS should include questions on their compatibility with international law including international humanitarian law as well the impact of their possible dissemination on international security. Our aim should be to strengthen the CCW in terms of its objectives and purposes through increased systemic controls on international armed conflict in a manner that does not widen the technology gap amongst states or encourage the increased resort to military force in the expectation of lesser causalities or that use of lethal force can be shielded from the dictates of public conscience." (Way ahead statement) (2015) |
|---|---|---|---|
| **Reference to accountability** | Cuba: "Tampoco podría hacerse una evaluación efectiva de la responsabilidad del Estado por hechos internacionalmente." (General statement and Position Paper/National Document) (2015)[*]<br><br>Ecuador: "Nos preocupan aspectos fundamentales que merecen ser analizados y discutidos en profundidad como… responsabilidad legal en cuanto a la delegación de autoridad y a la toma de decisiones [y] incumplimiento del Derecho Internacional Humanitario en cuanto a la secuencia ininterrumpida de responsabilidad." (General statement) (2015)[†]<br><br>Pakistan: "LAWS create an accountability and transparency vacuum and provide impunity to the user due to the inability to attribute responsibility for the harm that they cause. If the | Argentina: "La falta o baja frecuencia de control humano significativo de los SALA conducirá a decisiones sin intervención humana que podrían provocar consecuencias humanitarias impredecibles. La determinación de responsabilidades y rendición de cuentas por las consecuencias del empleo de los SALA se hace difuso y hasta impracticable, con lo cual ante esa situación se podrían considerar armas ilegales." (General statement) (2015)[‡]<br><br>Chile: "Creemos que el punto actual en que se encuentra la evolución del Derecho Internacional Humanitario aún no da respuestas solidas a los desafíos que plantea un sistema autónomo que llegara a tomar la decisión de quitar la vida, con independencia completa de la orden de un humano. Este es un desafío legal que contiene vacíos necesarios de llenar | Poland: "What if we accept MHC [meaningful human control] as a starting point for developing national strategies towards LAWS? We could view MHC from the standpoint of state's affairs, goals and consequences of its actions. In that way this concept could also be regarded as the exercise of "meaningful state control" (MSC). A state should always be held accountable for what it does, especially for the responsible use of weapons which is delegated to the armed forces. The same goes also for LAWS…. Looking at the present level of technological advancement, however, there are reasons for concern that the existing systems will not be able to meet [IHL] principles. Hence the importance of developing further the MHC concept and its institutional extension - the idea of MSC. The presence of human control in the form of institutional framework guarantees itself a reference to certain standards - legal and related |

[*] "Neither would it be possible to make an effective evaluation of State responsibility for international acts [translator's note: or perhaps "incidents"]."

[†] "We are worried about fundamental aspects that deserve to be analyzed and discussed in depth like…legal responsibility in regard to the delegation of authority and the taking of decisions [and] violation of International Humanitarian Law in relation to the interrupted sequence of responsibility."

[‡] "The lack or low amount of significant human control in LAWS will bring decisions without human intervention that would provoke unpredictable humanitarian consequences. The determination of responsibility and accountability for the consequences of the use of LAWS will be diffuse and almost impracticable, which in that situation would make it possible to consider them illegal weapons."

nature of a weapon renders responsibility for its consequences impossible, its use should be considered unethical and unlawful. Also, in the event of a security breach or a compromised system, who would be held responsible; the programmer, the hardware manufacturer, the commander who deploys the system or the user state?" (General statement) (2015)

Pakistan: "LAWS create an accountability vacuum and provide impunity to the user due to the inability to attribute responsibility for the harm that they cause. If the nature of a weapon renders responsibility for its consequences impossible, its use should be considered unethical and unlawful." (General Statement) (2016) como la responsabilidad final en caso de error…. o la posibilidad real de la rendición de cuentas" (General statement) (2015)*

France: "Je veux citer également la question de la responsabilité, qui est naturellement centrale. A ce stade, rien ne permet de définir avec certitude les contours de la responsabilité de chaque acteur, qui dépendra de leur rôle dans l'utilisation du SALA. La possibilité d'identifier un acteur responsable est cruciale pour savoir si les principes existants du DIH demeurent suffisants ou non." (General statement) (2015)†

France: "Même si un SALA s'avérait capable de respecter le DIH, il resterait toutefois un certain nombre de problèmes. Un premier problème est celui de la dilution de la responsabilité, qui serait plus difficile mais peut-être pas impossible à établir." (Ethics/Overarching issues statement) (2015)‡

Korea: "[W]e are wary of fully autonomous weapons systems that remove meaningful human control from the operation loop, due to the risk of malfunctioning, potential accountability gap and ethical concerns." (General statement) (2015)

Netherlands: "We see the notion of meaningful human control as an important concept for the discussion on LAWS. Command responsibility is an issue here." (General statement) (2015) international customs. Human or institutional oversight upholds accountability, the rule of law and supports procedures through which our decisions may be verified." (Characteristics of LAWS/Meaningful Human Control statement) (2015)

United Kingdom: "Turning now to the issue of the accountability chain, the UK's position is that there must always be human oversight and control in the decision to deploy weapons. It is in this person or with these people that responsibility must initially be vested. Responsibility will flow up through the Chain of Command, which is so important in military structures. These chains of command are vital not just for accountability and compliance with the law, but also in order for decisions to be made and for military judgement to be exercised. 9. Inherent in that individual and chain of command responsibility is not just individual criminal responsibility, both nationally and potentially internationally, but also State responsibility." (Challenges to IHL statement) (2015)

Israel: "Another related question that has been raised during this session addresses the issue of accountability. In Israel's view, it is safe to assume that human judgment will be an integral part of any process to introduce LAWS, and will be applied throughout the various phases of the research, development, programming, testing, review, approval, and decision to employ them. LAWS will operate as designed and programmed by humans. In cases where employment of LAWS would involve a violation of the law, individual

---

* "We believe that at the current moment in its evolution, International Humanitarian Law still does not give solid answers to the challenges laid out by an autonomous system that will be become able to take the decision to take a life, with complete independence from a human order. This is a legal challenge that contains lacunae that must be filled like the final responsibility in case of error…or the real possibility of accountability."

† "I would like to cite equally the question of responsibility, that is naturally central. At this stage, nothing allows one to define with certainty the contours of the responsibility of each actor, that will depend on their role in the utilization of LAWS. The possibility of identifying a responsible actor is crucial to know if the existing principals of IHL remain sufficient or not."

‡ "Even if LAWS proved to be capable of respecting IHL, a certain number of problems would nevertheless remain. A first problem is that of the dilution of responsibility, that would be difficult but perhaps not impossible to establish."

Spain: "[E]s necesaria la capacidad de control y supervisión humana en la fase de selección del blanco militar, incluida la capacidad de abortar el proceso de lanzamiento del arma de que se trate. Esta imperativa intervención humana en el proceso de activación del sistema y su posterior supervisión, al mismo tiempo, y en toda lógica, deberá permitir una atribución clara y precisa de responsabilidad jurídica personal." (General statement) (2015)[*]

Sri Lanka: "[T]he use of LAWS could open up new challenges on compliance with IHL principles such as distinction, proportionality, precaution and military necessity. Left unanswered this will also lead to a crucial accountability gap." (General statement) (2015)

Switzerland: "An uninterrupted accountability chain is essential for the implementation of international law, incl. IHL. This holds true not only for cases where LAWS would be used in an unlawful manner, but especially also in cases where such systems malfunction or cause unintended harm. We see a wide range of legal mechanisms, national and international, that could come into play to ensure accountability in the use of LAWS. On the international level, the primary enforcement mechanisms would be the rules governing State responsibility and, in case of individual culpability, international criminal law. We all have the responsibility to ensure that legal liability for violations of international law cannot be evaded through the use of LAWS. In this respect questions seem to arise primarily with regard to establishing the intent required for holding a person criminally liable for the use of LAWS. Perhaps our discussions could benefit from considerations in related fields where increasingly automatic or autonomous systems already exist, such as in the automobile industry. We would welcome more in-depth analysis on possible accountability

accountability would be sought in accordance with the law." (Challenges to IHL Statement) (2016)

Italy: "Apart from systems entirely controlled by humans, we could first consider weapons systems that act on the basis of criteria pre-programmed by human operators. Such criteria determine the type of target to be selected and potentially engaged, together with the geographical area and amount of time in which the search for targets will be carried out. These systems – which have also been called "highly automated"[§§]– could be characterized by high degrees of autonomy in several functions, even some critical ones, but their behavior and actions can still be attributed to the human operator, who remains accountable… There may not be any accountability gap in this case, given that the effects of these weapons could be ascribed to the human operators who decided to deploy and activate them. Obviously, people in charge of weapons deployment and activation decisions will need to take due account of the environment in which they would operate." ("Towards a Working Definition of LAWS" Statement) (2016)

Netherlands: "We do not foresee an accountability gap arising as long as humans exercise meaningful human control in the wider loop of the decision-making process for deploying autonomous weapon systems. In that case the existing legal regime is adequate to hold offenders accountable, as there is no change in the accountability of commanders, subordinates or those in positions of political or administrative responsibility who make the decisions. Likewise, state responsibility remains unchanged in the event of deployment of autonomous weapon systems under human control, according to the advisory committee." (General Statement) (2016)

---

[*] "The capacity to control and human supervision in the phase of selection of a military target, including the capacity to abort the launch process of the weapon in question, is necessary. It is necessary to have human intervention in the process of activation of the system and its later supervision, at the same time, and in all logic, it should allow a clear and precise attribution of personal legal responsibility."
[§§] Note that when Italy refers to certain weapons systems as "highly automated" here, they mean ones that have some important automatic functions but are not fully autonomous.

gaps and ways to close them." (Challenges to IHL statement) (2015)

Zambia: "Zambia also takes note of the challenges the increasing degree of autonomy would present to International Humanitarian Law and therefore would not advocate for any such weapons systems that would water down the aspects of responsibility and accountability in armed conflict. Our focus should instead be on strengthening such norms." (Way ahead statement) (2015)

Germany: "The use [of any weapons system] must always observe an unequivocal accountability chain. This is of crucial importance for the use of any weapons system." (General Statement) (2016)

Sierra Leone: "We should bear in mind that any weapons systems that are developed might fall into the wrong hands, including non-state actors for which accountability would not be easily established and could be a good reason why these weapons should not be developed in the first place." (General Statement) (2016)

Sri Lanka: "The challenges to be addressed during such a dialogue ranges from the need for a definition for lethal autonomous weapon systems, to clarity on 'meaningful human control,' the accepted degree of autonomy that enables compliance with international human rights and humanitarian law that can successfully address the void in accountability issues, and moral and ethical concerns in usage, including selecting targets." (General Statement) (2016)

Sri Lanka: "Another important concept emerging in relation to the issues of definition as well as on the accountability of the use of LAWS is 'Meaningful Human Control' (MHC). Given that issues such as the exact level of human control and the

Poland: "Also, from the military perspective, it is important to satisfy the need to both introduce the latest technologies into warfare and create environments where humans may be held accountable for their decisions. In our opinion, such a need can be satisfied through exercising Meaningful Human Control (MHC) over the critical functions of LAWS. Therefore, we see rationale in continuing the analysis of LAWS against the concept of Meaningful Human Control where further exploration of such a concept may significantly facilitate the discussion on the definitions." ("Towards a Working Definition of LAWS" Statement) (2016)

Poland: "If robots are designed to act autonomously, who is to control them, and hence, to be held accountable for robot actions? To help answering this question, we would like to propose to look at the possibility of human control over the robotic systems rather than the actual execution of such control. Following this logic, a person accountable for robot actions is the user who has a possibility to take over control over a robotic system at every moment of the robot conduct, without necessarily executing such control. This refers not only to taking over manual control over the system but also the decisions we make that influence robots' goals." (emphasis in original) (Human Rights and Ethical Issues Statement) (2016)

Sri Lanka: "The debate on how and what provisions of IHL should be applied in the case of LAWS and who should be held accountable in the event of unlawful use are some off the fundamental issues that need an answer. We need to take into consideration how the existing international legal regimes could effectively address the future forms of warfare and weapons, in particular lethal autonomous weapons. The challenge of addressing the accountability gap in this context means to what extent an individual, organizations or a State could be held liable for a crime committed by a fully autonomous weapon. As the ICRC notes* under the law of

---

necessary parameters of 'meaningfulness' are yet to be defined, we encourage states to continue the dialogue on this concept, focusing on further defining its context and application with a view to contributing to a working definition on LAWS and to regulate the increasing autonomy of weapons." (General Statement) (2016)

Turkey: "We, as others, attach importance to the humanitarian aspect of the matter. Therefore, we support the notions like need for human control and accountability for such weapon systems. Nevertheless, taking into consideration that yet such weapon systems do not exist and we are working on an issue which is still hypothetical, we hesitate on the accuracy of a general prohibition pre-emptively" (General Statement) (2016)

United Kingdom: "Turning now to the issue of the accountability chain, the UK's position is that there must always be human oversight in the decision to deploy weapons. It is with this person/people that responsibility lies. Responsibility will flow up through the Chain of Command, which is so important in military structures. This chain of command is vital not just for accountability and compliance with the law, but also in order for decisions to be made and communicated, for forces to be controlled and for military judgement to be exercised. Both state and individual criminal responsibility are inherent in this concept of command responsibility. If in the future LAWS that could comply with an Article 36 Review were ever to exist, we do not believe that accountability would or should be any different from what has already been outlined above. The person who decides to deploy the weapon would ultimately be responsible for the consequences of its use. Accountability might even be improved if we assume that the automated record systems that an autonomous system would need in

State responsibility, in addition to accountability for violations of IHL committed by its armed forces, a State could also be held liable for violations of IHL caused by an autonomous weapon system that it has not, or has inadequately tested or reviewed prior to deployment. Further, under the laws of product liability, manufacturers and programmers could also be held accountable for errors in programming or for the malfunction of an autonomous weapon system. However, establishing evidence that the operator or manufacturers knew or should have known the possibility of the crime committed by a complicated artificial intelligence system fed into the weapon will be a difficult task. Therefore, we recommend this aspect also be given due attention when discussing Article 36 implementation, to ensure a clear accountability chain with regard to autonomous weapons." (General Statement) (2016)

Switzerland[*]: "Another important issue arising with regard to AWS is that of accountability, namely in terms of individual criminal responsibility and of state responsibility. Given that AWS possess no agency or legal personality of their own, the question of individual criminal responsibility focuses entirely on the responsibility of humans that are involved as operators, commanding officers, programmers, engineers, technicians or in other relevant functions. If the deployment of an AWS results in a serious violation of IHL, and if that violation is the consequences of culpable fault on the part of a human being the latter may be subjected to criminal prosecution for war crimes or, depending on the circumstances of the case, also for crimes against humanity or genocide. Criminal culpability is self-evident in the case of deliberate and premeditated intent. It is less so in the case of recklessness or (advertent) negligence, or of simple acceptance of a risk that violations will or may occur. With regard to war crimes, article 85(3) of Additional Protocol I requires "willfulness", with national practices varying as to

---

https://www.icrc.org/en/document/international-humanitarian-law-and-challenges-contemporary-armed-conflict.
[*] This is a fairly long excerpt but has many important points. Italics here are intended to emphasize the most important parts and are not included in the original.

order to operate may provide better evidence to support subsequent investigation." (Challenges to IHL Paper) (2016)

Spain: "Consideramos siempre necesaria la participación de un operador humano, así como el establecimiento de principios de atribución clara de responsabilidad jurídica personal sobre los criterios de uso de cualquier tipo de arma." (General Statement) (2016)[*]

Costa Rica : "Las armas autónomas letales pueden llevar a modificar la naturaleza de los conflictos armados. Su existencia aumentará el riesgo de operaciones encubiertas y vulneraciones deliberadas del derecho internacional humanitario, exacerbaría la asimetría de ciertos conflictos armados y conduciría a la impunidad debido a la imposibilidad de atribuir la autoría de los ataques." (General Statement) (2016)[†]

Costa Rica: "La responsabilidad de la persona y el Estado es fundamental para garantizar la rendición de cuentas, tanto en el derecho internacional humanitario como en el derecho internacional de los derechos humanos. Sin la rendición de cuentas se reducen la disuasión y la prevención, lo que tiene como consecuencia una menor protección de los civiles y las posibles víctimas de crímenes de guerra. Los robots no tienen capacidad de discernimiento moral, por lo que si causan pérdidas de vida no se les puede exigir ningún tipo de responsabilidad, como sería normalmente el caso si las decisiones hubieran sido tomadas por seres humanos. ¿En quién recaería entonces la responsabilidad? Si no hay the meaning to be given to this requirement. The International Tribunal for the Former Yugoslavia has stated that, as a matter of customary law, indirect intent would be sufficient to fulfil the mental requirement (mens rea).[†††] Conversely, the Rome Statute of the International Criminal Court does not foresee criminal liability for indirect intent,[‡‡‡] except in the case of command responsibility. for the conduct of subordinates.[§§§] As a matter of concept, command responsibility does not entail the commander's direct criminal responsibility for crimes committed by his subordinates, but for his or her culpable failure to prevent, suppress or repress crimes committed by persons (i.e. not machines) under his or her command and control. *Strictly speaking, therefore, a commander's failure to duly control AWS operating under his command is not a case of command responsibility within the contemporary understanding of this concept, but may constitute a direct violation of the duties of precaution, distinction, proportionality or any other obligation imposed by IHL.* This does not exclude that, as the functions of human soldiers are increasingly "delegated" to AWS, it may become appropriate de lege ferenda to extend the commander's supervisory duty, mutatis mutandis and by analogy, also to AWS operating under his direct command and control. *Overall, under current international law, whether or not there is an "accountability gap" for operators, commanders and other humans involved in the operation of AWS depends on the applicable mens rea standard. As a general assumption, the more significant human involvement in a specific AWS operation is (such as humans "in the loop"), the easier it is to assign individual responsibility. This* assumption may be relevant with a view to the general

---

[*] "We consider the participation of a human operator as necessary [/requisite], alongside the establishment of principles of clear attribution of personal legal responsibility, as among the criteria for the use of any type of weapon."

[†] "Lethal autonomous weapons can lead to changing the nature of armed conflict. Their existence will increase the risk of covert operations and deliberate violations of international humanitarian law, exacerbate the asymmetry of certain armed conflicts and lead to impunity because of the impossibility of attributing responsibility for attacks."

[†††] Cf. ICTY, Prosecutor v. Tihomir Blaskic, Judgement of 29 July 2004, Appeals Chamber, para. 42.

[‡‡‡] Article 30 of the Rome Statute of the International Criminal Court.

[§§§] Article 28 of the Rome Statute of the International Criminal Court.

responsabilidad hay impunidad." (Human Rights and Ethical Issues Statement) (2016)[*]

Mexico: "Otro desafío se presenta en el momento de determinar responsabilidades legales, particularmente penales, derivadas del uso de estas armas, ya que es evidente que a un arma no pueden atribuírsele responsabilidades. En su situaciones en las que el uso de estas armas pudiera derivar en posibles violaciones al derecho internacional, no existe actualmente un marco jurídico suficientemente claro que permita fácilmente la atribución de responsabilidades." (General Statement) (2016)[†]

Ecuador: "Rendición de cuentas y asunción de responsabilidades: Definitivamente existiría un vacío jurídico al respecto ya que tanto el DIH como el Derecho Penal Internacional juzga violaciones de la ley cometidas por seres humanos. En el caso de los SALA, no se podría juzgarlos como máquinas, y para establecer responsabilidades existiría una larga cadena que va desde el comandante y toda la cadena de mando hasta el programador, el ingeniero y el productor o fabricante." (General Statement) (2016)[‡]

Zimbabwe: "[H]ow will these systems determine their targets? Who will be accountable for violations of international humanitarian law? Who will be criminally liable for war crimes, where such crimes are committed by fully obligation of States to respect and ensure respect for IHL. The second dimension of accountability derives from general international law governing the responsibility of States for internationally wrongful acts. States remain legally responsible for unlawful acts and resulting harm caused by AWS they employ, including due to malfunction or other undesired or unexpected outcomes. The rules governing attribution of conduct to a State are pertinent in relation to AWS as with any other means and methods of warfare. Given that AWS lack legal personality in the first place, they cannot become agents in a human sense, whether state agents or non-state actors. The question of State responsibility therefore does not turn on the nature or capability of the AWS, but of legal and factual status of the person or entity deciding on its employment. A decision of a person or entity exercising public powers or governmental authority (e.g. the armed forces) to employ an AWS in a given situation certainly would be attributable to the State.[‡‡‡] *The result is that States cannot escape international responsibility by a process of "delegating" certain tasks to AWS.*" (Working Paper) (2016)

United Kingdom: "I turn now to the phrase meaningful human control. This is not a concept that the UK actively uses in its doctrine, principally because what may or may not be meaningful is almost an entirely subjective judgment: therefore, any system based on this concept would be open to a wide range of interpretation. This level of ambiguity

---

[*] "The responsibility of the individual and the State is essential to ensure accountability, both in international humanitarian law and international law of human rights. Without accountability, deterrence and prevention are reduced, with the consequence of reduced protection of civilians and potential victims of war crimes. Robots do not have capacity for moral discernment, so that you cannot demand any responsibility from them for causing loss of life, as would normally be the case if the decisions had been taken by human beings. With whom then would responsibility fall? If there is no responsibility, there is impunity."

[†] "Another challenge presents itself when trying to determine the legal responsibilities, particularly criminal, arising from the use of these weapons, since it is clear that responsibility cannot be attributed to a weapon. In those situations where the use of these weapons could lead to possible violations of international law, there is currently no sufficiently clear legal framework that would easily facilitate the attribution of responsibilities."

[‡] "Accountability and the assumption of responsibilities: A legal vacuum would definitively exist with this respect, since both IHL and the ICL judge violations of the law committed by human beings. In the case of LAWS, you could not judge them as machines, and to establish responsibilities there would exist a long chain running from the commander and the entire chain of command to the programmer, engineer and producer or manufacturer."

[‡‡‡] Article 4 et seq. of the Articles on State Responsibility for Internationally Wrongful Acts (2001). See also Article 91 of Additional Protocol I to the Geneva Conventions.

automated machines? Of course one could argue that overall responsibility lies with the military commanders who make the decisions to deploy such weapons. However, this is a whole new and complex area that we are entering, which will be very difficult to fathom as far as international humanitarian law is concerned. Consider that, unmanned aerial military vehicles, more commonly referred to as 'drones', which are remotely controlled by human operators, are already wreaking havoc on civilians and the environment. Thus completely autonomous weapon systems can only be much worse on the accountability scales. These are, some among the many questions for which we have no answers. In the absence of such answers, my delegation is of the view that we should maintain meaningful human control over military weapons or weapons with a dual use." (Speech) (2016)* would not be helpful in agreeing definitions. Furthermore, some of the terms and phrases used to define MHC will themselves also need to be defined, for example, full situational awareness in order to have an informed understanding. Variances in definitions and criteria of MHC, particularly with regard to accountability, do not align with the current UK doctrine. Therefore, the UK believes it would be useful to research relevant doctrine when trying to define accountability. In essence, MHC describes the relationship between weapons technology (that can in part function autonomously) and the operator. It is suggested that the phrase MHC is changed to more accurately reflect the premise of human-machine interaction, for example intelligent partnership." (Towards a Working Definition of LAWS Paper) (2016)

France: "Le caractère autonome d'un système d'armement létal soulèverait également la question des modalités de recherche des responsabilités des peronnes ayant participé à sa mise en œuvre à son déploiment. La France estime que le DIH, là aussi, debrait servir de base utile à la recherche de la responsabilité des décideurs politiques et militaires, industriels, programmeurs, ou opérateurs, sera néanmoins ppossible en cas d'infraction au droit international humanitaire commise par le biais de ces systèmes." (Challenges to IHL Paper) (2016)***

Chile: "En lo relative a la rendición de cuentas, y ante la imposibilidad de aplicar la justicia a un SAL, en virtud de los artículos ya citados se hace necesario que el derecho internacional determine quienes asumirían la

---

| | | | responsabilidad penal y civil en la cadena de mando, incluido el nivel político ante actos ilegales de los SAL, así como tambi´n la eventual responsabilidad de los fabricantes y programadores de los SAL. Ignorar esta necesidad sería una violación al Artículo 86 del Protocolo Adiciona I de los Convenciones de Ginebra, el que establece la responsabilidad penal de los mandos superiores por actos de sus subordinados." (Human Rights and Ethical Issues Statement) (2016)* |
|---|---|---|---|
| **Reference to humanity** | Pakistan: "The use of LAWS will make war even more inhumane… Whilst automated weapons and automatic weapons have to some degree a 'human in the loop', autonomous implies no scope for such 'interference' by any human, calling into question the principles of IHL [including] humanity and military necessity." (General statement) (2015)<br><br>Pakistan: "LAWS are by nature unethical, because there is no longer a human in the loop and the power to make life and death decisions are delegated to machines which inherently lack compassion and intuition. This will make war more inhumane. Regardless of the level of sophistication and programming, machines cannot replace humans in making the vital decision of taking another human's life." (General Statement) (2016)<br><br>Pakistan: "Based on these considerations, the introduction of LAWS would be illegal, unethical, inhumane and unaccountable as well as destabilizing for international peace and security with grave consequences. Therefore, their further development and use must ideally be pre-emptively banned through a dedicated Protocol of the CCW." (General Statement) (2016) | Ecuador: "[E]stas nuevas tecnologías [LAWS]…pueden estar reñidas con el Derecho Internacional Humanitario, la ética, [y] los principios de humanidad." (General statement) (2015)†<br><br>Austria: "The underlying unity of international humanitarian law is grounded on the basic values of humanity shared by all civilizations. The idea of humanity plays a crucial role and is reflected in the Martens clause, which is a binding rule under IHL and demands the application of "the principle of humanity" in armed conflict. In the context of LAWS, an interesting parallel is sometimes drawn to landmines, which were banned because of the delegation of the decision to initiate lethal force from humans." (Working paper) (2015)<br><br>Sri Lanka: "Rapid advancement of Artificial Intelligence (AI) and the possibility of fully autonomous functioning in weapon systems devoid of human control have created unprecedented risks and challenges to humanity and human values, demanding undelayed global attention." (General Statement) (2016)<br><br>Sri Lanka: "While acknowledging the important contribution by the experts in the field over the past two years, time is now opportune for this body to move further and to initiate a dialogue on this issue among States, who must eventually | Ireland: "The decisive questions may well be whether such weapons are acceptable under the principles of humanity, and if so, under what conditions." (General statement) (2015)<br><br>Ghana: "Our ultimate objective as States remains the preservation of human dignity and respect for basic sanctity of humanity at all times and most, especially, during armed conflicts. The laws of war must in this regard remain at the forefront of all our efforts and ahead of technological developments. Technology must not be allowed to overtake our commitment to these goals." (Way ahead statement) (2015)<br><br>Poland: "The main principles of IHL which are of interest to us would be: humanity, military necessity, discrimination and proportionality…. Looking at the present level of technological advancement, however, there are reasons for concern that the existing systems will not be able to meet those principles. Hence the importance of developing further the MHC [meaningful human control] concept and its institutional extension - the idea of MSC [meaningful state control]. The presence of human control in the form of institutional framework guarantees itself a reference to certain standards - legal and related international customs. Human or institutional oversight upholds accountability, the |

---

* "In relation to accountability, and in light of the impossibility of applying justice to a LAWS, in virtue of the aforementioned Articles it becomes necessary that international law determine who assumes the criminal and civil responsibility in the chain of command, including at the political level, for illegal acts of LAWS, as well as the eventual responsibility of the manufacturers and programmers [/developers] of the LAWS. Ignoring this necessity would be a violation of Article 86 of API, which establishes the criminal responsibility of superior commanders for the acts of subordinates."

† "These new technologies [LAWS]…can be in conflict with International Humanitarian Law, ethics, and the principles of humanity."

| | | | |
|---|---|---|---|
| | | make the ultimate call. We hope that such an intergovernmental process will help in ensuring clarity on the concerns of States, as well as to create a matrix of common elements. If we fail to live up to this expectation, it would result in denying the 2016 Review Conference which meets only once in five years, a historic opportunity to address this pressing issue decisively and frame the work that the CCW proposes to undertake in this connection. Our failure today, could result in the intensity of the development of LAWS in an unregulated environment, to the detriment of humanity." (General Statement) (2016)<br><br>Costa Rica: "Debe ser un ser humano quien siempre tome la decisión de emplear la fuerza letal, y en consecuencia, interiorizar el costo de cada vida perdida en las hostilidades, o asumir la responsabilidad por ello, como parte de un proceso deliberativo de interacción humana. No es aceptable La la delegación de este proceso, que deshumanizaría aún más los conflictos armados, pero sobre todo, bajo ninguna circunstancia debe permitir la comunidad internacional que se diluyan las responsabilidades institucionales, jurídicas y políticas que subyacen en el uso de la fuerza y que son inherentes al ser humano y al pacto social." (Human Rights and Ethical Issues Statement) (2016)* | rule of law and supports procedures through which our decisions may be verified." (Characteristics of LAWS/Meaningful human control statement) (2015)<br><br>Australia: "We have observed and considered the various ways of framing the question [including] an ethical approach, which raises the fundamental question whether the principles of humanity and dictates of public conscience can ever allow machines to select, attack and kill human beings, entirely outside of human control." (General Statement) (2016)<br><br>Sri Lanka: "It is important to consider safeguards that can help avoid the abuse and unintended consequences of the AI technology while reaping its benefits for the betterment of humanity." (General statement) (2016) |
| **Reference to international human rights law[†]** | Cuba: "se deben respetar los principios y propósitos de la Carta de las Naciones Unidas, la Declaración Universal de los Derechos Humanos, las obligaciones jurídicamente vinculantes en materia de derechos humanos que defienden el derecho a la vida, las libertades fundamentales y el respeto a la dignidad humana. Asimismo, aspectos del derecho consuetudinario basados en la ética." (Way ahead | Chile: "[T]here are arguments that consider that Human Rights are applicable and should be respected in the case of use of force at any time. They are complementary to IHL in case of armed conflict and, where there is no such conflict, Human Rights norms should apply exclusively. This has also been underscored by the International Court of Justice, which considers that both branches of International Law are to be taken into consideration for the protection of the human being and that the protection offered by human rights | Ireland: "Ireland also has concerns regarding eventual use of these technologies outside of traditional combat situations, for example in law enforcement, and this is one reason why we also see value in discussing these questions in other relevant fora such as, for example, the Human Rights Council, as the issue of autonomy is weapons systems is also relevant for International Human Rights Law." (General statement) (2015) |

---

* "It must always be a human being who makes the decision to use lethal force, and in consequence, [rough translation] internalizing the cost of each life lost in hostilities, or assuming responsibility for it, as part of a deliberative process of human interaction. It is unacceptable for this delegation that this process further dehumanize armed conflict; but above all, under no circumstances should the international community allow the dilution of the institutional, legal [/juridical] and political responsibilities that underlie the use of force and are inherent in humanity [/being human] and the social contract."

[†] General references to "international human rights law" were excluded, in favor of only including more specific statements, e.g., to specific human rights, specific bodies, specific questions of application, etc, as the former provided little substantive guidance (particularly since some states were or may have been referencing law enforcement, when international human rights law is generally understood to be in operation).

statement)[*] (2015) (Though this excerpt does not explicitly state AWS violate these laws, in the context of the statement, that meaning seems to be intended.)

Ecuador: "[Con respeto a LAWS, hay una] incongruencia con...los derechos humanos como el derecho a la vida y a la dignidad." (General statement) (2015)[†]

Mexico: "México considera que...[la] potencial uso [de los sistemas de armas plenamente autónomos] representa un riesgo en contra de los derechos humanos más fundamentales, como son el derecho a la vida y la dignidad." (Way ahead statement) (2015)[‡]

Pakistan: "The standards of International Human Rights Law are even more stringent [than the IHL rules which Pakistan believes LAWS violate]. These rules can be complex and entail subjective decision making requiring human judgment." (General statement) (2015) conventions does not cease in case of armed conflict.... it has become clearer that the battlefield use of LAWS would potentially affect Human Rights, including the right to life, the right to dignity, the right to freedom and security and the prohibition of torture and other forms of cruel, inhumane or degrading treatment. It should be noted that Human Rights are based on the principle of universality and timelessness, as established in the Universal Declaration of Human Rights and reiterated at the Vienna World Conference on Human Rights in 1993, which agreed that all States have the duty, independently of their political, economic and cultural systems, to promote and protect all human rights and fundamental freedoms as the birthright of all human beings." (Paper) (2015)

Sierra Leone: "Under no circumstances should the taking of the life of human beings be entrusted to machines, however well programmed. Sierra Leone therefore believes that the Human Rights Council should remained seized on the human rights aspects of LAWS, while respecting the mandate of CCW." (General Statement) (2016)

Sri Lanka: "While the primary focus has been on autonomous weapons usage in armed conflicts, once developed, there would be no guarantee that the same would not be used in the domestic law enforcement activities, with lethal or less-lethal force. As pointed out by the Special Rapporteur on

South Africa: "The use of such a weapon systems would need to comply with the fundamental rule of International Humanitarian Law, including those of distinction, proportionality and military necessity, as well as their potential impact on human rights." (General statement) (2015)

Spain: "Nuestro principal punto de partida en este empeño debe fundamentarse, como debe hacerlo además en relación con cualquier otro tipo de armas, en la necesidad del respeto más escrupuloso del Derecho Internacional Humanitario y del Derecho Internacional de los Derechos Humanos, cuya primacía entendemos irrenunciable." (General statement) (2015)[‡‡]

Sierra Leone: "We expect that [attending states will] provide a clear path for...taking decisions that would respect the human rights, including the right to life, of concerned persons." (General statement) (2015)

Sweden: "The review process may also include the use of non-lethal weapons by the armed forces or the use of lethal weapons by law enforcement agencies. In these instances the legality of the new weapon and its use needs to comply with human rights law which include the right to life and principles of necessity and proportionality as these principles are understood in the legal framework of human rights law." (Challenges to IHL Statement) (2016)[§§]

---

[*] "The principles and purposes of the United Nations Charter, the Universal Declaration of Human Rights, the legally binding human rights obligations that defend the right to life, fundamental liberties, and respect for human dignity must be respected. Additionally, aspects of customary law [translator's note: or "common law"] based on ethics."

[†] "[With respect to LAWS, there is an] incongruency between human rights like the right to life and to dignity."

[‡] "Mexico considers that...the potential use [of LAWS] represent a threat to the most fundamental human rights, like the right to life and to dignity."

[‡‡] "Our principle point of departure in this effort must be based, as must be done in relation to whatever type of weapons, on the need for the most scrupulous respect for International Humanitarian Law and International Human Rights Law, whose primacy we understand to be irrenounceable."

[§§] Sweden here is referring to its own Article 36 (Geneva Conventions Optional Protocol I) review process in order to emphasize that it is not limited to military-grade weapons. Apparently, they are also concerned with how the weapons they develop (even ones that may not be employed in a strictly military setting) would affect international human rights norms like the principles of necessity and proportionality. Thus, they seem to imply that any autonomous weapon systems that Sweden develops would be subject to an internal Article 36 review, regardless of whether it was intended to be employed for military purposes or for internal policing purposes.

| | | extrajudicial, summary or arbitrary executions, such use, both in a military context, and a law enforcement context could pose serious violation of human rights, in particular the right to life and dignity.[*] Given this non-derogable human rights dimension of the subject, we encourage that the matter continues to be pursued in the Human Rights Council as well, under relevant agenda items." (General Statement) (2016)<br><br>Costa Rica: "La cuestión ética que subyace en el debate sobre las armas autónomas letales es la dependencia creciente de la capacidad de las computadoras al tomar una decisión sobre si utilizar o no la fuerza contra seres humanos. Las utilización de estas armas podría tener repercusiones para el derecho a la vida, el derecho a la integridad física, el derecho a la dignidad humana y el derecho a la reparación." (Human Rights and Ethical Issues Statement) (2016)[†]<br><br>Chile: "El uso de SAL puede dejar eventualmente en estado de desprotección e indefensión legal a los civiles y combatientes que pudieran ser víctimas de actos cometidos por un SAL, violando de esa manera el derecho a la protección de la ley y de la presentación de de recursos legales de acuerdo a los artículos 7 y 8 de la Declaración Universal de los Derechos Humanos." (Human Rights and Ethical Issues Statement) (2016)[‡] | |

---

[*] Christof Heyns (Special Rapporteur) Report of the Special Rapporteur on extrajudicial, summary or arbitrary executions A/69/265 of 6 August 2014.

[†] "The ethical issue underlying the debate over lethal autonomous weapons is the growing dependence on the ability of computers to make a decision on whether to use force against human beings. The use of these weapons could have implications for the right to life, the right to physical integrity, the right to human dignity and the right to reparation."

[‡] "The use of LAWS could eventually lead to a state of unprotected and defenseless civilians and combatants who might become victims of acts committed by a LAWS, thus violating the right to protection under the law and the availability of legal recourse in accordance with Articles 7 and 8 of the UDHR."

# APPENDIX II

## STATES' POSITIONS AS REFLECTED BY THEIR STATEMENTS AT THE 2015 AND 2016 CCW MEETINGS OF EXPERTS ON LAWS

NB: When italics are used in the table below, they are for the purpose of highlighting particularly noteworthy portions of states' remarks that express their views on the law, especially when excerpts are lengthy; italics are not from the original statements.

| State[1] | Currently Unacceptable, Unallowable, or Unlawful | Need to monitor or continue to discuss | Need to regulate | Need to ban (or favorably disposed towards the idea) | Need for meaningful human control[2] | AP I Article 36 review necessary | Refers to legal principles while remaining undecided on per se legality of AWS |
|---|---|---|---|---|---|---|---|
| **Algeria** | | "Nous sommes confiants que les débats que nous aurons cette semaine permettront d'affiner notre compréhension sur les développements en cours concernant les systèmes autonomes en general, les approaches technique et légale envue de définir les SALA, les défis que posent ces systèmes au DIH ainsi que sur les questions relatives à léthique, aux droits de l'homme et à la sécurité." (General Statement) (2016)[3] | | "Aussi, la delegation algérienne est en faveur de la prohibition de l'acquisition, conception, développement, essais, déploiment, transfert, et utilisation des systèmes d'armes d'armes létaux autonomes «robots tueurs» par l'etablissement d'un instrument international juridiquement contraignant. Il serait, également, judicieux de prendre des mesures immédiates, par le biais d'un moratoire, en vue de surseoir au développement de ces systèmes." (General Statement) (2016)[4] | | | "L'introduction des SALA dans des conflits armés soulèverait de sérieux problems quant au respect des principes du DIH liés, d'une part, aux capacités de « jugement» et d'adaptation de ces systems aux environnements dynamiques dans lesquels ils opéreraient, et d'autre part, a leur prévisibilité et fiabilité. Les aspects moral et éthique de l'emploi de tells systems cotnre des êtres humains, en situation de guerre ou de paix, viennent ajouter de la complexité quant à un traitement approprié de cette question." (General Statement) (2016)[5] |
| **Argentina** | | | | | "La falta o baja frecuencia de control humano significativo de los SALA conducirá a decisiones sin intervención humana que podrían provocar consecuencias humanitarias impredecibles." (General statement) (2015)[6] | | "Resulta obvio que los principios humanitarios del DIH sobre proporcionalidad y discriminación aplicado a los SALA se encuentran visiblemente comprometidos y con numerosas alternativas de incumplimiento según se comporten un número de variables que intervienen en su uso. Es conveniente que cualquier desarrollo de los SALA, este sujeto a que se demuestre en forma indubitable que las mismas |

[1] Brazil, Canada China, Russia, and Palestine also offered statements in 2015, but their text is not available online. See UN Office at Geneva, 2015 Meeting of Experts on LAWS, Disarmament, http://www.unog.ch/80256EE600585943/(httpPages)/6CE049BE22EC75A2C1257C8D00513E26?OpenDocument (last visited March 13, 2016). Algeria, a non-member of CCW, see Chairperson of the Informal Meeting of Experts, Advaced Copy of the Report of the 2015 Informal Meeting of Experts on Lethal Autonomous Weapons Systems (LAWS) 1 (2015), http://www.genf.diplo.de/contentblob/4567632/Daten/5648986/201504berichtexpertentreffenlaws.pdf., also offered a statement, see UN Office at Geneva, above, and it too is unavailable online.

[2] The lack of clarity about what "meaningful human control" (MHC) and even "AWS"/"LAWS" actually mean could be obscuring a variety of views about the legality of AWS. In some cases, a state may believe MHC is achievable, so even if it opposes AWS without MHC on legal, moral, policy, or some other grounds, it might not believe a full ban is necessary; it may believe a version of AWS with MHC is possible and acceptable. Other states sometimes seem to think that no meaningful MHC is possible, with their statements about the necessity of MHC therefore implying AWS are not acceptable (though even then the states do not always explicitly call for a ban or stake their opposition on legal grounds). Adding to the confusion is the fact that some states seem to refer to civilian uses of LAWS which at first glance are not necessarily lethal, indicating they are more likely considering the underlying technology, not lethal autonomous weapons systems.

[3] "We trust that the debates of the forthcoming week will allow us to refine our comprehension of the ongoing developments in the field of autonomous systems in general, of the technical and legal approaches to define the lethal autonomous weapons systems, of the challenges created by these systems for international humanitarian law, and of the issues related to ethics, human rights and security."

[4] "The Algerian delegation is in favor of prohibiting the acquisition, design, development, test, deployment, transfer, and use of the lethal autonomous weapons systems "killing robots" by means of a legally binding international instrument. It would also be wise to take immediate measures, by means of a moratorium, with a view to delay the development of such systems."

[5] "Introducing lethal autonomous weapons systems into armed conflicts would raise serious problems with regard to compliance with the principles of international humanitarian law, linked, on one side, with the ability of these systems to "discern" and adapt to the dynamic environments where they would operate, and, on the other side, with their predictability and reliability. The moral and ethical features of using such systems against human beings, in situations of war or peace, make an appropriate treatment of this issue even more difficult."

[6] "The lack or low amount of significant human control in LAWS will bring decisions without human intervention that would provoke unpredictable humanitarian consequences."

| State[1] | Currently Unacceptable, Unallowable, or Unlawful | Need to monitor or continue to discuss | Need to regulate | Need to ban (or favorably disposed towards the idea) | Need for meaningful human control[2] | AP I Article 36 review necessary | Refers to legal principles while remaining undecided on per se legality of AWS |
|---|---|---|---|---|---|---|---|
| | | | | | | | poseen la capacidad de discriminar y de diferenciar la proporcionalidad conforme las instrumentos legales existentes. La falta o baja frecuencia de control humano significativo de los SALA conducirá a decisiones sin intervención humana que podrán provocar consecuencias humanitarias impredecibles. *La determinación de responsabilidades y rendición de cuentas por las consecuencias del empleo de los SALA se hace difuso y hasta impracticable, con lo cual ante esa situación se podrían considerar armas ilegales.*" (General statement) (2015)[7] |
| Austria | "To take just the example of the IHL principle of distinction: today, *clearly only humans are capable to distinguish reliably between civilians and combatants in a real combat situation,* thereby ensuring observance of the principle. Whether technology will be able to create at some future point machines with an equivalent capability seems to be a matter of speculation at this stage. In any case, the blurring of the fundamental distinction between the military and civilian spheres, between front and rear, as an ever more prominent feature of modern warfare, does not make this an easy task." (General statement) (2015) | "The two preceding Geneva expert meetings... provided a forum for the presentation and discussion of expert knowledge, and they offered an opportunity for political dialogue among governments with the participation of civil society... All of this needs to be continued." (General Statement) (2016) | "We are keenly aware that technology is moving fast, outpacing diplomatic deliberations. Let me therefore repeat Austria's call on States from last year's meeting. In order not to create undesirable faits accompli, states should decide immediately to refrain from, or suspend, activities which risk to prejudge the outcome of the international political discussion on LAWS." (General Statement) (2016) | | "At the same time, IHL of course will continue to require human control over armed attacks. The question therefore is how technological change can be managed so that human control can continue to be exercised in a meaningful way. To take an example: It is doubtful whether a single human actor surveilling the pre-programmed activity of a swarm of LAWS from a distance would be able to exercise meaningful, as opposed to purely formal control over the situation. Rather, we tend therefore to believe that the deployment of lethal force would have to be decided upon in a conscious and informed manner on a case by case | "The basis for the lawfulness of new weapons can be found in Article 36 of the First Additional Protocol to the Geneva Conventions which stipulates the obligation of every state party "to determine whether its employment would, in some or all circumstances, be prohibited by this Protocol or by any other rule of international law applicable to the High Contracting Party". This obligation covers obligations under treaties and customary | "Whether technology will be able to create at some future point machines with an equivalent capability [to humans' capability to distinguish reliably between civilians and combatants in a real combat situation] seems to be a matter of speculation at this stage. In any case, the blurring of the fundamental distinction between the military and civilian spheres, between front and rear, as an ever more prominent feature of modern warfare, does not make this an easy task....Austria acknowledges that much of the advanced technology associated with the development of LAWS has applications in the civil but also in the military fields that are perfectly acceptable. The essential point is to ensure that technology is applied in a responsible way. In particular, States should pay utmost attention that the pursuit of a particular technological development does not increase political and strategic risks, that it is fully compatible with the universal legal |

---

[7] "It is obvious that the humanitarian principles of IHL about proportionality and discrimination applied to LAWS would find themselves visibly compromised and with numerous other options in violation depending on the involvement of a number of variables that could be part of their use. It is advisable that whatever the development of LAWS, it is shown in indubitable form that LAWS have the capacity to discriminate and distinguish proportionality in conformance with existing legal instruments. It is obvious that the The lack or low amount of significant human control in LAWS will bring decisions without human intervention that would provoke unpredictable humanitarian consequences. The determination of responsibility and accountability for the consequences of the use of LAWS will be diffuse and almost impracticable, which in that situation would make it possible to consider them illegal weapons."

| State[1] | Currently Unacceptable, Unallowable, or Unlawful | Need to monitor or continue to discuss | Need to regulate | Need to ban (or favorably disposed towards the idea) | Need for meaningful human control[2] | AP I Article 36 review necessary | Refers to legal principles while remaining undecided on per se legality of AWS |
|---|---|---|---|---|---|---|---|
| | | | | | basis by a human actor, who could be held responsible under international law." (General statement) (2015)<br><br>"The perspective of weapons that may in the future take decisions about the use of force without human intervention poses a challenge to international humanitarian law. The concept of "meaningful human control" was brought up in this context, which should not be seen as introducing a new legal norm, but as evaluating LAWS on the basis of the existing standards in international humanitarian law. The use of these weapons has to be assessed on the basis of existing norms and principles of international humanitarian law, from which the necessity of a certain 'human control' can be derived.... The assessment of compliance with the existing standards and rules under IHL has to be taken in a contextual manner in the light of concrete circumstances. Circumstances in the battlefield are shifting and human control of a weapon is a necessary prerequisite. IHL does require that combatants can make an objective assessment of the facts when applying force | international law." (Working paper) (2015) | framework, and that it is handled in as transparent a way as possible. Therefore Austria calls on States to stop, or refrain from, any developments which do not clearly satisfy these criteria" (General statement) (2015) |

| State[1] | Currently Unacceptable, Unallowable, or Unlawful | Need to monitor or continue to discuss | Need to regulate | Need to ban (or favorably disposed towards the idea) | Need for meaningful human control[2] | AP I Article 36 review necessary | Refers to legal principles while remaining undecided on *per se* legality of AWS |
|---|---|---|---|---|---|---|---|
|  |  |  |  |  | and targeting an objective. This assessment must be made on a case-by-case basis, in view of the concrete circumstances. In this context, the concurrence and inter-action of the three principles of proportionality, distinction and precaution can be seen as the basis for what can be considered under IHL as a requirement to consider until when human control needs to be maintained. Such a concept implies that States have to use particular restrain before deciding about the development and the deployment of new weapons, even if the evaluation of each of these principles on their own may not necessarily lead to a negative compliance assessment." (Working paper) (2015)

"The principles of international humanitarian law imply the need for human control over the use of armed force, which is also reflected in the concept of 'meaningful human control' currently discussed in the context of laws." (General Statement) (2016) |  |  |
| **Australia** |  | "We must work harder in our collaborative examination of the issues, looking through all the relevant frames: technological, |  |  | "Over the coming week, we look forward to hearing more on meaningful human control; predictability; | "May I reaffirm at the outset that Australia takes seriously our responsibilities under | "We have observed and considered the various ways of framing the question [including] a legal approach, which asks how IHL applies to weaponisation of |

| State[1] | Currently Unacceptable, Unallowable, or Unlawful | Need to monitor or continue to discuss | Need to regulate | Need to ban (or favorably disposed towards the idea) | Need for meaningful human control[2] | AP I Article 36 review necessary | Refers to legal principles while remaining undecided on per se legality of AWS |
|---|---|---|---|---|---|---|---|
| | | legal and ethical. And we must work with the aim not just of stating our own positions, but of seeking common ground." (General Statement) (2016) | | | human judgement and critical functions." (General Statement) (2016) | the existing legal framework for reviewing new weapons under Article 36 of Additional Protocol I of 1977 to the Geneva Conventions of 1949. We fully support and adhere to the obligation to undertake a review of any proposed new weapon, means or method of warfare to determine whether its employment would, in some or all circumstances, be prohibited by international humanitarian law or other international law applicable to Australia. We encourage others also to undertake Article 36 reviews and look forward to hearing about the processes other States undertake to conduct a Review." (General Statement) (2016) | increasingly autonomous systems, whether lethal autonomous weapons systems would function in conformity with IHL rules, whether clarification or interpretation of existing law is required, or whether new rules need to be developed." (General Statement) (2016)

"If we were to settle, ultimately, on an agreement that there were limits to the autonomy that lethal weapons may possess, or that there were limits to the weaponisation of autonomous systems, we would also have to design ways, not just of defining, but of implementing, such limits, and of verifying compliance. We should not underestimate the complexity of this task. Common understandings and universal acceptance are essential, indeed foundational, to any effective and lasting agreement. *As an international community, we remain some way from common understandings and universal acceptance of the potential use of LAWS, and a long way from being able to set enforceable standards for their use.*" (General Statement) (2016) |
| Bolivia | | | | According to the Campaign to Stop Killer Robots, Bolivia has "called for a ban on fully autonomous weapons systems, citing concerns that the right to life should not be delegated and doubts that international humanitarian and human rights law is sufficient | | | |

| State[1] | Currently Unacceptable, Unallowable, or Unlawful | Need to monitor or continue to discuss | Need to regulate | Need to ban (or favorably disposed towards the idea) | Need for meaningful human control[2] | AP I Article 36 review necessary | Refers to legal principles while remaining undecided on per se legality of AWS |
|---|---|---|---|---|---|---|---|
| | | | | to deal with the challenges posed."[8] | | | |
| Canada | | "Canada continues to believe that International Humanitarian Law is sufficiently robust to regulate emerging technologies. That said, we also recognize that LAWS may raise unique challenges with regards to the weapons review process, such as related to testing and evaluation. There may also be unique challenges to ensuring the lawful use of LAWS generally, and in light of the significant impact a host of contextual factors could have upon their potential use. Such complexities could rightfully be the subject of further international discussion." (General Statement) (2016) | | | "Better fleshing out conceptual notions of 'meaningful human control' or appropriate human judgement is a "concrete, pragmatic and useful way in which we as an international community can continue to grapple with the challenges and possibilities posed by LAWS." (General Statement) (2016) | "Increased transparency and information sharing around guidelines and best practices for weapons reviews could play an important role in assisting States fulfill their Article 36 obligations with regards to LAWS." (General Statement) (2016) | "Working to promote and implement existing mechanisms for ensuring compliance with international law" is a "concrete, pragmatic and useful way in which we as an international community can continue to grapple with the challenges and possibilities posed by LAWS." (General Statement) (2016)<br><br>"Canada continues to believe that International Humanitarian Law is sufficiently robust to regulate emerging technologies. That said, we also recognize that LAWS may raise unique challenges with regards to the weapons review process, such as related to testing and evaluation. There may also be unique challenges to ensuring the lawful use of LAWS generally, and in light of the significant impact a host of contextual factors could have upon their potential use. Such complexities could rightfully be the subject of further international discussion." (General Statement) (2016) |
| Chile[9] | "En esta oportunidad quisiera hacer en primer lugar un comentario general: la posición de Chile relativa al desarrollo y posible uso de sistemas de armas sin control humano es de prohibición preventive… Nuestro país tiene la certeza que el despliegue de los SAL irá contra las disposiciones actuales de | "Por ello, el ejercicio que está llevando bajo el amparo de la Convención sobre Ciertas Armas Convenciales es un acierto político y esta delegación considera que debe seguir como tema permanente de la Agenda de la Convención." (Human Rights and Ethical Issues) (2016)[11] | "Si bien el avance de la tecnología podría desarrollar una inteligencia artificial capaz de distinguir entre objetos civiles y militares, elegir medios proporcionales para enfrentar a un adversario e incluso planificar al nivel estratégico una campaña militar | "En esta oportunidad quisiera hacer en primer lugar un comentario general: la posición de Chile relativa al desarrollo y posible uso de sistemas de armas sin control humano es de prohibición preventive… Nuestro país tiene la certeza que el despliegue de los SAL irá contra las disposiciones actuales de DDHH y de Derecho Internacional | "En este sentido, cabe señalar que en último término el único freno al daño indiscriminado por cualquier tipo de arma es la identificación de aquellos que manejan las armas – cualquiera sea su lado en el conflicto- con lo humano recíproco, es decir el bien ulterior de la humanidad. Este elemento identitario, si | | "Creemos que el punto actual en que se encuentra la evolución del Derecho Internacional Humanitario aún no da respuestas solidas a los desafíos que plantea un sistema autónomo que llegara a tomar la decisión de quitar la vida, con independencia completa de la orden de un humano. Este es un desafío legal que contiene vacíos necesarios de llenar como la responsabilidad final en caso de error, la merma en la dignidad humana o la |

---

| State[1] | Currently Unacceptable, Unallowable, or Unlawful | Need to monitor or continue to discuss | Need to regulate | Need to ban (or favorably disposed towards the idea) | Need for meaningful human control[2] | AP I Article 36 review necessary | Refers to legal principles while remaining undecided on per se legality of AWS |
|---|---|---|---|---|---|---|---|
| | DDHH y de Derecho Internacional Humanitaro. Más aun, Chile cree que ambos sistemas jurídicos no están preparados ni tienen jurisprudencia específica relativa a este tema. En tanto ello suceda, es de toda lógica apoyar la prohibición de los SAL antes de su despliegue. El CICR ha sido claro en mencionar que no existen disposiciones relativas al uso de SAL." (Human Rights and Ethical Issues Statement) (2016)[10] | | minimizando bajas civiles, el juicio, el discernimiento y el criterio político son atributos humanos esenciales para enfrentar con éxitor una crisis o conflict armado. Si bien los SAL podrían tener una serie benficios, especialmente el de evitar exponer a seres humanos a peligros mortales, su uso sin control humano significativo es un riesgo innecesario cuyos costos podrían ser muy altos en términos de sufrimiento humano. A manera de conclusión, es claro que la falta de control humano significativo los SAL también conlleva riesgos evidentes para todo el Sistema Internacional de los DDHH, el Derecho Internacional y la paz y por lo tanto su uso de desarrollo debe ser claramente normado por la comunidad internacional." (Human | Humanitaro. Más aun, Chile cree que ambos sistemas jurídicos no están preparados ni tienen jurisprudencia específica relativa a este tema. En tanto ello suceda, es de toda lógica apoyar la prohibición de los SAL antes de su despliegue. El CICR ha sido claro en mencionar que no existen disposiciones relativas al uso de SAL." (Human Rights and Ethical Issues Statement) (2016)[13] | entregado a un "efecto espejo", limita la capacidad de volición y de acción del ser humano. Creemos que el punto actual en que se encuentra la evolución del Derecho Internacional Humanitario aún no da respuestas solidas a los desafíos que plantea un sistema autónomo que llegara a tomar la decisión de quitar la vida, con independencia completa de la orden de un humano. Este es un desafío legal que contiene vacíos...[E]l trabajo en este foro debería ir acercándose hacia la definición de los imperativos éticos, por todos aceptados, para la mantención de un control humano significativo sobre cualquier sistema de armas, y su traducción en parámetros legales internacionales. Desde nuestra perspectiva nacional, se hace inaceptable que meros artefactos pudiesen comenzar a estar en condiciones de tomar decisiones autónomas sobre la vida y la muerte de | | posibilidad real de la rendición de cuentas."[15] (General statement) (2015) |

---

[10] "On this occasion, I would like to first make a general comment: the position of Chile relating to the development and possible use of weapons systems without human control is a preventative prohibition... *Our country is confident that the deployment of LAWS will go against the existing provisions of IHRL and IHL.* Moreover, Chile believes that both systems are not prepared nor do they have the specific jurisprudence relative to this subject. To the extent that this remains the case, it is entirely logical to support the prohibition of LAWS before their deployment. The ICRC has been clear in mentioning that there are no provisions relating to the use of LAWS." (Italics added)

[13] "On this occasion, I would like to first make a general comment: the position of Chile relating to the development and possible use of weapons systems without human control is a preventative prohibition... Our country is confident that the deployment of LAWS will go against the existing provisions of IHRL and IHL. Moreover, Chile believes that both systems are not prepared nor do they have the specific jurisprudence relative to this subject. To the extent that this remains the case, it is entirely logical to support the prohibition of LAWS before their deployment. The ICRC has been clear in mentioning that there are no provisions relating to the use of LAWS."

[15] "We believe that at the current moment in its evolution, International Humanitarian Law still does not give solid answers to the challenges laid out by an autonomous system that will be become able to take the decision to take a life, with complete independence from a human order. This is a legal challenge that contains lacunae that must be filled like the final responsibility in case of error, the reduction in human dignity, or the real possibility of accountability."

| State[1] | Currently Unacceptable, Unallowable, or Unlawful | Need to monitor or continue to discuss | Need to regulate | Need to ban (or favorably disposed towards the idea) | Need for meaningful human control[2] | AP I Article 36 review necessary | Refers to legal principles while remaining undecided on per se legality of AWS |
|---|---|---|---|---|---|---|---|
| | | | Rights and Ethical Issues) (2016)[12] | | las personas." (General statement) (2015)[14]<br><br>"Si bien los SAL podrían tener una serie benficios, especialmente el de evitar exponer a seres humanos a peligros mortales, su uso sin control humano significativo es un riesgo innecesario cuyos costos podrían ser muy altos en términos de sufrimiento humano. A manera de conclusión, es claro que la falta de control humano significativo los SAL también conlleva riesgos evidentes para todo el Sistema Internacional de los DDHH, el Derecho Internacional y la paz y por lo tanto su uso de desarrollo debe ser claramente normado por la comunidad internacional." (Human Rights and Ethical Issues) (2016)[15] | | |
| Colombia | | "A mi país le resulta fundamental contar con definiciones ampliamente aceptables internacionalmente. En ese contexto, valoramos la | "Entendemos que…el concepto de autonomía se refiere no sólo a la capacidad de operar por sí solo, sino también a la | | "De particular interés para mi delegación es observar la diferenciación que se establece entre los conceptos de | | |

[12] "While the advancement of technology could develop an AI capable of distinguishing between civilian and military objects, choose proportional means to confront an adversary and even plan at the strategic level, minimizing civilian casualties, the judgment, discrimination and the political viewpoint are essential human attributes for successfully confronting a crisis or armed conflict. While LAWS may have a series of benefits, especially in avoiding exposing human beings to mortal dangers, their use without significant [/meaningful] human control is an unnecessary risk whose costs could be very high in terms of human suffering. In conclusion, it is clear that the absence of significant human control over the LAWS also carries with it evident risks for the entire system of IHRL, international law and peace and therefore their use and development ought to be clearly regulated by the international community."

[14] "In this regard, it should be noted that ultimately the last brake on indiscriminate damage by whatever type of weapon is the identification of those who handle the weapons--whatever their side in the conflict--with their fellow human, that isto say the ultimate good of humanity. This identity element, if it introduces a "mirror effect," limits the volition and action capacity of the human being. We believe that the actual point at which one finds the evolution of International Humanitarian Law still does not give solid answers to the challenges laid out by an autonomous system that will be become able to take the decision to take a life, with complete independence from a human order. This is a legal challenge that contains lacunae…The work in this forum should be to work towards a definition of the ethical imperatives, accepted by all, for the maintenance of significant human control over whatever system of weapons, and its translation into international legal parameters. From our national perspective, it is unacceptable that mere devices could begin to be able to make autonomous decisions about the life and death of people."

[15] "While LAWS may have a series of benefits, especially in avoiding exposing human beings to mortal dangers, their use without significant [/meaningful] human control is an unnecessary risk whose costs could be very high in terms of human suffering. In conclusion, it is clear that the absence of significant human control over the LAWS also carries with it evident risks for the entire system of IHRL, international law and peace and therefore their use and development ought to be clearly regulated by the international community."

| State[1] | Currently Unacceptable, Unallowable, or Unlawful | Need to monitor or continue to discuss | Need to regulate | Need to ban (or favorably disposed towards the idea) | Need for meaningful human control[2] | AP I Article 36 review necessary | Refers to legal principles while remaining undecided on *per se* legality of AWS |
|---|---|---|---|---|---|---|---|
| | | oportunidad de participar en estas reuniones y beneficiarnos del nutrido intercambio de puntos de vista. Reconociendo el avance temático alcanzado, vemos que persisten áreas en las que aún se requiere mayor examen." (Way ahead statement) (2015)[17] | capacidad de tomar decisiones sin la intermediación de un ser humano. Para mi delegación es indudable que la regulación de este último tipo de armas es requerida a nivel multilateral con el fin de garantizar que persista en todo momento un control por parte de los seres humanos, para evitar que sea una máquina la que tome decisiones de vida o muerte sobre las personas. Para mi país, resulta igualmente importante establecer diferencias entre el concepto de armas ofensivas, y el de aquellas tecnologías, que pueden contar con diferentes grados de autonomía y que puedan ser utilizadas en aplicaciones de tipo militar, como herramientas para contribuir a la preservación de la vida o en labores de tipo humanitario." (Way | | automatización y autonomía. Entendemos que el primer concepto, el de automatización involucra la capacidad de un instrumento para operar por sí solo, mientras que el concepto de autonomía se refiere no sólo a la capacidad de operar por sí solo, sino también a la capacidad de tomar decisiones sin la intermediación de un ser humano. Para mi delegación es indudable que la regulación de este último tipo de armas es requerida a nivel multilateral con el fin de garantizar que persista en todo momento un control por parte de los seres humanos, para evitar que sea una máquina la que tome decisiones de vida o muerte sobre las personas." (Way ahead statement) (2015)[19] | | |

---

[17] "To my country it is fundamental to have definitions that are widely acceptable internationally. In that context, we appreciate the opportunity to participate in these meetings and benefit from the considerable exchange in points of view. Recognizing the thematic progress made, we see that areas which still require further examination persist."

[19] "Of particular interest to my delegation is observing the differentiation that is established between the concepts of automation and autonomy. We understand that the first concept, that of automation involves the ability of an instrument to operate by itself, while the concept of autonomy refers not only to the ability to operate by itself, but also the ability to make decisions without the intermediation of a human being. For my delegation it is unquestionable that regulation of the latter kind of weapon is required at the multilateral level in order to ensure that control on the part of human beings persists at all times , to prevent it from being a machine that make decisions of life or death about people."

| State[1] | Currently Unacceptable, Unallowable, or Unlawful | Need to monitor or continue to discuss | Need to regulate | Need to ban (or favorably disposed towards the idea) | Need for meaningful human control[2] | AP I Article 36 review necessary | Refers to legal principles while remaining undecided on *per se* legality of AWS |
|---|---|---|---|---|---|---|---|
| | | | ahead statement) (2015)[18]  (Note that in 2015 Colombia was not necessarily endorsing regulation in contrast to a ban, but rather was focused on the need for some sort of control over autonomous weapons, in contrast to automized weapons.) | | | | |
| Costa Rica | "A nuestro criterio, de llegar a desarrollarse, los sistemas de armas autonómas letales podrían tomar decisions de vida o muerte sin la intervención de un ser humano, por lo que serían contrarios al derecho internacional humanitario y el derecho internacional de los derechos humanos. Como lo han señalado muchas delegaciones, cualquier tipo de arma que se desarrolle debe respetar los principios de distinción, proporcionalidad y precaución en el ataque. El uso de la fuerza debe ser el ultimo recurso y conlleva, inexorablemente, una responsabilidad humana, politica, jurídica e institucional que es | "Empezamos esta semana con el deseo de seguir debatiendo sobre los aspectos técnicos de las armas autonómas letales. Estos debates serán de mucha utilidad para seguir conociendo, un poco méas a fondo, este tema tan complejo." (General Statement) (2016)[21] | | "Las pasadas reuniones de expertos han servido para ir formando un entendimiento común. Mi delegación es del criterio que estas armas deberían prohibirse antes de que lleguen a construirse, de la misma forma que se hizo con los láseres cegadores. Por ello, creemos conveniente que la próxima Conferencia de Revisión estudie la posibilidad de convocar una reunion de expertos gubernamentales que pueda identificar elementos necesarios para elaborar una convención internacional." (General Statement) (2016)[22] | | | |

[18] "We understand that … the concept of autonomy refers not only to the ability to operate by itself, but also the ability to make decisions without the intermediation of a human being. For my delegation it is clear that regulation of the latter type of weapon is required at the multilateral level in order to ensure that control by human beings persists at all times, to prevent it from being a machine that make decisions of life or death about people. For my country, it is equally important to differentiate between the concept of offensive weapons, and those technologies, which can have varying degrees of autonomy and can be used in military applications, as tools to contribute to the preservation of life or in work of a humanitarian nature."

[21] "We set out this week with the desire to continue discussing the technical aspects of lethal autonomous weapons. These discussions will be of great use for understanding this complex topic a little more thoroughly."

[22] "Past meetings have helped in the formation of a common understanding. My delegation is of the opinion that these weapons ought to be prohibited before they are built, in the same manner as with blinding laser weapons. Therefore, we find it fitting that the next Review Conference consider convening a meeting of government experts who can identify the elements needed to develop an international convention."

| State[1] | Currently Unacceptable, Unallowable, or Unlawful | Need to monitor or continue to discuss | Need to regulate | Need to ban (or favorably disposed towards the idea) | Need for meaningful human control[2] | AP I Article 36 review necessary | Refers to legal principles while remaining undecided on per se legality of AWS |
|---|---|---|---|---|---|---|---|
| | indelegable." (General Statement) (2016)[20] | | | | | | |
| Croatia | | "Given the complexity of the subject we are dealing with, Croatia is in favor of establishing a Group of Governmental Experts to lead the way forward (in exploring the issue of LAWS)." (Final statement) (2015) | | In 2015, Croatia did not necessarily endorse a ban on all LAWS, but it seemed to at least indicate it would be favorably inclined towards efforts to ban any LAWS that did not involve "meaningful human control" (presumably, at least a meaningful opportunity to override an act).

" [W]e as mankind are ethically obliged to ensure meaningful control with regard to the lethal use of force… [E]ven the idea of developing an international prohibition of weapons systems operating without meaningful human control should not be something unthinkable, particularly given the calls for a moratorium on the development of such weapons." (General statement) (2015)

"[O]ur position is that fundamental questions of life and death cannot be assigned to armed autonomous weapons systems…[E]fforts conducted by [a] possible GGE might not prove sufficient to assure that humanity retains full control over its own fate. Thus, the possibility of (creating) a future legally-binding instrument which will set clear rules on the issue of | " [W]e as mankind are ethically obliged to ensure meaningful control with regard to the lethal use of force… [E]ven the idea of developing an international prohibition of weapons systems operating without meaningful human control should not be something unthinkable, particularly given the calls for a moratorium on the development of such weapons." (General statement) (2015)

"[O]ur position is that fundamental questions of life and death cannot be assigned to armed autonomous weapons systems…[E]fforts conducted by [a] possible GGE might not prove sufficient to assure that humanity retains full control over its own fate. Thus, the possibility of (creating) a future legally-binding instrument which will set clear rules on the issue of weaponized autonomous systems should not be left completely out of sight." (Final statement) (2015)

In 2015, Croatia did not speak to whether the need | | |

| State[1] | Currently Unacceptable, Unallowable, or Unlawful | Need to monitor or continue to discuss | Need to regulate | Need to ban (or favorably disposed towards the idea) | Need for meaningful human control[2] | AP I Article 36 review necessary | Refers to legal principles while remaining undecided on *per se* legality of AWS |
|---|---|---|---|---|---|---|---|
| | | | | weaponized autonomous systems should not be left completely out of sight." (Final statement) (2015) | for meaningful human control was based on law; instead, its claims seemed to rest on ethics. | | |
| Cuba[23] | | | | "Cuba favorece la adopción de un instrumento internacional jurídicamente vinculante que disponga la prohibición total de las armas letales autónomas, especialmente las antipersonales. La letalidad, además de la autonomía, es un patrón básico que debe guiar la prohibición o regulación de las armas letales autónomas u otras categorías de armas. Mientras mayor sea la letalidad, más estricto debe ser el marco que las regule. No se podría emplear estas armas con plenas garantías de cumplimiento y observancia de las normas y principios del Derecho Internacional Humanitario (DIH). No podría garantizarse la distinción entre civiles y combatientes, ni la evaluación de proporcionalidad, entre otros principios básicos del DIH. Tampoco podría hacerse una evaluación efectiva de la responsabilidad del Estado por hechos internacionalmente. Cuba considera que los beneficios tácticos que aparentemente resultarían del empleo de las armas letales autónomas, pudieran derivar en que los Estados que las | | "Las nuevas tecnologías tienen que acogerse a lo ya dispuesto en el artículo 36 del Protocolo I Adicional a los Convenios de Ginebra de 1977." (General statement) (2015)[26]

"Reafirmamos que debe respetarse el artículo 36 del Protocolo Adicional I de los Convenios de Ginebra de 1949, relativo a la protección de las víctimas de los conflictos armados, el cual establece que "Cuando una Alta Parte contratante estudie, desarrolle, adquiera o adopte una nueva arma, o nuevos medios o métodos de guerra, tendrá la obligación de determinar si su empleo, en ciertas condiciones o en todas las circunstancias, estaría prohibido por el presente Protocolo o por cualquier otra norma de derecho internacional aplicable | |

---

[23] Cuba apparently also delivered a statement on Transparency and the Way Ahead, but its text is not available online. See UN Office at Geneva, 2015 Meeting of Experts on LAWS, Disarmament, http://www.unog.ch/80256EE600585943/(httpPages)/6CE049BE22EC75A2C1257C8D00513E26?OpenDocument (last visited March 13, 2016).

[26] "New technologies must accept what is already arranged in Article 36 of Additional Protocol I to the Geneva Conventions of 1977."

| State[1] | Currently Unacceptable, Unallowable, or Unlawful | Need to monitor or continue to discuss | Need to regulate | Need to ban (or favorably disposed towards the idea) | Need for meaningful human control[2] | AP I Article 36 review necessary | Refers to legal principles while remaining undecided on *per se* legality of AWS |
|---|---|---|---|---|---|---|---|
| | | | | poseen dejen de considerar el conflicto armado como un último recurso. También resulta preocupante el riesgo de que estas armas caigan en manos de actores no estatales no autorizados. El alto costo de la tecnología requerida por las armas autónomas solo puede ser asumido por los países desarrollados, lo que incrementa aún más la asimetría entre países ricos y pobres. Pensamos que los cuantiosos recursos humanos y financieros que se dedican a la investigación y desarrollo de las armas letales autónomas, deberían utilizarse en beneficio del desarrollo social y económico de la humanidad. *Cuba reafirma que en tanto no exista una norma internacional que prohíba estas armas*, las mismas deben regirse por las disposiciones del Derecho Internacional." (General statement; generally repeated in Position paper/National document) (2015)[24] | | a esa Alta Parte contratante". (Way ahead statement) (2015)[27] | |
| | | | | "Reafirmamos que debe haber una prohibición total y completa de las armas autónomas letales, | | | |

| State[1] | Currently Unacceptable, Unallowable, or Unlawful | Need to monitor or continue to discuss | Need to regulate | Need to ban (or favorably disposed towards the idea) | Need for meaningful human control[2] | AP I Article 36 review necessary | Refers to legal principles while remaining undecided on *per se* legality of AWS |
|---|---|---|---|---|---|---|---|
| | | | | especialmente las antipersonales, mediante un instrumento jurídicamente vinculante. Pensamos que debe trabajarse en una definición internacionalmente acordada de armas autónomas letales como aquellas armas programadas por el hombre para que, una vez activadas, puedan seleccionar y atacar objetivos sin necesidad de otra intervención humana para cumplir las tareas que se les asigna." (Characteristics of LAWS statement) (2015)[25] | | | |
| Czech Republic | | "There are obvious risks associated with introduction of weapons with autonomous capabilities, but as with any other weapon there are undoubtedly certain benefits as well. The risks would be mitigated by the obligation of states to review these new weapons against the requirements of international humanitarian law or any rule of international law applicable to the reviewing party to acceptable level. The Czech Republic remains convinced that there is already an obligation of High Contracting Parties of the Additional Protocol I to the Geneva Conventions to review whether new weapon, means or method of warfare would comply with international humanitarian law or not. The benefits of these weapons could be increased by developing | "There are obvious risks associated with introduction of weapons with autonomous capabilities, but as with any other weapon there are undoubtedly certain benefits as well. The risks would be mitigated by the obligation of states to review these new weapons against the requirements of international humanitarian law or any rule of international law applicable to the reviewing party to acceptable level. The Czech Republic remains convinced that there is already an obligation of High Contracting Parties of the Additional | | "The Czech Republic is of view [sic] that the *ultimate decision to end somebody's life must remain under meaningful human control.* This principle should be a common understanding in the international community and *we believe it is already implicitly inherent to international humanitarian law.* The challenging part is to establish what precisely 'meaningful human control' would entail." (General statement) (2015) | | "The Czech Republic remains convinced that there is already an obligation of High Contracting Parties of the Additional Protocol I to the Geneva Conventions to review whether new weapon, means or method of warfare would comply with international humanitarian law or not." (General statement) (2015) |

| State[1] | Currently Unacceptable, Unallowable, or Unlawful | Need to monitor or continue to discuss | Need to regulate | Need to ban (or favorably disposed towards the idea) | Need for meaningful human control[2] | AP I Article 36 review necessary | Refers to legal principles while remaining undecided on *per se* legality of AWS |
|---|---|---|---|---|---|---|---|
| | | autonomous capabilities that can lead to better protection of non-combatants lives. We should be mindful of all the pros and cons and should not jump to premature conclusions such as that the development, production and use of these weapons should be absolutely and pre-emptively prohibited....In any way, given the complexity of these matters, it would be useful to have at least some key definitions as soon as possible in order to ensure common understanding of what we are actually talking about" (General statement) (2015) | Protocol I to the Geneva Conventions to review whether new weapon, means or method of warfare would comply with international humanitarian law or not. The benefits of these weapons could be increased by developing autonomous capabilities that can lead to better protection of non-combatants lives. We should be mindful of all the pros and cons and should not jump to premature conclusions such as that the development, production and use of these weapons should be absolutely and pre-emptively prohibited.... From humanitarian point of view it might be more reasonable to concentrate on certain critical autonomous features of weapons that could be regulated or prohibited, rather than pursue absolute ban of these weapons." (General statement) (2015) | | | | |
| Denmark | | | | | "[T]he use of autonomous weapons systems...must be in compliance with international humanitarian law...and must remain under 'meaningful human control.'" (General statement) (2015) | | "[T]he use of autonomous weapons systems...must be in compliance with international humanitarian law...and must remain under 'meaningful human control.'" (General statement) (2015) |

| State[1] | Currently Unacceptable, Unallowable, or Unlawful | Need to monitor or continue to discuss | Need to regulate | Need to ban (or favorably disposed towards the idea) | Need for meaningful human control[2] | AP I Article 36 review necessary | Refers to legal principles while remaining undecided on *per se* legality of AWS |
|---|---|---|---|---|---|---|---|
| Ecuador | "Estas nuevas tecnologías...pueden estar reñidas con el Derecho Internacional Humanitario, la ética, los principios de humanidad y los dictados de la conciencia pública, por lo que su desarrollo debería ser prohibido para prevenir su uso futuro....[Con respecto a estas tecnologías, hay una] incongruencia con el Derecho Internacional Humanitario y con los derechos humanos como el derecho a la vida y a la dignidad. Nos preocupan aspectos fundamentales que merecen ser analizados y discutidos en profundidad como el uso dual de los sistemas autónomos para fines pacíficos y para fines bélicos; la ausencia de infalibilidad de tales sistemas y posibilidad cierta de errores de programación y de despliegue; su vulnerabilidad ante los ataques cibernéticos; responsabilidad legal en cuanto a la delegación de autoridad y a la toma de decisiones; incumplimiento del Derecho Internacional Humanitario en cuanto a la secuencia ininterrumpida de responsabilidad y las normas de la distinción, la proporcionalidad y las | "Por todas estas consideraciones, Ecuador apoya lo siguiente: 1. Avanzar en esta Reunión de Expertos hacia una definición de trabajo y caracterización de los SALA, que permita concretar la materia de nuestras deliberaciones y el futuro camino a seguir." (2016)[29] | | "Estas nuevas tecnologías...que pueden estar reñidas con el Derecho Internacional Humanitario, la ética, los principios de humanidad y los dictados de la conciencia pública, por lo que su desarrollo debería ser prohibido para prevenir su uso futuro....[Con respecto a estas tecnologías, hay una] incongruencia con el Derecho Internacional Humanitario y con los derechos humanos como el derecho a la vida y a la dignidad. Nos preocupan aspectos fundamentales que merecen ser analizados y discutidos en profundidad como el uso dual de los sistemas autónomos para fines pacíficos y para fines bélicos; la ausencia de infalibilidad de tales sistemas y posibilidad cierta de errores de programación y de despliegue; su vulnerabilidad ante los ataques cibernéticos; responsabilidad legal en cuanto a la delegación de autoridad y a la toma de decisiones; incumplimiento del Derecho Internacional Humanitario en cuanto a la secuencia ininterrumpida de responsabilidad y las normas de la distinción, la proporcionalidad y las precauciones en los ataques; la inobservancia de la ética y de los derechos humanos fundamentales, en particular | "Algunas opiniones expresadas en anteriores reuniones de expertos han . señalado la posibilidad de mantener o establecer un control humano significativo en las funciones críticas de estos sistemas, en lo que se refiere a la identificación de objetivos y uso de fuerza letal. Pero otras opiniones de expertos en la materia indican que con el aumento de la autonomía, el control humano no es posible y la decisión de uso de la fuerza Letal pasaría a las máquinas. Al parecer, el meollo de la discusión debería centrarse en la autonomía en las "funciones críticas" de los sistemas de armas existentes y emergentes y contestar la pregunta clave "¿en qué punto y en cuáles circunstancias corremos el riesgo de perder el control humano significativo sobre el uso de la fuerza?" ¿Estamos dispuestos a correr ese riesgo? Creemos que los Estados y la Comunidad Internacional, debemos actuar de manera oportuna y eficaz para adecuar el Derecho Internacional a fin de que responda con mayor agilidad a los retos y desafíos de carácter ético, | | "Distinción: Es improbable que los SALA puedan ser programados para que puedan distinguir entre los combatientes y los civiles. Su modo mecánico de inteligencia hace imposible aplicar la regla de distinción no sólo de los civiles, sino de los combatientes fuera de combate por estar heridos o enfermos, de aquellos que se rinden y desertores. Proporcionalidad: Los SALA al no tener el razonamiento humano para aplicar la regla de proporcionalidad en el complejo ambiente de la guerra |

| State[1] | Currently Unacceptable, Unallowable, or Unlawful | Need to monitor or continue to discuss | Need to regulate | Need to ban (or favorably disposed towards the idea) | Need for meaningful human control[2] | AP I Article 36 review necessary | Refers to legal principles while remaining undecided on *per se* legality of AWS |
|---|---|---|---|---|---|---|---|
|  | precauciones en los ataques; la inobservancia de la ética y de los derechos humanos fundamentales, en particular de la cláusula de Martens... La Constitución del Ecuador...prohíbe y condena el desarrollo y uso de armas de destrucción masiva y de armas de efectos indiscriminados violatorias del Derecho Internacional Humanitario como es el caso de los Drones armados y sería el caso de las Armas Letales Autónomas. Consideramos inaceptable, inadmisible y anti-ético permitir que las máquinas decidan sobre la vida o muerte de seres humanos. Estamos abiertos a la discusión y esperamos, al final de esta reunión, tener respuestas satisfactorias a un sinnúmero de interrogantes, como por ejemplo: la distinción entre civiles y combatientes; identificación de los |  |  | de la cláusula de Martens...En ausencia de respuestas satisfactorias a muchas preguntas y de la falta de garantía de cumplimiento con el Derecho Internacional Humanitario, Ecuador considera que los Estados debemos tomar acciones a tiempo para prevenir la creación y desarrollo de los Sistemas de Armas Letales Autónomas a través de normas y leyes nacionales que los prohíban y de un instrumento internacional jurídicamente vinculante que prohíba el desarrollo, uso e inversiones en tales sistemas" (General statement) (2015)[30]<br><br>(See also "Currently unlawful" box for further relevant 2015 excerpts.)<br><br>"Por todas estas consideraciones, Ecuador apoya lo siguiente: 1. Avanzar en esta Reunión de Expertos hacia una definición de trabajo y caracterización de los SALA, que permita concretar la materia de nuestras | jurídico y humanitario, que imponen las investigaciones y nuevos adelantos científicos y tecnológicos para uso bélico en el presente y futuro, como es el caso de los Drones armados y de los Sistemas de Armas Letales Autónomas." (General statement) (2015)[32] |  |  |

[30] "These new technologies ... can be at odds with international humanitarian law, ethics, principles of humanity and the dictates of public conscience, so that their development should be banned to prevent future use. ... [With respect to these technologies there is an] incongruity with international humanitarian law and human rights like the right to life and dignity. We are concerned about fundamental aspects that deserve to be analyzed and discussed in depth like the dual use of autonomous systems for peaceful purposes and for military purposes; the absence of infallibility in such systems and the true possibility of errors in programming and deployment; their vulnerability to cyber attacks; legal responsibility regarding the delegation of authority and decision-making; breach of international humanitarian law regarding the uninterrupted sequence of responsibility and rules of distinction, proportionality and precautions in attack; failure to observe ethics and fundamental human rights, in particular the Martens Clause...In the absence of satisfactory answers to many questions and the failure to ensure compliance with international humanitarian law, Ecuador considers that States must take action in time to prevent the creation and development of Lethal Autonomous Weapons Systems through norms and national laws that ban them and a legally binding international instrument prohibiting the development, use and investment in such systems."

[32] "Some opinions expressed in previous meetings of experts have pointed out the possibility of maintaining or establishing meaningful human control in the critical functions of these systems, in what is referred to as the identification of targets and use of lethal force. But other opinions of experts in the field indicate that with increasing autonomy, human control is not possible and the decision to use lethal force would pass to the machines. Apparently, the crux of the discussion should focus on autonomy in "critical functions" of existing and emerging weapons systems and answer the key question "to what extent and in what circumstances we we run the risk of losing significant human control over the use of force?" Are we willing to take that risk? We believe that States and the international community, must act in a timely and effective manner to tailor international law to the purpose of responding with greater agility to the challenges of ethical, legal and humanitarian character, imposed by research and new scientific and technological advances for military use in the present and future, as in the case of armed drones and Autonomous Lethal Weapon Systems."

| State[1] | Currently Unacceptable, Unallowable, or Unlawful | Need to monitor or continue to discuss | Need to regulate | Need to ban (or favorably disposed towards the idea) | Need for meaningful human control[2] | AP I Article 36 review necessary | Refers to legal principles while remaining undecided on *per se* legality of AWS |
|---|---|---|---|---|---|---|---|
| | objetivos militares; la capacidad de cancelar un ataque ante el riesgo de efectos mortales y desproporcionados en civiles; la distinción entre combatientes activos y aquellos fuera de combate o que se han rendido; la distinción entre civiles que participan en las hostilidades y aquellos armados que no participan, como los agentes de seguridad pública o cazadores; ausencia de sentimientos como la compasión y perdón ante una rendición; la responsabilidad y rendición de cuentas ante crímenes de guerra y violaciones del Derecho Internacional Humanitario" (General statement) (2015)[28] | | | deliberaciones y el futuro camino a seguir; 2. Acordar en esta Reunión recomendar a la Conferencia de Examen de la Convención de Armas Convencionales el establecimiento de un Grupo de trabajo Intergubernamental para la elaboración de un Tratado Internacional de Prohibición Completa del Desarrollo, Producción y Uso de los SALA; 3. Mientras el Tratado de Prohibición es negociado y entra en vigor, establecer una moratoria sobre la inversión, investigación, ensayo, producción, ensamblaje, transferencia, adquisición, emplazamiento y uso de los SALA; 4. Fortalecer los mecanismos nacionales para la revisión legal y la implementación del DIH para asegurar que nuevos tipos de armas puedan ser usados de conformidad con el DIH. Finalmente, adoptar las normas jurídicas necesarias a nivel nacional para prohibir la inversión, el desarrollo, producción y uso de los SALA." (General Statement) (2016)[31] | | | |

---

[28] "These new technologies ... can be at odds with international humanitarian law, ethics, principles of humanity and the dictates of public conscience, so that their development should be banned to prevent future use. ... [With respect to these technologies there is an] incongruity with international humanitarian law and human rights like the right to life and dignity. We are concerned about fundamental aspects that deserve to be analyzed and discussed in depth like the dual use of autonomous systems for peaceful purposes and for military purposes; the absence of infallibility in such systems and the true possibility of errors in programming and deployment; their vulnerability to cyber attacks; legal responsibility regarding the delegation of authority and decision-making; breach of international humanitarian law regarding the uninterrupted sequence of responsibility and rules of distinction, proportionality and precautions in attack; failure to observe ethics and fundamental human rights, in particular the Martens Clause... Ecuador's Constitution ... prohibits and condemns the development and use of weapons of mass destruction and indiscriminate weapons in violation of international humanitarian law, as in the case for armed drones and would be the case for Autonomous Lethal Weapons. We consider unacceptable, impermissible and unethical to allow machines to decide the life or death of human beings. We are open to discussion and we hope at the end of this meeting, to have satisfactory answers to countless questions, such as: the distinction between civilians and combatants; identification of military objectives; the ability to cancel an attack faced with the risk of fatal and disproportionate effects on civilians; the distinction between active combatants and those outside of combat or who have surrendered; the distinction between civilians that participate in hostilities and those who do not, like public security officers or hunters; absence of feelings such as compassion and forgiveness in the face of surrender; responsibility and accountability for war crimes and violations of international humanitarian law."

[31] "For all of these considerations, Ecuador supports the following: 1. In this Expert Meeting, advance towards a working definition and characterization of LAWS, that would permit the concretization of the material [/outcomes] of our deliberations and the future path to follow. 2. Agree in this Meeting to recommend to the CCW Conference the establishment of an Intergovernmental Working Group for the elaboration of an International Treaty for the Complete [/Comprehensive] Prohibition of the Development, Production and Use of LAWS. 3. While the

| State[1] | Currently Unacceptable, Unallowable, or Unlawful | Need to monitor or continue to discuss | Need to regulate | Need to ban (or favorably disposed towards the idea) | Need for meaningful human control[2] | AP I Article 36 review necessary | Refers to legal principles while remaining undecided on *per se* legality of AWS |
|---|---|---|---|---|---|---|---|
| Egypt | | | | "International attention to subject of lethal autonomous weapons has grown rapidly over the past year. Such weapons have generated widespread concern about their impacts, including with respect to distinction, proportionality, and their lack of accountability. At present there is no treaty body governing such technologies, but there is overarching rules governing this field via international humanitarian law. The need for evaluation is urgent and timely. Experience shows that it is necessary to ban a weapon system that is found to be excessively injurious or indiscriminate before they are deployed, as we have seen with blinding lasers and non-detectable fragments. We look forward to the convening of the experts meeting and hopes it works as an eye-opener. There are ramifications for the value of human life. We are concerned about the possibility of acquisition by terrorists and armed groups. A ban could prevent this, but until that is achieved, we support the calls for a moratorium on development of such technology to allow for meaningful debate and to reach greater international consensus. It might be too late | | | |

Treaty of Prohibition is being negotiated and entering into force, establishing a moratorium over the investment, research, testing, production, assembly, transfer, acquisition, deployment [/installation?], and use of LAWS. 4. Strengthen national mechanisms for legal review and the implementation of IHL to ensure that new types of weapons can be used in conformance with IHL. Finally, adopting the necessary legal standards at the national level to prohibit the investment, design, production and use of LAWS."

| State[1] | Currently Unacceptable, Unallowable, or Unlawful | Need to monitor or continue to discuss | Need to regulate | Need to ban (or favorably disposed towards the idea) | Need for meaningful human control[2] | AP I Article 36 review necessary | Refers to legal principles while remaining undecided on *per se* legality of AWS |
|---|---|---|---|---|---|---|---|
| | | | | after they are developed to work on an appropriate response. Technology should not overtake humanity. This technology raises many concerns that need to be fully addressed" (CCW Intervention) (2013)[33] | | | |
| Finland | | "The discussions have demonstrated the complexity of the issue. What especially makes LAWS a challenging issue is the fact that we are discussing characteristics of a system instead of a particular clearly defined weapon. This makes the discussion uniquely distinct from any other discussion on the disarmament fora. When thinking about LAWS we are in fact discussing whether autonomy may be used within a specific task namely using lethal force. Since the issue of LAWS is so multifaceted, we will need some clearly defined definitions at some stage. Compared with last year's deliberations it seems that in this year's deliberations the concept of Meaningful Human Control proved not to be as clear a concept as we had originally thought. It remains to be seen if this concept would serve future discussions in a way that would help us in clearly defining LAWS. As we are also speaking about special characteristics of a system that has not yet been developed we are inevitably | | | | "Finland will, for our part, review the national implementation of article 36 during this year and we are also open to the idea of creating international standards for the implementation of this norm." (General Statement) (2016) | "Finland highlights the importance of adhering to the rule of international humanitarian law in all situations. In our opinion IHL is fully applicable also in a situation where LAWS would be used as a means of warfare on the battleground. We further underline, that each and every state has the ultimate responsibility in every situation where norms of international humanitarian or human rights law are breached." (General Statement) (2016) |

---

[33] See Campaign to Stop Killer Robots, Country Statements on Killer Robots 14-15 (2014), http://www.stopkillerrobots.org/wp-content/uploads/2013/03/KRC_CountryStatus_14Mar2014.pdf (quoting Egypt's November 15, 2013 CCW intervention). As explained above, though Egypt did not express a desire for a ban via a written statement at the 2015 or 2016 CCW Meeting of Experts, it did orally indicate the cited preference for a moratorium on the development of AWS until more debate has occurred. Egypt's position has been included here to more fully represent states' attitudes on an important issue.

| State[1] | Currently Unacceptable, Unallowable, or Unlawful | Need to monitor or continue to discuss | Need to regulate | Need to ban (or favorably disposed towards the idea) | Need for meaningful human control[2] | AP I Article 36 review necessary | Refers to legal principles while remaining undecided on *per se* legality of AWS |
|---|---|---|---|---|---|---|---|
| | | facing a situation where speculation still plays a major role. Taking into consideration the rapid technological development it is almost impossible to foresee how technology will evolve in the coming decades and how this technology will be used. Thus it is impossible to say with certainty weather future systems could fully comply with IHL. Instead of speculating how technology will evolve in the future, it might be better to concentrate on certain critical functions or how the interaction between the system and humans would be addressed. We should be asking ourselves what is left when we strip the speculations away from the discussions. What we will be left with is not the question whether LAWS can comply with IHL or not. Based on our moral and ethical considerations we will rather have to address the fundamental questions on whether we want an autonomous weapon to become a reality or not. As High Contracting Parties to the CCW, it is our responsibility and obligation to protect current and future generations from excessive harm. But whether this is best done by banning or by allowing the development of LAWS is not a simple question. Human beings are not perfect, we make mistakes and our judgment can be easily affected. At the same time it is also fair to assume that machines, as human creations, are subject to flaws and imperfections as well. The question is really whether we | | | | | |

| State[1] | Currently Unacceptable, Unallowable, or Unlawful | Need to monitor or continue to discuss | Need to regulate | Need to ban (or favorably disposed towards the idea) | Need for meaningful human control[2] | AP I Article 36 review necessary | Refers to legal principles while remaining undecided on *per se* legality of AWS |
|---|---|---|---|---|---|---|---|
| | | foresee that human kind will cause less harm to itself and coming generations by relying on machines or relying on humans and their judgment. This is where we need to converge our opinions further. We belong to those that would see it possible to continue the discussions on LAWS in one format or another. A more formal mode of discussions might serve a purpose when we consider the path towards the 2016 Review Conference. Not prejudging the outcome of any discussions we would see benefits in a more defined process and a more focused discussion. As some others have indicated one possibility to take the discussions forward would include the establishment of a Group of Governmental Experts as the chairman has also put forward as an option in his food-for-thought paper. We also stand ready to consider different ways to improve transparency concerning LAWS." (Way ahead statement) (2015)<br><br>"We are looking forward to a vivid and fruitful exchange during the upcoming week. We believe that the framework of the CCW is the right place to continue the discussions also in the future. We greatly appreciate and support the chairman's effort to steer these discussions towards finding a common working definition of LAWS, as we believe that this is the key to deepening our deliberations. We see that this work would merit from continued | | | | | |

| State[1] | Currently Unacceptable, Unallowable, or Unlawful | Need to monitor or continue to discuss | Need to regulate | Need to ban (or favorably disposed towards the idea) | Need for meaningful human control[2] | AP I Article 36 review necessary | Refers to legal principles while remaining undecided on per se legality of AWS |
|---|---|---|---|---|---|---|---|
| | | discussions within the framework of a Governmental Group of Experts and we thus support that the establishment of a GGE would be one of the recommendations coming out of this meeting." (General Statement) (2016) | | | | | |
| France | | "La France est persuadée de l'utilité de des réunions informelles d'experts et de la nécessité d'approfondir notre réflexion sur les SALA. Il appartient à la CCAC de rester saisie d'un sujet qui entre pleinement dans son mandat. Pour conclure, je souhaite rappeler la suggestion française, introduite l'année dernière, d'un réexamen périodique de la question des SALA, si l'état de nos connaissances ne nous permet pas de faire aboutir rapidement nos réflexions." (General statement) (2015)[34]<br><br>"S'agissant maintenant de l'avenir, et sans préjuger des décisions qui seront prises par la réunion des Etats-parties en novembre prochain, la France est favorable à la poursuite de nos travaux, dans un cadre informel tant que nous n'aurons pas une compréhension partagée de ce dont nous parlons. Nous souhaitons donc le renouvellement du mandat de notre groupe à l'identique. Pour | | | | "L'Article 36 du Protocole I additionnel aux Conventions de Genève, qui prévoit l'évaluation des nouveaux systèmes d'armes au regard du DIH, peut constituer un cadre de réflexion pertinent." (General Statement) (2016)[37]<br><br>"La France considère que l'examen de licéité prévu par l'article 36 du 1er protocole additionnel aux conventions de Genève constitue une base essentielle pour répondre aux défis poses par les technologies émergentes en matière de systèmes d'armes, y compris celles visant au renforcement de leur autonomie." | "D'un point de vue juridique, je crois que de nombreuses délégations ont souligné l'importance du respect du DIH dans les phases de développement et d'emploi des SALA. La France estime que les principes du DIH s'appliquent pour encadrer le développement et l'emploi des SALA. Je veux citer également la question de la responsabilité, qui est naturellement centrale. A ce stade, rien ne permet de définir avec certitude les contours de la responsabilité de chaque acteur, qui dépendra de leur rôle dans l'utilisation du SALA. La possibilité d'identifier un acteur responsable est cruciale pour savoir si les principes existants du DIH demeurent suffisants ou non." (General statement) (2015)[39]<br><br>"Il est aujourd'hui trop tôt pour savoir si l'on pourra un jour développer des SALA conformes dans leur emploi aux principes de discrimination et de proportionnalité du DIH, mais nous ne pouvons pas prévoir les progrès techniques à venir. Par ailleurs, comme cela a été rappelé plusieurs fois dans cette enceinte, tout dépend du milieu dans lequel ces systèmes seront déployés : l'incapacité présumée de ces systèmes à distinguer un civil d'un combattant ne pose problème que dans un |

[34] "France is convinced of the value of informal meetings of experts and the need to deepen our reflection on the SALA. It is up to the CCW to remain seized of a subject fully within its mandate. To conclude, I wish to remind you of the French suggestion, introduced last year, of a periodic review of the question of LAWS, if the state of our knowledge does not allow us to swiftly conclude our reflections."

[37] "Article 36 of Protocol I additional to the Geneva Conventions, which provides for the evaluation of new weapons in light of international humanitarian law, can constitute a pertinent framework for reflection."

[39] "From a legal point of view, I believe that numerous delegations have underlined the important of respect for IHL in the phases of development and employment of LAWS. France believes that IHL principles apply to frame [or regulate] the development and employment of LAWS. I would like to cite equally the question of responsibility, that is naturally central. At this stage, nothing allows one to define with certainty the contours of the responsibility of each actor, that will depend on their role in the utilization of LAWS. The possibility of identifying a responsible actor is crucial in order to know if the existing principals of IHL remain sufficient or not."

| State[1] | Currently Unacceptable, Unallowable, or Unlawful | Need to monitor or continue to discuss | Need to regulate | Need to ban (or favorably disposed towards the idea) | Need for meaningful human control[2] | AP I Article 36 review necessary | Refers to legal principles while remaining undecided on *per se* legality of AWS |
|---|---|---|---|---|---|---|---|
| | | autant, nous estimons également nécessaire d'avoir, en novembre prochain, une discussion sur l'évolution du processus, notamment dans la perspective de la conférence d'examen de la CCAC en 2016. Nous pourrions notamment, dans le cadre de la conférence d'examen, envisager de passer à un format permettant d' adopter des conclusions négociées par consensus. Je saisis cette opportunité en conclusion, pour rappeler la suggestion française, introduite l'année dernière, d'un réexamen périodique de la question des SALA, si l'état de nos connaissances ne nous permet pas de faire aboutir rapidement nos réflexions." (Review processes/Way ahead statement) (2015)[35]<br><br>"La France est persuadée de la nécessité d'approfondir notre réflexion sur les SALA et de la légitimé de CCAC à rester saisie d'un sujet qui correspond pleinement à son mandat." (General Statement) (2016)[36] | | | | (Challenges to IHL Paper) (2016)[38] | environnement où la machine aura à faire cette distinction entre civils et combattants, ce qui n'est pas toujours le cas. Tous les champs de bataille ne comprennent pas de civils. Les SALA sont donc soumis à une forte logique de milieu et leur déploiement dans les milieux spatiaux et sous-marins, par exemple, semble a priori poser moins de problems... Même si un SALA s'avérait capable de respecter le DIH, il resterait toutefois un certain nombre de problèmes. Un premier problème est celui de la dilution de la responsabilité, qui serait plus difficile mais peut-être pas impossible à établir. Une autre question serait cellé de savoir si la prolifération des SALA – et les nouveaux moyens ou méthodes de guerre qu'il pourrait impliquer – satisferait aux objectifs de maintien de la paix et de la sécurité internationale de la Charte des Nations Unies." (Ethics/Overarching issues statement) (2015)[40]<br><br>"D'un point de vue juridique, la France estime que les principes du DIH s'appliquent pour encadrer le développement et l'emploi des SALA. A ce stade, il est impossible de determiner si un SALA pourrait ou non respecter le DIH, mais de développer de tells systèmes que |

[35] "Turning now to the future, and without prejudice to the decisions that will be taken by the Meeting of States Parties this coming November, France supports the continuation of our work in an informal setting until we have a shared understanding of what we're talking about. Identically, we wish the renewal of the mandate of our group. However, we also consider it necessary to have, in November, a discussion on the evolution of the process, particularly in view of the CCW Review Conference in 2016. We could include, as part of the conference review, a consideration of switching to a format to adopt conclusions negotiated by consensus. I take this opportunity in conclusion, to remind you of the French suggestion, introduced last year, to a periodic review of the issue of LAWS, if the state of our knowledge does not allow us to swiftly conclude our thoughts."

[36] "France is convinced of the need to deepen our reflection on lethal autonomous weapons systems as well as of the legitimacy of the CCWC to remain seized of a matter that fully corresponds to its mandate."

[38] "France considers that the test of lawfulness set forth in Article 36 of Protocol I additional to the Geneva Conventions represents an essential starting point to address the challenges raised by the emerging technologies in the field of weapons, including those technologies that aim at strengthening the autonomy of weapons."

[40] "Today it is too early to know if we will one day be able to develop LAWS that conform in their use with the principles of discrimination and proportionality in IHL, but we cannot predict the technological progress to come. Moreover, as has been reiterated numerous times in this chamber, it all depends on the environment in which these systems will be deployed: the presumed incapacity of these systems to distinguish a civilian from a combatant will only pose a problem in an environnement in which a machine will have to make that distinction between civilians and combatants, which is not always the case. All battlefields do not include civilians. LAWS are then subject to a strong logic of environment and their use in space [ie, outer space] and undersea environments, for example, seems a priori to pose fewer problems... Even if LAWS proved to be capable of respecting IHL, a certain number of problems would nevertheless remain. A first problem is that of the dilution of responsibility, that would be difficult but perhaps not impossible to establish. Another question would be whether the proliferation of LAWS—and new means or methods of warfare that they would implicate—would satisfy the United Nations' Charter's objectives of maintaining peace and international security."

| State[1] | Currently Unacceptable, Unallowable, or Unlawful | Need to monitor or continue to discuss | Need to regulate | Need to ban (or favorably disposed towards the idea) | Need for meaningful human control[2] | AP I Article 36 review necessary | Refers to legal principles while remaining undecided on per se legality of AWS |
|---|---|---|---|---|---|---|---|
| | | | | | | | si leur capacité à s'y conformer était prouvée." (General Statement) (2016)[43] |
| Germany | "Germany will not accept that the decision over life and death is taken solely by an autonomous system without any possibility for human intervention....Germany is of the opinion that given the actual state of artificial intelligence and other components of LAWS, a legal weapons review for the time being inevitably would lead to the result of LAWS being illegal, as they are not able to meet the requirements set out by Article 36 AP I." (Final statement) (2015)

"We will not accept that the decision to use force, in particular the decision over life and death, is taken solely by an autonomous system without any possibility for a human intervention in the selection and engagement of targets....[Such a use would be a] red line [that] should not be crossed." (General statement) (2015)

Though on their own the first part of the first 2015 excerpt and the entirety of the second 2015 excerpt could be mere reflections of | "If there are no LAWS yet and nobody seems to have the intention to cross the line where we would lose human control over a given weapon system then we should take care to closely monitor the development and introduction of any new weapon system to guarantee that there will be no transgression....Germany would...welcome the establishment of a GGE [Group of Governmental Experts] in the framework of the CCW to discuss and propose transparency measures." (Final statement) (2015)

"We reaffirm our will to contribute in pushing this urgent issue forward within the Convention on Certain Conventional Weapons which is the appropriate forum for further discussions and settlements in regard to this emerging technology." (General Statement) (2016) | | | "We will not accept that the decision to use force, in particular the decision over life and death, is taken solely by an autonomous system without any possibility for a human intervention in the selection and engagement of targets....[Such a use would be a] red line [that] should not be crossed." (General statement) (2015)

"We have reiterated on Monday the two pillars of the German position with respect to LAWS: Unconditional respect for-international law and the necessity to exercise appropriate levels of human control over the use of force." (Weapons reviews/Possible Challenges to IHL statement) (2015)

"Germany will not accept that the decision over life and death is taken solely by an autonomous system without any possibility for human intervention." (Final statement) (2015)

"In our view a working definition of LAWS should start with identifying a | "The development of any autonomous weapons system, and LAWS in particular, would clearly require [Article 36] legal reviews." (Possible Challenges to IHL/Weapons Reviews Statement) (2015)

"Without LAWS being a reality yet and without even a definition of LAWS transparency and confidence building measures especially with regard to the introduction of new weapons systems are of crucial importance. We should therefore make full use of the process of Legal Weapons Review in accordance to Art 36 AP I in sharing national regulations, looking for common standards and specific procedures for early detecting developments where human control risks to get lost." (General Statement) (2016)

"To implement the obligation pursuant to Article 36 of the 1977 | "Independent of the above is the question whether future LAWS will be able to live up to the discussed requirements. in order to be lawful. In our view, this considerable 'technical challenge' that developers face is nevertheless not deemed to put the sufficiency [sic] of the existing law into question." (Weapons reviews/Challenges to IHL statement) (2015)

Germany affirmed "the principle of unconditional respect for International Law and International humanitarian law. The use of possible future weapons systems, also LAWS are subject to International Law without restrictions." (General Statement) (2016) |

[43] "From a legal viewpoint, France considers that the principles of international humanitarian law apply in framing the development and use of lethal autonomous weapons systems. At this stage, it is impossible to determine whether a lethal autonomous weapons system could or not comply with international humanitarian law, but France, faithful to its international undertakings, could not envisage developing such systems unless their ability to comply with it were proven."

| State[1] | Currently Unacceptable, Unallowable, or Unlawful | Need to monitor or continue to discuss | Need to regulate | Need to ban (or favorably disposed towards the idea) | Need for meaningful human control[2] | AP I Article 36 review necessary | Refers to legal principles while remaining undecided on *per se* legality of AWS |
|---|---|---|---|---|---|---|---|
| | policy and could be limited only to AWS wherein no human override is possible, the second part of the former indicates Germany's opposition was at least partially based on a belief all AWS are currently unlawful. That does leave open the possibility that, for Germany, a *future form* of AWS that was more technologically advanced and allowed for human overrides *could be lawful*.<br><br>"As already underlined in the Meetings of Experts before, Germany will certainly adhere to the principle that it is not acceptable, that the decision to use force, in particular the decision over life and death, is taken solely by an autonomous system without any possibility for a human intervention." (General Statement) (2016)[42] | | | | common understanding of at least minimum requirements regarding the necessity to exercise appropriate levels of human control over the use of force, especially the decision over life and death." (General Statement) (2016) | Additional Protocol I to the 1949 Geneva Conventions, Germany established a permanent Steering Group within the Federal Ministry of Defence (MOD) entitled "Review of New Weapons and Methods of Warfare… Obviously, other states may use different methods of examination for the Article 36 review process. We believe that international trust and confidence-building could be furthered by increasing transparency regarding these review mechanisms. A first step could be to make public the national procedures. The CCW could provide the adequate framework." (Statement on the Implementation of Weapons Reviews under Article 36 Additional Protocol I) (2016) | |
| Ghana[43] | | | | "Ghana is very much concerned about the possible use of lethal autonomous weapon systems at any time in the future, for the many | | | |

[42] It's important to note that Germany doesn't actually use the word "unlawful." Nevertheless, it does seem as though they are indicating that they consider the use of lethal force by fully autonomous weapon systems to be illegitimate. Not only do they explicitly state that it is "not acceptable" for a weapon system to have control over life and death, but they portray their current stance as a repetition of the stance that they took in last year's meeting. (In last year's meeting, they unequivocally stated that they considered LAWS to be unlawful.)

[43] Ghana offered statements at the meeting but is not a party to the CCW (other non-signatory states attended but did not offer statements). See Chairperson of the Informal Meeting of Experts, Advaced Copy of the Report of the 2015 Informal Meeting of Experts on Lethal Autonomous Weapons Systems (LAWS) 1 (2015), http://www.genf.diplo.de/contentblob/4567632/Daten/5648986/201504berichtexpertentreffenlaws.pdf.

| State[1] | Currently Unacceptable, Unallowable, or Unlawful | Need to monitor or continue to discuss | Need to regulate | Need to ban (or favorably disposed towards the idea) | Need for meaningful human control[2] | AP I Article 36 review necessary | Refers to legal principles while remaining undecided on *per se* legality of AWS |
|---|---|---|---|---|---|---|---|
| | | | | reasons and fears that theses systems present to us by their very nature. It is obvious that proponents of these systems believe that they will not be the victims but others will. We need to avoid moving in this direction of self perfection to the promotion and preservation of human dignity för humanity as a whole. History confirms that todays victim can become tomorrows perpetrator, especially, when we take into consideration the ever increasing development and spread of technology. Won't we be heading towards a potential quagmire in the near future. In our view fully automated lethal systems must be proscribed before they are fully developed because of the concerns aforesaid and shared by a larger number of delegations represented here in this meeting." (Ethics/Challenges to IHL statement) (2015)<br><br>"We join the recommendation made by other delegations that Member States commit to engage in discussions that should enable us establish a weapons review mechanism to ensure the prevalence of transparency and enable the international community monitor weapon developments. This must constitute a part of an overall drive towards the promulgation of a convention that regulates and proscribes the production | | | |

| State[1] | Currently Unacceptable, Unallowable, or Unlawful | Need to monitor or continue to discuss | Need to regulate | Need to ban (or favorably disposed towards the idea) | Need for meaningful human control[2] | AP I Article 36 review necessary | Refers to legal principles while remaining undecided on *per se* legality of AWS |
|---|---|---|---|---|---|---|---|
| | | | | of those weapons that cannot meet the basic standards set for us by the IHL and IHRL....We also need to consider by the next meeting a mandate to commence negotiations on the promulgation of the aforementioned Convention." (Way ahead statement) (2015) | | | |
| Greece | | | | | "The discussion [about prohibition], however, takes a very different dimension when it is addressed ethically or politically, bringing to the fore the question of 'meaningful human control', but this is not a legal norm. Hence, we should in our view be clear about what it is we are discussing and avoid a conflation which makes things even more complicated." (Challenges to IHL statement) (2015) | | "For the sake of argument, let us suppose that in the future autonomous weapon systems are developed which can fully comply with IHL and its cardinal principles, such as distinction, proportionality and precautions in attack; a weapon operating with better precision than being under human control. We are not there yet; indeed we are far from that juncture, however, for the sake of our debate, let us envisage such a hypothetical scenario. In such a case, one may ask oneself what would the legal basis be to justify their prohibition. Some have argued that we should draw parallels from the blinding lasers precedent when we banned a weapon that did not yet exist. However, blinding lasers were prohibited because they violated the rule that a weapon should not be of a nature to cause superfluous injury or unnecessary suffering. Again though, let us suppose that this criterion is also fulfilled by a future autonomous weapon....[T]o argue that LAWs comply or do not comply with IHL at this stage would amount to an oracle of Delphi. What is left then is basically an ethical question, not a legal one. It boils down to the fundamental question of whether humans should delegate life and death decisions to machines and definitely Greece, like others, does not feel comfortable with such a prospect. Or as Germany stated on Monday, full autonomy is a line that should not be crossed, the line being when there is no longer any human oversight, as the |

| State[1] | Currently Unacceptable, Unallowable, or Unlawful | Need to monitor or continue to discuss | Need to regulate | Need to ban (or favorably disposed towards the idea) | Need for meaningful human control[2] | AP I Article 36 review necessary | Refers to legal principles while remaining undecided on *per se* legality of AWS |
|---|---|---|---|---|---|---|---|
| | | | | | | | delegate from the United Kingdom remarked earlier. The question which then arises is how does one operationalize this ethical concern into a legal provision. The only legal principle which comes to mind is the Martens Clause, given its dependence on the dictates of public conscience. Does though such a general principle suffice to lead to the codification in the future of a new set of legally binding rules? We have our doubts. Indeed, should we isolate this issue to its legal parameters, then- in our view- there is no other logical conclusion than the one made by Dr. Boothby earlier, that is, that a thorough and systematic weapons review is the only practical solution, at least at the present stage, to address the issue of LAWS from a legal angle. The discussion, however, takes a very different dimension when it is addressed ethically or politically, bringing to the fore the question of 'meaningful human control', but this is not a legal norm. Hence, we should in our view be clear about what it is we are discussing and avoid a conflation which makes things even more complicated." (Challenges to IHL statement) (2015) |
| India | | "In our view, there continue to be wide divergences on issues such as "meaningful human control". It is also not clear whether distinctions can be drawn between oversight, review, control or judgement or how they would apply to a new weapon system from the time of its conception, design and development to production, deployment and use or for that matter when does a weapon system cross the line to become a new weapon or its use constitute a new method of warfare. These are complex questions with no easy answers. | | | | | "[A] discussion on LAWS should include questions on their compatibility with international law including international humanitarian law as well as the impact of their possible dissemination on international security. Our aim should be to strengthen the CCW in terms of its objectives and purposes through increased systemic controls on international armed conflict in a manner that does not widen the technology gap amongst states or encourage the increased resort to military force in the expectation of lesser causalities or that use of lethal force can be shielded from the dictates of public conscience." (Way ahead statement) (2015) |

| State[1] | Currently Unacceptable, Unallowable, or Unlawful | Need to monitor or continue to discuss | Need to regulate | Need to ban (or favorably disposed towards the idea) | Need for meaningful human control[2] | AP I Article 36 review necessary | Refers to legal principles while remaining undecided on *per se* legality of AWS |
|---|---|---|---|---|---|---|---|
| | | In these circumstances, it may be prudent not to jump to definitive conclusions. At the same time, we cannot ignore the inexorable march of technology, in particular of dual use nature, expanding the autonomous dimension of lethal weapon systems, while keeping in mind the CCW remains a relevant and acceptable framework for addressing such issues of concern to the international community. Hence, there may be merit in continued consideration of LAWS on the basis of an agreed mandate to be adopted by the Meeting of States Parties in November this year." (Way ahead statement) (2015)  "In our view, there continue to be wide divergences on key issues- definitional issues, mapping autonomy- whether distinctions can be drawn between oversight, review, control or judgment or how they would apply to a new weapon system from the time of its conception, design and development to production, deployment and use or for that matter when does a weapon system cross the line to become a new weapon or its use constitute a new method of warfare." (General Statement) (2016 | | | | | "In our view, a discussion on LAWS should include questions on their compatibility with international law including international humanitarian law." (General Statement) (2016) |
| Ireland[44] | | "My delegation has noted a range of overlapping nuances and assumptions made by states during interventions throughout this process. For example some | | | "Ireland's starting position in relation to the development of Lethal Autonomous Weapons Systems is that weapons | | "The decisive questions may well be whether such weapons are acceptable under the principles of humanity, and if so, under what conditions. Ireland also has concerns regarding eventual use of these |

---

[44] Ireland is listed as having made a "Characteristics of LAWS" statement, but the text of this statement is not available online; instead, a duplicate copy of its general statement seems to have been uploaded. See UN Office at Geneva, 2015 Meeting of Experts on LAWS, Disarmament, http://www.unog.ch/80256EE600585943/(httpPages)/6CE049BE22EC75A2C1257C8D00513E26?OpenDocument (last visited March 13, 2016). Also, its statement on "Transparency and the Way Ahead" is not available online. See id.

| State[1] | Currently Unacceptable, Unallowable, or Unlawful | Need to monitor or continue to discuss | Need to regulate | Need to ban (or favorably disposed towards the idea) | Need for meaningful human control[2] | AP I Article 36 review necessary | Refers to legal principles while remaining undecided on *per se* legality of AWS |
|---|---|---|---|---|---|---|---|
| | | delegations have stated that Lethal Autonomous Weapons Systems do not exist at present and others say that fully Autonomous Weapons Systems do not exist and may even never exist. This flexible terminology has been helpful in allowing states to engage in the process from their own perspective but if we are to move on to more substantive discussions we will need an agreed basis for that work." (Draft Statement on Definitions) (2016) | | | should remain under effective Human Control." (General statement) (2015) | | technologies outside of traditional combat situations, for example in law enforcement, and this is one reason why we also see value in discussing these questions in other relevant fora such as, for example, the Human Rights Council, as the issue of autonomy is weapons systems is also relevant for International Human Rights Law." (General statement) (2015) |
| Israel | | "The first assumption relates to the necessity to maintain an open mind regarding both potential risks as well as possible positive capabilities of future LAWS. It is difficult to foresee today how autonomous capabilities may look like in ten, twenty or fifty years from now. As a consequence, any responsible discussion of future LAWS, should be undertaken in a cautious and prudent fashion. The second assumption is that an assessment of such systems and of their employment should be conducted on a case by case basis. Future LAWS could take on a variety of forms, have a wide array of capabilities and nuances, and may be intended to operate in a range of operational environments, from the simplest ones to more complicated ones. Consequently, a serious deliberation on legal aspects of LAWS should take these factors into account." (General statement) (2015) | | | | | "[T]he use of future LAWS, as any other means of warfare, must comply with the applicable rules of IHL. In fact, prudent employment of LAWS may even promote compliance with IHL. In this context, it seems that states should, when considering a lethal autonomous weapon system, subject the system in question to an internal legal review." (General statement) (2015)

"Several States mentioned the phrase "meaningful human control". Several other States did not express support for this phrase. Some of them thought that it was too vague... We have also noted, that even those who did choose to use the phrase "meaningful human control", had different understandings of its meaning....In our view, it is safe to assume that human judgment will be an integral part of any process to introduce LAWS, and will be applied throughout the various phases of research, development, programming, testing, review, approval, and decision to employ them. LAWS will not actually be making decisions or exercising judgment by themselves, but will operate as designed and programmed by humans. Humans who intend to develop and |

| State[1] | Currently Unacceptable, Unallowable, or Unlawful | Need to monitor or continue to discuss | Need to regulate | Need to ban (or favorably disposed towards the idea) | Need for meaningful human control[2] | AP I Article 36 review necessary | Refers to legal principles while remaining undecided on *per se* legality of AWS |
|---|---|---|---|---|---|---|---|
| | | "These discussions have also highlighted that LAWS do not currently exist and that, as the technologies are rapidly developing, it would be difficult, if at all possible, at this stage, to predict how future LAWS would look like, and what their characteristics, capabilities and limitations will be. At the same time, fundamental questions were left open. For example, there seemed to be no agreement as to the exact definition of LAWS, and there were clearly divergent views on questions relating to the appropriate level of human judgment, or involvement / intervention, over LAWS. In this regard, many states - including Israel - were not supportive of the call made by some states for a preemptive ban on LAWS. Considering the divergent views on these questions, it is our view that an incremental, step by step approach, is not only preferable but inevitable. There is much work still ahead of us in order to effectively assess the various aspects of LAWS and potentially forge shared understandings in this regard." (General Statement) (2016) | | | | | employ a lethal autonomous weapon system, are responsible to do so in a way that ensures the system's operation in accordance with the rules of IHL. In this regard, the context − referring to the specific system and the specific scenario of use − is of utmost importance. The characteristics and capabilities of each system must be adapted to the complexity of its intended environment of use. Where deemed necessary, the system's operation would be limited by, for example, restricting the system's operation to a specific perimeter, during a limited timeframe, against specific types of targets, to conduct specific kinds of tasks, or other such limitations which are all set by a human. Likewise, for example, if necessary, a system could be programmed to refrain from action when facing complexities it cannot resolve." (Characteristics of LAWS statement) (2015)<br><br>"There seemed to be a general understanding that the use of LAWS, like other weapon systems, is subject to the Law of Armed Conflict and that LAWS should undergo legal review before they are deployed. Israel shares these understandings and will further elaborate on the issue of legal review in the course of the session dedicated to IHL, later this week." (General Statement) (2016)<br><br>"We should also be aware of the military and humanitarian advantages that may be associated with LAWS, both from operational as well as legal and ethical aspects. These may include better precision of targeting which would minimize collateral damage and reduce risk to combatants and non-combatants." (General Statement) (2016) |

| State[1] | Currently Unacceptable, Unallowable, or Unlawful | Need to monitor or continue to discuss | Need to regulate | Need to ban (or favorably disposed towards the idea) | Need for meaningful human control[2] | AP I Article 36 review necessary | Refers to legal principles while remaining undecided on *per se* legality of AWS |
|---|---|---|---|---|---|---|---|
| | | | | | | | "On the issue of human machine interface, it is safe to assume that human judgment will be an integral part of any process to introduce LAWS, and will be applied throughout the various phases of research, development, programming, testing, review, approval, and decision to employ them. LAWS will only operate as designed and programmed by humans." (General Statement) (2016)<br><br>"Notwithstanding that Israel is not a party to the First Additional Protocol to the Geneva Conventions, and as such is not bound by Article 36 of that Protocol, Israel is of the view that applying legal reviews to new weapons is the best instrument for a State to ensure that it uses only lawful means of warfare during armed conflicts." (Challenges to IHL Statement) (2016)<br><br>"In order to determine the legality of the weapon under consideration, the legal review focuses on examining three questions: (a) Whether the weapon in question is capable of being used discriminately; (b) Whether the weapon is calculated to cause superfluous injury or unnecessary suffering; (c) Whether the weapon falls within a category of weapons that has been specifically prohibited or restricted by an international. In some cases the outcome of the review may be a finding that the weapon is not unlawful per se, but that its legal use is subject to specific restrictions arising out of the applicable rules of international law." (Challenges to IHL Statement) (2016) |
| Italy | | "Among them we agree to have further focused discussions on the issues concerning the definition of LAWS; on the crucial question of dual-use technology and the challenges regarding increasingly complex technology | | | | | "A different group of weapons systems includes those able to make autonomous decisions based on their own learning and rules, and that can adapt to changing environments independently of any pre-programming. Such systems, which could select targets and decide when to use |

| State[1] | Currently Unacceptable, Unallowable, or Unlawful | Need to monitor or continue to discuss | Need to regulate | Need to ban (or favorably disposed towards the idea) | Need for meaningful human control[2] | AP I Article 36 review necessary | Refers to legal principles while remaining undecided on *per se* legality of AWS |
|---|---|---|---|---|---|---|---|
| | | in the military sphere, as well as on the issues concerning the guarantee and full respect for international humanitarian law in the development, acquisition and deployment of increasingly complex weapon systems…[A]llow me to underline that the additional expertise and knowledge that academic and research institutions, think-tanks, and NGOs can bring on the LAWS issue, would certainly have a positive impact on our work. We therefore support the continued participation of civil society in this debate." (General statement) (2015)

"Italy has started an in-depth inter-agency analysis on LAWS, which also involves representatives from the private sector industry and is set to continue for some time. Such a review process takes account of the reflections and knowledge emerged during the meetings of the Group of Experts and, in return, aims to elaborate and provide a valid contribution to discussions in the CCW framework. In this regard, let me anticipate that Italy, in cooperation with the International Institute of Humanitarian Law of Sanremo and the ICRC, intends to organize a round table on "Weapons and the International Rule of Law" to be held next September. The event will be devoted, inter alia, to the "legal review of new weapons" and will be attended by academics, experts, and diplomats. We are | | | | | force, would be entirely beyond human control… We cannot exclude that those systems – in particular offensive ones – may pose issues of compliance with IHL and raise ethical dilemmas. However, *we believe that existing IHL rules already provide relevant parameters to assess the legality also of this second group of weapons…* Once again, at this stage we believe that the adoption of a total ban or other kinds of general limitations on fully autonomous technologies would be premature, given that the field is in constant, dynamic evolution and that such restrictions would hinder the development of technologies with very useful civilian applications." ("Towards a Working Definition of LAWS" Statement) (2016) |

| State[1] | Currently Unacceptable, Unallowable, or Unlawful | Need to monitor or continue to discuss | Need to regulate | Need to ban (or favorably disposed towards the idea) | Need for meaningful human control[2] | AP I Article 36 review necessary | Refers to legal principles while remaining undecided on per se legality of AWS |
|---|---|---|---|---|---|---|---|
| | | confident that the outcome of this round table will offer useful guidance for our upcoming work in Geneva." (General Statement) (2016)<br><br>"We consider it very valuable to continue discussions in the framework of the CCW, allowing us to keep close attention to current developments and make relevant decisions, should the need arise." ("Towards a Working Definition of LAWS" Statement) (2016) | | | | | |
| Japan | | "Japan would like to reiterate that it is important to clarify the definition of LAWS, but at the same time we recognize that reaching a consensus is not easy at this stage considering the deliberations at the Meeting of Experts in 2014. Therefore, we consider it useful to conduct in-depth discussions on the main elements of LAWS, such as autonomy and meaningful human control, which will be discussed at this meeting." (General statement) (2015)<br><br>"Japan would like to reiterate that it is important to clarify the definition of LAWS… Japan is willing to engage in such discussions in a constructive manner… Taking into account that the High Contracting Parties acknowledge that LAWS do not exist at present, we believe it is crucial for the Parties to develop a common understanding of LAWS. Should such a common understanding be deepened as a result of the deliberations at the | | | | | "Japan is of the view that Lethal Autonomous Weapons Systems (LAWS) should be discussed with a focus on various aspects of technology, ethics, law and military affairs, and that it is not appropriate to draw conclusions from any one of them." (Working Paper) (2016) |

| State[1] | Currently Unacceptable, Unallowable, or Unlawful | Need to monitor or continue to discuss | Need to regulate | Need to ban (or favorably disposed towards the idea) | Need for meaningful human control[2] | AP I Article 36 review necessary | Refers to legal principles while remaining undecided on *per se* legality of AWS |
|---|---|---|---|---|---|---|---|
| | | Third Informal Meeting, we believe it is possible to engage in further considerations. Japan, for its part, will actively contribute to discussions at this Informal Meeting in a constructive manner." (General Statement) (2016) | | | | | |
| Korea, Republic of | | "[W]e need to further clarify the concept and scope of autonomy as well as the legality regarding the use of autonomous weapons systems. We need to closely examine LAWS, not in isolation but in conjunction with the tasks it performs, the types of targets it engages and the contexts in which it is employed. [CCW Parties need to] deepen their understanding of LAWS technology and its related implications…. [I]dentical technology is used for LAWS and civilian robots. In our view, the discussions on LAWS should not be carried out in a way that can hamper research and development of robotic technology for civilian use. In this regard, we note the advantages of looking into the dual-use characteristics of LAWS technology under a separate and dedicated session." (General statement) (2015) | | | "The 2014 CCW Meeting of Experts on LAWs [sic]…have led to a broad consensus on the importance of 'meaningful human control' over the critical functions of selecting and engaging targets…. [W]e are wary of fully autonomous weapons systems that remove meaningful human control from the operation loop, due to the risk of malfunctioning, potential accountability gap and ethical concerns." (General statement) (2015) | | |
| Mexico[45] | "México considera que los sistemas de armas plenamente autónomos no podrían cumplir con los | "[M]i delegación considera conveniente que el debate de este tema continúe y se profundice. Seguimos creyendo que las | | In favor of "the negotiation of a legally-binding instrument to preemptively ban fully autonomous weapons"[46] | | | "Los desarrollos tecnológicos bélicos, incluidos los Sistemas de Armas Letales Autónomas (SALAS) deben cumplir con las normas del Derecho Internacional |

[45] Mexico apparently also delivered a General Statement in 2015, but its text is not available online. See id.

[46] Mexico did not express its desire for a ban via a written statement at the 2015 or 2016 CCW Meeting of Experts. It did, however, orally indicate a preference for a ban during the 2016 meeting. See Campaign to Stop Killer Robots, Report on Activities: Convention on Conventional Weapons Third Informal Meeting of Experts on Lethal Autonomous Weapons Systems 16 (2016), http://www.stopkillerrobots.org/wp-content/uploads/2013/03/KRC_CCWx2016_Jun27upld-1.pdf (reporting that "Mexico announced that it favors 'the negotiation of a legally-binding instrument to preemptively ban fully autonomous weapons.'…It affirmed that negotiations 'should not necessarily be done through CCW.'"). Mexico's position in favor of a ban has been included here to more fully represent states' attitudes on an important issue.

| State[1] | Currently Unacceptable, Unallowable, or Unlawful | Need to monitor or continue to discuss | Need to regulate | Need to ban (or favorably disposed towards the idea) | Need for meaningful human control[2] | AP I Article 36 review necessary | Refers to legal principles while remaining undecided on per se legality of AWS |
|---|---|---|---|---|---|---|---|
| | principios de derecho internacional humanitario y que su potencial uso también representa un riesgo en contra de los derechos humanos más fundamentales, como son el derecho a la vida y la dignidad." (Way ahead statement) [46] (2015) | nociones de control humano significativo ("meaningful human control"), así como la de autonomía en funciones críticas ("critical functions") pueden ser vías que debemos continuar explorando. Igualmente, como muchos otros lo hicieron durante esta semana, insistimos en la necesidad de que las discusiones sobre los sistemas de armas autónomas no se restrinjan al marco de la CCAC, dados los potenciales efectos que podrían tener los SALAS en materia de derechos humanos y en la estabilidad global, lo que requiere expandir el enfoque y los foros en los que se atienda este tema." (Way ahead statement) [47] (2015) | | | | | Humanitario (DIH), normas convencionales y consuetudinarias; en particular las normas de distinción, proporcionalidad y precauciones en el ataque." (General Statement) (2016) [49]<br><br>"Mi país considera que para cumplir con los requerimientos del DIH, los SALAS deben tener además la capacidad de distinguir entre combatientes activos y personal de las fuerzas armadas fuera de combate, civiles que participan directamente en las hostilidades, fuerzas de seguridad públicas, personal sanitario, entre otros." (General Statement) (2016) [50]<br><br>"Los Sistemas de Armas Letales Autónomas también deben cumplir el principio de proporcionalidad previsto en los Convenios de Ginebra, que exige que, cuando llegaren a ocurrir daños civiles como consecuencia incidental derivada de un ataque contra un objetivo militar, éstos no resulten excesivos en relación con la ventaja militar directa y concreta prevista. Asimismo, los SALAS deben tener la capacidad de adoptar las precauciones razonables en sus operaciones con el fin de reducir al mínimo possible el número de víctimas y daños a persona o objetos civiles." (General Statement) (2016) [51]<br><br>"México considera que la capacidad de los Sistemas de Armas Letales Autónomas, de cumplir con las normas y los principios |

[46] "Mexico believes that the fully autonomous weapons systems would not be able to comply with the principles of international humanitarian law and that their potential use also represents a risk against the most fundamental human rights, like the right to life and dignity."

[47] "My delegation considers it appropriate that the debate on this issue continue and deepen. We continue to believe that the notions of significant human control ('meaningful human control'), as well as autonomy in critical functions ('critical functions') may be routes that we should continue exploring. Also, as many others did during this week, we insist on the necessity that discussions on autonomous weapons systems are not restricted to the framework of the CCW, given the potential effects that LAWS could have on human rights and global stability, which requires expanding the focus and forums in which this issue is addressed."

[49] "The development of military technology, including LAWS, must comply with the norms of IHL, conventional and customary law; in particular, the rules of distinction, proportionality and precautions in attack."

[50] "My country believes that in order to meet the requirements of IHL, LAWS must also have the capacity to distinguish between active combatants and [hors de combat], civilians who directly participate in hostilities, public security forces, health [/medical] personnel, among others."

[51] "LAWS must also comply with the principle of proportionality provided for in the Geneva Conventions, which requires that, when anticipating incidental civilian harm resulting from an attack on a military object, they will not be excessive in relation to the concrete and direct military advantage anticipated. Also, LAWS must have the capacity of adopting reasonable precautions in their operations with an eye to reducing to the minimum possible the number of victims and harms to civilian persons and object."

| State[1] | Currently Unacceptable, Unallowable, or Unlawful | Need to monitor or continue to discuss | Need to regulate | Need to ban (or favorably disposed towards the idea) | Need for meaningful human control[2] | AP I Article 36 review necessary | Refers to legal principles while remaining undecided on per se legality of AWS |
|---|---|---|---|---|---|---|---|
| | | | | | | | del DIH, se encuentra fuertemente vinculada a su nivel de autonomia y se dependencia operacional de un ser humano en sus diferentes capacidades y escenarios en los que pudiera emplearse, lo anterior en razón de que es evidente que a mayor autonomia se dificulta reconocer el grado de responsabilidad humana en su operación. En este context, existen Fuertes preocupaciones de que los SALAS totalmente autónomos e independientes de control humano puedan cumplir cabalmente con las exigencias de las normas y los principios del DIH." (General Statement) (2016)[52] |
| Morocco | | "Nous partageons le point de vue que ce genre d'exercice permettra de créer les conditions propices pour developer une compréhension commune sur la question des SALA. Parmi l'un des sujets cruciaux soulevés dans votre document à réflexion, la definition des SALA, nous semble-t-il demeure un chantier important à investor afin de mieux cerner le champ et la portée du développement et de l'usage de ces nouvelles armes, qui posent des questions d'éthiques, de droit international et de droit international humanitaire. A cet égard, nous souhaitons qu'une approche constructive puisse être adoptee en vue de pouvoir ensemble jeter les bases nécessaires à l'élaboration d'une definition precise qui pourrait nous aider à mieux comprendre la nature de ces nouvelles armes | | | "Il semble que la majorité s'accodait, lors des deux reunions informelles précédentes, pour souligner la nécessité de maintenir sous contrôle humain les fonctions essentielles des systems d'armes létaux. En effect, les SALA, comme l'indique leur nom, sont des systems d'armes qui une fois actives, peuvent sélectionner et attaquer des cibles sans intervention humaine. Or, ma delegation est d'avis que les SALA doivent être concus de facon a impliquer des responsables humains et insiste que le contrôle humain effectif doit etre toujours humaine demeure centrale et mérite une attention particuliere, car, du point de vue moral, il est | | |

| State[1] | Currently Unacceptable, Unallowable, or Unlawful | Need to monitor or continue to discuss | Need to regulate | Need to ban (or favorably disposed towards the idea) | Need for meaningful human control[2] | AP I Article 36 review necessary | Refers to legal principles while remaining undecided on *per se* legality of AWS |
|---|---|---|---|---|---|---|---|
| | | dont l'emploi doit se conformer avec les règles fondamentales du droit international." (General Statement) (2016)[53] | | | inconceivable d'admettre que des armed autonomes auraient le pouvoir de decider de la vie ou de la mort d'un être humain ou qu'on confie la decision de vie ou de mort a un systeme autonome." (General Statement) (2016)[54] | | |
| **Netherlands** | | "Although it seems somewhat early for regulation, we can and should look into the need. In our opinion, regulation should be effective, proportional and we should be able to implement it in a verifiable way. The dual use nature of Artificial Intelligence technologies is playing a role here." (General statement) (2015)<br><br>"[M]any of the issues that were raised deserve further consideration. Also many questions remain. We therefore see greet [sic] value in continuing our discussions and will support a new mandate to that effect during the CCW meeting in November. More in particular we think future discussions could focus on 3 topics in particular: 1. we should continue to focus our discussions on the human role, including on meaningful human control and work towards a common understanding; 2. command and | In 2015, not yet, but should begin to think about it (see "Need to monitor or continue to discuss")<br><br>"It is essential to differentiate between an autonomous weapon, in which humans play a crucial role in the wider loop of human control, and a fully autonomous weapon, in which humans are beyond the wider loop and human control no longer plays any role. The Netherlands firmly rejects the development and deployment of such fully autonomous weapon systems that have no meaningful human control at all. However, we currently do not support a | | "We see the notion of meaningful human control as an important concept for the discussion on LAWS." (General statement) (2015)<br><br>"As long as autonomous weapon systems are under meaningful human control there is no reason to assume that they will by definition fall into one of the categories of weapons that are banned under international humanitarian law… The Netherlands believes meaningful human control should be exercised within the 'wider loop.' This means that human control within the wider targeting process should be meaningful. The wider targeting process includes target selection, weapon selection and implementation planning, | "All weapon systems (and their eventual use in armed conflicts) should meet the rules and regulations of international law. Part of that is Art. 36 of the 1st Protocol to the Geneva Conventions…. Another important element is the exchange of best practices, in particular concerning Art. 36, Art. 36 Commissions and the development of policies in addressing this issue. The mandate is there, there is a lot of information and there is common ground to move forward." (General statement) (2015)<br><br>"It is important that when procuring | |

---

[53] "We share the point of view that this kind of exercise will create the conditions favorable for the development of a common understanding of the question of LAWS. Among one of the crucial topics raised in your working paper, the definition of LAWS seems to us as an important work to be done in order to better understand the scope and reach of the development and use of these new weapons, which pose questions of ethics, international law and international humanitarian law. In this regard, we wish that a constructive approach be adopted in order to lay down together the required basis for the drafting of a precise definition which could help us better understand the nature of these new weapons, the use of which must comply with the fundamental rules of international law."

[54] "It seems that the majority agreed during the two preceding informal meetings to underline the need to maintain under human control the essential functions of lethal weapons systems. For LAWS, as their name shows, are weapons systems which once activated can select and attack targets without human intervention. Thus, my delegation's opinion is that LAWS must be designed so that individuals are involved and it emphasizes that effective human control must always be implemented with LAWS. In my delegation's view, the notion of human responsibility remains central and deserves a particular attention, as from a moral standpoint it is inconceivable to admit that autonomous weapons would have the authority to decide between the life and death of a human being or that such decision be entrusted with an autonomous system."

| State[1] | Currently Unacceptable, Unallowable, or Unlawful | Need to monitor or continue to discuss | Need to regulate | Need to ban (or favorably disposed towards the idea) | Need for meaningful human control[2] | AP I Article 36 review necessary | Refers to legal principles while remaining undecided on *per se* legality of AWS |
|---|---|---|---|---|---|---|---|
| | | control issues are an important part of meaningful human control and could be looked at in greater detail; 3. we also see value to further discuss the idea of a peer review process on Article 36. Furthermore we should continue to identify the defining elements of LAWS. We see this as an essential step to take the process further. However, it should not be used to slow us down." (Way ahead statement) (2015)<br><br>"Though in our opinion fully autonomous weapons systems, without human control, might not be expected within a short period, we re-iterate the importance of continued monitoring of the rapid technological developments within the field of artificial intelligence." (emphasis in original) (General Statement) (2016)<br><br>"The AIV/CAVV believes that there are various practical objections to a moratorium or a ban. Much of the relevant technology is being developed for peaceful purposes in the civilian sector and has both civilian and military (dual-use) applications. It is therefore difficult to draw a clear distinction between permitted and prohibited technologies. In | moratorium on the development of fully autonomous weapon systems for practical reasons. Such a moratorium would be inexpedient and unfeasible, mainly due to the fact that most Artificial Intelligence technology comes from civilian developments, e.g. autonomous car developments. That technology progress should not suffer from a moratorium, especially when the effectiveness of such a moratorium is very doubtful at the least." (General Statement) (2016)[56] | | including an assessment of potential collateral damage. Furthermore, it includes decisions like the programming of conditions and parameters of the autonomous weapon and the decision of the autonomous weapon's deployment." (General Statement) (2016) | autonomous weapons, the government should ensure that… the procedure relating to Article 36 of the First Additional Protocol to the Geneva Conventions is strictly applied. With respect to Article 36 we believe that the concept of meaningful human control should serve as a benchmark for this purpose." (General Statement) (2016)<br><br>"The Netherlands strongly calls for… a more widespread implementation by States of the Article 36 procedure at the national level, greater transparency concerning the outcomes of these procedures and more and better international information sharing." (General Statement) (2016)<br><br>"The Netherlands believes that, in assessing whether autonomous weapon systems are under meaningful human control, there is an | |

[56] The Netherlands position here is complex and it requires a little elaboration. They are strongly opposed to what they call "fully autonomous" weapon systems, in which humans are entirely out of the loop. But note that they do not actually indicate that such weapons are per se unlawful. They also reject the idea that even such fully autonomous weapons systems ought to be banned, largely because they are concerned that such a ban would interfere with civilian uses of autonomous AI technology. So while they never actually use the word "regulate," their position fits most closely into the "Need to Regulate" column because they seem to be advocating that states regulate the development of autonomous AI technologies, allowing partially autonomous weapon systems, while discouraging fully autonomous weapons systems from being developed. They also indicate that their current position is subject to change, and that they will re-evaluate in five years.

| State[1] | Currently Unacceptable, Unallowable, or Unlawful | Need to monitor or continue to discuss | Need to regulate | Need to ban (or favorably disposed towards the idea) | Need for meaningful human control[2] | AP I Article 36 review necessary | Refers to legal principles while remaining undecided on per se legality of AWS |
|---|---|---|---|---|---|---|---|
| | | addition, there is no international consensus on the definition of the relevant concepts. The question thus becomes: a moratorium on what? A non-proliferation regime would also be hard to enforce, as it would be difficult to establish the existence of 'weapons' in the case of dual-use technology and readily available programming languages. Countries would not be able to trust that other countries were respecting the agreement. During the CCW's informal meetings of experts in April 2015, it became apparent that there was no support among states for a moratorium or a ban. Only five countries (Cuba, Ecuador, Egypt, the Holy See and Pakistan) indicated that they would support such an initiative. A treaty establishing a moratorium or a ban is not viable without widespread support. For these reasons, the AIV/CAVV currently regards this option as inexpedient and unfeasible. However, it cannot rule out that advances in the field of artificial intelligence and robotics might necessitate revision of this position in the future." (Advisory report) (2016)[55] | | | | important role for the Article 36 procedure of the First Additional Protocol to the Geneva Conventions... The Netherlands also advocates greater transparency at the international level concerning the national Article 36 procedures and encourages more information sharing on procedures, best practices and outcomes of Article 36 reviews. Therefore the Netherlands calls for the formulation of an interpretative guide that clarifies the current legal landscape with regard to the deployment of weapons with autonomous functions. Such a document could list best practices on issues such as the role of meaningful human control in relation to the deployment of autonomous weapons. We are of the opinion that this could be a useful step to a better common understanding of this complex topic." (Challenges to IHL Statement) (2016) | |

---

[55] Though this table is mostly confined to statements at the 2015 and 2016 CCW Meetings of Experts, this excerpt from a recent advisory report provided to the Dutch government, see ADVISORY COUNCIL ON INTERNATIONAL AFFAIRS, AUTONOMOUS WEAPON SYSTEMS: THE NEED FOR MEANINGFUL HUMAN CONTROL (2016), http://aiv-advice.nl/8gr, and used to formulate the government's position, has been included as it addresses the possibility of a ban, an important issue. Though the excerpt quoted above is from the summary, it reflects the conclusions of the report in its entirety (which, for example, states, "the AIV/CAVV currently regards a moratorium as inexpedient and unfeasible. However, it cannot rule out that developments in the field of artificial intelligence and robotics might necessitate revision of this position in the future. It is therefore important to closely monitor such developments and ensure that the government actively participates in international discussions on the legal, ethical, technological and policy implications of autonomous weapons," id. at 47).

| State[1] | Currently Unacceptable, Unallowable, or Unlawful | Need to monitor or continue to discuss | Need to regulate | Need to ban (or favorably disposed towards the idea) | Need for meaningful human control[2] | AP I Article 36 review necessary | Refers to legal principles while remaining undecided on *per se* legality of AWS |
|---|---|---|---|---|---|---|---|
| New Zealand | | "We are particularly conscious of the need for progress in the determination of a working definition of LAWS. We are well aware that this task is not an easy one with the 2015 MEX session already having highlighted very different views among States on this critical issue." (General Statement) (2016)<br><br>"New Zealand would certainly consider it desirable that the CCW now intensify its work on LAWS. Discussions to date have already identified a number of serious challenges posed by LAWS to international humanitarian law as well, more broadly, to a number of other standards. It is therefore timely that we move into a more intensive and sustained format for engagement on these issues" (General Statement) (2016) | | | | | "We also look forward to an informed debate on the challenges posed by LAWS for compliance with the norms and dictates of international humanitarian law. For New Zealand, the absolutely essential requirement is that the development and subsequent usage of any weapon system – including LAWS – must take place only in accordance with IHL. Compliance with IHL, and, as applicable, other aspects of international law, remains of the highest priority for New Zealand and will continue to be the determining factor in our approach to these issues." (General Statement) (2016) |
| Nicaragua | | | | "Put us on the list."[3] | | | |
| Pakistan | "LAWS *would not distinguish between combatants and non-combatants; they lack morality, mortality and judgement. The use of LAWS will make war even more inhumane... autonomous implies no scope for such 'interference' by any human, calling into question the principles of IHL: distinction, proportionality, precaution, humanity and military necessity. The standards of* | "The suggestion to limit the evaluation of the development and employment of LAWS to a purely national exercise including national reviews under Article-36, while appealing and convenient for the developers of such technologies, is not convincing or satisfactory for us and the large majority present here. Therefore, as a practical step forward, without prejudice to our preference for a comprehensive and pre-emptive ban on LAWS, we find merit in the establishment of a Group of Governmental Experts | | "The introduction of LAWS would be illegal, unethical, inhumane and unaccountable as well as destabilizing for international peace and security with grave consequences. Therefore, their further development and use must be pre-emptively banned through a dedicated Protocol of the CCW. Pending the negotiations and conclusions of a legally binding Protocol, the states currently developing such weapons should place an immediate moratorium on their | "The question of definitional clarity for the word, 'autonomous' is pertinent and requires immediate attention. Whilst automated weapons and automatic weapons have to some degree a 'human in the loop', autonomous implies no scope for such 'interference' by any human, calling into question the principles of IHL: distinction, proportionality, precaution, humanity and military necessity. The | | |

---

[57] Nicaragua did not express its desire for a ban via a written statement at the 2015 or 2016 CCW Meeting of Experts. It did, however, orally indicate a preference for a ban during the 2016 meeting. *See* Stop Killer Robots (@BanKillerRobots), Twitter (April 14, 2016, 9:10 AM), https://twitter.com/bankillerrobots/status/720645378895454208; Campaign to Stop Killer Robots, at 5. Nicaragua's position has been included here to more fully represent states' attitudes on an important issue.

| State[1] | Currently Unacceptable, Unallowable, or Unlawful | Need to monitor or continue to discuss | Need to regulate | Need to ban (or favorably disposed towards the idea) | Need for meaningful human control[2] | AP I Article 36 review necessary | Refers to legal principles while remaining undecided on *per se* legality of AWS |
|---|---|---|---|---|---|---|---|
| | *International Human Rights Law* are more stringent. These rules can be complex and *entail subjective decision making requiring human judgment....* Should a machine programmed on a complex set of algorithms, which is devoid of the notions of morality and humanity, be allowed to decide who should live and who should die? We are convinced that the answer is a firm NO....Faced with no loss or injury to their "human" combatants, the States employing LAWS would resort to use of force on a frequent basis – thus undermining the very basis of the restraints on the use of force that international law seeks to maintain. LAWS will lower the threshold of going to war, resulting in armed conflict no longer being a measure of last resort, but a recurrent "low-cost" affair instead... Like drones, civilians could be targeted and killed with LAWS through so-called signature strikes. The breaches of State sovereignty – in addition to breaches of International Humanitarian Law and International Human Rights Law – | (GGE) by the CCW Meeting of States Parties this year (2015), with a mandate to formally consider this issue and present a report to the CCW Review Conference next year (2016)." (Way ahead statement) (2015) | | production and use." (General statement) (2015)  "We remain convinced that the introduction of LAWS would be illegal, unethical, inhumane and unaccountable as well as destabilizing for international peace and security with grave consequences. Therefore, their further development and use must be pre-emptively banned through a dedicated Protocol of the CCW. Pending the negotiations and conclusions of a legally binding Protocol, the states currently developing such weapons should place an immediate moratorium on their production and use" (Way ahead statement) (2015)  "Based on these considerations, the introduction of LAWS would be illegal, unethical, inhumane and unaccountable as well as destabilizing for international peace and security with grave consequences. Therefore, their further development and use must ideally be pre-emptively banned through a dedicated Protocol of the CCW. Pending the negotiations and conclusions of a legally binding Protocol, the states currently developing such weapons should place an immediate moratorium on their | standards of International Human Rights Law are even more stringent. These rules can be complex and entail subjective decision making requiring human judgment. The question is simple: Should a machine programmed on a complex set of algorithms, which is devoid of the notions of morality and humanity, be allowed to decide who should live and who should die? We are convinced that the answer is a firm NO." (General statement) (2015)  "Although the concept of "meaningful human control" has gained some currency and traction in the context of LAWS, we are of the view that the concept of "meaningful human control" only provides an approach to discussing the weaponization of increasingly autonomous technologies; it does not provide a solution to the technical, legal, moral and regulatory questions that they pose." (General Statement) (2016)[58] | | |

| State[1] | Currently Unacceptable, Unallowable, or Unlawful | Need to monitor or continue to discuss | Need to regulate | Need to ban (or favorably disposed towards the idea) | Need for meaningful human control[2] | AP I Article 36 review necessary | Refers to legal principles while remaining undecided on *per se* legality of AWS |
|---|---|---|---|---|---|---|---|
| | associated with targeted killing programmes risk making the world less secure, with LAWS in the equation. LAWS create an accountability and transparency vacuum and provide impunity to the user due to the inability to attribute responsibility for the harm that they cause. *If the nature of a weapon renders responsibility for its consequences impossible, it's [sic] use should be considered unethical and unlawful.* Also, in the event of a security breach or a compromised system, who would be held responsible; the programmer, the hardware manufacturer, the commander who deploys the system or the user state?... The use of LAWS in the battlefield would amount to a situation of one-sided killing. Besides *depriving the combatants of the targeted state, the protection offered to them by the international law of armed conflict, LAWS would* also risk the lives of civilians and non-combatants on both sides. *It remains unclear as to how "combatants" will be defined in case of LAWS.* Will targets be chosen based on an algorithm that recognizes certain physical characteristics, for example, "beards and turbans"? Also, there are | | | production and use." (General Statement) (2016) | | | |

| State[1] | Currently Unacceptable, Unallowable, or Unlawful | Need to monitor or continue to discuss | Need to regulate | Need to ban (or favorably disposed towards the idea) | Need for meaningful human control[2] | AP I Article 36 review necessary | Refers to legal principles while remaining undecided on *per se* legality of AWS |
|---|---|---|---|---|---|---|---|
| | questions of the protection of those who are not; or no longer, taking part in fighting: "hors de combat". *How will LAWS distinguish between noncombatants from combatants or hors de combat?* Can a machine be *trusted to have the same or better discerning abilities as a human?* These questions remain unanswered... Like any other complex machine, *LAWS can never be fully predictable or reliable.* They could fail for a wide variety of reasons including human error, malfunctions, degraded communications, software failures, cyber attacks, jamming and spoofing, etc. There will always be a level of uncertainty about the way an autonomous weapon system will interact with the external environment.... *The introduction of LAWS would be illegal,* unethical, inhumane and unaccountable as well as destabilizing for international peace and security with grave consequences." (General statement) (2015)<br><br>Though many of the concerns Pakistan highlighted in 2015 could be argued to merely indicate certain uses of AWS would be illegal or that the use of AWS would have negative effects of the legal system, | | | | | | |

| State[1] | Currently Unacceptable, Unallowable, or Unlawful | Need to monitor or continue to discuss | Need to regulate | Need to ban (or favorably disposed towards the idea) | Need for meaningful human control[2] | AP I Article 36 review necessary | Refers to legal principles while remaining undecided on *per se* legality of AWS |
|---|---|---|---|---|---|---|---|
| | Pakistan also made clear its belief that AWS are *per se* illegal. "We remain convinced that the introduction of LAWS would be illegal, unethical, inhumane and unaccountable as well as destabilizing for international peace and security with grave consequences....[LAWS] pos[e] a number of technical, ethical, moral and legal challenges, including compliance with the International Humanitarian Law and the International Human Rights Law." (Way ahead statement) (2015) "LAWS cannot be programmed to comply with International Humanitarian Law (IHL), in particular its cardinal rules of distinction, proportionality, and precaution. These rules can be complex and entail subjective decision making requiring human judgment." (Paper/inputs) (2015) "LAWS cannot be programmed to comply with International Humanitarian Law (IHL), in particular its cardinal rules of distinction, proportionality, and precaution. These rules can be complex and entail subjective decision making requiring human judgment. The introduction of fully | | | | | | |

| State[1] | Currently Unacceptable, Unallowable, or Unlawful | Need to monitor or continue to discuss | Need to regulate | Need to ban (or favorably disposed towards the idea) | Need for meaningful human control[2] | AP I Article 36 review necessary | Refers to legal principles while remaining undecided on per se legality of AWS |
|---|---|---|---|---|---|---|---|
| | autonomous weapons in the battlefield would be a major leap backward on account of their profound implications on norms and behaviour that the world has painstakingly arrived at after centuries of warfare. We firmly believe that developments in future military technologies should follow the established law and not vice versa." (General Statement) (2016) | | | | | | |
| | "LAWS create an accountability vacuum and provide impunity to the user due to the inability to attribute responsibility for the harm that they cause. If the nature of a weapon renders responsibility for its consequences impossible, its use should be considered unethical and unlawful." (General Statement) (2016) | | | | | | |
| | "Besides depriving the combatants of the targeted state the protection offered to them by the international law of armed conflict, LAWS would also risk the lives of civilians and non-combatants. The unavailability of a legitimate human target of the LAWS user State on the ground could lead to reprisals on its civilians including through terrorist acts." (General Statement) (2016) | | | | | | |

| State[1] | Currently Unacceptable, Unallowable, or Unlawful | Need to monitor or continue to discuss | Need to regulate | Need to ban (or favorably disposed towards the Idea) | Need for meaningful human control[2] | AP I Article 36 review necessary | Refers to legal principles while remaining undecided on per se legality of AWS |
|---|---|---|---|---|---|---|---|
| Palestine[59] | | | | Yes | | | |
| Poland | "It seems that there is a broad agreement on that the development of fully autonomous weapon systems shall not be allowed. The question is how to allow the robotic systems to act autonomously to some extent, and at the same time never fully give up human control over such systems?" (Human Rights and Ethical Issues Statement) (2016)[60] | "[A]t this stage of the discussion, the CCW remains to be an appropriate forum to try to reach a common understanding of the main elements of this problem and possible ways forward....Generally speaking, technology in itself is neither good nor bad. What matters is how it is applied and used....[W]e are glad that part of our discussion will be devoted to the dual use [that is, civilian and military] issues." (General statement) (2015)<br><br>On the subject of exporting AWS technology: "At the present stage it would seem advisable to be able to prevent transfers of such systems and their components to undesirable end-users, whether states or non-state actors....A possible set of 'best practices' in export control might be a complementary and useful tool. The scope of such measures would be decided by states themselves....Furthermore, similar solutions could be extended to other areas: handling, transportation, testing or retransfer of LAWS....A joint effort of governmental experts to elaborate a set of 'best practices' might undoubtedly provide the | On the subject of exporting AWS technology: "At the present stage it would seem advisable to be able to prevent transfers of such systems and their components to undesirable end-users, whether states or non-state actors....A possible set of 'best practices' in export control might be a complementary and useful tool. The scope of such measures would be decided by states themselves....Furthermor e, similar solutions could be extended to other areas: handling, transportation, testing or retransfer of LAWS." (Best practices/way ahead statement) (2015)<br><br>The above statement would seem to be calling for national, not international, regulation, and is focused on export of LAWS technology. | | "The main principles of IHL which are of interest to us would be: humanity, military necessity, discrimination and proportionality. Looking at the present level of technological advancement, however, there are reasons for concern that the existing systems will not be able to meet those principles. Hence the importance of developing further the MHC [meaningful human control] concept and its institutional extension - the idea of MSC [meaningful state control]. The presence of human control in the form of institutional framework guarantees itself a reference to certain standards - legal and related international customs. Human or institutional oversight upholds accountability, the rule of law and supports procedures through which our decisions may be verified." (Characteristics of laws/meaningful human control statement) (2015)<br><br>"[I]t is too early in the process to make final | | "A state should always be held accountable for what it does, especially for the responsible use of weapons which is delegated to the armed forces. The same goes also for LAWS. The responsibility of states for such weapons should also be extended to their development, production, acquisition, handling, storage or international transfers. The proper application of state's power to control the development of increasingly autonomous weapons may play an important role in binding this process with instruments of International Humanitarian Law (IHL) and International Human Rights Law. The main principles of IHL which are of interest to us would be: humanity, military necessity, discrimination and proportionality. Looking at the present level of technological advancement, however, there are reasons for concern that the existing systems will not be able to meet those principles. Hence the importance of developing further the MHC [meaningful human control] concept and its institutional extension - the idea of MSC [meaningful state control]. The presence of human control in the form of institutional framework guarantees itself a reference to certain standards - legal and related international customs. Human or institutional oversight upholds accountability, the rule of law and supports procedures through which our decisions may be verified." |

---

[55] Palestine apparently offered a written statement for the 2015 CCW Meeting of Experts, but it is unavailable online. The Campaign to Stop Killer Robots does report that "[d]uring the meeting... [f]or the first time... Palestine said [it] supported a preemptive ban." *See* Campaign to Stop Killer Robots, at 5.

[60] It's important to make two points about this statement. First, Poland here is indicating only that *fully* autonomous weapons systems "shall not be allowed." According to its General Statement, Poland does not believe that fully autonomous weapon systems are currently in use. Therefore, Poland is not indicating that any autonomous weapon systems currently in use are unlawful. They are also not saying that any semi-autonomous weapon systems that might be developed later are necessarily not allowed. (Additionally, it is unclear whether "shall not be allowed" is equivalent to "illegal under international law." Poland does not explicitly use the word "unlawful." "Shall not be allowed" could possibly mean that their use would be discouraged or that individual states would independently choose to ban them, even if they were permitted under international law.) Second, based on the other states' General Statements, it seems highly doubtful that the "broad" agreement to which this statement refers actually exists. It is possible that this broad agreement emerged from the discussions at the conference that occurred after the states gave their General Statements, in which case a thorough review of the General Statements would not provide a complete picture.

| State[1] | Currently Unacceptable, Unallowable, or Unlawful | Need to monitor or continue to discuss | Need to regulate | Need to ban (or favorably disposed towards the idea) | Need for meaningful human control[2] | AP I Article 36 review necessary | Refers to legal principles while remaining undecided on per se legality of AWS |
|---|---|---|---|---|---|---|---|
| | | impetus towards a more conclusive discussion on definition issues. It would be an occasion to share national know-how and policies and exchange experiences with nonproliferation and arms-control regimes." (Best practices/way ahead statement) (2015)<br><br>"The actions related to the possible future deployment of such systems would require: close monitoring of their development; building an expert knowledge for better understanding of the nature of the systems; introduction of necessary control, in addition to already existing mechanisms." (General Statement) (2016) | | | judgements regarding the development and introduction of autonomy into certain parts of weapon systems, but it is of utmost importance to make sure that human beings remain accountable for use of their crucial functions.... There is a wide array of wrongdoings defined in the International Humanitarian Law for which a member of armed forces can be held accountable. But what is accountability? In terms of military activity, accountability is acknowledgment and assumption of responsibility for decisions, actions and their consequences, needed to achieve an authorized military objective. This is, however strongly associated with the level of authority and autonomy an individual is given. This is clearly visible when looking at the level of autonomy transferred down the chain of command. A commander of higher rank transfers part of his powers to a commander of a lower level. As a result we end up in a situation where an individual whose actions should be judged and penalized, if necessary, can always be identified" (Military perspective on accountability/Possible challenges to IHL statement) (2015) | | (Characteristics of laws/meaningful human control statement) (2015)<br><br>"[I]t is too early in the process to make final judgements regarding the development and introduction of autonomy into certain parts of weapon systems, but it is of utmost importance to make sure that human beings remain accountable for use of their crucial functions.... There is a wide array of wrongdoings defined in the International Humanitarian Law for which a member of armed forces can be held accountable. But what is accountability? In terms of military activity, accountability is acknowledgment and assumption of responsibility for decisions, actions and their consequences, needed to achieve an authorized military objective. This is, however strongly associated with the level of authority and autonomy an individual is given. This is clearly visible when looking at the level of autonomy transferred down the chain of command. A commander of higher rank transfers part of his powers to a commander of a lower level. As a result we end up in a situation where an individual whose actions should be judged and penalized, if necessary, can always be identified" (Military perspective on accountability/Possible challenges to IHL statement) (2015)<br><br>In the above 2015 statement, Poland also states, "can a machine be allowed to decide whether to kill or not? The military answer to that question is simply NO, we want and have to be in control." However, this response is framed in terms of military policy, not legality.<br><br>"Compliance with the fundamental rules and principles of international |

| State[1] | Currently Unacceptable, Unallowable, or Unlawful | Need to monitor or continue to discuss | Need to regulate | Need to ban (or favorably disposed towards the idea) | Need for meaningful human control[2] | AP I Article 36 review necessary | Refers to legal principles while remaining undecided on per se legality of AWS |
|---|---|---|---|---|---|---|---|
| | | | | | In the above 2015 statement, Poland also states, "can a machine be allowed to decide whether to kill or not? The military answer to that question is simply NO, we want and have to be in control." However, this response is framed in terms of military policy, not legality.<br><br>"Also, from the military perspective, it is important to satisfy the need to both introduce the latest technologies into warfare and create environments where humans may be held accountable for their decisions. In our opinion, such a need can be satisfied through exercising Meaningful Human Control (MHC) over the critical functions of LAWS. Therefore, we see rationale in continuing the analysis of LAWS against the concept of Meaningful Human Control where further exploration of such a concept may significantly facilitate the discussion on the definitions." ("Towards a Working Definition of LAWS" Statement) (2016) | | humanitarian law in the conduct of hostilities, that is distinction, proportionality and precautions in attack, poses formidable challenges, especially as future weapons with autonomy in their critical functions will be assigned more complex tasks and deployed in more dynamic environments than has been the case until now." (General Statement)<br><br>"In cluttered, dynamic and populated areas where civilian objects are close to military objectives and fighters are intermingled with civilians, LAWS would need to have highly sophisticated recognition abilities. In such a case, the system would be expected to distinguish between combatants and civilians and between military and civilian objects. This can be a challenging task even for human soldiers, let alone robotic systems that have only limited capabilities. *This is why, there should be always a human being involved in the targeting process to recognize situations of doubt that would cause a human being to hesitate before attacking.* In such circumstances States are obliged to refrain from attacking objects and persons." (Challenges to IHL Statement) (2016)<br><br>"To comply with the principle of proportionality, LAWS would at a minimum need to be able to estimate the expected amount of collateral harm that might come to civilians from an attack. However, the difficulty of LAWS to apply the proportionality principle lies not so much in the evaluation of the risks for civilians and civilian objects as in the evaluation of military advantage anticipated... We should remember that the concrete and direct military advantage anticipated resulting from an attack against a legitimate target constantly changes according to the plans |

| State[1] | Currently Unacceptable, Unallowable, or Unlawful | Need to monitor or continue to discuss | Need to regulate | Need to ban (or favorably disposed towards the idea) | Need for meaningful human control[2] | AP I Article 36 review necessary | Refers to legal principles while remaining undecided on *per se* legality of AWS |
|---|---|---|---|---|---|---|---|
| | | | | | | | and the development of military operations of both sides. Therefore the system must be constantly updated, taking into account factors resulting from the changing operational environment and battlefield in which such a system is deployed. The decisions and update in question may be provided only by States... Question to panelists: Do you think it is possible to program the autonomous weapon system on the basis of clear criteria to identify objective indicators and make assessments objectively?" (Challenges to IHL |
| Sierra Leone[61] | | "My delegation is optimistic that with the cooperation that has so far been enjoyed by this initiative, we are moving towards a fruitful outcome - an outcome that would address all the concerns that are being raised in relation to the development and deployment of Lethal Autonomous Weapons Systems. In this connection, Sierra Leone supports the establishment of a Governmental Group of Experts to more comprehensively address this issue and the forthcoming Fifth Review Conference will provide an opportunity to put it in place." (General Statement) (2016) | | Not explicitly in favor of a ban, but seemingly against any AWS not under human control: "Under no circumstances should the taking of the life of human beings be entrusted to machines, however well programmed. Sierra Leone therefore believes that the Human Rights Council should remained seized on the human rights aspects of LAWS, while respecting the mandate of CCW." (General Statement) (2016) | | "When an existing system gets to the stage when it could be considered as autonomous, it should be also subjected to the provisions of Article 36 of the Additional Protocol of the IHL." (General Statement) (2016)  "We have heard conflicting views as to whether LAWS could be in conformity with IHL.. My delegation trusts that all States would like to operate within the provisions of International Humanitarian Law and would take steps to respect Article 36 of the Additional Protocol 1 to the Geneva Conventions of 1949, relating to the | "[T]he full human rights and humanitarian impacts of the use of LAWS must be determining factor in decisions on their use or in their prohibition." (General statement) (2015) |

---

[61] Sierra Leona apparently also delivered a statement on Transparency and the Way Ahead, but its text is not available online. See id.

| State[1] | Currently Unacceptable, Unallowable, or Unlawful | Need to monitor or continue to discuss | Need to regulate | Need to ban (or favorably disposed towards the idea) | Need for meaningful human control[2] | AP I Article 36 review necessary | Refers to legal principles while remaining undecided on *per se* legality of AWS |
|---|---|---|---|---|---|---|---|
| | | | | | | Protection of Victims of International Armed Conflicts, in relation to new weapons. This article states, and I quote, 'In the study, development, acquisition or adoption of a new weapon, means or method of warfare, a High Contracting Party is under an obligation to determine whether its employment would, in some or all circumstances, be prohibited by this Protocol or by any other rule of international law applicable to the High Contracting Party.' The emphasis here is on international law and not just IHL." (emphasis in original) (General Statement) (2016) | |
| South Africa | | "Defining the characteristics of LAWS will help bring us closer to a definition, which is essential in reaching a common understanding as to the very nature of these weapons...Should LAWS also be regulated for their possible dual use applications? In this regard, the various components that make up these weapon systems probably have a wide range of peaceful applications. In addition, controlling such components would probably have a large measure of overlap with existing | | | "The concept of 'meaningful human control' is something that my delegation is supportive of. In our view, there should always be meaningful human control in the question of life and death." (General statement) (2015)

"After two informal Meeting of Experts, the concept of 'meaningful human control' or rather 'necessary human control' is a requirement that my delegation is | "Article 36 of Additional Protocol I of the Geneva Convention states that 'In the study, development, acquisition or adoption of a new weapon, means or method of warfare, a High Contracting Party is under an obligation to determine whether its employment would, in some or all circumstances, be prohibited by this | "The concept of 'meaningful human control' is something that my delegation is supportive of. In our view, there should always be meaningful human control in the question of life and death. …The use of [LAWS] would need to comply with the fundamental rule of International Humanitarian Law, including those of distinction, proportionality and military necessity, as well as their potential impact on human rights." (General statement) (2015)

"South Africa does not wish to see the development of legitimate, commercial robotics technology curtailed in any way |

| State[1] | Currently Unacceptable, Unallowable, or Unlawful | Need to monitor or continue to discuss | Need to regulate | Need to ban (or favorably disposed towards the idea) | Need for meaningful human control[2] | AP I Article 36 review necessary | Refers to legal principles while remaining undecided on per se legality of AWS |
|---|---|---|---|---|---|---|---|
| | | control regimes, such as the Missile Technology Control Regime. In our view, this is an issue that would require further study... The questions posed by the Chair in his working paper raise some pertinent issues and challenges -that will be crucial when - it comes to making a decision on how best to move forward. In this regard/ States Parties probably. need more time and information to come to a conclusion. For instance, should the call by some experts and States for a moratorium until a decision is taken be answered? However, given that moratoria are generally unilateral in nature, would a political declaration work better? Or should we rather pursue some rules of the road which could allow for flexibility to deal with an as yet undefined weapon system. In essence, the question would be whether we should rather put in place some rules of the road until there is a better understanding of the various concepts?" (General statement) (2015)<br><br>"In the absence of an agreed definition, States' continued engagement will lead to a common understanding of what these weapons are and move this debate forward towards a shared understanding. Mapping out the characteristics of LAWS and the concepts that are surrounding them will help bring us closer to a | | | supportive of. In the final analysis, there is a necessity for human control in the selection of targets to enforce accountability." (General Statement) (2016) | Protocol or by any other rule of international law applicable to the High Contracting Party.'" (General Statement) (2016)[62] | through any restrictive and prohibitive framework that would affect technology development or advancement for peaceful application. However, my delegation, once again reaffirms that all new means and methods of warfare should comply with the law of armed conflict. The use of such weapon systems would need to comply with the fundamental rule of International Humanitarian Law, including those of distinction proportionality and military necessity, as well as their potential impact on human rights." (General Statement) (2016) |

---

[62] Note that South Africa does not explicitly say that an Article 36 review is necessary, but by directly quoting it in its discussion of compliance with international law, it strongly implies that Article 36 is a relevant consideration when dealing with LAWS.

| State[1] | Currently Unacceptable, Unallowable, or Unlawful | Need to monitor or continue to discuss | Need to regulate | Need to ban (or favorably disposed towards the idea) | Need for meaningful human control[2] | AP I Article 36 review necessary | Refers to legal principles while remaining undecided on per se legality of AWS |
|---|---|---|---|---|---|---|---|
| | | definition, which is essential in reaching a common understanding on the very nature of these weapons." (General Statement) (2016)<br><br>"My delegation would be supportive of convening a GGE to take this discussion forward in a formal setting." (General Statement) (2016) | | | | | |
| Spain | | "Quisiera destacar, en esta primera oportunidad que nos brinda para expresar nuestros puntos de vista, la importancia de un primer criterio que entendemos fundamental para acotar de modo oportuno y justo la definición de estos sistemas de armas, como es el carácter ofensivo o defensivo de los mismos y su letalidad inherente, de modo que queden excluidos sistemas con diferentes niveles de automatismo que sean eminentemente defensivos, así como aquellos que no proyecten una fuerza letal, como por ejemplo los que establezcan contramedidas electrónicas…. Para ayudar a acotar las definiciones necesarias, además, deberemos tener en cuenta cuestiones como las normas de procedimiento previas a la activación. Adicionalmente, considerar el entorno también resulta relevante, ya que la aplicación de los principios del | | | "Nuestro principal punto de partida en este empeño debe fundamentarse, como debe hacerlo además en relación con cualquier otro tipo de armas, en la necesidad del respeto más escrupuloso del Derecho Internacional Humanitario y del Derecho Internacional de los Derechos Humanos, cuya primacía entendemos irrenunciable, en particular en relación con los principios de necesidad, proporcionalidad, distinción y precaución. Para lograr este objetivo, es necesaria la capacidad de control y supervisión humana en la fase de selección del blanco militar, incluida la capacidad de abortar el proceso de lanzamiento del arma de que se trate. Esta imperativa intervención humana en el proceso de activación del sistema y su | | "Debemos partir para ello del máximo respeto a la legalidad internacional, fundamentada en el Derecho Internacional Humanitario y el Derecho Internacional de los Derechos Humanos, contando con los principios de necesidad, distinción, proporcionalidad y precaución." (General Statement) (2016)[66]<br><br>"Finalmente, para una correcta aplicación de los principios del DIH en cada caso concreto, deberá considerarse también el entorno o ambiente en el que se desarrolle la acción, ya que la aplicación de dichos principios tiene lugar, generalmente, de forma individualizada." (General Statement) (2016)[67]<br><br>"Consideramos siempre necesaria la participación de un operador humano, así como el establecimiento de principios de atribución clara de responsabilidad jurídica personal sobre los criterios de uso de cualquier tipo de arma." (General Statement) (2016)[68]<br><br>"Será relevante, en general, considerar el carácter ofensivo o defensivo del sistema, |

[66] "We should begin with maximum respect for international law based on IHL and IHRL, the principles of necessity, distinction, proportionality and precaution."

[67] "Finally, the correct application of IHL principles in each concrete case [/case-by-case basis], ought also take into account the environment and surroundings in which the system would be used, and that the application of these principles are to be, generally, applied individually."

[68] "We consider the participation of a human operator as necessary [/requisite], alongside the establishment of principles of clear attribution of personal legal responsibility, as among the criteria for the use of any type of weapon."

| State[1] | Currently Unacceptable, Unallowable, or Unlawful | Need to monitor or continue to discuss | Need to regulate | Need to ban (or favorably disposed towards the idea) | Need for meaningful human control[2] | AP I Article 36 review necessary | Refers to legal principles while remaining undecided on per se legality of AWS |
|---|---|---|---|---|---|---|---|
| | | Derecho Internacional Humanitario es generalmente individualizada y dependerá por tanto del entorno en el que se produzca la acción. Estas precisiones, entre otras,...nos permitirán garantizar con pleno respeto a la legalidad internacional el legítimo derecho de autodefensa, especialmente en circunstancias en que los tiempos de reacción son críticos, y evitar interpretaciones erróneas o excesivamente generales sobre estos sistemas, resultando inoportuno el posible planteamiento de nuevas iniciativas jurídicas en el plano internacional antes de que establezcamos una definición clara y oportuna de los mismos. Además, teniendo en cuenta que el desarrollo de estas tecnologías afecta asimismo a ámbitos no militares, considerar posibles nuevas iniciativas jurídicas con ambigüedades o lagunas técnicas podría afectar negativamente a futuros desarrollos civiles." (General statement) (2015)[63]<br><br>"Estamos convencidos de que la Convención sobre Ciertas Armas Convencionales, es el marco | | | posterior supervisión, al mismo tiempo, y en toda lógica, deberá permitir una atribución clara y precisa de responsabilidad jurídica personal." (General statement) (2015)[65] | | así como su grado de letalidad inherente y las normas de procedimiento previas a su activación." (General Statement) (2016)[69] |

---

| State[1] | Currently Unacceptable, Unallowable, or Unlawful | Need to monitor or continue to discuss | Need to regulate | Need to ban (or favorably disposed towards the idea) | Need for meaningful human control[2] | AP I Article 36 review necessary | Refers to legal principles while remaining undecided on per se legality of AWS |
|---|---|---|---|---|---|---|---|
| | | idóneo para continuar con estas reflexiones, y confiamos en que nuestros debates contribuirán a establecer un consenso útil sobre el concepto de estos sistemas y su consideración a la luz del DIH. También esperamos que nuestros trabajos contribuyan a evitar posibles impactos negativos en el desarrollo de tecnologías para usos civiles." (General Statement) (2016)[64] | | | | | |
| Sri Lanka | | "The use of LAWS could open up new challenges on compliance with IHL principles such as distinction, proportionality, precaution and military necessity. Left unanswered this will also lead to a crucial accountability gap...As the Convention stipulates 'the civilian population and the combatants shall at all times remain under the protection and the authority of the principles of international law derived from established custom, from the principles of humanity and from the dictates of public conscience,' and we therefore need to be wary of allowing any level of autonomy in the use of weapons systems....[W]e encourage the continuation of the deliberations on this issue, to develop common understanding on the future challenges of use of LAWS and ways and means to overcome them, including the option of pre-emptive ban, for it is believed that prevention is always better than | "It is our understanding that the debate on LAWS is not merely a question to ban or not to ban autonomous technology in weapons systems, but rather a question of the acceptable threshold of the degree of autonomy in weapon systems that is in compliance with international law. In deciding so, it is necessary to be mindful of the fact that implications of the use of LAWS can vary substantially depending on the circumstances, the context that it is being used, the type and usage of the weapons, etc. Therefore, the debate should be an exercise to explore how we can take pre-emptive regulatory actions taking into account all above aspects, while preserving | | | "The challenge of addressing the accountability gap in this context means to what extent an individual, organizations or a State could be held liable for a crime committed by a fully autonomous weapon. As the ICRC notes under the law of State responsibility, in addition to accountability for violations of IHL committed by its armed forces, a State could also be held liable for violations of IHL caused by an autonomous weapon system that it has not, or has inadequately tested or reviewed prior to deployment. Further, under the laws of product liability, | "The issue of IHL compatibility has centrality in our deliberations towards developing an international legal instrument on regulating autonomous technology in weapons. The debate on how and what provisions of IHL should be applied in the case of LAWS and who should be held accountable in the event of unlawful use are some of the fundamental issues that need an answer." (General Statement) |

---

| State[1] | Currently Unacceptable, Unallowable, or Unlawful | Need to monitor or continue to discuss | Need to regulate | Need to ban (or favorably disposed towards the idea) | Need for meaningful human control[2] | AP I Article 36 review necessary | Refers to legal principles while remaining undecided on *per se* legality of AWS |
|---|---|---|---|---|---|---|---|
| | | the cure." (General statement) (2015)

"More robust engagement in the discussion from the global South is also vital, for it is these countries who are disadvantaged in the access to such technologies, and are likely to be more vulnerable during any potential warfare involving LAWS." (General Statement) (2016) | *the space for the peaceful use of the autonomous technology,* including non-lethal military and defensive purposes. Protocol IV of the CCW provides an example to this end, where the use of laser technology in a specific context was pre-emptively banned, but the same technology continues to be in use for various other peaceful purposes. The concept of 'dual-use technology' in the nuclear field also has relevance to the issue of LAWS. Therefore, it is important to consider safeguards that can help avoid the abuse and unintended consequences of the AI technology while reaping its benefits for the betterment of humanity." (General statement) (2016)

"While acknowledging that States have limited understanding on this subject, views continue to evolve and States are paying special attention in developing its own policy in this area. In this context, the voluntary measures for self-regulation at national levels, if any, may provide a valuable | | | manufacturers and programmers could also be held accountable for errors in programming or for the malfunction of an autonomous weapon system. However, establishing evidence that the operator or manufacturers knew or should have known the possibility of the crime committed by a complicated artificial intelligence system fed into the weapon will be a difficult task. Therefore, we recommend this aspect also be given due attention when discussing Article 36 implementation, to ensure a clear accountability chain with regard to autonomous weapons." (General Statement) (2016) | |

| State[1] | Currently Unacceptable, Unallowable, or Unlawful | Need to monitor or continue to discuss | Need to regulate | Need to ban (or favorably disposed towards the idea) | Need for meaningful human control[2] | AP I Article 36 review necessary | Refers to legal principles while remaining undecided on *per se* legality of AWS |
|---|---|---|---|---|---|---|---|
| | | | insight into the larger issue of addressing the international framework. State Parties may therefore announce such measures, as means of Transparency and Confidence Building Measures (TCBMs), in fulfilment of their moral obligation as the global efforts intensify towards establishing legal norms through a consensual approach." (General Statement) (2016)[70]<br><br>"While noting the positive commitments expressed by many States to not develop 'unpredictable autonomous weapons,' within their respective national security doctrines," we believe that national regulations themselves would not be sufficient to guarantee that these weapons will not be developed or used, as national military doctrines tend to evolve with 'potential risks' from outside. Furthermore, given the repeated emphasis on the danger of a possible military AI arms race, it is of utmost importance that the international community | | | | |

---

[70] Note that this particular excerpt from the Sri Lankan General Statement is not actually suggesting an international ban, but rather pointing out the value of voluntary bans adopted by individual states. However, also note that the excerpt immediately following this one indicates that a merely voluntary national regulation, absent more robust international regulations, would be insufficient.

| State[1] | Currently Unacceptable, Unallowable, or Unlawful | Need to monitor or continue to discuss | Need to regulate | Need to ban (or favorably disposed towards the idea) | Need for meaningful human control[2] | AP I Article 36 review necessary | Refers to legal principles while remaining undecided on *per se* legality of AWS |
|---|---|---|---|---|---|---|---|
| | | | understands the urgent and serious need for regulation of the use of artificial intelligence in weapons systems, which if not acted upon swiftly can be beyond any control. Therefore, while welcoming voluntary national measures, *Sri Lanka wishes to stress the need for negotiating a legally binding international instrument that regulates the use of autonomous technology in weapon systems.* We stand ready to support action towards that end." (General Statement) (2016) | | | | |
| Sweden | | "We are still lacking a clear definition of the term LAWS. There has been a tendency in this discussion to focus on technical issues, such as the dual-use nature of the technology involved and its application in both civilian and military systems. As some parties have suggested, it may be more fruitful to focus on identifying the critical functions of concern, with due consideration for the context in which a particular weapons system would be operating, as well as its effects, and take the discussion on definitions further from there.... The necessary level of human control... is an area that we expect to be explored further and we would be happy to do so in cooperation with others." (General statement) (2015) | "Exploring ways and means of regulating LAWS may at some point become desirable. As a step forward, at this stage, we would encourage transparency and propose information-sharing measures among interested states." (General statement) (2015) | | "[W]e have stated before our belief that humans' should not delegate to machines the power to make life-and-death decisions... It follows from our starting point of not delegating power of life and death to machines that Sweden would support the principle of applying Meaningful Human Control which has already been put forward by many parties. The necessary level of human control would depend on the particular situation and the requirements of international law in each case." (General statement) (2015) | "However, at the bottom of the issue lies the fact that a legal review of new weapons, means and methods of warfare is crucial. Sweden set up a Delegation for International law monitoring of Weapons projects already a long time ago. This delegation acts as an independent authority and is not part of the government. If the weapon projects assessed by the Delegation do not meet requirements within international law, the Delegation shall encourage or urge the authority that | "However, at the bottom of the issue lies the fact that a legal review of new weapons, means and methods of warfare is crucial. Sweden set up a Delegation for International law monitoring of Weapons projects already a long time ago. This delegation acts as an independent authority and is not part of the government. If the weapon projects assessed by the Delegation do not meet requirements within international law, the Delegation shall encourage or urge the authority that submitted the matter for examination to take appropriate measures. The Delegation reviews all weapons used by Swedish authorities. In this way we ensure that we fulfil the requirements in IHL – in particular art. 36 of Additional Protocol I – on implementing legal weapons reviews. Any possible autonomous weapons systems, as well as any other new weapon, or means or method of warfare, would be scrutinized in accordance with these |

| State[1] | Currently Unacceptable, Unallowable, or Unlawful | Need to monitor or continue to discuss | Need to regulate | Need to ban (or favorably disposed towards the idea) | Need for meaningful human control[2] | AP I Article 36 review necessary | Refers to legal principles while remaining undecided on per se legality of AWS |
|---|---|---|---|---|---|---|---|
| | | "As a step forward, at this stage, we believe that developing information-sharing measures among interested states is a useful way to go. We have listened carefully to suggestions on this from the expert panel and to comments from States and believe that it would be worthwhile developing some of the measures proposed, such as establishing points of contact and exchanging information on procedures and best practices on weapons reviews." (Way ahead statement) (2015)

"We believe that the CCW and its protocols present an effective means to respond in a flexible way to future developments in the field of weapons technology. We also look forward to continue working closely with the ICRC, and note that civil society has many valuable contributions to make to our work. Sweden continues to attach great importance to NGO participation in the CCW meetings. Against this background, we welcome the opportunity to continue our discussions on lethal autonomous weapon systems, LAWS, in this forum." (General Statement) (2016)

"An important question for our present meeting is how to move forward beyond this discussion. Like many other Parties, we would support a decision to create, at the Review Conference in December this year, a Governmental Group of Experts to | | | "From this week's discussions on LAWS, it appears that few if any delegations seem to actually advocate the use of fully autonomous weapons. The exchanges especially on the question of Meaningful Human Control point to a need to explore this concept further." (Way ahead statement) (2015)

"As our Foreign Minister has previously underlined, we believe that humans should always bear the ultimate responsibility when dealing with questions of life and death. As States we have an obligation to assess the legality of new weapons, and we will therefore welcome a continued discussion not least of these issues within the framework of the CCW." (General Statement) (2016) | submitted the matter for examination to take appropriate measures. The Delegation reviews all weapons used by Swedish authorities. In this way we ensure that we fulfil the requirements in IHL – in particular art. 36 of the Additional Protocol I – on implementing legal weapons reviews. Any possible autonomous weapons systems, as well as any other new weapon, or means or method of warfare, would be scrutinized in accordance with these procedures and this legal framework." (General Statement) (2016)

"We would like to contribute to the discussion with a brief account of the Swedish experiences of the Article 36 review process. The Swedish Delegation for International Humanitarian Law Monitoring of the Arms Projects (the Delegation) was established in 1974. The Delegation is an independent authority and not part of the Government or the Swedish Armed Forces. | procedures and this legal framework." (General Statement) (2016)
"Sweden has also recently re-appointed a National Commission for International Law and Disarmament. Similar Commissions in the past have proven to be very useful to advance our thinking on issues of international law, not least international humanitarian law, and the Government expects the new Commission to consider some topical issues where IHL and disarmament issues cross paths." (General Statement) (2016) |

| State[1] | Currently Unacceptable, Unallowable, or Unlawful | Need to monitor or continue to discuss | Need to regulate | Need to ban (or favorably disposed towards the idea) | Need for meaningful human control[2] | AP I Article 36 review necessary | Refers to legal principles while remaining undecided on *per se* legality of AWS |
|---|---|---|---|---|---|---|---|
| | | further examine the issue of LAWS. We hope that we might arrive at some common understanding so that the Expert Meeting could make a recommendation in this direction. Given the uncertainties in relation to many of the questions and issues regarding LAWS, we see that one promising issue for exploring in a GGE could well be the implementation of weapons review processes, including identification of best practices or benchmarks for such reviews. In this context, let me say that we agree with the direction suggested by Switzerland in their working-paper "Towards a 'compliance based' approach to LAWS", submitted to this meeting. Sweden remains open to discussing several possible ways forward, with a view to finding one that can enjoy consensus." (General Statement) (2016) | | | | Its organization and working methods are regulated in an ordinance... The Delegation's monitoring follows Article 36 of Additional Protocol I to the Geneva Conventions. It is thus examining whether the employment of a new weapon, means or method of warfare would, in some or all circumstances, be prohibited by the Additional Protocol I or by any other rule of international law applicable to Sweden- including human right and disarmament law. The Delegation reviews the characteristics of the weapon, how the weapon is planned to be used and other relevant aspects. In many cases the focus is on the usage of a new weapons or ammunition, and the Delegation thus reviews how the planned usage will adhere to the requirements of international law. There is also a need to control how the applying authority secure compliance with the legal requirements through training and | |

| State[1] | Currently Unacceptable, Unallowable, or Unlawful | Need to monitor or continue to discuss | Need to regulate | Need to ban (or favorably disposed towards the idea) | Need for meaningful human control[2] | AP I Article 36 review necessary | Refers to legal principles while remaining undecided on *per se* legality of AWS |
|---|---|---|---|---|---|---|---|
| | | | | | | education of users, the use of manuals or other types of instructions or regulations. The review is based on the law as it currently stands but the constant development of international law needs also to be taken into account. The Delegation has for these purposes denied a request by an applying authority to use a list of standardized requirements for future purchase of ammunition… In case the presented project does not fulfil the legal requirements the Delegation can recommend the applying authority to make the necessary modifications or issue limitations for tis use but it cannot halt the production of a weapon." (Challenges to IHL Statement) (2016) | |
| Switzerland | | "Ceci ne veut cependant pas dire que nous ne devrions pas envisager de déjà développer des mesures pratiques s'il apparaît que ceci peut contribuer à nos travaux et à répondre aux défis posés par les SALA. Il convient en effet de garder à l'esprit que la CCAC a devant elle une nouvelle fois l'opportunité de fournir une | | | "Accordingly, given the current state of robotics and artificial intelligence, it is difficult today to conceive of an AWS that would be capable of reliably operating in full compliance with all the obligations arising from existing IHL without any human control | "[M]y delegation considers the duty to conduct legal reviews in the study, development, acquisition or adoption of a new weapon, means or methods of warfare as an important element in | "IHL imposes manifold obligations which would have to be respected when using LAWS, in particular the principles governing the conduct of hostilities. For example, in order for LAWS to be lawfully employed in an armed conflict, challenging assessments are required to distinguish between civilian and military objectives or in evaluating whether the causation of unavoidable incidental harm to the civilian |

| State[1] | Currently Unacceptable, Unallowable, or Unlawful | Need to monitor or continue to discuss | Need to regulate | Need to ban (or favorably disposed towards the idea) | Need for meaningful human control[2] | AP I Article 36 review necessary | Refers to legal principles while remaining undecided on per se legality of AWS |
|---|---|---|---|---|---|---|---|
| | | réponse aux défis posés par un système d'arme en amont de son déploiement, et de ne pas attendre à devoir prendre des mesures correctives une fois les conséquences constatées sur le terrain. Le domaine de la transparence pourrait constituer un domaine idéal pour le développement de premières mesures concrètes, et auraient en sus l'avantage de contribuer à éclairer les discussions futures sur la thématique. La question de l'examen au niveau national de la licéité de nouvelles armes, y inclus tout nouveau système d'arme, au regard du droit international, soit des pratiques à suivre en la matière afin que les défis posés par les SALA soient pleinement pris en compte, constitue un autre domaine à approfondir concernant le développement de mesures pratiques." (General statement) (2015)[71]<br><br>"While there seems to be widespread agreement that the interplay between engagement-related functions and human-machine interaction should take center stage, discussions about what critical functions are and what constitutes an appropriate degree of control are ongoing and complex. At this stage, it would therefore appear premature to aim for a definition that seeks to | | | in the use of force, notably in the targeting cycle. On this basis, the question, therefore, is not whether States have a duty to control or supervise the development and/or employment of AWS, but how that control or supervision ought to be usefully defined and exerted. Would it be sufficient, for example, to rely on superior programming and strict reliability testing to make an AWS predictably compliant with IHL for its intended operational parameters? If so, would it be permissible to restrict human involvement to the proper activation of such an AWS? This working paper does not seek to prejudge these questions. However, it is useful to recognize that control can be exercised in various different ways, both independently and in combination. Arguably, in the future, a significant level of control can already be exerted in the development and programming phase. Through testing and evaluating AWS in the course of weapons reviews, predictability and reliability | preventing or restricting the employment of new weapons that would violate international law in all or some circumstances. The conduct of such reviews is an explicit treaty obligation for States Parties to the first Additional Protocol to the Geneva Conventions as expressed in its Art. 36. However, it would appear that the obligation to assess the legality of new weapons, also applies to all other States, as it flows directly from the general obligation of States to respect and ensure respect for IHL in all circumstances and from the general prohibition of using unlawful weapons or of using them in an unlawful manner." (Challenges to IHL statement) (2015)<br><br>"As with any other weapon, means or method of warfare, States have the positive obligation to determine, in the study, | population can be justified in view of the concrete and direct military advantage anticipated from that particular attack. These fundamental principles must not be circumvented by the use of LAWS. These and other legal requirements are derived directly from longstanding principles of IHL and allow for no compromise. It is therefore clear that existing IHL sets the bar very high in terms of technological prerequisites for the lawful use of LAWS in armed conflict." (Challenges to IHL statement) (2015)<br><br>"Of particular relevance for AWS is the prohibition of indiscriminate weapons. A weapons system would have to be regarded as indiscriminate if it cannot be directed at a specific military objective or if its effects cannot be limited as required by IHL and if, in either case, it is of a nature to strike military objectives and civilians or civilian objects without distinction. In other words, in order for an AWS to be lawful under this rule, it must be possible to ensure that its operation will not result in unlawful outcomes with respect to the principle of distinction." (Working Paper) (2016)<br><br>"With regard to the lawful use of a weapons system, the principles governing the conduct of hostilities need to be considered. Most notably, in order to lawfully use an AWS for the purpose of attack, belligerents must: (1 - Distinction) distinguish between military objectives and civilians or civilian objects and, in case of doubt, presume civilian status; (2 - Proportionality) evaluate whether the |

| State[1] | Currently Unacceptable, Unallowable, or Unlawful | Need to monitor or continue to discuss | Need to regulate | Need to ban (or favorably disposed towards the idea) | Need for meaningful human control[?] | AP I Article 36 review necessary | Refers to legal principles while remaining undecided on per se legality of AWS |
|---|---|---|---|---|---|---|---|
| | | draw a line between desirable, acceptable or unacceptable systems. There is merit in an inclusive discussion that does not prejudge the question of appropriate regulatory response for current and future systems, and that does not preclude an examination of the implications of a system for compliance with IHL." (Working Paper) (2016)

"Of course, over time, the discussions on autonomous weapons systems will advance. Once the very finality of this process becomes clearer, the purpose of the definition will shift. This would of course require us to reassess, and perhaps adapt this working definition to emerging needs." (Towards a Working Definition of LAWS Paper) (2016) | | | of such systems can also be reinforced. Predictability and reliability can also be increased by restricting the AWS' parameters of engagement in line with the system's compliance capabilities. Depending on operational requirements and system capabilities, further control can be exercised through real-time supervision, or through an autonomous or human operated override mechanism aimed at avoiding malfunction or, alternatively, ensuring safe failure." (Working Paper) (2016)

"This working paper has also put forward the notion that – given the current state of robotics and artificial intelligence – the relevant question is not whether a certain level of human control is called for, but what kind and level of human involvement in each of the different phases ranging from conceptualization, development and testing, to operational programming, employment and target engagement. At the heart of the issue is the question: what is the right quality of the human-machine | development, acquisition or adoption of any AWS, whether their employment would, in some or all circumstances, contravene existing international law. In this regard *the duty to conduct legal reviews, as specified in article 36 of Additional Protocol I, constitutes an important element in preventing or restricting the development and employment of new weapons that would not meet the obligations listed above.*[72] Moreover, adequate testing and reviews may also have implications on the level of State responsibility, including for malfunction of approved AWS. The legal review of AWS may present a number of challenges distinct from traditional weapons reviews. Specifically, the question is how such systems and their specific characteristics can be meaningfully tested. Beyond the purely technical challenge of assessing | incidental harm likely to be inflicted on the civilian population or civilian objects would be excessive in relation to the concrete and direct military advantage anticipated from that particular attack; (3 - Precaution) take all feasible precautions to avoid, and in any event minimize, incidental harm to civilians and damage to civilian objects; and cancel or suspend the attack if it becomes apparent that the target is not a military objective, or that the attack may be expected to result in excessive incidental harm." (Working Paper) (2016)

"The employment of AWS in the conduct of hostilities also raises particular challenges with regard to the prohibition of the denial of quarter and the protection of persons hors de combat, i.e. the protection from attack of the wounded and sick and those intending to surrender.2 Any reliance on AWS would need to preserve a reasonable possibility for adversaries to surrender. A general denial of this possibility would violate the prohibition of ordering that there shall be no survivors or of conducting hostilities on this basis (denial of quarter).[73]" (Working Paper) (2016)

"The present working paper has sought to map the most relevant IHL obligations applicable to the development and employment of AWS on the basis of a broad, inclusive working definition that allows for a discussion of different types of AWS. On one end of the spectrum of the proposed working definition, some types of AWS would be already unlawful under existing IHL, while on the other end of the |

---

[72] ICRC, A Guide to the Legal Review of New Weapons, Means and Methods of Warfare: Measures to Implement Article 36 of Additional Protocol I of 1977 (2006), available at https://www.icrc.org/eng/assets/files/other/icrc_002_0902.pdf.

[73] Article 40 of Additional Protocol I to the Geneva Conventions

| State[1] | Currently Unacceptable, Unallowable, or Unlawful | Need to monitor or continue to discuss | Need to regulate | Need to ban (or favorably disposed towards the idea) | Need for meaningful human control[2] | AP I Article 36 review necessary | Refers to legal principles while remaining undecided on per se legality of AWS |
|---|---|---|---|---|---|---|---|
| | | | | | interaction to ensure and facilitate compliance with IHL?" (Working Paper) (2016)<br><br>"Switzerland is of the view that *given the current state of robotics and artificial intelligence, it is difficult today to conceive of an autonomous weapons systems that would be capable of reliably operating in full compliance with all the obligations arising from existing IHL without any human control in the use of force, notably in the targeting cycle.* On this basis, the question, therefore, is not whether States have a duty to control or supervise the development and/or employment of autonomous weapons systems, but how that control or supervision ought to be usefully defined and exerted." (emphasis in original) (Challenges to IHL Paper) (2016) | IHL compliance of an AWS, there is also a conceptual challenge related to the fact that an autonomous system will assume an increasing number of determinations in the targeting cycle which traditionally are being taken care of by a human operator. For example, in traditional systems, the principle of proportionality was to be respected by the operator. It consequently fell outside the scope of an article 36 review. However, if an AWS is expected to perform this proportionality assessment by itself, that aspect will need to be added to legal reviews of these systems. New evaluation and testing procedures may need to be conceptualized and developed to meet this particular challenge."(Working Paper) (2016) | spectrum, some types of AWS can be readily qualified as unproblematic." (Working Paper) (2016) |
| **Turkey[74]** | | "How international humanitarian law and international human rights law would apply to lethal autonomous systems will need to be addressed while the technology develops. Establishing a common understanding among | | | "We, as others, attach importance to the humanitarian aspect of the matter. Therefore, we support the notions like need for human control and | | "How international humanitarian law and international human rights law would apply to lethal autonomous systems will need to be addressed while the technology develops." (General statement) (2015) |

[74] Turkey apparently also delivered a statement in the 2015 session on Transparency and the Way Ahead, but the text is not available online. See id.

| State[1] | Currently Unacceptable, Unallowable, or Unlawful | Need to monitor or continue to discuss | Need to regulate | Need to ban (or favorably disposed towards the idea) | Need for meaningful human control[2] | AP I Article 36 review necessary | Refers to legal principles while remaining undecided on *per se* legality of AWS |
|---|---|---|---|---|---|---|---|
| | | states would need considerable efforts. It would be fair to say we still have more questions than answers at this stage. Therefore, we look forward to hearing from delegations, academics and civil society in this discussion to allow us further understand the technical issues, characteristics of LAWS, legal and overarching issues." (General statement) (2015)<br><br>"In our opinion, prohibiting such systems before a broad agreement on a definition would not be pragmatic. In that regard, we consider that discussing terminologies like autonomous, fully autonomous and lethal autonomous might be useful." (General Statement) (2016)<br><br>"Finally, we are of the opinion that current international law and international humanitarian law provide the necessary basis regarding possible development of LAWS, but yet again we do not disregard any need to study the sufficiency of them on this matter." (General Statement) (2016) | | | accountability for such weapon systems. Nevertheless, taking into consideration that yet such weapon systems do not exist and we are working on an issue which is still hypothetical, we hesitate on the accuracy of a general prohibition pre-emptively" (General Statement) (2016) | | "Finally, we are of the opinion that current international law and international humanitarian law provide the necessary basis regarding possible development of LAWS, but yet again we do not disregard any need to study the sufficiency of them on this matter." (General Statement) (2016) |
| **United Kingdom** | | "To legislate now, without a clear understanding of the potential opportunities as well as dangers of a technology that we cannot fully appreciate, would risk leading to the use of generalised and unclear language which would be counter-productive. IHL has successfully accommodated previous evolutions in military technology such as the aeroplane and submarine. There is no | | | "The UK's clear position is that IHL is the applicable legal framework for the assessment and use of all weapons systems in armed conflict. Distinction, proportionality, military necessity and humanity are fundamental to compliance with IHL. Any LAWS, no matter what its specific technical characteristics, | "As required by Additional Protocol 1 to the Geneva Convention, the UK conducts legal reviews of weapons in accordance with Article 36 of the Protocol. The UK is aware that despite the large numbers of States being signatories to the first Protocol, not all | "From our perspective, to discuss LAWS is to discuss means and methods of warfare. As such, international humanitarian law provides the appropriate paradigm for discussion. To that end, we look forward to. sharing our views on the process of Legal Weapons Review. That is a process which has been developed exactly for situations like the one we are now facing, where the legality of new and novel weapons technologies need to be |

| State[1] | Currently Unacceptable, Unallowable, or Unlawful | Need to monitor or continue to discuss | Need to regulate | Need to ban (or favorably disposed towards the idea) | Need for meaningful human control[2] | AP I Article 36 review necessary | Refers to legal principles while remaining undecided on *per se* legality of AWS |
|---|---|---|---|---|---|---|---|
| | | reason to believe that IHL will not be capable of dealing with an evolution in automation." (Challenges to IHL statement) (2015)<br><br>"We look forward to further discussion on these issues and, in the longer term, to agreement on the applicability of IHL to this discussion, and the need to enhance compliance." (Challenges to IHL Paper) (2016)<br><br>"With regard to the recommendations that we might put forward at the end of the week, the UK would like to see agreement that future discussions should work towards: [1] Agreeing the relevance and importance of existing international humanitarian law to this debate. [2] Reaffirming key principles of IHL and commitment to increase compliance with those principles, including the importance of Legal Weapons Reviews. [3] Commitment to working towards a definition of LAWS consistent with the mandate of the discussion." (General Statement) (2016) | | | would have to comply with those principles to be capable of being used lawfully. However, the UK position is that those principles, and the requirement for precautions in attack, are best assessed and applied by a human. Within that process, a human may of course be supported by a system that has the appropriate level of automation to assist the human to make informed decisions." (Challenges to IHL statement) (2015)<br><br>"Article 36 Weapons Reviews are the correct means to assess a weapon, means, or method of warfare and its use, as required by Additional Protocol 1 to the Geneva Convention. The UK is aware that despite the large numbers of States being signatories to the first Protocol, not all formally conduct legal weapons reviews. Conversely there are States that are not signatories to the Protocol which conduct Article 36-style legal weapons reviews. The UK is committed to transparency where possible in this area, and so has published its weapons review procedures online. I will include the link in | formally conduct legal weapons reviews. Conversely there are States who are not signatories to the Protocol who conduct Article 36 style legal weapons review as a matter of good practice. We would like to encourage others by sharing UK practice and joining the debate in this area." (Challenges to IHL statement) (2015) | thoroughly assessed." (General statement) (2015)<br><br>"The UK's clear position is that IHL is the applicable legal framework for the assessment and use of all weapons systems in armed conflict. Distinction, proportionality, military necessity and humanity are fundamental to compliance with IHL. Any LAWS, no matter what its specific technical characteristics, would have to comply with those principles to be capable of being used lawfully." (Challenges to IHL statement) (2015)<br><br>"We believe that existing international humanitarian law is sufficient in assessing whether any future weapon system including LAWS would be capable of legal use." (Towards a Working Definition of LAWS Paper) (2016)<br><br>"The UK's clear position is that IHL is the applicable legal framework for the assessment and use of all weapons systems in armed conflict. Distinction, proportionality, military necessity and humanity are fundamental to compliance with IHL. Any weapon system, no matter what its specific technical characteristics or which or how many of its critical functions are autonomous, would have to comply with those principles to be capable of being used lawfully." (Challenges to IHL Paper) (2016 |

| State[1] | Currently Unacceptable, Unallowable, or Unlawful | Need to monitor or continue to discuss | Need to regulate | Need to ban (or favorably disposed towards the idea) | Need for meaningful human control[2] | AP I Article 36 review necessary | Refers to legal principles while remaining undecided on *per se* legality of AWS |
|---|---|---|---|---|---|---|---|
| | | | | | | the version of this statement which will be uploaded to the CCW website[75]." (Towards a Working Definition of LAWS Paper) (2016) | |
| | | | | | | "The details of individual UK Article 36 reviews are confidential due to factors including the classified nature of the equipment reviewed, the accompanying legal advice and the sensitive commercial and contractual nature of the related procurement processes. However, we can describe the five main areas considered in the reviews: 1. Whether the weapon is prohibited, or whether its use is restricted by any specific treaty provision or other applicable rule of international law; 2. Whether the weapon is of a nature to cause superfluous injury or unnecessary suffering; 3. Whether it is capable of being used discriminately; 4. Whether it may be expected to cause widespread, long-term | |

| State[1] | Currently Unacceptable, Unallowable, or Unlawful | Need to monitor or continue to discuss | Need to regulate | Need to ban (or favorably disposed towards the idea) | Need for meaningful human control[2] | AP I Article 36 review necessary | Refers to legal principles while remaining undecided on *per se* legality of AWS |
|---|---|---|---|---|---|---|---|
| | | | | | | and severe damage to the natural environment; and 5. Whether it is likely to be affected by current and possible future trends in the development of International Humanitarian Law. Any system, whether it displays any level of autonomy or not, would have to meet the required standards for all five of the areas of consideration. Assessing weapons systems with increasing levels of automation or autonomy does not require another process. The requirement for Article 36 Reviews is already prescribed in International Humanitarian Law. So we do not see the need for additional legislation, in the form of a pre-emptive ban. Instead, we would like to see greater compliance with existing IHL." (Challenges to IHL Paper) (2016) | |
| United States of America | | "[W]e believe that it is important to focus on increasing our understanding versus trying to decide possible outcomes. It remains our view that it is premature to try and determine | | | | | "We have consistently heard in the CCW interest expressed on the weapons review process and about the requirement to conduct a legal review of all new weapons systems, including LAWS. We believe that this is an area on which we should focus |

| State[1] | Currently Unacceptable, Unallowable, or Unlawful | Need to monitor or continue to discuss | Need to regulate | Need to ban (or favorably disposed towards the idea) | Need for meaningful human control[2] | AP I Article 36 review necessary | Refers to legal principles while remaining undecided on *per se* legality of AWS |
|---|---|---|---|---|---|---|---|
| | | where these discussions might or should lead" (General statement) (2015)<br><br>"We are prepared to continue to contribute to the robust discussions about these issues so that our collective understanding can grow further… we also continue to welcome contributions from civil society and technical experts to inform our discussion." (General Statement) (2016)<br><br>"The U.S. Delegation also looks forward to more in depth discussions with respect to human-machine interaction and about the phrase 'meaningful human control.' Turning first to the phrase 'meaningful human control,' we have heard many delegations and experts note that the term is subjective and thus difficult to understand. We have expressed these same concerns about whether 'meaningful human control' is a helpful way to advance our discussions. We view the optimization of the human/machine relationship as a primary technical challenge to developing lethal autonomous weapon systems and a key point that needs to be reviewed from the start of any weapon system development. Because this human/machine relationship extends throughout the development and employment of a system and is not limited to the moment of a decision to engage a target, we consider it more useful to talk about 'appropriate levels of | | | | | as an interim step as we continue our consideration of LAWS in CCW. The United States would like to see the Fifth Review Conference agree to begin work, as part of the overall mandate on LAWS, on a non-legally binding outcome document that describes a comprehensive weapons review process, including the policy, technical, legal and operational best practices that states could consider using if they decide to develop LAWS or any other weapon system that uses advanced technology. To be clear, the United States believes that the existence of such a document would not endorse the development of LAWS; it would assist a State in conducting a thorough weapons review if that State is considering developing LAWS or any new weapon system. It would also help ensure consistency and quality in the weapons review process by all States, regardless of the particular weapon being reviewed. It is also an opportunity for the CCW to take a concrete step related to LAWS in the near term, even while we continue to develop our common understanding of what constitutes LAWS." (General Statement) (2016) |

| State[1] | Currently Unacceptable, Unallowable, or Unlawful | Need to monitor or continue to discuss | Need to regulate | Need to ban (or favorably disposed towards the idea) | Need for meaningful human control[2] | AP I Article 36 review necessary | Refers to legal principles while remaining undecided on per se legality of AWS |
|---|---|---|---|---|---|---|---|
| | | human judgment.'" (General Statement) (2016) | | | | | |
| Zambia | "With regard to the relation between lethal autonomous weapons system and human rights, my delegations supports calls by other delegations that to delegate the decision to decide over life and death to machines, will be against human rights." (Way ahead statement) (2015) | "[I]t is clear that there are diverging views regarding the development and use of autonomous weapons. This suggests a possibility of further discussions that may take the current form or indeed as suggested in the summary that you presented to this meeting this morning. While the later could be a possible means, broad membership would be important to get as many views as possible. In this regard Zambia supports the views highlighted by other delegations for mobilization of financial resources that would enable participation of as many states as possible especially those from developing countries. This will enable common understanding of the nature and characteristics of such weapons and their likely impact on peace and security. With regard to the relation between lethal autonomous weapons system and human rights, my delegations supports calls by other delegations that to delegate the decision to decide over life and death to machines, will be against human rights. To this effect, while the CCW framework remains the appropriate forum to debate this subject the cooperation between the CCW and the Human Rights | "We believe that these procedures should be designed under the framework of IHL, however, currently international law may not have specific provisions on review mechanism as such these contradictions are making us go round in circles, thus there should be room for crafting additional regulations on the review of potential LAWS and the applicable procedures which would help with coming up with compliance measures. This therefore, implies coming up with additional legislation to cater for the review procedure under the provisions of IHL to be created." (Challenges to IHL Statement) (2016) | "In addition, the current debate is unable to justify a legal basis under the International Humanitarian Law or indeed the CCW protocols over the humane application of Autonomous Weapons in armed conflict. A prohibition on their (LAWS) use and to an acceptable degree their proliferation by member states should be on the CCW agenda as a preliminary measure to a conclusive endpoint where a basis shall be developed which should be capable of mitigating our current fears of releasing Lethal Autonomous Weapon Systems in armed conflict." (General Statement) (2016)[76] | "Zambia also takes note of the challenges the increasing degree of autonomy would present to International Humanitarian Law and therefore would not advocate for any such weapons systems that would water down the aspects of responsibility and accountability in armed conflict. Our focus should instead be on strengthening such norms." (Way ahead statement) (2015) | "Further, we see a grave weakness in exclusively looking to IHL to spell out whether LAWS are acceptable in armed conflict and how they fit in without causing apprehension. As many have observed and commented before on the subject, the CCW is the right platform for the creation of regulations to uphold current rules such as article 36 of Additional Protocol I to the Geneva Convention and any other international laws on conflict." (General Statement) (2016)[77] | "It is against this background that my delegation submits that considering a multilateral agreement at this point is most critical in providing clear margins for all states on the use of LAWS in armed conflict. This direction brings to the fore earlier calls by the ICRC (1987 Commentary 1466 on Article 36) for High Contracting Parties to collectively determine the possibly unlawful nature of a new weapon, both with regard to the provisions of the Protocol, and with regard to any other applicable rule of international law. Where resultant measures from such a forum are not taken, the State becomes liable for any wrongful outcomes in the use of LAWS." (General Statement) (2016) |

[76] The phrasing here is somewhat convoluted and difficult to fully decipher. However, it appears that Zambia is calling for some kind of temporary ban on LAWS until the international community gains a better understanding of them.

[77] This excerpt appears to endorse an Article 36 review, but later in the General Statement, Zambia says that "while a number of states have suggested that action around national legal reviews of weapons, under the framework of article 36 of Additional Protocol I to the Geneva Convention, could constitute a basis for addressing the serious concerns that states have raised in relation to autonomous weapons, our observation is that national reviews are insufficient to deal with LAWS." (General Statement) Zambia's position, therefore, seems to be that Article 36 reviews are a necessary, but not sufficient step in the international community's attempt to appropriately regulate LAWS.

| State[1] | Currently Unacceptable, Unallowable, or Unlawful | Need to monitor or continue to discuss | Need to regulate | Need to ban (or favorably disposed towards the idea) | Need for meaningful human control[2] | AP I Article 36 review necessary | Refers to legal principles while remaining undecided on *per se* legality of AWS |
|---|---|---|---|---|---|---|---|
| | | Council is therefore inevitable. The discussion in a forum of this nature restricts our thinking to more technical issues rather than consideration of ethical and moral issues. We stand to benefit more a lot more from a forum that is inclusive considering the multidisciplinary nature of the subject….My delegation therefore welcomes further discussions on the subject but wish to reiterate that such discussions should rather be inclusive. It is therefore important that during the Meeting of States Parties in November, a decision on how we should proceed on this subject should be made." (Way ahead statement) (2015)<br><br>"My delegation's view is that we find ourselves in a paradox as we debate the theme on ethics, human rights and the law of armed conflict; if International Humanitarian Law is based on the principal "making armed conflict as humane as possible," to what extent or at what point will this autonomous system assume humanity in order for IHL to justify their adoption for use in armed conflict? The fact that high contracting parties have not yet devised international legislation on the international approach towards AWs is enough reasons why states parties should trade slowly in allowing their use in armed conflict. Without established international rules, national reviews are currently inadequate to guide the use of LAWS in war. We stand ready to | | | | | |

| State[1] | Currently Unacceptable, Unallowable, or Unlawful | Need to monitor or continue to discuss | Need to regulate | Need to ban (or favorably disposed towards the idea) | Need for meaningful human control[2] | AP I Article 36 review necessary | Refers to legal principles while remaining undecided on *per se* legality of AWS |
|---|---|---|---|---|---|---|---|
| | | get more insight in this meeting and hope to move together with the same understanding with others." (General Statement) (2016) | | | | | |
| Zimbabwe[78] | | | | "We have been caught napping before, and if past experience can be our guide for the present and future, we join like-minded delegations in calling for a pre-emptive ban on lethal autonomous weapons systems. My delegation believes the time to act on this issue is now and that it is imperative to avoid a situation where a pre-emptive ban becomes a moot point." (CCW Speech) (2016) | "In situations where autonomous weapon systems are deployed to select and engage human targets in armed conflict, my delegation holds the view that there is need to maintain meaningful human control to ensure full observance of international humanitarian law." (CCW Speech) (2016) | | |

---

[78] Zimbabwe did not express its desire for a ban via a written statement at the 2015 or 2016 CCW Meeting of Experts. It did, however, indicate a preference for a ban during the 2016 CCW meeting (*not* the Meeting of Experts). *See* Speech for the Meeting of High Contracting Parties to the Convention on the Prohibitions or Restrictions on the Use of Certain Conventional Weapons which may be Deemed to be Excessively Injurious or to have Indiscriminate Effects (CCW) (Nov. 12-13, 2015), http://www.unog.ch/80256EDD006B8954/(httpAssets)/842EF3CB3B61A2FBC1257F0F003B9521/$file/zimbabwe.pdf. Zimbabwe's position has been included here to more fully represent states' attitudes on an important issue.

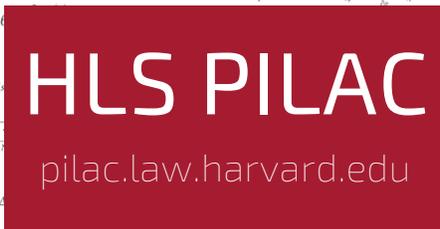

HLS PILAC

pilac.law.harvard.edu